\documentstyle[12pt,epsfig,newcom,artmod]{article}     
     
\def\beq{\begin{equation}}
\def\eeq{\end{equation}}
\def\beqar{\begin{eqnarray}}
\def\eeqar{\end{eqnarray}}
\def\barr#1{\begin{array}{#1}}
\def\earr{\end{array}}
\def\bfi{\begin{figure}}
\def\efi{\end{figure}}
\def\btab{\begin{table}}
\def\etab{\end{table}}
\def\bce{\begin{center}}
\def\ece{\end{center}}

\def\nl{\nonumber\\}

\def\ga{\gamma}
\def\de{\delta}
\def\la{\lambda}
\def\si{\sigma}

\def\refeq#1{\mbox{(\ref{#1})}}

\def\refse#1{\mbox{Section~\ref{#1}}}

\renewcommand{\ri}{{\mathrm{i}}}
\renewcommand{\rd}{{\mathrm{d}}}

\renewcommand{\rR}{{\mathrm{R}}}
\renewcommand{\rT}{{\mathrm{T}}}
\renewcommand{\rL}{{\mathrm{L}}}

\renewcommand{\M}{{\cal{M}}}

\def\mathswitch#1{\relax\ifmmode#1\else$#1$\fi}
\def\mathswitchr#1{\relax\ifmmode{\mathrm{#1}}\else$\mathrm{#1}$\fi}
\renewcommand{\PW}{\mathswitchr W}
\newcommand{\PB}{\mathswitchr B}
\renewcommand{\PZ}{\mathswitchr Z}
\renewcommand{\PA}{\mathswitchr A}
\renewcommand{\PH}{\mathswitchr H}
\renewcommand{\Pf}{\mathswitch f}

\renewcommand{\Pe}{\mathswitchr e}
\renewcommand{\Pd}{\mathswitchr d}
\renewcommand{\Pu}{\mathswitchr u}
\renewcommand{\Ps}{\mathswitchr s}
\renewcommand{\Pc}{\mathswitchr c}
\renewcommand{\Pb}{\mathswitchr b}
\renewcommand{\Pt}{\mathswitchr t}

\renewcommand{\PWp}{\mathswitchr {W^+}}
\renewcommand{\PWm}{\mathswitchr {W^-}}

\renewcommand{\MW}{\mathswitch {M_\PW}}
\renewcommand{\MZ}{\mathswitch {M_\PZ}}
\renewcommand{\MH}{\mathswitch {M_\PH}}

\newcommand{\thw}{\mathswitch {\theta_{\mathrm{w}}}}
\renewcommand{\cw}{\mathswitch {c_{\mathrm{w}}}}
\renewcommand{\sw}{\mathswitch {s_{\mathrm{w}}}}

\newcommand{\SUtwo}{\mathrm{SU(2)}}
\newcommand{\Uone}{\mathrm{U}(1)}
\def\ie{i.e.\ }

\def\cf{cf.\ }

\newcommand{\elm}{{\mathrm{em}}}
\renewcommand{\ew}{{\mathrm{ew}}}

\newcommand{\SC}{{\mathrm{LSC}}}
\renewcommand{\SS}{{\mathrm{SSC}}}

\newcommand{\cew}{C^{\ew}}

\newcommand{\sNB}{\tilde{\NB}}
\newcommand{\NB}{N}
\newcommand{\GB}{V}

\newcommand{\ls}{l(s)}

\newcommand{\lrM}{l(r_{kl},M^2)}

\newcommand{\lsM}{l(s,M^2)}
\newcommand{\lsW}{l(s,\MW^2)}

\newcommand{\lWla}{l(\MW^2,\la^2)}

\newcommand{\lWa}{l(\MW^2,M_{\GB_a}^2)}

\newcommand{\Ls}{L(s)}
\newcommand{\LrM}{L(|r_{kl}|,M^2)}
\newcommand{\Lrs}{L(|r_{kl}|,s)}
\newcommand{\LrMa}{L(|r_{kl}|,M_{\GB_a}^2)}
\newcommand{\LsM}{L(s,M^2)}
\newcommand{\LsW}{L(s,\MW^2)}

\newcommand{\Lkla}{L(m_k^2,\la^2)}

\newcommand{\LWla}{L(\MW^2,\lambda^2)}
\newcommand{\Lemk}{L^\elm(s,\lambda^2,m_k^2)}

\newcommand{\lrs}{\log{\frac{|r_{kl}|}{s}}}

\begin{document}     
     
\vspace*{-2cm} \renewcommand{\thefootnote}{\fnsymbol{footnote}}     
\begin{flushright}     
\end{flushright}     
\vskip 65pt     
     
\begin{center}     
\vspace{2cm}
{\Large {\bf Electroweak radiative corrections in \\
\vspace{0.2cm}
high energy processes
}}\\[0pt]     

\vspace{1.2cm}

{Michael Melles\footnote{{\bf 
Michael.Melles@psi.ch}} }\\     
\vspace{10pt}   
  
{Paul Scherrer Institute (PSI), CH-5232 Villigen, Switzerland. }  
\end{center}  
     
\vspace{60pt}     
\begin{abstract}     
Experiments at future colliders will attempt to unveil the origin of electroweak symmetry
breaking in the TeV range. At these energies the Standard Model (SM) predictions
have to be known precisely in order to disentangle various viable scenarios
such as supersymmetry and its manifestations. In particular, large logarithmic
corrections of the scale ratio $\sqrt{s}/M$, where $M$ denotes the gauge boson
masses, contribute significantly up to and including the two loop level. 
In this paper we review recent
progress in the theoretical understanding of the electroweak Sudakov corrections
at high energies up to subleading accuracy in the SM and the minimal supersymmetric
SM (MSSM). 
We discuss the symmetric part of the SM Lagrangian at high energies yielding
the effective theory employed in the framework of the infrared evolution equation
(IREE) method. 
Applications are presented for
important SM and MSSM processes relevant for the physics program of future linear colliders
including higher order purely electroweak angular dependent corrections.
The size of the higher order subleading electroweak corrections is found to 
change cross sections in the several percent regime at TeV energies and
their inclusion is thus mandatory for predictions of high energy processes
at future colliders. 
\end{abstract}

\vskip12pt     
     
\setcounter{footnote}{0} \renewcommand{\thefootnote}{\arabic{footnote}}     
     
\vfill     
\clearpage     
\pagestyle{plain}     

\tableofcontents

\clearpage
     
\section{Introduction} \label{sec:int}    

The Standard Model (SM) of particle physics \cite{gg,fgl,gla,sal,wei} 
has enjoyed unprecedented success over the last decades.
The discovery of the top quark at the Tevatron \cite{tev,tev2} leaves the Higgs particle 
\cite{higgs,higgs2,higgs3,kibble} as the last undiscovered
ingredient to complete the SM. While it is possible that the SM remains valid up to energies far beyond
experimental reaches, most theorists view the SM as an effective theory which is embedded in a larger
theory usually containing unification of the gauge interactions at a high scale $M_{GUT}$.

This expectation seems well motivated due to the presence of light neutrino masses established
at Super-Kamiokande \cite{kam} in connection with a seesaw mechanism involving $M_{GUT}$. Also coupling
unification in the minimal supersymmetric SM (MSSM) points to the existence of a higher scale
in nature where the forces unify. 

If, however, the SM is the effective low energy theory of a more complete and unified theory at 
$M_{GUT}$, the hierarchy problem must be taken seriously. Supersymmetry is able to stabilize 
the quadratic divergences in the scalar sector by canceling these terms with the corresponding superpartner
loop divergences if, and only if, the superpartner mass splittings are not much larger than the weak scale. 
Another possibility currently discussed is that there are large extra dimensions at the
TeV scale \cite{add,rs}, however, such a scenario only trades one problem (the existence of a large scale 
$M_{GUT}$)
for another (the existence of large extra dimensions of the ``right'' size). 

In any case, while the SM works extremely well, it does not explain electroweak symmetry breaking
(EWSB). A negative mass squared is introduced by hand in the SM, but in the larger theory the reason for EWSB
is expected to be dynamical such as in typical SUGRA models \cite{sugra}.
While many possible extensions of the SM exist, only experiments at future colliders will shed light on
the origin of EWSB expected to lie in the TeV regime.

At this point a few general remarks about the usage of the expression EWSB are appropriate in
order to not be misleading. It has been known for some time now that the Higgs mechanism does
not lead to a breaking of the local gauge invariance on the lattice \cite{el}. In general,
all vacuum expectation values (v.e.v.'s) of gauge dependent operators (such as $ \langle
0| \phi(x) | 0 \rangle $) can be shown to vanish. As was pointed out in Ref. \cite{froh},
the crucial point about the continuum version in the conventional perturbative formulation
of the Higgs mechanism is not as much the existence of a v.e.v. $v$, but rather the existence
of a non-trivial orbit minimizing the Higgs-potential. The apparent breaking of the original 
symmetry by $v$ is due to it being a gauge choice (which always breaks the gauge symmetry).
In other words,
if we were to reformulate the full theory in terms of only gauge invariant operators, then no
symmetry breaking would be visible (but of course new operators would occur describing, for 
instance, the different masses of the electroweak gauge bosons). Since it was also shown in
Ref. \cite{froh} that the difference between the manifestly gauge invariant picture and the 
conventional perturbative formulation vanishes for observables 
in the small coupling limit, we prefer to
use the standard terminology and operators in which the original symmetry is hidden. 
It is in this sense, that we use the expression ``broken gauge theory'' below.
 
The high precision measurements of SLC/LEP have limited the room for extensions
of the SM considerably and in general, they cannot deviate from the SM to a large extent
without evoking so-called conspiracy effects. It would therefore be very desirable to
have a leptonic collider at hand in the future in order to answer questions posed by
discoveries made at the LHC and possibly the Tevatron. In particular, if only a light Higgs
is discovered, say at 115 GeV, then it is mandatory to investigate all its properties
in detail to experimentally establish the Higgs mechanism including a possible reconstruction
of the potential and of course of the Yukawa couplings. In addition one would have to
look for additional heavy Higgs-bosons which could easily escape detection at the hadronic
machines, but can be discovered at the $\gamma \gamma$-option at TESLA
\cite{kmsz,gko,mks} up to masses reaching 80 \% of the c.m. energy.
If any supersymmetric particle would be found in addition, it is necessary to
clarify and/or test the relations between couplings and properties of all new particles in
as much detail as possible in a complementary way to what would already be known by that
time.
The overall importance of leptonic colliders would thus be to clarify the physics responsible
for the EWSB which in turn means it must be a high precision
machine.

On the theory side this means that effects at the 1 \% level should be under control in both
the SM as well as all extensions that are viable at that point.
The purpose of the present work is to summarize the recent activities and results relevant on 
this level of precision from electroweak radiative corrections at energies much larger than the
gauge boson masses and to apply these corrections to processes relevant to the linear collider
program. This does not mean that the corrections are negligible for hadronic machines, however,
for the high precision illustrations we focus here on $e^+ e^-$ machines in the TeV range. 

At the expected level of precision required to disentangle new physics effects from the SM
in the ${\cal O} \left( \leq 1 \% \right)$ regime, higher order {\it electroweak} radiative
corrections cannot be ignored at energies in the TeV range. 
As a consequence, there has been a lot of interest recently in the high energy behavior of the SM
\cite{cc1,bccrv1,bccrv2,brv1,cc,kp,kps,flmm,dp}.
The largest contribution
is contained in electroweak double logarithms (DL) of the Sudakov type and a comprehensive
treatment of those corrections is given in Ref. \cite{flmm} to all orders.
The effects of the mass-gap between the photon and Z-boson has been
considered in recent publications \cite{m2,bw} since spontaneously broken gauge theories lead
to the exchange of massive gauge bosons. In general one expects the
SM to be in the unbroken phase at high energies.
There are, however, some important differences of the electroweak theory with respect to an unbroken
gauge theory. Since the physical cutoff of the massive gauge bosons is the weak scale $M\equiv
M_{\rm W} \sim M_{\rm Z} \sim M_{\rm H}$,
pure virtual corrections lead to physical cross sections depending on the infrared ``cutoff''.
Only the photon needs to be treated in
a semi-inclusive way.
Additional complications arise due to the mixing involved to make the mass eigenstates and the fact
that at high energies, the longitudinal degrees of freedom are not suppressed.
Furthermore, since the asymptotic states are not group singlets, it is expected
that fully inclusive cross sections contain Bloch-Nordsieck violating electroweak corrections
\cite{ccc1}.

It has by now been established that the
exponentiation of the electroweak Sudakov DL calculated in Ref.
\cite{flmm} via the infrared evolution equation method (IREE) \cite{kl,kl2} with the fields of
the unbroken phase is indeed reproduced by explicit
two loop calculations with the physical SM fields \cite{m2,bw,hkk}. One also understands now
the origin of previous disagreements. The results of Ref. \cite{kp}, based on fully
inclusive cross sections in the photon, is not gauge invariant as already
pointed out in Ref. \cite{flmm}. The factorization used in Ref. \cite{cc} is based
on QCD and effectively only takes into account contributions from ladder diagrams. In the electroweak
theory, the three boson vertices, however, do not simply cancel the corresponding
group factors of the crossed ladder diagrams (as is the case in QCD) and thus, infrared
singular terms survive for left handed fermions (right handed ones are effectively
Abelian) in the calculation of Ref. \cite{cc}. The IREE method does
not encounter any such problems since all contributing diagrams are automatically
taken into account by determining the kernel of the equation in the effective regime
above and below the weak scale $M$.
It is then possible to calculate corrections in the effective high energy theory
in each case yielding the same result as calculations in the physical basis.
Thus, the
mass gap between the Z-boson and the photon can be included in a natural way
with proper matching conditions at the scale $M$.
For longitudinally polarized gauge bosons it was shown in Ref. \cite{m1, m3} that
the leading and subleading (SL) kernel can be obtained from the Goldstone boson equivalence theorem.

We specify in the next section how the high energy effective theory is obtained from
the SM and illustrate the approach followed in the main part of this work.

\subsection{The Standard Model} \label{sec:sm}

\newcommand{\ubar}{\bar{u}}
\newcommand{\dbar}{\bar d \mkern2mu}
\newcommand{\lbar}{\bar l \mkern2mu}

The complete classical Lagrangian $\L_{\mathrm{class}}$ of
the electroweak SM (EWSM) reads in terms of the physical fields, \ie the mass and charge
eigenstates $\FA_\mu$, $\FZ_\mu$, $\FW^\pm_\mu$, $\FH$, $l$, $\nu$,
$u$, and $d$, the would-be Goldstone fields $\phi^\pm$ and $\chi$, and
the physical parameters $e$, $\MW$, $\MZ$, $\MH$, $\Mf$, and $\VKM$, as follows \cite{bdj}:
\beqar \label{42Lphys}
\lefteqn{ \L_{\mathrm{class}} =
\sum_{f=l,\nu,u,d}\sum_i [ \bar f_i (\ri\slash{\partial} - \Mf)f_i
                          - e \Qf \bar f_i \ga^\mu f_i  \FA_\mu ] } \quad
\nl
&&{} + \sum_{f=l,\nu,u,d}\sum_i \frac{e}{\sw\cw}
                         [ \If \bar f_i^{\rL} \ga^\mu f_i^{\rL}
                          - \sw^2 \Qf \bar f_i \ga^\mu f_i] \FZ_\mu \nl
&&{} + \sum_{i,j} \frac{e}{\sqrt2\sw}
                      [ \bar u_i^{\rL} \ga^\mu \VKM_{ij} d_j^{\rL} \FW_\mu^+
                      + \dbar_i^{\rL} \ga^\mu \VKM_{ij}^{\dagger} u_j^{\rL}
                         \FW_\mu^-] \nl
&&{} + \sum_{i} \frac{e}{\sqrt2\sw}
                      [ \bar \nu_i^{\rL} \ga^\mu l_i^{\rL} \FW^+_\mu
                      + \lbar_i^{\rL} \ga^\mu \nu_i^{\rL} \FW^-_\mu] \nl
&&{} - \frac{1}{4} \left| \partial_\mu \FA_\nu - \partial_\nu \FA_\mu
            - \ri e(\FW^-_\mu \FW^+_\nu - \FW^-_\nu \FW^+_\mu) \right|^2 \nl
&&{} - \frac{1}{4} \left| \partial_\mu \FZ_\nu - \partial_\nu \FZ_\mu
    + \ri e\frac{\cw}{\sw}(\FW^-_\mu \FW^+_\nu - \FW^-_\nu \FW^+_\mu) \right|^2 \nl
&&{} - \frac{1}{2} \left| \partial_\mu \FW^+_\nu - \partial_\nu \FW^+_\mu
    - \ri e(\FW^+_\mu \FA_\nu - \FW^+_\nu \FA_\mu)
\right.  \nl &&{} \qquad \left.
    + \ri e\frac{\cw}{\sw}(\FW^+_\mu \FZ_\nu - \FW^+_\nu \FZ_\mu) \right|^2 \nl
&&{} + \frac{1}{2} \left| \partial_\mu (\FH + \ri\chi)
                  - \ri  \frac{e}{\sw}\FW^-_\mu \phi^+ + \ri \MZ\FZ_\mu
        + \ri \frac{e}{2\cw\sw} \FZ_\mu(\FH + \ri\chi) \right|^2 \nl
&&{} + \left| \partial_\mu \phi^+ + \ri e \FA_\mu \phi^+
        - \ri e \frac{\cw^2 - \sw^2}{2\cw\sw} \FZ_\mu \phi^+ - \ri \MW\FW^+_\mu
\right.  \nl &&{} \qquad \left.
        - \ri \frac{e}{2\sw} \FW^+_\mu(\FH + \ri\chi) \right|^2 \nl
&&{} - \frac{1}{2} \MH^2 \FH^2 - e\frac{\MH^2}{2\sw\MW} \FH
         \left(\phi^-\phi^+ + \frac{1}{2} |\FH + \ri \chi|^2 \right)
\nl
&&{}  - e^2\frac{\MH^2}{8\sw^2\MW^2}
         \left(\phi^-\phi^+ + \frac{1}{2} |\FH + \ri \chi|^2 \right)^2 \nl
&&{} - \sum_{f=l,\nu,u,d}\sum_i e\frac{m_{\Pf,i}}{2\sw\MW}
            ( \bar f_i f_i \FH -2\If  \ri\bar f_i \ga_5 f_i\chi) \nl
&&{} + \sum_{i,j} \frac{e}{\sqrt2\sw}
       \left(
       \frac{m_{u,i}}{\MW}\left(\bar u_i^{\rR} \VKM_{ij} d_j^{\rL} \phi^+
       + \dbar_i^{{\rL}} \VKM_{ij}^{\dagger} u_j^{\rR} \phi^-\right)
       \right. \nl &&{} \qquad \left.
      -\frac{m_{d,j}}{\MW}\left(\bar u_i^{\rL} \VKM_{ij} d_j^{\rR} \phi^+
       + \dbar_i^{\rR} \VKM_{ij}^{\dagger} u_j^{\rL} \phi^-\right)
       \right) \nl
&&{} - \sum_{i} \frac{e}{\sqrt2\sw}
           \frac{m_{l,j}}{\MW} \left(\bar \nu_i^{\rL} l_i^{\rR} \phi^+
           +\lbar_i^{\rR} \nu_i^{\rL}\phi^- \right) 
\eeqar

The quantization of the EWSM requires the introduction of a {\it
  gauge-fixing term} and of {\it Faddeev--Popov fields}.  We introduce
a gauge-fixing term of the form
\begin{equation} \label{44Lfix}
  {\cal L}_{\mathrm{fix}} = - \frac{1}{2\xiA} (C^{\PA})^{2}
  - \frac{1}{2\xiZ} (C^{\PZ})^{2} - \frac{1}{\xiW} C^{+} C^{-}
\end{equation}
with linear gauge-fixing operators
\begin{eqnarray}\label{44gf}
C^{\pm} &=& \partial^{\mu } W^{\pm}_{\mu } \mp \ri \MW \xiW' \phi ^{\pm} \nlc
C^{\PZ} &=& \partial ^{\mu} Z_{\mu } - \MZ \xiZ' \chi \nlc
C^{\PA} &=& \partial^{\mu} A_{\mu }
\end{eqnarray}
This general linear gauge contains five independent gauge parameters $\xia$,
$a=A,Z,\pm$, and $\xia'$, $a=Z,\pm$, where $\xi_{\pm}^{(\prime)}\equiv
\xiW^{(\prime)}$.

For $\xiW'=\xiW$ and $\xiZ'=\xiZ$  the terms involving the would-be Goldstone
fields in \refeq{44gf} cancel the mixing terms $V_\mu\,\partial^{\mu}
\phi$ in the classical Lagrangian \refeq{42Lphys} up to irrelevant
total derivatives. This gauge is called
{\it 't~Hooft-Feynman gauge}
\index{t Hooft gauge@'t~Hooft gauge}%
and is used in the following if not stated otherwise.

The corresponding Faddeev--Popov ghost-field Lagrangian reads
\begin{eqnarray} \label{44Lghost}
\lefteqn{\L_{\mathrm{ghost}} = \biggl\{ -
  \ubar^+(\partial^\mu\partial_\mu + \xiW' \MW^2) u^+ + \ri e
  (\partial^\mu\ubar^+) \left(A_\mu - \frac{\cw}{\sw}Z_\mu \right)u^+ }\quad
\nl &&{} - \ri e (\partial^\mu\ubar^+) W^+_\mu \left(u^A -
\frac{\cw}{\sw}u^Z\right) \nl &&{} - e\MW\xiW' \ubar^+
\left[\frac{1}{2\sw}(H+\ri \chi) u^+ - \phi^+\left(u^A -
\frac{\cw^2-\sw^2}{2\cw\sw}u^Z\right)\right] \nl &&{} - (u^+ \to u^-,\, W^+
\to W^-,\, \phi^+ \to \phi^-,\, \ri \to -\ri)\biggr\} \nl &&{} -
\ubar^Z(\partial^\mu\partial_\mu + \xiZ' \MZ^2) u^Z - \ri
e\frac{\cw}{\sw} (\partial^\mu\ubar^Z) \left( W^+_\mu u^- -W^-_\mu u^+\right)
\nl &&{} - e\MZ\xiZ' \ubar^Z \left[\frac{1}{2\cw\sw} \FH u^Z -
\frac{1}{2\sw} \left(\phi^+ u^- + \phi^- u^+ \right)\right] \nl &&{} - \ubar^A
\partial^\mu\partial_\mu u^A + \ri e (\partial^\mu\ubar^A) \left( W^+_\mu
u^- - W^-_\mu u^+\right)
\end{eqnarray}

Adding up all terms \refeqs{42Lphys}, \refeqf{44Lfix} and
\refeqf{44Lghost} we obtain the complete Lagrangian of the EWSM
suitable for higher-order calculations,
\begin{equation}
{\cal L}_{\mathrm{GSW}} = {\cal L}_{\mathrm{class}} + {\cal L}_{\mathrm{fix}} + {\cal L}_{\mathrm{ghost}}
\label{eq:LSM}
\end{equation}
The Feynman rules which can be derived from the Lagrangian in Eq. (\ref{eq:LSM}) are
given in appendix \ref{sec:fr} in the 't Hooft-Feynman gauge for the physical fields
of the broken gauge theory.

\begin{figure}
\centering
\epsfig{file=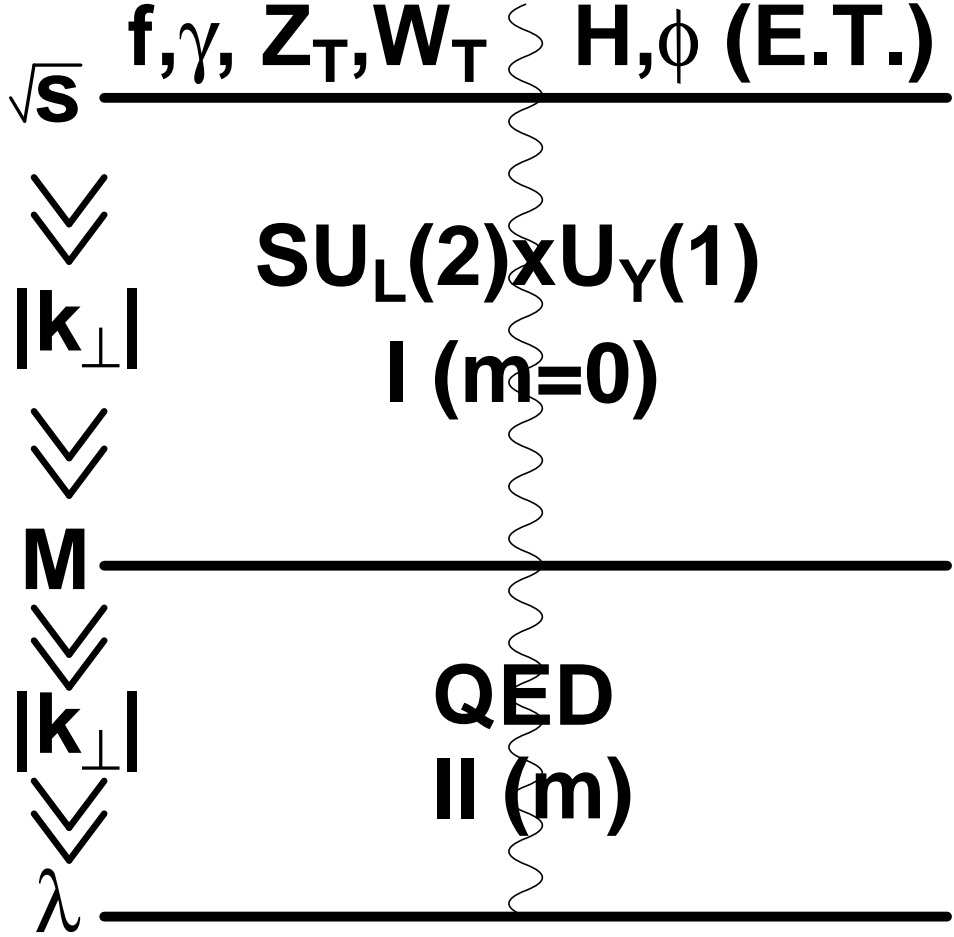,width=12.5cm}
\caption{The schematic depiction of the effective high energy regimes ($\sqrt{s}
\gg M \gg \lambda$) in the
framework of the infrared evolution equation method. In region I), the high
energy corrections are obtained effectively in the unbroken $SU_L(2)\times
U_Y(1)$ theory described by ${\cal L}_{\mathrm{symm}}$ in Eq. (\ref{eq:lsym})
where all terms connected to the v.e.v. can be neglected to SL accuracy.
For external fermions and transverse
gauge bosons this picture contains at the subleading level Yukawa interactions
and contributions from Higgs doublets to the anomalous scaling violations.
For external longitudinal gauge bosons ($\phi = \{ \phi^+, \phi^-, \chi \}$)
the equivalence theorem (E.T.) is
employed yielding effectively a scalar theory charged under the unbroken
gauge group. Again this scenario contains at the subleading level 
Yukawa terms introduced by the spontaneous symmetry breaking. For all charged
particles, the soft photon effects, regulated here by a fictitious photon mass
$\lambda$, are included by integrating in region II)
which incorporates pure QED effects including mass terms. 
In the calculation $\lambda$ is replaced
by a cutoff $\mu$ on the exchanged $|{\mbox{\boldmath $k$}_{\perp}}|$. 
The matching condition is given by the
requirement that the high energy solution in region I) is obtained if the
infrared cutoff $\mu$ is chosen to be the gauge boson mass $M$.}
\label{fig:su2u1}
\end{figure}

At high energies and for processes that are not mass suppressed or dominated by resonances
we can neglect particle masses and terms connected to the vacuum expectation value
(v.e.v.) of the broken gauge theory to the level of SL accuracy \cite{dp2}. Thus, instead
of the Lagrangian in Eq. (\ref{eq:LSM}) we use a high energy approximation
of $\L_{\mathrm{symm}}$ which is based on the fields of the unbroken phase in
the symmetric basis and neglect all terms with a mass dimension.
It is composed of a Yang-Mills part, a Higgs and a fermion
part which are given by \cite{dhab}:
\begin{equation}
{\cal L}_{\mathrm{YM}}=- \frac{1}{4} \left( \partial_\mu W^a_\nu-\partial_\nu W^a_\mu + g
\varepsilon^{abc} W^b_\mu W^c_\nu \right)^2  
-\frac{1}{4} \left( \partial_\mu B_\nu-\partial_\nu B_\mu \right)^2
\end{equation}
where $\varepsilon^{abc}$ is the totally antisymmetric tensor of SU(2). The Higgs
part consists of a single complex scalar SU(2) doublet field with hypercharge $Y=1$:
\begin{equation}
\Phi (x) = \left( \begin{array}{lc} \phi^+ (x) & \\ \phi^0 (x) & \end{array} \right)
\end{equation}
with $\phi^0 (x) = \frac{1}{\sqrt{2}} \left( H(x)+i \chi (x) \right)$ and where the
v.e.v. is neglected. $\phi^+$, $\phi^-$ and
$\chi$ denote the would-be Goldstone bosons and $H$ the physical Higgs field. 
They couple to the gauge fields via
\begin{equation}
{\cal L}_{\mathrm{H}} = \left( D_\mu \Phi \right)^\dagger \left( D^\mu \Phi \right)
\end{equation}
where we omit the self coupling part (the potential) and the covariant derivative is
given by
\begin{equation}
D_\mu = \partial_\mu - i g T^a W^a_\mu + i g^\prime \frac{Y}{2} B_\mu
\end{equation}
The left handed fermions transform as doublets and the right handed ones as singlets under
the gauge group. The fermionic part of the symmetric Lagrangian is then given by
\begin{equation}
{\cal L}_{\mathrm{F}} = \sum_j \left( \overline{L}_j i \gamma^\mu D_\mu L_j \right)
+\sum_j \left( \overline{R}_j i \gamma^\mu D_\mu R_j \right)
-\sum_{j,l} \left( \overline{L}_j G_{jl} R_l  \Phi  + h.c. \right)
\end{equation}
The covariant derivative acting on right handed
fields contains no term proportional to $g$. The Yukawa coupling matrices are denoted by $G_{jl}$
noting that for up-quarks, the charge conjugated Higgs field must be used.
The high energy effective symmetric part of the Lagrangian is then given by
\begin{equation}
{\cal L}_{\mathrm{symm}} = {\cal L}_{\mathrm{YM}} + {\cal L}_{\mathrm{H}} + {\cal L}_{\mathrm{F}}
+ {\cal L}_{\mathrm{fix}} + {\cal L}_{\mathrm{FP}}
\label{eq:lsym}
\end{equation}
where the corresponding ghost and gauge fixing terms are given by
\begin{equation}
{\cal L}_{\mathrm{fix}} = - \frac{1}{2} \left[ \left( F^W \right)^2
+\left( F^B \right)^2 \right]
\end{equation}
with
\begin{equation}
F^W =  \frac{1}{\sqrt{\xi_W}} \partial^\mu W^a_\mu \;, \;\;
F^B = \frac{1}{\sqrt{\xi_B}} \partial^\mu B_\mu 
\end{equation}
and
\begin{equation}
{\cal L}_{\mathrm{FP}} = \overline{u}^\alpha (x) \frac{ \delta F^\alpha}{\delta \theta^\beta (x)} u^\beta (x)
\end{equation}
where $ \frac{ \delta F^\alpha}{\delta \theta^\beta (x)}$ is the variation of the gauge fixing operators
$F^\alpha$ under the infinitesimal gauge transformations characterized by $\theta^\beta (x)$. The 
Faddeev Popov ghosts are denoted by $u^\alpha (x)$.
The corresponding Feynman rules are thus analogous to a theory with an unbroken $SU_L(2)\times U_Y(1)$
and fermions or scalars in the fundamental representation respectively. The new ingredient in
${\cal L}_{\mathrm{F}}$ is the Yukawa term. In addition we have in the gauge boson sector the coupling
of the gauge bosons to scalars through the covariant derivative in ${\cal L}_{\mathrm{H}}$. This
effective regime corresponds to region I) in Fig. \ref{fig:su2u1} where the wavy line
separates the transverse sector (analogous to an unbroken gauge theory) and the scalar sector,
where for the would-be Goldstone bosons the equivalence theorem (E.T.) must be used. Note that all gauge bosons 
contained in ${\cal L}_{\mathrm{symm}}$ are massless and an infrared cutoff will treat 
$W^a_\mu$ and $B_\mu$ fields in the same way.

In the following we always use the 't Hooft-Feynman gauge.
For the high energy regime, the Lagrangian in Eq. (\ref{eq:lsym}) is convenient since it allows
for an approach via the IREE method \cite{kl,kl2} described in section \ref{sec:bgt}
and thus, for a consistent treatment of higher order SL corrections at high energies. In Ref.
\cite{dp2} it was proven that Eq. (\ref{eq:lsym}) at one loop to SL accuracy
gives the same results as calculations
based on the physical Lagrangian in Eq. (\ref{eq:LSM}). The approach in Ref. \cite{dp2} uses collinear
Ward identities to show that SL contributions from the v.e.v. part of the Lagrangian (\ref{eq:LSM})
do not contribute additional terms not already contained in ${\cal L}_{\mathrm{symm}}$.
In particular this means that for longitudinal degrees of freedom at high energy we can employ the Goldstone
boson equivalence theorem and to SL accuracy, we treat the would-be Goldstone bosons $\phi^+$, $\phi^-$
and $\chi$ as physical degrees of freedom in the ultrarelativistic limit. 
At higher orders, all terms related to the renormalization
of the Goldstone bosons are sub-subleading (SSL).

Fig. \ref{fig:su2u1} also indicates that this approach is only
valid in the high energy regime $\sqrt{s} \gg M$ and that the QED corrections from below the weak scale
must be included by appropriate matching conditions at $M$. 

Thus the overall approach 
consists of identifying the relevant degrees of freedom in region I) and II), integrating out
the contributions to SL accuracy and by matching the solution found in II in such a way that at the weak scale
$M$ the solution in region I) is reproduced. 

\subsection{Organization of the paper} \label{sec:org}

The paper is organized as follows. In section \ref{sec:ugt} we summarize the various ingredients needed
to calculate SL virtual corrections in unbroken gauge theories. While these corrections do not lead to
physical observables in those theories, the IREE approach allows for an application of the results
of section \ref{sec:ugt} to broken gauge theories in the high energy limit in section \ref{sec:bgt}. 
The QED effects from the region below the weak scale are implemented with the appropriate matching
conditions as indicated above. As mentioned above, we
use the term ``broken gauge theories'' in the sense that
the local symmetry is hidden due to the degeneracy of the vacuum ground state
and thus not evident in the physical states.
The associated local BRST relations, however, still hold \cite{el,froh}. 

In section \ref{sec:app} the results summarized in section \ref{sec:bgt} are applied to specific
processes relevant to a future linear collider program. In particular the importance of the higher
than one loop corrections is emphasized. We present our concluding remarks in section \ref{sec:out}
and discuss lines of future work needed for precision prediction at future TeV colliders.

\section{Unbroken gauge theories} \label{sec:ugt}

In this section we summarize the results obtained for virtual corrections in
unbroken gauge theories at high energies.
These contributions will be crucial for the high energy regime of the SM in section \ref{sec:bgt}.

\subsection{Sudakov double logarithms} \label{sec:sdl}

The high energy asymptotics of electromagnetic processes was calculated many
years ago within the framework of QED \cite{s}. In particular
the amplitude for $e^{+}e^{-}$ elastic scattering at a fixed angle ($s\sim
|t|\sim |u|\gg m^{2}\gg \lambda ^{2}$, where $m$ is the electron and $%
\lambda $ a fictitious \footnote{$\lambda$ plays the role of the infrared
cut-off. In physical cross sections the divergence in $\lambda$ of the
elastic amplitude is canceled with the analogous divergences in processes
with soft photon emissions.} photon mass) in the DL approximation has the
form 
\begin{equation}
{\cal M}={\cal M}_{\rm Born}\; \Gamma ^{2}\left( \frac{s}{m^{2}},\frac{m^{2}}{%
\lambda ^{2}}\right) 
\end{equation}
where ${\cal M}_{\rm Born}$ is the Born amplitude for $e^{+}e^{-}$ scattering
and $\Gamma $ is the Sudakov form factor. The DL approximation applies in
the energy regime
\begin{equation}
\frac{e^2}{4 \pi^2} \log ^{2}\frac{s}{m^{2}}\sim \frac{e^2}{4 \pi^2} \log \frac{s}{m^{2}}\log \frac{%
m^{2}}{\lambda ^{2}}\sim 1
\end{equation}
where the QED coupling $e^{2}/4\pi \ll 1$. Thus each charged
external particle effectively contributes $\sqrt{\Gamma }$ to the total
amplitude. The Sudakov form factor appears in the elastic scattering of an
electron off an external field \cite{s}. It is of the form:
\begin{equation}
\Gamma \left( \frac{s}{m^{2}},\frac{m^{2}}{\lambda ^{2}}\right) =\exp \left(
-\frac{e^2}{8\pi^2 }R\left( \frac{s}{m^{2}},\frac{m^{2}}{\lambda ^{2}}%
\right) \right)
\end{equation}
To specify $R$ it is convenient to use the Sudakov parametrization of the
momentum of the exchanged virtual photon :
\begin{equation}
k=vp_{1}+up_{2}+k_{\perp } \label{eq:kpdef}
\end{equation}
for massless fermions and
\begin{equation}
k=v \left(p_{1}-\frac{m^2}{s} p_2\right)+u \left( p_{2}-\frac{m^2}{s} p_1 \right)+k_{\perp }
\label{eq:kpdefm}
\end{equation}
\begin{figure}[t]
\centering
\epsfig{file=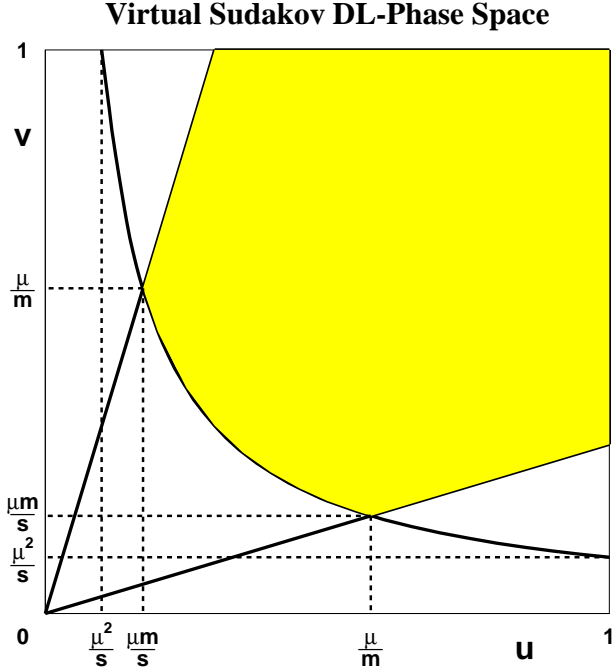,width=10cm}
\caption{The virtual Sudakov DL-phase space in massive QED for the function $R$ in the $\{ u,v \}$
representation. The cutoff $\mu$ plays the role of $\lambda$ for $\mu \ll m$. The shaded area is the region
of integration and is symmetric with respect to $u$ and $v$. For $\mu \geq m$ the relevant phase space
is mass independent.}
\label{fig:suv}
\end{figure}              
\noindent for massive fermions. $p_{1}$ and $p_{2}$ are the initial and final momenta of the scattered
electron and in the following we denote the Euclidean component
\begin{equation}
{\mbox{\boldmath $k$}_{\perp}}^2=-k_\perp^2>0 \label{eq:kpeuc}
\end{equation}
 $R\left( \frac{s}{m^{2}},\frac{m^{2}}{\lambda ^{2}}\right)$ can then be written as the integral over $u$ and $v$ 
 after rewriting the measure as $d^4k=d^2k_\perp d^2k_\parallel$ with
 \begin{eqnarray}
 d^2k_\perp &=& |{\mbox{\boldmath $k$}_{\perp}}| d |{\mbox{\boldmath $k$}_{\perp
 }}| d \phi =
 \frac{1}{2} d {\mbox{\boldmath $k$}^2_{\perp}} d \phi = \pi d {\mbox{\boldmath
 $k$}^2_{\perp}}
 \\
 d^2k_\parallel &=& | \partial (k^0,k^x)/ \partial (u,v)| d u d v \approx \frac{
 s}{2} du dv
 \end{eqnarray}
 where we turn the coordinate system such that the $p_1,p_2$ plane corresponds to
$0,x$ and the
 $y,z$ coordinates to the $k_\perp$ direction so that it is purely spacelike (see Eq.
(\ref{eq:kpeuc})).
 The last equation follows from $p_i^2=0$, i.e. $p_{i_x}^2\approx p_{i_0}^2$ and
 \begin{equation}
 (p_{1_0}p_{2_x}-p_{2_0}p_{1_x})^2 \approx (p_{1_0}p_{2_0}-p_{2_x}p_{1_x})^2=(p_1
 p_2)^2=(s/2)^2
 \end{equation}
Integrating according to
the DL phase space of Fig. \ref{fig:suv} (where $\mu$ plays the role of $\lambda$):
\begin{eqnarray}
R\left( \frac{s}{m^{2}},\frac{m^{2}}{\lambda ^{2}}\right)
&=&\int_{0}^{1}du\int_{0}^{1}dv\,\left( \frac{1}{u+m^{2} \; v/s}\right)
\left( \frac{1}{v+m^{2} \; u/s}\right) \,\,\theta (suv-\lambda ^{2}) \nonumber \\
&\approx&\int_{0}^{1}\frac{du}{u}\int_{0}^{1} \frac{dv}{v} \, \theta \left( u-m^{2} \; v/s \right)
\,\, \theta \left( v-m^{2} \; u/s\right) \,\,\theta (suv-\lambda ^{2}) 
\label{eq:R}
\end{eqnarray}
where $s\sim |t|\sim 2p_{1}p_{2}$. The first two factors in the integrand
correspond to the propagators of the virtual fermions which occur in the
one-loop triangle Sudakov diagram. The $\theta $ - function appears as a
result of the integration of the propagator of the photon over its
transverse momentum $k_{\perp }$: 
\begin{equation}
\frac{i}{k^2- \lambda^2+i\varepsilon}=
\frac{i}{s u v - \lambda^2-{\mbox{\boldmath $k$}^2_{\perp}}+i\varepsilon}= {\cal P}
\frac{i}{s u v - \lambda^2-{\mbox{\boldmath $k$}^2_{\perp}}} + \pi \delta ( s u v -
\lambda^2-{\mbox{\boldmath $k$}^2_{
\perp}}) \label{eq:propid}
\end{equation}
writing it in form of the real and imaginary parts (the principle value is indicated
by ${\cal P}$).
The latter does not contribute to the DL asymptotics and at higher orders gives
subsubleading
contributions.                                                                                               
We note that the main contribution comes
from the region near the photon mass shell:
\begin{equation}
suv=\lambda ^{2}+{\mbox{\boldmath $k$}_{\perp }^{2}}\,.  \label{eq:kp}
\end{equation}
To DL accuracy Eq. (\ref{eq:R}) gives for $\lambda \ll m$:
\begin{equation}
R\left( \frac{s}{m^{2}},\frac{m^{2}}{\lambda ^{2}}\right) =\frac{1}{2}\ln
^{2}\frac{s}{m^{2}}+\ln \frac{s}{m^{2}}\ln \frac{m^{2}}{\lambda ^{2}}
\end{equation}
where the result comes equally from two different kinematical regions, $v\gg
u$ and $u\gg v$ as is evident from Fig. \ref{fig:suv}. Therefore one can write $R=2r$.

\begin{figure}[t]
\centering
\epsfig{file=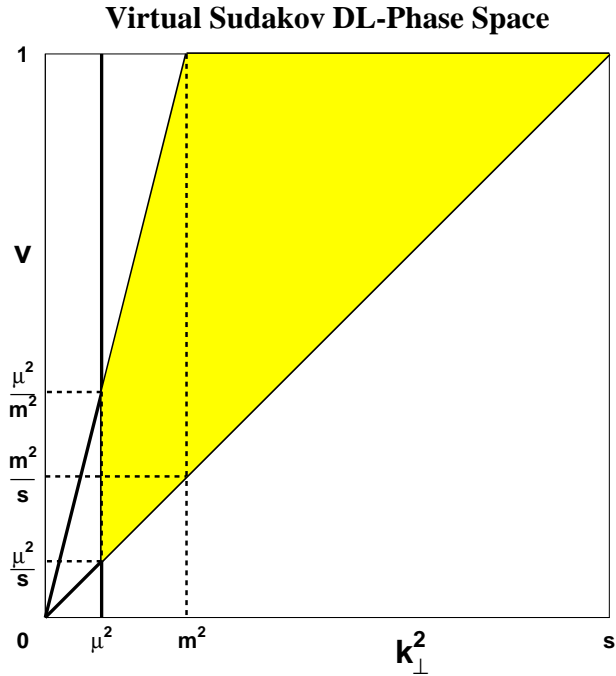,width=10cm}
\caption{The virtual Sudakov DL-phase space in massive QED for the function $R$ in the $\{
{\mbox{\boldmath $k$}_{\perp }^2}, v \}$ representation. The shaded area is the region
of integration. For $\mu \ll m$ the cufoff plays the role of $\lambda$ in the text.
For $\mu \geq m$ the relevant phase space
is mass independent as in the $\{u,v\}$ representation of Fig. \ref{fig:suv}.}
\label{fig:skp}
\end{figure}              
We can obtain physical insight by presenting the two equal contributions
separately. In the first region, with $v\gg u$, the virtual photon is
emitted along $p_{1}$ and the parameter $v$ is given by the ratio of
energies of the photon and the initial electron. Here instead of $u$, it is
convenient to use Eq. (\ref{eq:kp}) to replace it by the square of the
transverse momentum component of the photon. Then integrating over $v$ and ${%
\mbox{\boldmath
$k$}_{\perp }^{2}}$ according to the DL phase space in Fig. \ref{fig:skp} gives
\begin{equation}
r\left( \frac{s}{m^{2}},\frac{m^{2}}{\lambda ^{2}}\right) =
\int_{\lambda /\sqrt{s}}^{1}\frac{dv}{v}\int_{\lambda ^{2}}^{sv^{2}}\frac{d{%
\mbox{\boldmath
$k$}_{\perp }^{2}}}{{\mbox{\boldmath $k$}_{\perp }^{2}}+m^{2}v^{2}}\simeq
\int_{\lambda ^{2}}^{s}\frac{d{\mbox{\boldmath $k$}_{\perp }^{2}}}{{%
\mbox{\boldmath
$k$}_{\perp }^{2}}}\int_{|{\scriptsize {\mbox{\boldmath $k$}}_{\perp }|/%
\sqrt{s}}}^{{\rm min} (|{\scriptsize {\mbox{\boldmath $k$}}_{\perp }|/m,\;1)}}%
\frac{dv}{v} \label{eq:r}
\end{equation}
in the DL approximation, which may be evaluated to give half of $R$. The
quantity $r$ is proportional to the probability $w_{i}$ of the emission of a
soft and almost collinear photon from an external particle with energy $%
\sqrt{s}$ and mass $m_{i}$, i.e.
\begin{equation}
w_{i}(s,\lambda ^{2})=\frac{e^2}{4\pi^2 }\,r\left( \frac{s}{m_{i}^{2}},%
\frac{m_{i}^{2}}{\lambda ^{2}}\right)  \label{eq:a10}
\end{equation}
If several charged particles participate in a process, for example $%
e^{+}e^{-}\rightarrow f\bar{f}f\bar{f}$, then analogous contributions appear
for each external line, provided all external invariants are large and of
the same order. This leads to the general result
\begin{equation}
{\cal M}={\cal M}_{{\rm Born}}\exp
\left( -\frac{1}{2}\sum_{i=1}^{n}w_{i}(s,%
\lambda^{2})\right)  \label{eq:a11}
\end{equation}
where $n$ is the number of external lines corresponding to charged
particles. In summary the soft emissions described by the Sudakov form
factor is a quasi-classical effect which does not depend on the hard
dynamics of the process. In particular there are no quantum mechanical
interference effects in the DL Sudakov corrections, for large scattering
angles.

\subsection{Gribov's factorization theorem} \label{sec:grib}

In this section we discuss a factorization theorem due to Gribov \cite{grib, grib2,grib3,grib4}.
It was originally derived in bremsstrahlung off hadrons at high energies 
in the context of QED but can appropriately be extended to non-Abelian
gauge theories. We follow the original derivation for real emission processes
noting that the form of factorization of virtual corrections must be analogous
due to the KLN theorem \cite{k,ln}.

Consider bremsstrahlung off a fermion with mass $m$ in the laboratory system. 
We denote the invariants
according to the notation depicted in Fig. \ref{fig:grib} as follows:
\begin{figure}
\centering
\epsfig{file=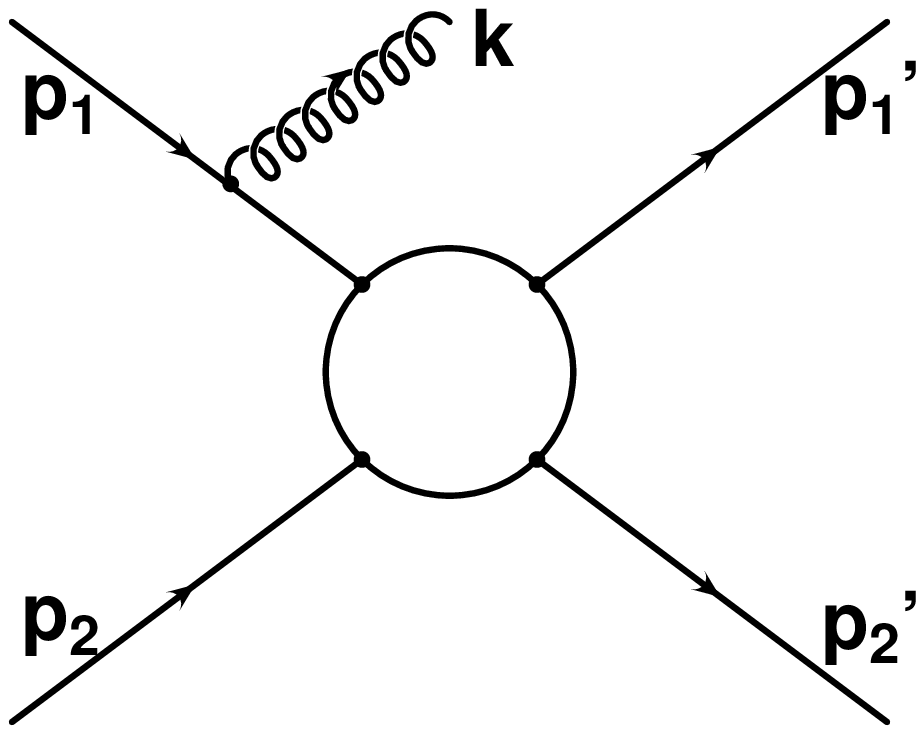,width=7cm}
\hspace{1cm}
\epsfig{file=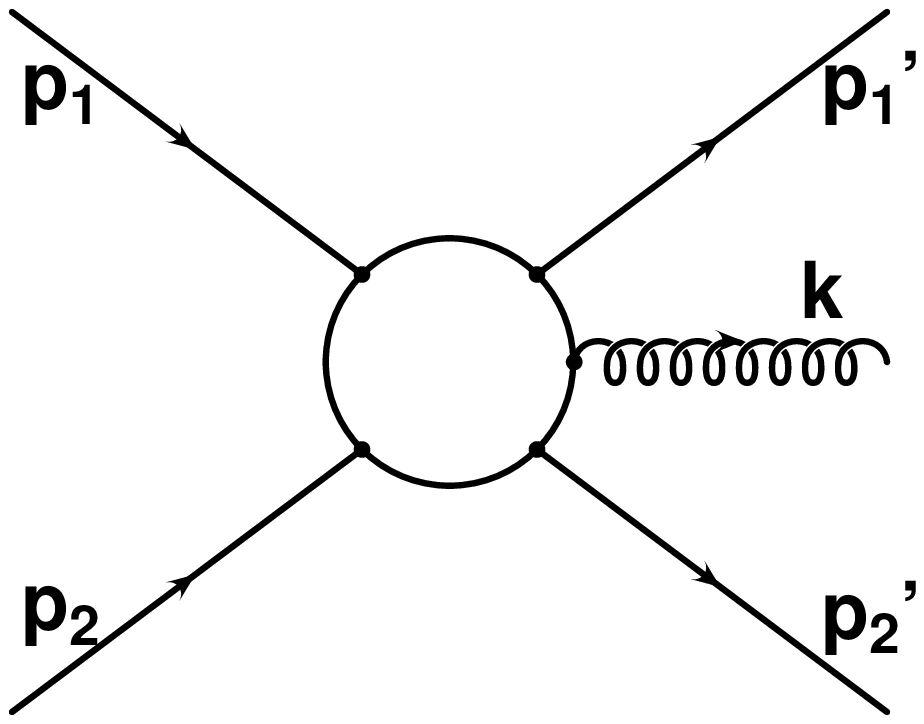,width=7cm} 
\caption{Bremsstrahlung in a process involving charged fermions. At high energies only the external
legs contribute to DL accuracy.}
\label{fig:grib}
\end{figure}
\begin{eqnarray}
s'&=& (p_1'+p_2')^2, \;\; t_2=(p_2'-p_2)^2 
\end{eqnarray}
The usual eikonal argument is that for $2p_1k \ll m^2$,
only the diagram on the l.h.s. is large,
yielding (neglecting $m$ in $M$):
\begin{eqnarray}
{\cal M} &=& - \frac{\varepsilon^* p_1}{p_1k} M(s',t_2) 
\end{eqnarray}
At large energies we have
\begin{eqnarray}
2p_1k &=& 2 | {\bf p_1}| \; | {\mbox{\boldmath $k$}} | (1-\cos \theta) + \frac{| {\mbox{\boldmath $k$}} |}{
| {\bf p}_1|} m^2 
\end{eqnarray}
Thus for ${\cal M}$ to be large we need in any case: $\frac{| {\mbox{\boldmath $k$}} |}{|{\bf p}_1|} \ll 1$.
It follows that the condition $| {\bf p}_1| \; | {\mbox{\boldmath $k$}} | \theta^2 \ll m^2$ 
should be fulfilled for small emission angles $\theta$.

Gribov observed, however, that ${\cal M}$ is large in broader region!
In the region $2p_1k > m^2$, $\frac{| {\mbox{\boldmath $k$}} |}{|{\bf p}_1|} \ll
1$ we have with $p_1=(p^0_1,0,0,|{\bf p}_1|)$, $k=(k^0,k^0 \sin \theta, 0 , k^0 \cos \theta)$ and
$\varepsilon^*=(0,\cos \theta, \pm i, -\sin \theta )$:
\begin{eqnarray}
{\cal M} = - \frac{2}{| {\mbox{\boldmath $k$}} | \theta} M(s',t_2) 
\end{eqnarray}
i.e. a sufficient condition is: $|{\mbox{\boldmath $k$}_{\perp }}| \approx | {\mbox{\boldmath $k$}} |
\theta \ll m$.

The proof proceeds as follows:

\noindent $M$ is taken on the mass shell in order to ensure gauge
invariance! In covariant form
the conditions read\footnote{We consider here only the case of initial state radiation in analogy
to Ref. \cite{grib,grib2}.}:
\begin{eqnarray}
\frac{2p_1k}{s} &\ll& 1, \; \frac{2p_2k}{s} \ll 1, \; {\mbox{\boldmath $k$}_{\perp }^2}
=\frac{4p_1kp_2k}{s} \ll m^2 \label{eq:cond}
\end{eqnarray}
We can then write the amplitude in a gauge invariant form:
\begin{eqnarray}
{\cal M} = \left( \frac{\varepsilon^* p_2}{p_2k}-\frac{\varepsilon^* p_1}{p_1k} \right) M(s,t)
\label{eq:Minv}
\end{eqnarray}
and write it in the following way:
\begin{eqnarray}
{\cal M}_\mu = {p_1}_\mu M_1 + {p_2}_\mu M_2 +q_\mu M_3, \;\;\;\; q=p_2-p_2'
\end{eqnarray}
Gauge invariance yields:
\begin{eqnarray}
p_1k M_1+p_2k M_2+qk M_3=0 \label{eq:ginv}
\end{eqnarray}
At high energies
$t_2-t_1=2qk = \frac{2kp_1}{s} 2 q p_2 + \frac{2kp_2}{s} 2 q p_1 +
2|{\mbox{\boldmath $k$}_\perp}| | {\mbox{\boldmath $q$}_\perp}| 
\approx 2|{\mbox{\boldmath $k$}_\perp}| | {\mbox{\boldmath $q$}_\perp}|$. 
Thus, $2qk$ is small, so that
there is no need to distinguish between $s, s^\prime$ and between $t_1,t_2$ in
the inner on-shell amplitude $M$ of Eq. (\ref{eq:Minv}). Eq. (\ref{eq:ginv}) then reads
\begin{eqnarray}
p_1k M_1=-p_2k M_2 
\end{eqnarray}
Thus
\begin{eqnarray}
M_1 &=& -\frac{M(s,t)}{p_1k}-p_2k \; T(s,t,p_1k,p_2k) \\
M_2 &=& \frac{M(s,t)}{p_2k}-p_1k \;T(s,t,p_1k,p_2k) 
\end{eqnarray}
where the only pole of $M_1$ is at $p_1k$,
that of $M_2$ at $p_2k$.
$M_3$ and $T$ have no singularity
at $p_1k$ or $p_2k$, and
\begin{eqnarray}
{\cal M}_\mu &=& \left( \frac{{p_2}_\mu}{p_2k}-\frac{{p_1}_\mu}{p_1k} \right)
[ M(s,t) + p_1kp_2k T] +q_\mu M_3  \\
{\cal M} &=& {\varepsilon^*}^\mu {\cal M}_\mu = - \frac{2}{|{\mbox{\boldmath $k$}_{\perp }}|} \left[ M(s,t)+
\frac{1}{4} {\mbox{\boldmath $k$}_{\perp }^2} s T \right] + \varepsilon^* q M_3 
\end{eqnarray}
Since $M_3$ and $T$  are functions of $p_1k$, $p_2k$
they could be of order of the first term.

\begin{figure}
\centering
\epsfig{file=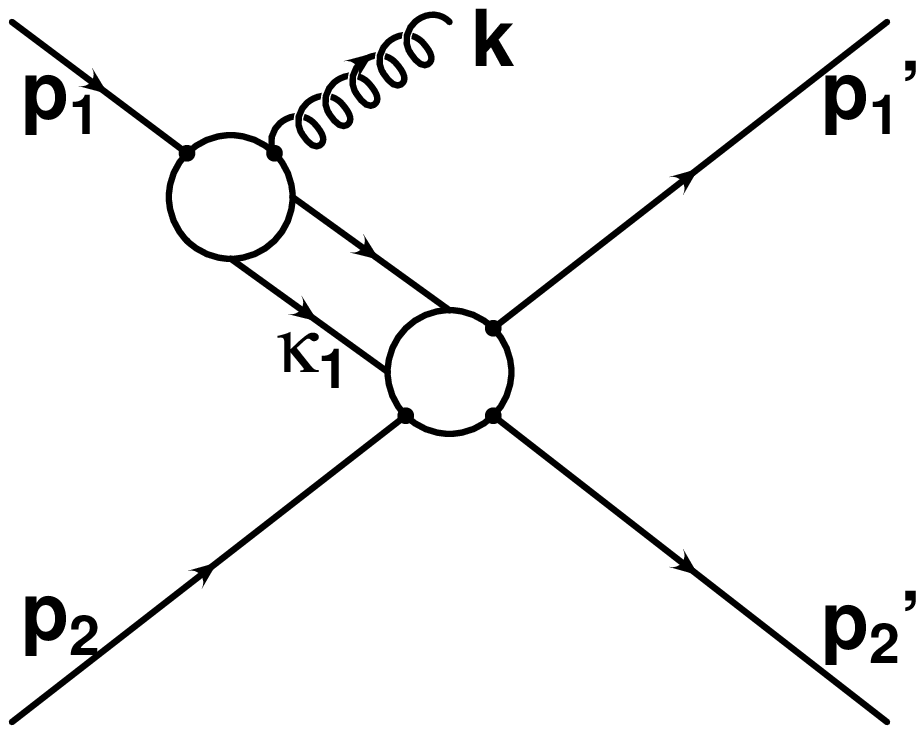,width=7cm}
\hspace{1cm}
\epsfig{file=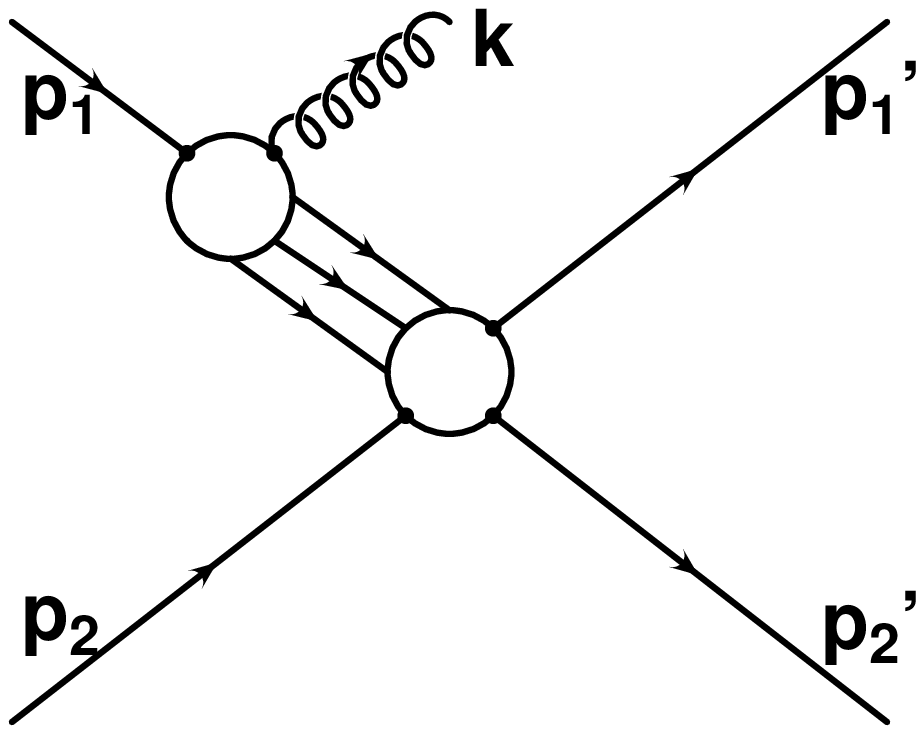,width=7cm} 
\caption{Higher order splittings determining the imaginary part of the scattering amplitude. At high
energies, Gribov proofed that the pole terms in the variables $2p_ik$ do not dominate the amplitude
and that the large terms factorize with respect to $1/|{\mbox{\boldmath $k$}_{\perp }}|$.}
\label{fig:sp}
\end{figure}

To show that this is {\it not}
the case, consider the
imaginary part (discontinuity) of ${\cal M}$
in $p_1k$, $p_2k$:
\begin{eqnarray}
{\mbox Im} \; {\cal M} = -\frac{1}{2} |{\mbox{\boldmath $k$}_{\perp }}| s \; {\mbox Im} \; T +
\varepsilon^* q \;{\mbox Im}
\; M_3 
\end{eqnarray}
It is determined by all possible splittings (not in $s,t_1,t_2$)
like the ones depicted in Fig. \ref{fig:sp}.
The simplest two particle intermediate state contains the amplitude
\begin{eqnarray}
{\cal M}_s = \left( \frac{\varepsilon^* \kappa_1}{\kappa_1k}-\frac{\varepsilon^* p_1}{p_1k} \right)
M_s((p_1+k)^2,{q'}^2) 
\end{eqnarray}
where $\kappa_1$ is the momentum of the charged particle.

As $p_1$ is large
and $(p_1+k)^2 \ll s$, $|{\mbox{\boldmath $\kappa$}}_1|$ is also large and along $
{\mbox{\boldmath $p$}}_1$:
\begin{eqnarray}
\frac{{\mbox{\boldmath $\kappa$}}_1}{\kappa_1k} \approx \frac{{\mbox{\boldmath $p$}}_1}{p_1k}
+\frac{{\mbox{\boldmath $q$}_{\perp
}}^\prime}{\kappa_1k} 
\end{eqnarray}
where ${\mbox{\boldmath $q$}_{\perp }}^\prime$ is the component of
 ${\mbox{\boldmath $\kappa$}}_1$
perpendicular to ${\mbox{\boldmath $p$}}_1$. Thus
\begin{eqnarray}
{\cal M}_s = \frac{\varepsilon^* q_\perp^\prime}{\kappa_1k} M_s ((p_1+k)^2,{q'}^2) 
\end{eqnarray}
We therefore observe that the large terms cancel! For higher splittings 
the cancellation proceeds analogously since in the intermediate states all
particles formed at high energies are parallel to the original particle momentum.

Thus, the large contributions to the original bremsstrahlung amplitude
are given by
\begin{eqnarray}
{\cal M} = - \frac{2}{|{\mbox{\boldmath $k$}_{\perp }}|} M (s,t) 
\end{eqnarray}
for $2p_1k \ll s$, $2p_2k \ll s$ and ${\mbox{\boldmath $k$}_{\perp }}^2 \ll \mu \leq m^2$.
The cutoff $\mu$ is introduced for later convenience.
An analogous factorization in $\frac{1}{{\mbox{\boldmath $k$}_{\perp }}^2}$
then holds for virtual corrections with ${\mbox{\boldmath $k$}_{\perp }}^2 \ge \mu^2$
since the sum of real and virtual corrections must be independent of the infrared cutoff.

In order to treat Non-Abelian gauge theories we need to
introduce a gauge invariant cutoff on all virtual
particles with momentum $\kappa_i$:
\begin{eqnarray}
{\mbox{\boldmath $\kappa_i$}_\perp}^2 \ge \mu^2, \; \kappa_i=v_ip_1+u_ip_2+{\kappa_i}_\perp 
\end{eqnarray}
and it is understood that $\mu \gg \Lambda_{\rm QCD}$ in order to remain in the perturbative
regime. The crucial point now is that
$\mu$ determines {\it both} the positions of the thresholds in the
variables $2p_1k$, $2p_2k$ {\it and} the minimum momentum transfers \cite{fad}:
\begin{eqnarray}
&& (p_1-k)^2=(\kappa_1+\kappa_2)^2 = \nonumber \\
&& {{\mbox{\boldmath $\kappa$}_1}_\perp}^2(1+\frac{u_2}{u_1})
+{\mbox{\boldmath $\kappa_2$}_\perp}^2(1+\frac{u_1}{u_2}) - {\mbox{\boldmath $k$}_{\perp }}^2 \ge 4\mu^2
\end{eqnarray}
for ${\mbox{\boldmath $k$}_{\perp }}^2 \ll \mu^2$. Thus the cut starts from
$4\mu^2$ and
\begin{eqnarray}
|(k-q_1)^2|=({{\mbox{\boldmath $q$}_1}_{\perp }}-{\mbox{\boldmath $k$}_{\perp }})^2-s(u_1-u)(v_1-v)
\ge \mu^2 
\end{eqnarray}
Now the same dispersive arguments are applicable to QCD
 as they
were in QED.

Thus we can consider again the simplest situation, when the additional
soft gauge boson is emitted in the process with all invariants $s_{lj}$
large. Of course, for the emission of a boson almost collinear to the
particle the direction of the particle with momentum $p_{i}$, the invariant
$%
2kp_{i}$ is small in comparison with $s$. In the case of non-Abelian gauge
theories the corresponding amplitude for the emission of a soft gauge boson
with small ${\mbox{\boldmath $k$}_{\perp }^{2}}\ll \mu ^{2}$ has, according
to the Gribov theorem as derived above, the following form in non-Abelian theories:
\begin{equation}
{\cal M}^{a}(p_{1},...,p_{n};k;\mu ^{2})=\sum_{j=1}^{n}g_s\;\frac{\varepsilon
^{*}p_{j}}{kp_{j}}\;
T^{a}(j)\;{\cal M}(p_{1},...,p_{n};\mu ^{2})\,.
\label{eq:rgrib}
\end{equation}
where $g_s$ denotes the QCD (or $SU(N)$) coupling.
The possible corrections to this factorized expression are of the order of
${%
\mbox{\boldmath $k$}_{\perp }^{2}}/\mu ^{2}$. However, to DL accuracy,
we can substitute $\mu ^{2}$ in the arguments of the scattering amplitudes
by its boundary value ${\mbox{\boldmath $k$}_{\perp }^{2}}$. Notice that the
amplitude on the r.h.s. of (\ref{eq:rgrib}) is taken on-the-mass shell, which
guarantees its gauge invariance. The result (\ref{eq:rgrib}) is highly
non-trivial in the Feynman diagram approach. It means, that the region of
applicability of the classical formulas for the Bremsstrahlung amplitudes is
significantly enlarged at high energies. 

The form of the virtual factorization and the subsequent resummation is the
topic of the following section.

\subsection{Infrared evolution equations} \label{sec:iree}

Sudakov effects have been widely discussed for non-Abelian gauge theories,
such as $SU(N)$ and can be calculated in various ways (see, for instance,
\cite{NA,NA2,NA3,NA4,NA5,NA6,NA7,NA8,NA9}). We consider here the scattering amplitude in the simplest
kinematics when all its invariants $s_{lj}=2p_{l}p_{j}$ are large and of the
same order $s_{lj}\sim s$. A general method of finding the DL asymptotics
(not only of the Sudakov type) is based on the infrared evolution equations
describing the dependence of the amplitudes on the infrared cutoff $\mu $ of
the virtual particle transverse momenta \cite{kl,kl2}. This cutoff plays the
same role as $\lambda $ in QED, but, unlike $\lambda $, it is not
necessary that it vanishes and it may take an arbitrary value. It can be introduced in
a gauge invariant way by working, for instance, in a finite phase space
volume in the transverse direction with linear size $l\sim 1/\mu $. Instead
of calculating asymptotics of particular Feynman diagrams and summing these
asymptotics for a process with $n$ external lines it is convenient to
extract the virtual particle with the smallest value of $|{%
\mbox{\boldmath
$k$}_{\perp }}|$ ($k_\perp \perp p_j,p_l$)
in such a way, that the transverse momenta $|{%
\mbox{\boldmath $k$}_{\perp }^{\prime }}|$ of the other virtual particles
are much bigger
\begin{equation}
\mbox{\boldmath $k$}_{\perp }^{\prime ^{2}}\gg {\mbox{\boldmath $k$}_{\perp
}^{2}}\gg \mu ^{2}\;.
\end{equation}
For the other particles ${\mbox{\boldmath $k$}_{\perp }^{2}}$ plays the role of
the initial infrared cut-off $\mu ^{2}$.

In particular, the Sudakov DL corrections are related to the exchange of
soft gauge bosons, see Fig.~1. For this case the integral over the
momentum $k$ of the
soft (i.e. $|k^0|\ll \sqrt{s}$) virtual boson with the smallest
${\mbox{\boldmath $k$}}_\perp$ can be factored off, which leads to the
following infrared evolution equation:
\begin{eqnarray}
{\cal M}(p_1,...,p_n;\mu^2) & = & {\cal M}_{\rm Born}(p_1,...,p_n) -\frac{i}{2}
\frac{g_s^2}{(2\pi)^4} \sum_{j,l=1, j \neq l}^n \int_{s \gg \mbox{{\scriptsize \boldmath $k$}}^2_\perp
\gg \mu^2} \frac{d^4k}{k^2+i \epsilon} \;\;
\frac{p_jp_l}{(kp_j)(kp_l)}  \nonumber \\
& & \times \; T^a(j) T^a(l) {\cal M} (p_1,...,p_n;{\mbox{\boldmath $k$}%
^2_\perp}) \,,  \label{eq:vem}
\end{eqnarray}
where the amplitude ${\cal M}(p_1,...,p_n;{\mbox{\boldmath $k$}^2_\perp})$
on the right hand side is to be taken on the mass shell, but with the
substituted infrared cutoff: $\mu^2 \longrightarrow {\mbox{\boldmath $k$}%
^2_\perp}$. 
From Eq. (\ref{eq:kpdef}) and the on-shell condition (\ref{eq:kp})
it is clear that $\frac{p_jp_l}{(kp_j)(kp_l)}=
\frac{2}{{\mbox{\boldmath $k$}^2_\perp}}$
and that Eq. (\ref{eq:vem}) has the required factorized form for the virtual corrections
according to
the discussion in section \ref{sec:grib}.

The generator $T^a(l) (a=1,...,N)$ acts on the color indices of
the particle with momentum $p_l$. The non-Abelian gauge coupling is $g$.
In Eq. (\ref{eq:vem}), and below, $\mbox{\boldmath $k$}_\perp$ denotes the component
of the gauge boson momentum $\mbox{\boldmath $k$}$ transverse to the particle
emitting this boson. Note that in Sudakov DL corrections there are no interference
effects, so that we can talk about the emission (and absorption) of a gauge
boson by a definite (external) particle, namely by a particle with momentum
almost collinear to $\mbox{\boldmath $k$}$. It can be expressed in invariant form
as $\mbox{\boldmath $k$}^2_\perp \equiv \min ( 2(kp_l)(kp_j)/(p_lp_j))$ for all $j \neq l$.
\begin{figure}
\centering
\epsfig{file=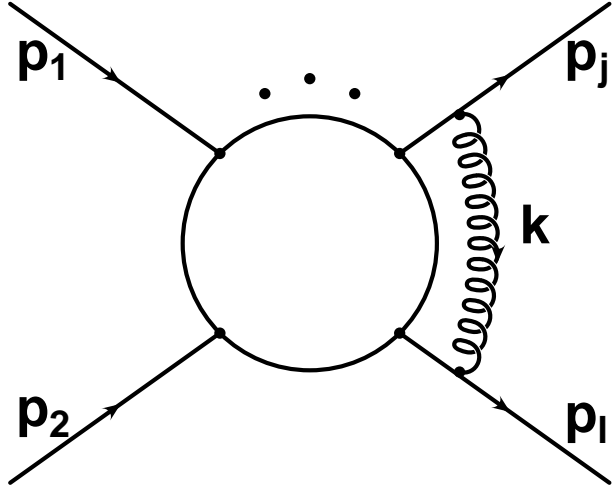,width=8cm}
\caption{Feynman diagrams contributing to the infrared evolution
equation (\ref{eq:vem}) for a process with $n$ external legs. In a general
covariant gauge the
virtual gluon with the smallest value of ${\mbox{\boldmath $k$}}_{\perp}$ is attached to
different external lines. The inner scattering amplitude is assumed to be
on the mass shell.}
\end{figure}
The above factorization is directly related to the non-Abelian generalization of the
Gribov theorem in Eq. (\ref{eq:rgrib}).

The form in which we present
Eq. (\ref{eq:vem}) corresponds to a covariant gauge
for the gluon with
momentum $k$. 
In this region
for $j\neq l$ we have $p_{j}p_{l}/kp_{j}\simeq E_{l}/\omega $, where $E_{l}$
is the energy of the particle with momentum $p_{l}$ and $\omega$ the
frequency of the emitted gauge boson. Using the conservation of the total non-Abelian
group charge:
\begin{equation}
\sum_{j=1}^n T^a(j) {\cal M}(p_1,...,p_j,...,p_n;{\mbox{\boldmath
$k$}_{\perp }^{2}})=0 \label{eq:cnag}
\end{equation}
we can reduce the double sum over the gauge boson insertions in Eq. (\ref{eq:vem})
to a single sum over external legs. In addition it is convenient to use the Sudakov
parametrization analogously to Eq. (\ref{eq:kpdef}) and to replace the variable
$u$ by ${\mbox{\boldmath$k$}_{\perp }^{2}}$ according to Eq. (\ref{eq:kp}). 
The infrared evolution equation then
takes on the form:
\begin{eqnarray}
{\cal M}(p_{1},...,p_{n};\mu ^{2}) &=&{\cal M}_{\rm Born}(p_{1},...,p_{n})-\frac{%
2g_s^{2}}{(4\pi )^{2}}\sum_{l=1}^{n}\int_{\mu ^{2}}^{s}\frac{d{%
\mbox{\boldmath $k$}_{\perp }^{2}}}{{\mbox{\boldmath $k$}_{\perp
}^{2}}}\int_{|\mbox{\scriptsize \boldmath $k$}_{\perp }|/\sqrt{s}}^{{\rm min}
(|\mbox{\scriptsize \boldmath $k$}_{\perp }|/m_l,\;1)}\frac{dv}{v}  \nonumber \\
&&\times \;C_{l}{\cal M}(p_{1},...,p_{n};{\mbox{\boldmath
$k$}_{\perp }^{2}}%
)\;\;,  \label{eq:dgvem}
\end{eqnarray}
where $C_{l}$ is the eigenvalue of the Casimir operator $T^{a}(l)T^{a}(l)$
($%
C_{l}=C_{A}$ for gauge bosons in the adjoint representation of the gauge
group $SU(N)$ and $C_{l}=C_{F}$ for fermions in the fundamental
representation).

The differential form of the infrared evolution equation follows immediately
from (\ref{eq:dgvem}):
\begin{equation}
\frac{\partial {\cal M}(p_{1},...,p_{n};\mu ^{2})}{\partial \log (\mu
^{2})}%
=K(\mu ^{2}){\cal M}(p_{1},...,p_{n};\mu ^{2})\,,  \label{eq:ee}
\end{equation}
where
\begin{equation}
K(\mu ^{2})\equiv -\frac{1}{2}\sum_{l=1}^{n}\frac{\partial W_{l}(s,\mu
^{2})%
}{\partial \log (\mu ^{2})}
\end{equation}
with
\begin{equation}
W_{l}(s,\mu ^{2})=\frac{g_s^{2}}{4\pi^{2}}C_{l}\,r\left( \frac{s}{m_{l}^{2}},
\frac{m_{l}^{2}}{\mu ^{2}}\right) \,.  \label{eq:wl}
\end{equation}
As in the Abelian case, $W_l$ is the probability to emit a soft and
almost collinear gauge boson from the particle $l$ with mass
$m_l$, subject to the infrared cut-off $\mu $ on the transverse momentum. Note
again that the cut-off $\mu $ is not taken to zero. The function $r$
is determined by (\ref{eq:r}) for arbitrary values of the ratio $m_l/\mu $. To
logarithmic accuracy, we obtain from (\ref{eq:wl}):
\begin{equation}
\frac{\partial W_{l}(s,\mu ^{2})}{\partial \log (\mu ^{2})}=-\frac{g_s^{2}}{%
8\pi ^{2}}C_{l}\log \frac{s}{\max (\mu ^{2},m_{l}^{2})}\,.
\end{equation}
The infrared evolution equation (\ref{eq:ee}) should be solved with an
appropriate initial condition. In the case of large scattering angles, if we
choose the cut-off to be the large scale $s$ then clearly there are no
Sudakov corrections. The initial condition is therefore
\begin{equation}
{\cal M}(p_{1},...,p_{n};s)={\cal M}_{\rm Born}(p_{1},...,p_{n}),
\end{equation}
and the solution of (\ref{eq:ee}) is thus given by the product of the Born
amplitude and the Sudakov form factors:
\begin{equation}
{\cal M}(p_{1},...,p_{n};\mu ^{2})={\cal M}_{\rm Born}(p_{1},...,p_{n})\exp
\left( -\frac{1}{2}\sum_{l=1}^{n}W_{l}(s,\mu ^{2})\right)
\end{equation}
Therefore we obtain an exactly analogous Sudakov exponentiation for the
gauge group $SU(N)$ to that for the Abelian case, see (\ref{eq:a11}).
Theories with semi-simple gauge groups can be considered in a similar way.

\subsection{Subleading corrections from splitting functions} \label{sec:sf}

At high energies, where particle masses can be neglected, the form of soft
and collinear divergences is universal. In this regime it is then appropriate
to employ the formalism of the Altarelli-Parisi approach \cite{ap} and to
calculate the corresponding splitting functions. We will do so below only 
for the virtual case. An important observation in this connection is that
at high energies, all subleading terms are either of the collinear or
the RG type. This can be seen as follows:

The types of soft, i.e. $|k^0| \ll \sqrt{s}$, divergences in loop
corrections with massless particles, unlike the collinear
logarithms, can be obtained by setting all $k$ dependent terms in the numerator of tensor
integrals to zero (since the terms left are of the order of the hard scale $s$).
Thus it is
clear that the tensor structure which emerges is that of the inner scattering amplitude in Fig.
\ref{fig:nll} taken on the mass-shell, times a scalar function of the given loop correction.
In the Feynman gauge, for instance, we find for the
well known vertex corrections the familiar three-point function $C_0$ and for higher point
functions we note that in the considered case all infrared divergent scalar integrals reduce to
$C_0$ multiplied by factors of $\frac{1}{s}$ etc.. The only infrared divergent three point function
is given by
\begin{equation}
C_0(s/\mu^2) \equiv \int_{{\mbox{\boldmath $k$}^2_{\perp}} > \mu^2} \frac{d^4k}{(2\pi)^4} \frac{1}{
(k^2+i\varepsilon)(k^2+2p_jk+i\varepsilon)(k^2-2p_lk+i\varepsilon)}
\label{eq:c0mudef}
\end{equation}
The function $C_0(s/\mu^2)$ is fastly converging for large ${\mbox{\boldmath $k$}^2_{\perp}}$ and we
are interested here in the region $\mu^2 \ll s$ in order to obtain large logarithms. Then
logarithmic corrections come from the region ${\mbox{\boldmath $k$}^2_{\perp}}\ll s|u|, s|v| \ll s$
(the strong inequalities give DL, the simple inequalities single ones)
and we can write to logarithmic accuracy:
\begin{eqnarray}
C_0(s/\mu^2) &=& \frac{s \pi}{2 (2 \pi)^4} \int^\infty_{-\infty}du \int^\infty_{-\infty} dv
\int^\infty_{\mu^2} d {\mbox{\boldmath
$k$}^2_{\perp}} \times \nonumber \\ &&
\frac{1}{(s u v - {\mbox{\boldmath $k$}^2_{\perp}}+i\varepsilon)(
s u v - {\mbox{\boldmath $k$}^2_{\perp}}+su +i\varepsilon) (
s u v - {\mbox{\boldmath $k$}^2_{\perp}}-s v +i\varepsilon)} \nonumber \\
&\approx& \frac{s i \pi^2}{2 (2 \pi)^4} \int^1_{-1} \frac{du}{su} \int^1_{-1}
\frac{dv}{sv} \int^\infty_{-\infty}
d {\mbox{\boldmath$k$}^2_{\perp}} \theta ( {\mbox{\boldmath$k$}^2_{\perp}}-
\mu^2 ) \delta ( suv-{\mbox{\boldmath$k$}}^2_{\perp}) \nonumber \\
&\approx&  \frac{i}{2(4\pi)^2s} \int^{1}_{-1} \frac{du}{u} \int^{1}_{-1} \frac{dv}{v} \theta (suv - \mu^2)
\nonumber \\
&=&  \frac{i}{(4\pi)^2s} \int^{1}_{0} \frac{du}{u} \int^{1}_{0} \frac{dv}{v} \theta (suv - \mu^2)
\nonumber \\
&=&  \frac{i}{(4\pi)^2s} \int^1_\frac{\mu^2}{s} \frac{du}{u} \int^1_\frac{\mu^2}{su} \frac{dv}{v}
\nonumber \\
&=&  \frac{i}{2(4\pi)^2s} \log^2 \frac{s}{\mu^2}
\end{eqnarray}
Thus, no single soft logarithmic corrections are present in $C_0(s/\mu^2)$.
In order to see that this result is not just a consequence of our regulator, we repeat the calculation
for a fictitious gluon mass\footnote{Note that this regulator spoils gauge invariance and leads to
possible inconsistencies at higher orders. Great care must be taken for instance when a three gluon
vertex is regulated inside a loop integral.}. In this case we have
\begin{equation}
C_0(s/\lambda^2) \equiv \int \frac{d^4k}{(2\pi)^4} \frac{1}{(k^2-\lambda^2+i\varepsilon)
(k^2+2p_jk+i\varepsilon)(k^2-2p_lk+i\varepsilon)}
\label{eq:c0def}
\end{equation}
It is clear that $C_0(s/\lambda^2)$ contains soft and collinear divergences ($k \parallel p_{j,l}$) and
is regulated with the cutoff $\lambda$, which plays the role of $\mu$ in this case.
Integrating over Feynman parameters we find:
\begin{equation}
C_0(s/\lambda^2)=\frac{i}{(4\pi)^2s} \left( \frac{1}{2} \log^2 \frac{\lambda^2-i \varepsilon}{-s} + \frac{\pi^2}{3}
\right) \label{eq:c0res}
\end{equation}
We are only interested here in the real part of loop corrections of scattering
amplitudes since they are multiplied by the Born amplitude and the imaginary pieces contribute
to cross sections at the next to next to leading level as mentioned above. In fact, the minus sign
inside the double logarithm corresponds precisely to the omitted principle value contribution
of Eq. (\ref{eq:propid}) in the previous calculation.
Thus, no single soft logarithmic correction is present in the case when particle masses can be
neglected.

This feature prevails to higher orders as well since it has been shown that also in non-Abelian
gauge theories
the one-loop Sudakov form factor exponentiates \cite{NA}-\cite{NA9}.

In case we would keep mass-terms, even two point functions, which in our scheme can only
yield collinear logarithms, would contain a soft logarithm due to the mass-renormalization which
introduces a derivative contribution (see for instance Ref. \cite{m}). In conclusion, all leading
soft corrections are contained in double logarithms (soft and collinear),
and subleading logarithmic corrections
in a massless theory, with all invariants large ($s_{j,l}=2p_jp_l\sim {\cal O}(s)$) compared to the
infrared cutoff,
are of the collinear type or renormalization group logarithms.

\begin{figure}
\centering
\epsfig{file=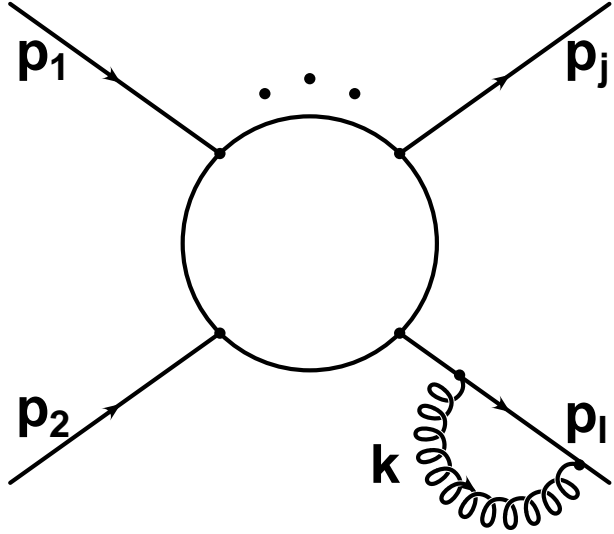,width=8cm}
\caption{In an axial gauge, all collinear logarithms come from corrections to a particular external line
(depending on the choice of the four vector $n^\nu$ satisfying $n^\nu A^a_\nu=0$) as
illustrated in the figure. In a covariant gauge, the sum over all possible
insertions is reduced to a sum over all $n$-external legs due to Ward identities. Overall, these
corrections factorize with respect to the Born amplitude.}
\label{fig:col}
\end{figure}

\begin{figure}
\centering
\epsfig{file=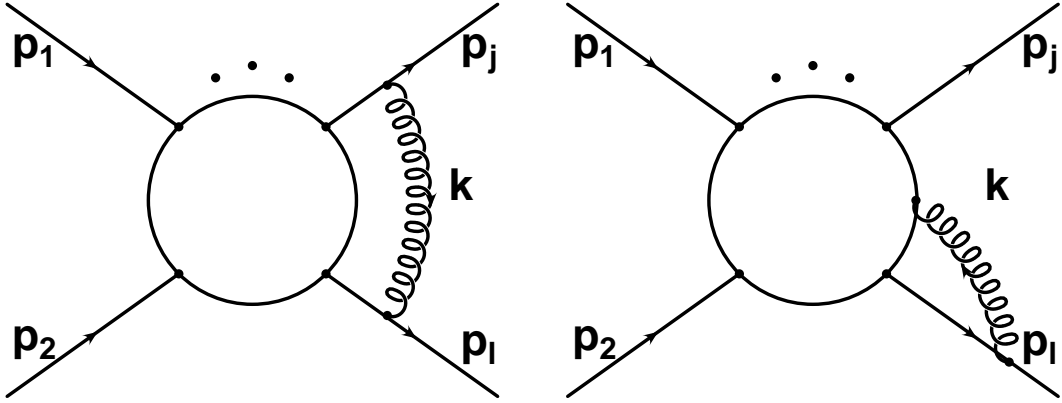,width=14cm}
\caption{Feynman diagrams contributing to the infrared evolution
equation (\ref{eq:mrg}) for a process with $n$ external legs. In a general
covariant gauge the
virtual gluon with the smallest value of ${\mbox{\boldmath $k$}}_{\perp}$ is attached to
different external lines. The inner scattering amplitude is assumed to be
on the mass shell.} \label{fig:nll}
\end{figure}

The universal nature of collinear type logarithmic corrections can then easily be seen 
in an axial gauge where collinear logarithms are related to corrections
on a particular external leg depending on the choice of the four vector $n_\nu$ \cite{col,col2,col3,col4}.
A typical diagram is depicted in Fig. \ref{fig:col}. In a general covariant
gauge this corresponds (using Ward identities)
to a sum over insertions in all $n$ external legs \cite{flmm}.
We can therefore adopt the strategy to extract the gauge invariant contribution from the external
line corrections on the invariant matrix element at the subleading level.
The results of the above discussion are thus
important in that they allow the use of the Altarelli-Parisi
approach to calculate the subleading contribution to the evolution kernel of Eq. (\ref{eq:ee}).
We are here only concerned with virtual corrections and use the universality of the splitting
functions to calculate the subleading terms. 
For brevity we discuss both scalar as well as conventional QCD simultaneously. In each case
one only needs to switch off the other type of field in the fundamental representation to obtain
the case of interest. The $\beta$-function in both cases differs in the non-glue part but
since this difference is of no consequence in our later discussion we don't distinguish between
the two.
For the purpose of calculating SL virtual corrections we use the virtual quark, scalar quark
and gluon contributions to the splitting functions
$P^V_{qq}(z)$, $P_{ss}^V$ and $P^V_{gg}(z)$ 
describing the probability to emit a soft and/or collinear virtual
particle with energy fraction $z$ of the original external line four momentum.
The infinite momentum frame corresponds to the Sudakov parametrization with lightlike vectors.
In general, the splitting functions $P_{BA}$
describe the probability of finding a particle $B$ inside a particle $A$
with fraction $z$
of the longitudinal momentum of $A$ with probability
${\cal P}_{BA}$ to first order \cite{ap}:
\begin{equation}
d {\cal P}_{BA}(z)=\frac{\alpha_s}{2\pi} P_{BA} d t
\end{equation}
where the variable $t=\log \frac{s}{\mu^2}$ for our purposes. It then follows \cite{ap} that
\begin{equation}
d {\cal P}_{BA}(z)=\frac{\alpha_s}{2\pi} \frac{z(1-z)}{2} \overline{\sum_{spins}} \frac{|V_{A
\longrightarrow B+C}|^2}{{\mbox{\boldmath$k$}^2_{\perp}}} d \log {\mbox{\boldmath$k$}^2_{\perp}} \label{eq:dpabres}
\end{equation}
where $V_{A\longrightarrow B+C}$ denotes the elementary vertices and
\begin{equation}
P_{BA}(z)=\frac{z(1-z)}{2} \overline{\sum_{spins}} \frac{|V_{A
\longrightarrow B+C}|^2}{{\mbox{\boldmath$k$}^2_{\perp}}}
\end{equation}
The upper bound on the integral over $d {\mbox{\boldmath$k$}^2_{\perp}}$ in Eq. (\ref{eq:dpabres})
is $s$ and it is thus
directly related to $d t$.
Regulating the virtual infrared divergences with the transverse momentum cutoff as described above,
we find the virtual contributions to the splitting functions for external quark, scalar quarks
and gluon lines:
\begin{eqnarray}
P^V_{qq}(z)&=& C_F \left( - 2 \log \frac{s}{\mu^2} + 3 \right) \delta(1-z) \label{eq:pqqv} \\
P^V_{ss}(z)&=& C_F \left( - 2 \log \frac{s}{\mu^2} + 4 \right) \delta(1-z) \label{eq:pssv} \\
P^V_{gg}(z)&=& C_A \left( - 2 \log \frac{s}{\mu^2} + \frac{4}{C_A} \beta^{\rm QCD}_0 \right) \delta(1-z) \label{eq:pggv}
\end{eqnarray}
The functions can be calculated directly from loop corrections to the elementary
processes \cite{aem,a,dot,m3} and the logarithmic term corresponds to the
leading kernel of section \ref{sec:iree}.
We introduce virtual distribution functions which include only the effects of loop computations.
These fulfill the Altarelli-Parisi equations\footnote{Note that the off diagonal
splitting functions $P_{qg}$ and $P_{gq}$  etc. do not contribute to the virtual probabilities to the order
we are working here. In fact, for virtual corrections there is no need to introduce off-diagonal terms
as the corrections factorize with respect to the Born amplitude. The normalization of the Eqs.
(\ref{eq:pqqv}), (\ref{eq:pssv}) and (\ref{eq:pggv}) corresponds to 
calculations in two to two processes on the cross section
level with the gluon symmetry factor $\frac{1}{2}$ included. The results, properly normalized,
are process independent.}
\begin{eqnarray}
\frac{\partial q(z,t)}{\partial t}&=& \frac{\alpha_s}{2\pi} \int^1_z \frac{dy}{y} q(z/y,t) P^V_{qq}(y)
\label{eq:apqq} \\
\frac{\partial s(z,t)}{\partial t}&=& \frac{\alpha_s}{2\pi} \int^1_z \frac{dy}{y} s(z/y,t) P^V_{ss}(y)
\label{eq:apss} \\
\frac{\partial g(z,t)}{\partial t}&=& \frac{\alpha_s}{2\pi} \int^1_z \frac{dy}{y} g(z/y,t) P^V_{gg}(y)
\label{eq:apgg}
\end{eqnarray}
The splitting functions are related by
$P_{BA}=P^R_{BA}+P^V_{BA}$, where
$R$ denotes the contribution from real gauge boson emission\footnote{$P_{qq}$ was first calculated by
V.N.~Gribov and L.N.~Lipatov in the context of QED \cite{gl,gl2}.}. $P_{BA}$ is free of logarithmic
corrections and positive definite.
The subleading term in Eq. (\ref{eq:pggv})
indicates that the only subleading corrections in the pure
glue sector are related to a shift in the scale of the coupling. These corrections enter with a
different sign compared to the conventional running coupling effects.
For fermion and scalar external lines there is an additional
subleading correction from collinear terms which is not related to a change in the scale of the
coupling.

Inserting the virtual probabilities of Eqs. (\ref{eq:pqqv}), (\ref{eq:pssv}) and (\ref{eq:pggv}) into the Eqs.
(\ref{eq:apqq}), (\ref{eq:apss}) and (\ref{eq:apgg}) we find:
\begin{eqnarray}
q(1,t)&=&q_0 \exp \left[ - \frac{ \alpha_s C_F}{2 \pi} \left( \log^2 \frac{s}{\mu^2}
- 3 \log \frac{s}{\mu^2} \right) \right] \label{eq:qsol} \\
s(1,t)&=&s_0 \exp \left[ - \frac{ \alpha_s C_F}{2 \pi} \left( \log^2 \frac{s}{\mu^2}
- 4 \log \frac{s}{\mu^2} \right) \right] \label{eq:ssol} \\
g(1,t)&=&g_0 \exp \left[ - \frac{ \alpha_s C_A}{2 \pi} \left( \log^2 \frac{s}{\mu^2}
- \frac{4}{C_A} \beta^{\rm QCD}_0 \log \frac{s}{\mu^2} \right) \right] \label{eq:gsol}
\end{eqnarray}
where $\beta^{\rm QCD}_0=\frac{11}{12}C_A-\frac{1}{3}T_Fn_f$ with $C_A=3$, $C_F=4/3$ and $T_F=\frac{1}{2}$.
These functions describe the total contribution for the emission of virtual particles (i.e. $z=1$),
with all invariants large compared to the cutoff $\mu$, to the
densities $q(z,t)$, $s(z,t)$ and $g(z,t)$. The normalization is on the level of the cross section.
For the invariant matrix element we thus find at the subleading level for processes with $n$
external lines:
\begin{eqnarray}
&& {\cal M} (p_1,...,p_n,g_s,\mu)={\cal M} (p_1,...,p_n,g_s) \times \nonumber \\
&& \exp \left(- \frac{1}{2}
\sum^{n_q}_{j=1} W_{q_j}(s,\mu^2)-\frac{1}{2} \sum^{n_s}_{i=1} W_{s_i}(s,\mu^2) 
-\frac{1}{2} \sum^{n_g}_{l=1} W_{g_l}(s,\mu^2) \right)
\label{eq:msol}
\end{eqnarray}
with $n_q+n_g+n_s=n$, and
\begin{eqnarray}
W_q(s,\mu^2)&=&\frac{ \alpha_s C_F}{4 \pi} \left( \log^2 \frac{s}{\mu^2}
- 3 \log \frac{s}{\mu^2} \right) \label{eq:wq} \\
W_s(s,\mu^2)&=&\frac{ \alpha_s C_F}{4 \pi} \left( \log^2 \frac{s}{\mu^2}
- 4 \log \frac{s}{\mu^2} \right) \label{eq:ws} \\
W_g(s,\mu^2)&=&\frac{ \alpha_s C_A}{4 \pi} \left( \log^2 \frac{s}{\mu^2} - \frac{4}{C_A}
\beta^{\rm QCD}_0 \log \frac{s}{\mu^2} \right)
\label{eq:wg}
\end{eqnarray}
The functions $W_q$, $W_s$ and $W_g$ correspond to the probability of emitting a virtual
soft and/or collinear gauge
boson from the particle $q$, $g$ subject to the infrared cutoff $\mu$. Typical diagrams contributing
to Eq. (\ref{eq:msol}) in a covariant gauge are depicted in Fig. \ref{fig:nll}.
In massless QCD there is no need for the label $W_{q_j}$, $W_{s_i}$ or $W_{g_l}$, 
however, we write it for later convenience.
The universality of the splitting functions is crucial in obtaining the above result.                          

\subsection{Anomalous scaling violations} \label{sec:asv}

\begin{figure}
\centering
\epsfig{file=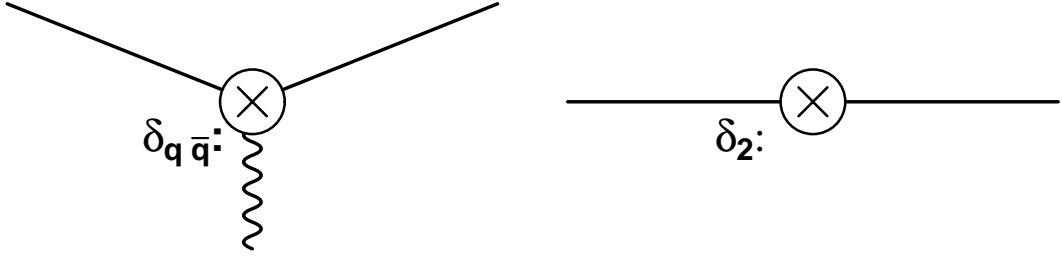,width=14cm}
\caption{The two counterterms contributing to the quark anomalous dimension $\gamma_q
= \frac{\partial}{\partial \log \overline{\mu}^2} \left( -\delta_{\overline{q}q}+\delta_2 \right)$
or $\gamma_s = \frac{\partial}{\partial \log \overline{\mu}^2} \left( -\delta_{s^*s}+\delta_2 \right)$.
Here $\overline{\mu}$ denotes the $\overline{\rm MS}$ dimensional regularization mass parameter.
Due to divergences in loop corrections there are scaling violations also in the massless theory.}
\label{fig:gqq}
\end{figure}

The solution presented in Eq. (\ref{eq:msol}) determines the evolution of the virtual scattering
amplitude ${\cal M} (p_1,...,p_n,g_s,\mu)$ for large energies at fixed angles and subject to
the infrared regulator $\mu$. In the massless case there is a one to one correspondence between the
high energy limit and the infrared limit as only the ratio $s/\mu^2$ enters as a dimensionless
variable \cite{pqz,p}.
Thus, we can generalize the Altarelli-Parisi equations (\ref{eq:apqq}), (\ref{eq:apss})
 and (\ref{eq:apgg}) to the invariant
matrix element in the language of the renormalization group. For this purpose, we define the
infrared singular (logarithmic) anomalous dimensions
\begin{equation}
\Gamma_q (t) \equiv \frac{C_F \alpha_s}{4 \pi} t \;\;;\;\; \Gamma_s (t) \equiv \frac{C_F \alpha_s}{4 \pi} t
\;\;;\;\; \Gamma_g (t) \equiv \frac{C_A \alpha_s}{4 \pi} t
\label{eq:irad}
\end{equation}
Infrared divergent anomalous dimensions have been derived in the context of renormalization properties
of gauge invariant Wilson loop functionals \cite{kr}. In this context they are related to undifferentiable
cusps of the path integration and the cusp angle $~p_jp_l/\mu^2$ gives rise to the logarithmic nature
of the anomalous dimension. In case we use off-shell amplitudes, one also has contributions from end
points of the integration \cite{kr}.
The leading terms in the equation below have also been discussed in Refs. \cite{ct}, \cite{ar} and
\cite{ku} in the context of QCD.
With these notations we
find that Eq. (\ref{eq:msol}) satisfies
\begin{eqnarray}
&& \left( \frac{\partial}{\partial t} + \beta^{\rm QCD} \frac{\partial}{\partial g_s} + n_g \left(
\Gamma_g(t)+ \frac{1}{2} \gamma_g \right) + n_q
\left( \Gamma_q(t) + \frac{1}{2} \gamma_q \right) \right. \nonumber \\
&& \left. + n_s \left( \Gamma_s(t) + \frac{1}{2} \gamma_s \right) \right)
  {\cal M} (p_1,...,p_n,g_s,\mu) =0 \label{eq:mrg}
\end{eqnarray}
to the order we are working here and
where ${\cal M}(p_1,...,p_n,g_s,\mu)$ is taken on the mass-shell.
The difference in the sign of the derivative term compared to Eq. (\ref{eq:ee}) is due to the fact
that instead of differentiating with respect to $\log \mu^2$ we use $\log s/\mu^2$.
The anomalous dimensions are given by $\gamma_g=-\frac{\alpha_s}{\pi} \beta^{\rm QCD}_0
=-\frac{\alpha_s}{\pi} \left( \frac{11}{12} C_A- \frac{n_f}{3} T_F \right)$, 
$\gamma_s=-C_F\frac{\alpha_s}{\pi} $ and $\gamma_{q}=-C_F \frac{3}{4} \frac{\alpha}{\pi}$. 
As mentioned above in pure scalar QCD the $\beta$-function differs in the non-glue part
from $\beta^{\rm QCD}$.
The quark-antiquark operator anomalous dimension $\gamma_{q}$ or in scalar QCD $\gamma_s$
enter even for massless theories as the quark antiquark
operator leads to
scaling violations through loop effects since the quark masslessness is not protected by gauge
invariance and a dimensionful infrared cutoff needs to be introduced. Thus, although the Lagrangian
contains no $m \overline{\psi}\psi$ or $m^2 \phi^* \phi$ term, 
quantum corrections lead to the anomalous scaling violations
in the form of $\gamma_q$ or $\gamma_s$. The factor $\frac{1}{2}$ occurs since we write Eq. (\ref{eq:mrg})
in terms of each external line
separately\footnote{In case of a massive theory, we could, for instance avoid the
anomalous dimension term $\gamma_q$ by adopting the pole mass definition. In this case, however, we
would obtain terms in the wave function renormalization, and in any case, the one to one correspondence
between UV and IR scaling, crucial for the validity of Eq. (\ref{eq:mrg}), is violated.}.
For the gluon, the scaling violations due to the infrared cutoff are manifest in
terms of an anomalous dimension proportional to the $\beta$-function since the
gluon mass is protected by gauge invariance from loop corrections.
Thus, in the bosonic sector
the subleading terms correspond effectively to a scale change of the
coupling.
Fig. \ref{fig:gqq} illustrates the corrections to the external quark-antiquark lines from loop effects.

Except for the infrared singular anomalous dimension (Eq. (\ref{eq:irad})), all other terms
in Eq. (\ref{eq:mrg}) are the standard contributions to the renormalization group equation for
S-matrix elements \cite{sterm}. In QCD, observables with infrared singular anomalous dimensions, regulated
with a fictitious gluon mass, are ill
defined due to the masslessness of gluons.
In the electroweak theory, however, we can legitimately
investigate only virtual corrections since the gauge bosons will require a mass.
Eq. (\ref{eq:mrg}) will thus be very useful in section \ref{sec:bgt}.

\subsection{Renormalization group corrections} \label{sec:rg}

\begin{figure}
\centering
\epsfig{file=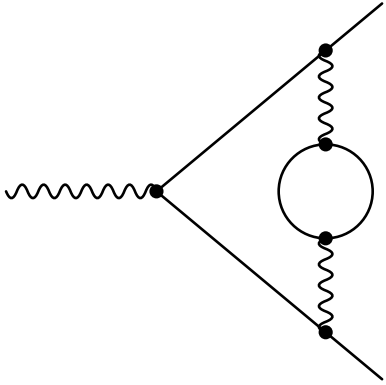,width=5cm}
\caption{A QED diagram at the two loop level yielding a SL-RG correction. The explicit result
obtained in Ref. \cite{bur} for the case
of equal masses relative to the Born amplitude
was $-\frac{1}{36} \frac{e^4}{16 \pi^4} \log^3 \frac{s}{m^2}
=\frac{1}{12} \beta_0^{\rm QED} \frac{e^4}{16 \pi^4} \log^3 \frac{s}{m^2}$. This result
is reproduced exactly by including a running coupling into the one loop vertex
correction diagram. The argument of the coupling must depend on the
component of the loop momentum (going into the fermion loop) which is perpendicular to the
external fermion momenta. In QCD, although more diagrams contribute, the net effect is just
to replace $\beta_0^{\rm QED} \longrightarrow \beta_0^{\rm QCD}$ in the above expression.}
\label{fig:SLRG}
\end{figure}

In this section we review the case of higher order RG-corrections in 
unbroken gauge theories like QCD following Ref. \cite{m4}. Explicit comparisons with
higher order calculations for the on-shell Sudakov form factor 
revealed that the relevant RG scale in the respective diagrams is
indeed the perpendicular Sudakov component \cite{b,ddt,ddt2,ddt3,cl}.
It should be noted, however, that in particular for the massive cases, the discussion below is
not valid close to any of the thresholds. In such cases it is useful to consider ``physical
renormalization schemes'' such as discussed in Ref. \c,mV2,mV3ite{mV,mV2,mV3}, which display a gauge
invariant, continuous
and smooth flavor threshold behavior with automatic decoupling of heavy particles.
For our purposes here,
we give correction factors for each external line below. The universal nature of the higher
order SL-RG corrections can be seen as follows. Consider the gauge invariant fermionic part
($\sim n_f$) as indicative of the full $\beta^{\rm QCD}_0$ term (replacing $n_f = \frac{3}{T_F} \left(
\frac{11}{12} C_A - \beta^{\rm QCD}_0 \right) $). In order to lead to subleading, i.e. ${\cal O} \left(
\alpha^n_s \log^{2n-1} \frac{s}{\mu^2} \right)$, this loop correction must be folded
with the exchange of a gauge boson between two external lines (producing a DL type contribution)
like the one depicted in Fig. \ref{fig:SLRG}.
Using the conservation of the total non-Abelian group charge, i.e. Eq. (\ref{eq:cnag}),
the double sum over all external insertions
$j$ and $l$ is reduced to a single sum over all $n$ external legs.
Thus these types of
corrections can be identified with external lines at higher orders. The same conclusion
is reproduced by the explicit pole structure of $\overline{\rm MS}$ renormalized scattering
amplitudes at the two loop level in QCD \cite{cat}.
The results presented in Ref. \cite{cat} have been confirmed recently by explicit massless two loop 
QCD calculations \cite{dix,dix2,glov,glov2,glov3,glov4}.
In addition, from the expression in Ref. \cite{cat} it can be seen that the SL-RG corrections are
independent of the spin, i.e. for both quarks and gluons the same running coupling argument is
to be used. This is a consequence of the fact that these corrections appear only in loops
which can yield DL corrections on the lower order level and as such, the available DL phase
space is identical up to group theory factors.
We begin with the virtual case.

\subsubsection{Virtual corrections}

The case of virtual SL-RG corrections for both massless and massive partons has been discussed in
Ref. \cite{ms3} with a different Sudakov parametrization. Below we show the identity of both
approaches.
The form of the corrections is given in terms of the probabilities $W_{i_V} \left( s,\mu^2
\right)$. To logarithmic accuracy, they correspond to the probability to emit a soft and/or
collinear virtual parton from particle $i$ at high energies subject to an infrared
cutoff $\mu$. At the amplitude level all expressions below are
universal for each external line and exponentiate according to
\begin{equation}
{\cal M} (p_1,...,p_n,g_s, \mu) = {\cal M}_{\rm Born} (p_1,...,p_n,g_s) \exp \left(- \frac{1}{2}
\sum_{i=1}^n W_{i_V} \left( s,\mu^2 \right) \right)
\end{equation}
where $n$ denotes the number of external lines. We begin with the massless case.

\subsubsection*{Massless QCD}

In the following we denote the running QCD-coupling by
\begin{equation}
\alpha_s ({\mbox{\boldmath $k$}_{\perp }^{2}})
 = \frac{\alpha_s(\mu^2)}{1+ \frac{\alpha_s(\mu^2)
 }{\pi} \beta^{\rm QCD}_0 \log \frac{{\mbox{\boldmath $k$}_{\perp }^{2}}}{\mu^2}}
 \equiv \frac{\alpha_s(\mu^2)}{1+c \; \log \frac{{\mbox{\boldmath $k$}_{\perp }^{2}}}{\mu^2}}
 \label{eq:rc}
 \end{equation}
 Up to two loops the massless $\beta$-function is independent of the chosen
 renormalization scheme and is gauge invariant in minimally subtracted schemes
 to all orders \cite{c}. These features will also hold for the derived
 renormalization group correction factors below in the high energy regime.
 The scale $\mu$ denotes the infrared cutoff on the exchanged ${\mbox{\boldmath $k$}_{\perp }}$
 between the external momenta $p_j,p_l$, where the Sudakov decomposition is given by
 $k=vp_l+up_j + k_\perp$, such that $p_jk_\perp =p_lk_\perp =0$. The cutoff $\mu$ serves
 as a a lower limit on the exchanged Euclidean component $\mbox{\boldmath $k$}^2_\perp=
 -k^2_\perp > 0$ as in the previous sections.
 In order to avoid the Landau pole we must choose $\mu > \Lambda_{\rm QCD}$.
 Thus, the expressions given in this section correspond for quarks to the case where $m \ll \mu$.
 For arbitrary external lines we then have
\begin{equation}
{\widetilde W}^{\rm DL}_{i_V} \left(s,\mu^2 \right) =  \frac{\alpha_s C_i}{2\pi}
\int^s_{\mu^2} \frac{d {\mbox{\boldmath $k$}_{\perp }^{2}}}{
{\mbox{\boldmath $k$}_{\perp }^{2}}} \int^1_{{\mbox{\boldmath $k$}_{\perp }^{2}}/s}
\frac{d v}{v} =  \frac{\alpha_s C_i}{4\pi} \log^2 \frac{s}{\mu^2}
 \end{equation}
The RG correction is then described by including the effect of the running coupling from the
scale $\mu^2$ to $s$ according to \cite{b,ddt,ddt2,ddt3,cl} (see also discussions in
Refs. \cite{ms3,mdur}):
\begin{eqnarray}
{\widetilde W}^{\rm RG}_{i_V}\left(s,\mu^2 \right) &=&  \frac{C_i}{2\pi}
\int^s_{\mu^2} \frac{d {\mbox{\boldmath $k$}_{\perp }^{2}}}{
{\mbox{\boldmath $k$}_{\perp }^{2}}} \int^1_{{\mbox{\boldmath $k$}_{\perp }^{2}}/s}
\frac{d v}{v}
 \frac{\alpha_s(\mu^2)}{1+c \; \log \frac{{\mbox{\boldmath $k$}_{\perp }^{2}}}{\mu^2}} \nonumber \\
&=& \frac{\alpha_s(\mu^2) C_i}{2 \pi } \left\{ \frac{1}{c} \log \frac{s}{\mu^2}
\left( \log \frac{\alpha_s(\mu^2)}{\alpha_s
(s)} - 1 \right) + \frac{1}{c^2}
\log \frac{\alpha_s(\mu^2)}{\alpha_s(s)} \right\} \label{eq:vrg}
 \end{eqnarray}
where $C_i=C_A$ for gluons and $C_i=C_F$ for quarks.
For completeness we also give the
subleading terms of the external line correction which is of course also
important for phenomenological applications. The terms depend on the external line
and the complete result
to logarithmic accuracy is given by:
\begin{eqnarray}
W^{RG}_{g_V}\left(s,\mu^2 \right)
&=& \frac{\alpha_s(\mu^2) C_A}{2 \pi } \left\{ \frac{1}{c} \log \frac{s}{\mu^2}
\left( \log \frac{\alpha_s(\mu^2)}{\alpha_s
(s)} - 1 \right) + \frac{1}{c^2}
\log \frac{\alpha_s(\mu^2)}{\alpha_s(s)} \right. \nonumber \\
&& \left.
 -\frac{2}{C_A} \beta^{\rm QCD}_0 \log \frac{s}{\mu^2}
\right\} \label{eq:gvrgex} \\
W^{RG}_{q_V}\left(s,\mu^2 \right)
&=& \frac{\alpha_s(\mu^2) C_F}{2 \pi } \left\{ \frac{1}{c} \log \frac{s}{\mu^2}
\left( \log \frac{\alpha_s(\mu^2)}{\alpha_s
(s)} - 1 \right) + \frac{1}{c^2}
\log \frac{\alpha_s(\mu^2)}{\alpha_s(s)} \right. \nonumber \\
&& \left.
 -\frac{3}{2} \log \frac{s}{\mu^2}
\right\} \label{eq:qvrgex}
 \end{eqnarray}
It should be noted that the subleading term in Eq. (\ref{eq:gvrgex}) proportional to $\beta^{\rm
QCD}_0$
is not a conventional renormalization group corrections but rather an anomalous scaling dimension,
and enters with the opposite sign \cite{m1} compared to the conventional RG contribution
(see section \ref{sec:asv}).

\subsubsection*{Massive QCD}

Here we give results for the case when the infrared cutoff $\mu \ll m$, where $m$ denotes
the external quark mass. We begin with the case of equal external and internal line masses:

\subsubsection*{Equal masses}

Following Ref. \cite{ms3}, we use the gluon on-shell condition
$suv={\mbox{\boldmath $k$}_{\perp }^{2}}$ to calculate the integrals.
We begin with the correction factor for each external massive quark line. Following the
diagram in Fig. \ref{fig:suv} we find:
\begin{eqnarray}
{\widetilde W}^{RG}_{q_V} \left(s,\mu^2 \right) &=&  \frac{C_F}{2 \pi} \int^1_0
\frac{d u}{u} \int^1_0 \frac{d v}
{v} \theta ( s uv - \mu^2) \theta ( u
- \frac{m^2}{s} v) \theta (v- \frac{m^2}{s} u
) \nonumber \\
&& \times \frac{\alpha_s(m^2)}{1+c \; \log \frac{s uv}{m^2}}
\nonumber \\ &=&
\frac{C_F}{2 \pi} \left\{ \int^\frac{\mu}{m}_\frac{\mu^2}{s}
\frac{d u}{u} \int^1_\frac{\mu^2}{
s u} \frac{d v}{v} +
\int^1_\frac{\mu}{m} \frac{d u}{u} \int^1_
{\frac{m^2}{s}u} \frac{d v}{v} \right. \nonumber
\\ && \left.  -\int^\frac{\mu m}{s}_\frac{\mu^2}{s} \frac{d
u}{u}
\int^1_\frac{\mu^2}{su} \frac{d v}{v}
- \int^\frac{m^2}{s}_\frac{\mu m}{s} \frac{d u}{u}
\int^1_{\frac{s}{m^2}u} \frac{d v}{v} \right\}
\frac{\alpha_s(m^2)}{1+c \;\log \frac{s uv}{m^2}} \nonumber \\
&=& \frac{\alpha_s(m^2) C_F}{2 \pi } \left\{ \frac{1}{c} \log \frac{s}{m^2}
\left( \log \frac{\alpha_s(\mu^2)}{\alpha_s
(s)} - 1 \right) + \frac{1}{c^2}
\log \frac{\alpha_s(m^2)}{\alpha_s(s)} \right\}
\label{eq:SvRG}
\end{eqnarray}
The $\mu$-dependent terms cancel out of any physical cross section (as they must) when
real soft Bremsstrahlung contributions are added
and $c=\alpha_s(m^2) \beta^{\rm QCD}_0 / \pi$ for massive quarks.
In order to demonstrate that the
result in Eq. (\ref{eq:SvRG}) exponentiates, we calculated in Ref. \cite{ms3} the explicit two loop
renormalization group improved massive virtual Sudakov corrections,
containing a different ``running scale'' in each loop.
It is of course also possible to use the scale ${\mbox{\boldmath $k$}_{\perp }^{2}}$ directly.
In this case we have according to the diagram in Fig. \ref{fig:skp}:
\begin{eqnarray}
{\widetilde W}^{RG}_{q_V}  \left(s,\mu^2 \right) &=&  \frac{C_F}{2\pi} \left[ \int^{m^2}_{\mu^2} \frac{d {\mbox{\boldmath $k$}_{\perp }^{2}}}{
{\mbox{\boldmath $k$}_{\perp }^{2}}} \int^{{\mbox{\boldmath $k$}_{\perp }^{2}}/m^2}_{
{\mbox{\boldmath $k$}_{\perp }^{2}}/s} \frac{d v}{v} +
\int^s_{m^2} \frac{d {\mbox{\boldmath $k$}_{\perp }^{2}}}{
{\mbox{\boldmath $k$}_{\perp }^{2}}} \int^1_{{\mbox{\boldmath $k$}_{\perp }^{2}}/s}
\frac{d v}{v} \right]
 \frac{\alpha_s(m^2)}{1+c \; \log \frac{{\mbox{\boldmath $k$}_{\perp }^{2}}}{m^2}} \nonumber \\
&=& \frac{\alpha_s(m^2) C_F}{2 \pi } \left\{ \frac{1}{c} \log \frac{s}{m^2}
\left( \log \frac{\alpha_s(\mu^2)}{\alpha_s
(s)} - 1 \right) + \frac{1}{c^2}
\log \frac{\alpha_s(m^2)}{\alpha_s(s)} \right\}
 \end{eqnarray}
which is the identical result as in Eq. (\ref{eq:SvRG}).
For completeness we also give the
subleading terms of the pure one loop form factor which is again
important for phenomenological applications. The complete result
to logarithmic accuracy is thus given by:
\begin{eqnarray}
W^{RG}_{q_V} \left(s,\mu^2 \right)
&=& \frac{\alpha_s(m^2) C_F}{2 \pi } \left\{ \frac{1}{c} \log \frac{s}{m^2}
\left( \log \frac{\alpha_s(\mu^2)}{\alpha_s
(s)} - 1 \right) + \frac{1}{c^2}
\log \frac{\alpha_s(m^2)}{\alpha_s(s)} \right. \nonumber \\
&& \left.
 -\frac{3}{2} \log \frac{s}{m^2} - \log \frac{ m^2}{
 \mu^2} \right\} \label{eq:mqvrgex}
 \end{eqnarray}
For $m=\mu$ Eq. (\ref{eq:mqvrgex}) agrees with
Eq. (\ref{eq:qvrgex}) in the previous section for massless quarks.

\subsubsection*{Unequal masses}

In this section we denote the external mass as before by $m$ and the internal mass
by $m_i$ and thus, the constant $c=\alpha_s(m^2_i) \beta_0^{\rm QCD} /\pi$.
We consider only the case at high energies taking the first two families
of quarks as massless. The
running of all light flavors is implicit in the $n_f$ term of the
$\beta_0^{\rm QCD}$ function. The result is then given by:
\begin{eqnarray}
{\widetilde W}^{RG}_{q_V}  \left(s,\mu^2 \right) &=&  \frac{C_F}{2\pi} \left[ \int^{m^2}_{\mu^2} \frac{d {\mbox{\boldmath $k$}_{\perp }^{2}}}{
{\mbox{\boldmath $k$}_{\perp }^{2}}} \int^{{\mbox{\boldmath $k$}_{\perp }^{2}}/m^2}_{
{\mbox{\boldmath $k$}_{\perp }^{2}}/s} \frac{d v}{v} +
\int^s_{m^2} \frac{d {\mbox{\boldmath $k$}_{\perp }^{2}}}{
{\mbox{\boldmath $k$}_{\perp }^{2}}} \int^1_{{\mbox{\boldmath $k$}_{\perp }^{2}}/s}
\frac{d v}{v} \right]
 \frac{\alpha_s(m_i^2)}{1+c \; \log \frac{{\mbox{\boldmath $k$}_{\perp }^{2}}}{m_i^2}} \nonumber \\
&=& \frac{\alpha_s(m_i^2) C_F}{2 \pi } \left\{ \frac{1}{c} \log \frac{s}{m^2}
\left( \log \frac{\alpha_s(\mu^2)}{\alpha_s
(s)} - 1 \right) \right. \nonumber \\ && \left. + \frac{1}{c}
\log \frac{\alpha_s(m^2)}{\alpha_s(s)} \left( \frac{1}{c}+  \log \frac{m^2}{m_i^2}
\right) \right\}
 \end{eqnarray}
It is evident that the effect of unequal masses is large only for a large mass splitting.
In QCD, we always assume scales larger than $\Lambda_{\rm QCD}$ and with our assumptions we
have only the
ratio of $m_t/m_b$ leading to significant corrections.

The full subleading expression is accordingly given by:
\begin{eqnarray}
W^{RG}_{q_V} \left(s,\mu^2 \right)
&=& \frac{\alpha_s(m_i^2) C_F}{2 \pi } \left\{ \frac{1}{c} \log \frac{s}{m^2}
\left( \log \frac{\alpha_s(\mu^2)}{\alpha_s
(s)} - 1 \right)  \right. \nonumber \\ && \left. + \frac{1}{c}
\log \frac{\alpha_s(m^2)}{\alpha_s(s)} \left( \frac{1}{c}+  \log \frac{m^2}{m_i^2}
\right) -\frac{3}{2} \log \frac{s}{m^2} - \log \frac{ m^2}{
 \mu^2} \right\} \nonumber \\
&=& \frac{\alpha_s(m_i^2) C_F}{2 \pi } \left\{ \frac{1}{c} \log \frac{s}{m^2}
\left( \log \frac{\alpha_s(\mu^2)}{\alpha_s
(s)} - 1 \right)  \right. \nonumber \\ && \left. + \frac{1}{c^2} \frac{\alpha_s(m_i^2)}{
\alpha_s(m^2)}
\log \frac{\alpha_s(m^2)}{\alpha_s(s)}
 -\frac{3}{2} \log \frac{s}{m^2} - \log \frac{ m^2}{
 \mu^2} \right\} \label{eq:miqvr}
 \end{eqnarray}
For $m=m_i$ Eq. (\ref{eq:miqvr}) agrees with
Eq. (\ref{eq:mqvrgex}) in the previous section for equal mass quarks.

If we want to apply the above result for the case of QED corrections later, then there
is no Landau pole (at low energies)
and we can have large corrections of the form $m_b/m_e$ etc.
In this case the running coupling term is given by
\begin{equation}
e^2({\mbox{\boldmath $k$}_{\perp }^{2}})= \frac{e^2}{1-
\frac{1}{3} \frac{e^2}{4 \pi^2} \sum_{j=1}^{n_f
} Q_j^2 N^j_C \log
\frac{{\mbox{\boldmath $k$}_{\perp }^{2}}}{m_j^2}} \label{eq:eeff}
\end{equation}
and instead of Eq. (\ref{eq:miqvr}) we have:
\begin{eqnarray}
&& W^{RG}_{f_V} \left(s,\mu^2 \right)
= \frac{e_f^2}{8 \pi^2 } \left\{ \frac{1}{c} \log \frac{s}{m^2}
\left( \log \frac{e^2(\mu^2)}{e^2
(s)} - 1 \right) + \right. \nonumber \\ && \left.  \frac{1}{c^2}
\log \frac{e^2(m^2)}{e^2(s)} \left( 1- \frac{1}{3}\frac{e^2}{4 \pi^2}
\sum_{j=1}^{n_f } Q_j^2 N^j_C \log \frac{m^2}{m_j^2}
\right) -\frac{3}{2} \log \frac{s}{m^2} - \log \frac{ m^2}{
 \mu^2} \right\} \label{eq:mifvr}
 \end{eqnarray}
and where $c=- \frac{1}{3}\frac{e^2}{4 \pi^2} \sum_{j=1}^{n_f } Q_j^2 N^j_C$.

\subsubsection{Real gluon emission}

We discuss the massless and massive case separately since the structure of the divergences
is different in each case. For massive quarks we discuss two types of restrictions on the
experimental requirements, one in analogy to the soft gluon approximation.
The expressions below exponentiate on the level of the cross section, i.e.
for observable scattering cross sections they are of the form
\begin{eqnarray}
d \sigma (p_1,...,p_n,g_s,\mu_{\rm expt}) &=& d \sigma_{\rm Born}
(p_1,...,p_n,g_s) \times \nonumber \\ &&
\exp \left\{ \sum_{i=1}^n \left[ W_{i,R} \left( s,\mu^2,\mu^2_{\rm expt}
\right)-W_{i,V} \left( s,\mu^2 \right) \right] \right\}
\end{eqnarray}
where the sum in the exponential is independent of $\mu$ and only depends on the cutoff
$\mu_{\rm expt}$ defining the experimental cross section.
We begin with the
massless case.

\subsubsection*{Emission from massless partons}

In this section we consider the emission of real gluons with a cutoff
${\mbox{\boldmath $k$}_{\perp }} \leq \mu_{\rm expt}$, related to the
experimental requirements. For massless partons we have at the DL level:
\begin{equation}
{\widetilde W}^{DL}_{i_R} \left(s,\mu^2,\mu^2_{\rm expt} \right) =  \frac{\alpha_sC_i}{\pi}
\int^{\mu^2_{\rm expt}}_{\mu^2} \frac{d {\mbox{\boldmath $k$}_{\perp }^{2}}}{
{\mbox{\boldmath $k$}_{\perp }^{2}}} \int^{\sqrt s}_{|{\mbox{\boldmath $k$}_{\perp }}|}
\frac{d \omega}{\omega}
= \frac{\alpha_sC_i}{4\pi}  \left\{ \log^2 \frac{s}{\mu^2} - \log^2 \frac{s}{\mu^2_{\rm expt}}
\right\}
 \end{equation}
and thus for the RG-improved correction:
\begin{eqnarray}
&& {\widetilde W}^{RG}_{i_R}\left(s,\mu^2,\mu^2_{\rm expt} \right) =  \frac{C_i}{\pi}
\int^{\mu^2_{\rm expt}}_{\mu^2} \frac{d {\mbox{\boldmath $k$}_{\perp }^{2}}}{
{\mbox{\boldmath $k$}_{\perp }^{2}}} \int^{\sqrt s}_{|{\mbox{\boldmath $k$}_{\perp }}|}
\frac{d \omega}{\omega}
 \frac{\alpha_s(\mu^2)}{1+c \; \log \frac{{\mbox{\boldmath $k$}_{\perp }^{2}}}{\mu^2}} \nonumber \\
&&= \frac{C_i\alpha_s(\mu^2)}{2\pi} \!\! \left\{ \frac{1}{c} \log \frac{s}{\mu^2} \left( \log \frac{
\alpha_s (\mu^2)}{\alpha_s (\mu^2_{\rm expt})} - 1 \! \right) - \frac{1}{c} \log \frac{\mu^2_{\rm expt}}{
s}
+ \frac{1}{c^2} \log \frac{\alpha_s (\mu^2)}{\alpha_s (\mu^2_{\rm expt})} \right\}
 \end{eqnarray}
This expression depends on $\mu$ as it must in order to cancel the infrared divergent virtual
corrections. In fact the sum of real
plus virtual corrections on the level of the cross section is given by
\begin{eqnarray}
&& W^{RG}_{i_R} \left(s,\mu^2,\mu^2_{\rm expt} \right)-W^{RG}_{i_V}\left(s,\mu^2 \right) =
\nonumber \\ &&
\frac{C_i}{2 \beta^{\rm QCD}_0} \!\! \left\{ \log \frac{s}{\mu^2} \log \frac{
\alpha_s (s)}{\alpha_s (\mu^2_{\rm expt})}  -  \log \frac{\mu^2_{\rm expt
}}{s}
+ \frac{1}{c} \log \frac{\alpha_s (s)}{\alpha_s (\mu^2_{\rm expt})} \right\} \nonumber \\
&=& \frac{C_i}{2 \beta^{\rm QCD}} \left( \frac{\pi}{\alpha(s) \beta^{\rm QCD}_0}
\log \frac{ \alpha_s (s)}{\alpha_s
(\mu^2_{\rm expt})} -  \log \frac{\mu^2_{\rm expt}}{s} \right)
 \end{eqnarray}
and thus independent of $\mu$.
The full expressions to subleading accuracy are thus:
\begin{eqnarray}
W^{RG}_{g_R}\left(s,\mu^2,\mu^2_{\rm expt} \right)
&=& \frac{C_A\alpha_s(\mu^2)}{2\pi} \!\! \left\{ \frac{1}{c} \log \frac{s}{\mu^2} \left( \log \frac{
\alpha_s (\mu^2)}{\alpha_s (\mu^2_{\rm expt})} - 1 \! \right) - \frac{1}{c} \log
\frac{\mu^2_{\rm expt}}{s} \right. \nonumber \\ && \left.
+ \frac{1}{c^2} \log \frac{\alpha_s (\mu^2)}{\alpha_s (\mu^2_{\rm expt})}
 -\frac{2}{C_A} \beta^{\rm QCD}_0 \log \frac{s}{\mu^2}
\right\} \label{eq:grrgex} \\
W^{RG}_{q_R} \left(s,\mu^2,\mu^2_{\rm expt} \right)
&=& \frac{C_F\alpha_s(\mu^2)}{2\pi} \!\! \left\{ \frac{1}{c} \log \frac{s}{\mu^2} \left( \log \frac{
\alpha_s (\mu^2)}{\alpha_s (\mu^2_{\rm expt})} - 1 \! \right) - \frac{1}{c}
\log \frac{\mu^2_{\rm expt}}{s} \right. \nonumber \\ && \left.
+ \frac{1}{c^2} \log \frac{\alpha_s (\mu^2)}{\alpha_s (\mu^2_{\rm expt})}
 -\frac{3}{2} \log \frac{s}{\mu^2}
\right\} \label{eq:qrrgex}
 \end{eqnarray}
All divergent ($\mu$-dependent) terms cancel when the full virtual corrections are added.

\subsubsection*{Emission from massive quarks}

In the case of a massive quark, i.e. $\mu \ll m$,
the overall infrared divergence is not as severe.
This means we can discuss different requirements which all have the correct divergent
pole structure canceling the corresponding terms from the virtual contributions.
We divide the discussion in two parts as above.

\subsubsection*{Equal masses}

The constant $c=\alpha_s(m^2) \beta^{\rm QCD}_0 / \pi$ below.
We have the following expression without a running coupling:
\begin{eqnarray}
\!\!\!\!\!\!\!\! && \!\! W_{q_R}\left(s,\mu^2,\mu^2_{\rm expt} \right) =  \frac{\alpha_s C_F}{\pi}
\int^{\mu^2_{\rm expt}}_{\mu^2} d {\mbox{\boldmath $k$}_{\perp }^{2}}
\int^{\sqrt s}_{|{{\mbox{\boldmath $k$}_{\perp }}}|}
\frac{d \omega}{\omega} \frac{{\mbox{\boldmath $k$}_{\perp }^{2}}}{\left(
{\mbox{\boldmath $k$}_{\perp }^{2}}+ m^2/s \; \omega^2 \right)^2} \nonumber \\
\!\!\!\!\!\!\!\! &\approx& \!\! \left\{ \! \begin{array}{lc} \frac{\alpha_s C_F}{2 \pi} \!\!\left(
\! \frac{1}{2} \log^2 \frac{s}{m^2}
+ \log \frac{s}{m^2} \log \frac{m^2}{\mu^2}
- \log \frac{m^2}{\mu^2} - \frac{1}{2} \log^2 \frac{s}{\mu_{\rm expt}^2} \right) &\!\! , \, m \ll \mu_{\rm expt} \\
\frac{\alpha_s C_F}{2 \pi} \!\!
\left( \! \log^2 \frac{s}{m^2}
+ \log \frac{s}{m^2} \log \frac{m^2}{\mu^2}
+ \log \frac{\mu^2}{\mu_{\rm expt}^2} - \log \frac{s}{m^2} \log
\frac{s}{\mu_{\rm expt}^2} \right)
&\!\! , \, \mu_{\rm expt} \ll m
\end{array} \right.
 \end{eqnarray}
If we want to employ a restriction analogously to the soft gluon approximation, we find
independently of the quark mass \cite{m1,m2}:
\begin{eqnarray}
\!\!\!\!\!\!\!\!\!\!&& W_{q_R}\left(s,\mu^2,\mu^2_{\rm expt} \right) =  \frac{\alpha_s C_F}{\pi}
\int^{\mu^2_{\rm expt}}_{\mu^2} d {\mbox{\boldmath $k$}_{\perp }^{2}}
\int^{\sqrt{\mu_{\rm expt}}}_{|{{\mbox{\boldmath $k$}_{\perp }}}|}
\frac{d \omega}{\omega} \frac{{\mbox{\boldmath $k$}_{\perp }^{2}}}{\left(
{\mbox{\boldmath $k$}_{\perp }^{2}}+ m^2/s \; \omega^2 \right)^2} \nonumber \\
\!\!\!\!\!\!\!\!\!\!&& \approx
\frac{\alpha_s C_F}{2 \pi}
\left( \frac{1}{2} \log^2 \frac{s}{m^2}
+ \log \frac{s}{m^2} \log \frac{m^2}{\mu^2}
- \log \frac{m^2}{\mu^2}+ \log \frac{s}{\mu_{\rm expt}^2} - \log \frac{s}{m^2} \log
\frac{s}{\mu_{\rm expt}^2} \right)
 \end{eqnarray}
In all cases above we have not taken into account all subleading collinear logarithms related
to real gluon emission. In order
to now proceed with the inclusion of the running coupling terms it is convenient to first
consider only the DL phase space in each case. Thus we find
\begin{eqnarray}
{\widetilde W}^{RG}_{q_R}\left(s,\mu^2,\mu^2_{\rm expt} \right) &=&  \frac{\alpha_s (m^2) C_F}{2\pi}
\left( \int^{m^2}_{\mu^2} \frac{d {\mbox{\boldmath $k$}_{\perp }^{2}}}{
{\mbox{\boldmath $k$}_{\perp }^{2}}}
\log \frac{s}{m^2} + \int_{m^2}^{\mu_{\rm expt}^2} \frac{d {\mbox{\boldmath $k$}_{\perp }^{2}}}{
{\mbox{\boldmath $k$}_{\perp }^{2}}} \log \frac{s}{{\mbox{\boldmath $k$}_{\perp }^{2}}} \right)
\frac{1}{1+c \log \frac{
{\mbox{\boldmath $k$}_{\perp }^{2}}}{m^2}} \nonumber \\
&\approx&  \frac{\alpha_s(m^2) C_F}{2 \pi} \!\left[
\frac{1}{c} \log \frac{s}{m^2} \left( \log \frac{ \alpha_s ( \mu^2 )}{ \alpha_s ( \mu_{\rm expt}^2 )}
-1 \right) + \frac{1}{c} \log \frac{s}{\mu_{\rm expt}^2 } \right. \nonumber \\
&& \left. \;\;\;\;\;\;\;\;\;\;\;\;\;\;\;\;\;+\frac{1}{c^2}
\log \frac{ \alpha_s (m^2)}{ \alpha_s ( \mu_{\rm expt}^2 )} \right] \; , \; m \ll
 \mu_{\rm expt}
 \end{eqnarray}
and
\begin{eqnarray}
{\widetilde W}^{RG}_{q_R} \left(s,\mu^2,\mu^2_{\rm expt} \right) &=&  \frac{\alpha_s (m^2) C_F}{2\pi}
\int^{\mu_{\rm expt}^2}_{\mu^2} \frac{d {\mbox{\boldmath $k$}_{\perp }^{2}}}{
{\mbox{\boldmath $k$}_{\perp }^{2}}}
\log \frac{s}{m^2}
\frac{1}{1+c \log \frac{
{\mbox{\boldmath $k$}_{\perp }^{2}}}{m^2}} \nonumber \\
&\approx& \frac{\alpha_s (m^2) C_F}{2\pi} \frac{1}{c}
\log \frac{s}{m^2}
\log \frac{ \alpha_s ( \mu^2 )}{ \alpha_s ( \mu_{\rm expt}^2 )}  \; , \; \mu_{\rm expt} \ll m
 \end{eqnarray}
The full subleading expressions are thus given by
\begin{eqnarray}
W^{RG}_{q_R} \left(s,\mu^2,\mu^2_{\rm expt} \right)
&\approx& \frac{\alpha_s (m^2)C_F}{2 \pi} \!\left[ \frac{1}{c}
\log \frac{s}{m^2} \left( \log \frac{ \alpha_s ( \mu^2 )}{ \alpha_s ( \mu_{\rm expt}^2 )}
-1 \right) + \frac{1}{c} \log \frac{s}{\mu_{\rm expt}^2 } \right. \nonumber \\ && \left.
+\frac{1}{c^2}
\log \frac{ \alpha_s (m^2)}{ \alpha_s ( \mu_{\rm expt}^2 )}
- \log \frac{m^2}{\mu^2}
\right] \; , \; m \ll
 \mu_{\rm expt}
 \end{eqnarray}
and
\begin{equation}
W^{RG}_{q_R} \left(s,\mu^2,\mu^2_{\rm expt} \right)
\approx \frac{\alpha_s (m^2) C_F}{2\pi} \left[ \frac{1}{c}
\log \frac{s}{m^2}
\log \frac{ \alpha_s ( \mu^2 )}{ \alpha_s ( \mu_{\rm expt}^2 )}
+  \log \frac{\mu^2}{\mu_{\rm expt}^2 } \right] \; , \; \mu_{\rm expt} \ll m
 \end{equation}
In case we also impose a cut on the integration over $\omega$ we have independently of the relation
between $m$ and $\mu_{\rm expt}$ assuming only $m^2 \ll s$:
\begin{eqnarray}
{\widetilde W}^{RG}_{q_R}\left(s,\mu^2,\mu^2_{\rm expt} \right) &=&  \frac{\alpha_s (m^2) C_F}{2\pi}
\left( \int^{\frac{m^2 \mu_{\rm expt}^2}{s}}_{\mu^2} \frac{d {\mbox{\boldmath $k$}_{\perp }^{2}}}{
{\mbox{\boldmath $k$}_{\perp }^{2}}}
\log \frac{s}{m^2} + \int_{\frac{m^2 \mu_{\rm expt}^2}{s}}^{\mu_{\rm expt}^2}
\frac{d {\mbox{\boldmath $k$}_{\perp }^{2}}}{
{\mbox{\boldmath $k$}_{\perp }^{2}}} \log \frac{\mu_{\rm expt}^2}{{\mbox{\boldmath $k$}_{\perp }^{2}}} \right)
\nonumber \\ && \times
\frac{1}{1+c \log \frac{
{\mbox{\boldmath $k$}_{\perp }^{2}}}{m^2}} \nonumber \\
&\approx&  \frac{\alpha_s (m^2) C_F}{2 \pi}  \left[ \frac{1}{c} \log \frac{s}{m^2}
 \left( \log \frac{ \alpha_s (\mu^2)}{\alpha_s (\mu_{\rm expt}^2)} - 1  \right) \right.
 \nonumber \\ && \left. + \frac{1}{c}
\log \frac{
s}{\mu_{\rm expt}^2} \log \frac{ \alpha_s(\mu^2_{\rm expt})}{ \alpha_s (\mu^2_{\rm expt}m^2/s)} +
 \frac{1}{c^2} \log \frac{
\alpha_s(\mu^2_{\rm expt}m^2/s)}{\alpha_s(\mu_{\rm expt}^2)} \right] \label{eq:rem}
 \end{eqnarray}
This expression agrees with the result obtained in Ref. \cite{ms3} where the gluon on-shell
condition ${\mbox{\boldmath $k$}_{\perp }^{2}}=suv$ was used and one integral over one Sudakov
parameter was done numerically. In Ref. \cite{ms3} it was also shown that the RG-improved
virtual plus soft form factor also
exponentiates by explicitly calculating the two loop RG correction with each loop containing
a running coupling of the corresponding ${\mbox{\boldmath $k$}_{\perp }^{2}}$.

The full subleading expression for the RG-improved soft gluon emission correction is thus given by
\begin{eqnarray}
\!\!\!&&  W^{RG}_{q_R} \left(s,\mu^2,\mu^2_{\rm expt} \right)
\approx  \frac{\alpha_s (m^2) C_F}{2 \pi}  \left[ \frac{1}{c} \log \frac{s}{m^2}
 \left( \log \frac{ \alpha_s (\mu^2)}{\alpha_s (\mu_{\rm expt}^2)} - 1  \right) \right.
 \nonumber \\ \!\!\!&& \left. + \frac{1}{c}
\log \frac{
s}{\mu_{\rm expt}^2} \log \frac{ \alpha_s(\mu^2_{\rm expt})}{ \alpha_s (\mu^2_{\rm expt}m^2/s)} +
 \frac{1}{c^2} \log \frac{
\alpha_s(\mu^2_{\rm expt}m^2/s)}{\alpha_s(\mu_{\rm expt}^2)}
- \log \frac{m^2}{\mu^2} + \log \frac{s}{\mu_{\rm expt}^2}
\right]
 \end{eqnarray}
for the equal mass case. The case of different external and internal masses is again important
for applications in QED and will be discussed next.

\subsubsection*{Unequal masses}

While the gluonic part of the $\beta$-function remains unchanged we integrate again only
from the scale of the massive fermion which is assumed to be in the perturbative regime.
For applications to QED, however, we need the full expressions below.
Here we discuss only the case analogous to the soft gluon approximation. Considering again
only the high energy scenario we have
for the case of an external mass $m$ and a fermion loop mass $m_i$:
\begin{eqnarray}
{\widetilde W}^{RG}_{q_R}\left(s,\mu^2,\mu^2_{\rm expt} \right) &=&  \frac{\alpha_s (m_i^2) C_F}{2\pi}
\left( \int^{\frac{m^2 \mu_{\rm expt}^2}{s}}_{\mu^2} \frac{d {\mbox{\boldmath $k$}_{\perp }^{2}}}{
{\mbox{\boldmath $k$}_{\perp }^{2}}}
\log \frac{s}{m^2} + \int_{\frac{m^2 \mu_{\rm expt}^2}{s}}^{\mu_{\rm expt}^2}
\frac{d {\mbox{\boldmath $k$}_{\perp }^{2}}}{
{\mbox{\boldmath $k$}_{\perp }^{2}}} \log \frac{\mu_{\rm expt}^2}{{\mbox{\boldmath $k$}_{\perp }^{2}}} \right)
\nonumber \\ && \times
\frac{1}{1+c \log \frac{
{\mbox{\boldmath $k$}_{\perp }^{2}}}{m^2_i}} \nonumber \\
&\approx&  \frac{\alpha_s (m^2_i) C_F}{2 \pi}  \left[ \frac{1}{c} \log \frac{s}{m^2}
 \left( \log \frac{ \alpha_s (\mu^2)}{\alpha_s (\mu_{\rm expt}^2m^2/s)} - 1  \right) \right.
 \nonumber \\ && \left. + \frac{1}{c}
\log \frac{
m_i^2}{\mu_{\rm expt}^2} \log \frac{ \alpha_s(\mu^2_{\rm expt})}{ \alpha_s (\mu^2_{\rm expt}m^2/s)} +
 \frac{1}{c^2} \log \frac{
\alpha_s(\mu^2_{\rm expt}m^2/s)}{\alpha_s(\mu_{\rm expt}^2)} \right] \label{eq:reMm}
 \end{eqnarray}
This expression agrees with the result obtained in Eq. (\ref{eq:rem}) for the case $m_i=m$.

The full subleading expression for the RG-improved soft gluon emission correction is thus given by
\begin{eqnarray}
\!\!\!&&  W^{RG}_{q_R} \left(s,\mu^2,\mu^2_{\rm expt} \right)
\approx  \frac{\alpha_s (m^2_i) C_F}{2 \pi}  \left[ \frac{1}{c} \log \frac{s}{m^2}
 \left( \log \frac{ \alpha_s (\mu^2)}{\alpha_s (\mu_{\rm expt}^2m^2/s)} - 1  \right) \right.
 \nonumber \\ \!\!\!&& \left. + \frac{1}{c}
\log \frac{m_i^2
}{\mu_{\rm expt}^2} \log \frac{ \alpha_s(\mu^2_{\rm expt})}{ \alpha_s (\mu^2_{\rm expt}m^2/s)} +
 \frac{1}{c^2} \log \frac{
\alpha_s(\mu^2_{\rm expt}m^2/s)}{\alpha_s(\mu_{\rm expt}^2)}
- \log \frac{m^2}{\mu^2} + \log \frac{s}{\mu_{\rm expt}^2}
\right] \nonumber \\
&& =  \frac{\alpha_s (m^2_i) C_F}{2 \pi}  \left[ \frac{1}{c} \log \frac{s}{m^2}
 \left( \log \frac{ \alpha_s (\mu^2)}{\alpha_s (\mu_{\rm expt}^2m^2/s)} - 1  \right) \right.
 \nonumber \\ \!\!\!&& \left.
 + \frac{1}{c^2} \frac{\alpha_s(m_i^2)}{\alpha_s(\mu^2_{\rm expt})} \log \frac{
\alpha_s(\mu^2_{\rm expt}m^2/s)}{\alpha_s(\mu_{\rm expt}^2)}
- \log \frac{m^2}{\mu^2} + \log \frac{s}{\mu_{\rm expt}^2}
\right] \label{eq:sg}
 \end{eqnarray}
As mentioned above, this expression is more useful for applications in QED or if the mass
ratios are very large. In QED we have again the running coupling of the form given in
Eq. (\ref{eq:eeff}), and Eq. (\ref{eq:sg}) becomes
\begin{eqnarray}
\!\!\!&&  W^{RG}_{f_R} \left(s,\mu^2,\mu^2_{\rm expt} \right)
\approx  \frac{e_f^2 }{8 \pi^2}  \left[ \frac{1}{c} \log \frac{s}{m^2}
 \left( \log \frac{ e^2 (\mu^2)}{e^2 (\mu_{\rm expt}^2m^2/s)} - 1  \right) \right.
 \nonumber \\ \!\!\!&&  + \frac{1}{c^2} \log \frac{e^2(\mu^2_{\rm expt}m^2/s)}{e^2(\mu_{\rm expt}^2)}
\left( 1-
 \frac{1}{3} \frac{e^2}{4 \pi^2} \sum^{n_f}_{j=1}
 Q^2_j N_C^j
\log \frac{
\mu_{\rm expt}^2}{m_j^2} \right)
\nonumber \\ && \left.
- \log \frac{m^2}{\mu^2} + \log \frac{s}{\mu_{\rm expt}^2}
\right] \label{eq:sp}
 \end{eqnarray}
where again $c=-\frac{1}{3} \frac{e^2}{4 \pi^2} \sum^{n_f}_{j=1}Q^2_j N_C^j$.
This concludes the discussion of SL-RG effects in QCD. As a side remark we mention that
for scalar quarks, the same function appears as for fermions since the DL-phase space
for both cases is identical. Only $\beta_0$ differs in each case.

\section{Broken gauge theories} \label{sec:bgt}

In the following we will apply the results obtained in the previous sections to the case of spontaneously
broken gauge theories. It will be necessary to distinguish between
transverse and longitudinal degrees of freedom. The physical motivation in this approach is that for
very large energies, $s \gg M_W^2 \equiv M^2$, the electroweak theory is in the unbroken phase,
with effectively an
$SU(2) \times U(1)$ gauge symmetry as described by the high energy symmetric part of the
Lagrangian in Eq. (\ref{eq:lsym}). We will calculate the corrections to this theory and use
the high energy solution as a matching condition for the regime for values of $\mu < M$.

We begin by considering some simple kinematic arguments for massive vector bosons.
A vector boson at rest has momentum $k^\nu=(M,0,0,0)$ and a polarization vector that is a linear
combination of the three orthogonal unit vectors
\begin{equation}
e_1\equiv (0,1,0,0)\;\;\;,\;\;\; e_1\equiv (0,0,1,0)\;\;\;,\;\;\;e_3\equiv (0,0,0,1) \;\;\;. \label{eq:edef}
\end{equation}
After boosting this particle along the $3$-axis, its momentum will be $k^\nu=(E_k,0,0,k)$. The three
possible polarization vectors are now still satisfying:
\begin{equation}
k_\nu \varepsilon^\nu_j = 0 \;\;\;,\;\;\; \varepsilon_j^2=-1 \;\;\;. \label{eq:econd}
\end{equation}
Two of these vectors correspond to $e_1$ and $e_2$ and describe the transverse polarizations. The
third vector satisfying (\ref{eq:econd}) is the longitudinal polarization vector
\begin{equation}
\varepsilon_L^\nu (k) = (k/M,0,0,E_k/M) \label{eq:lpv}
\end{equation}
i.e. $\varepsilon_L^\nu (k)=k^\nu/M + {\cal O}(M/E_k)$ for large energies. These considerations illustrate that
the transversely polarized degrees of freedom at high energies are related to the massless theory,
while the longitudinal degrees of freedom need to be considered separately.

Another manifestation of the different high energy nature of the two polarization states is
contained in the Goldstone boson equivalence theorem.
It states that the unphysical Goldstone boson that is ``eaten
up'' by a massive gauge boson still controls its high energy asymptotics. A more precise formulation
is given below in section \ref{sec:long}.

Thus we can legitimately use the results obtained in the massless non-Abelian theory
for transverse degrees of freedom at high energies and for longitudinal gauge bosons
by employing the Goldstone boson
equivalence theorem.

Another difference to the situation in an unbroken non-Abelian theory is the mixing of the physical fields
with the fields in the unbroken phase. These complications are especially relevant for the $Z$-boson
and the photon.

\subsection{Fermions and Transverse degrees of freedom} \label{sec:tr}

\begin{figure}
\centering
\epsfig{file=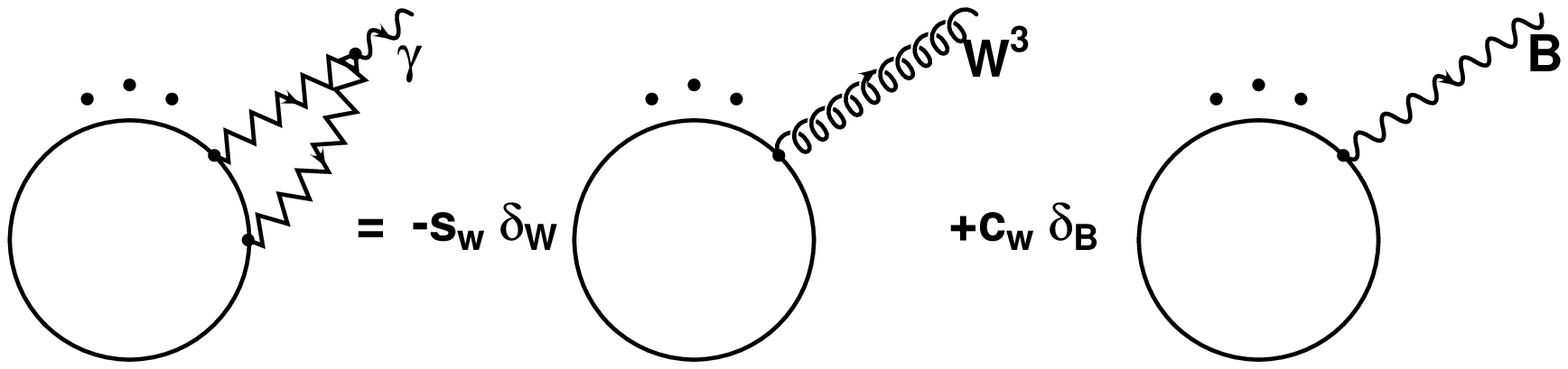,width=14cm}
\vspace{0.5cm} \\
\epsfig{file=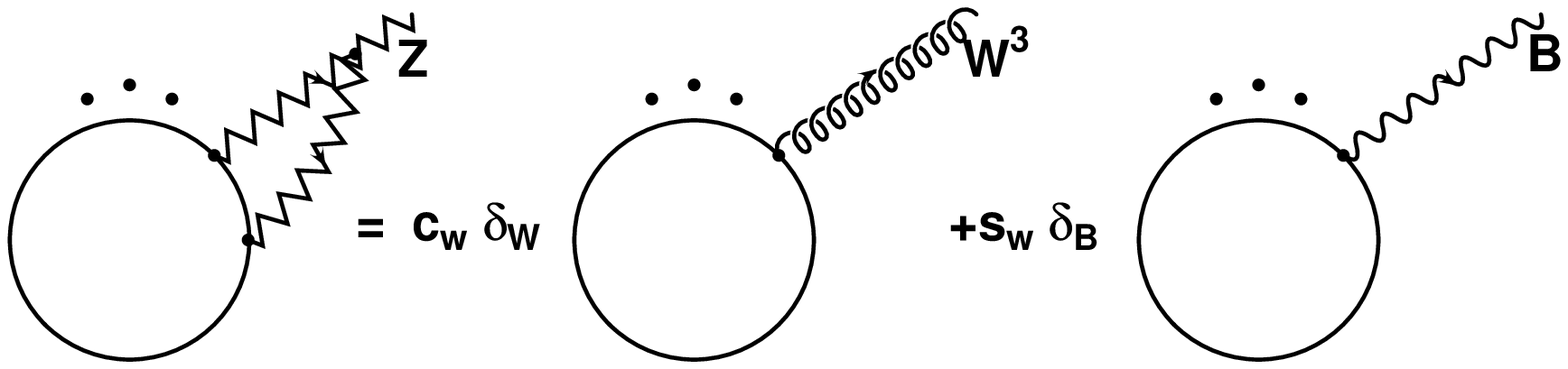,width=14cm}
\vspace{0.5cm} \\
\epsfig{file=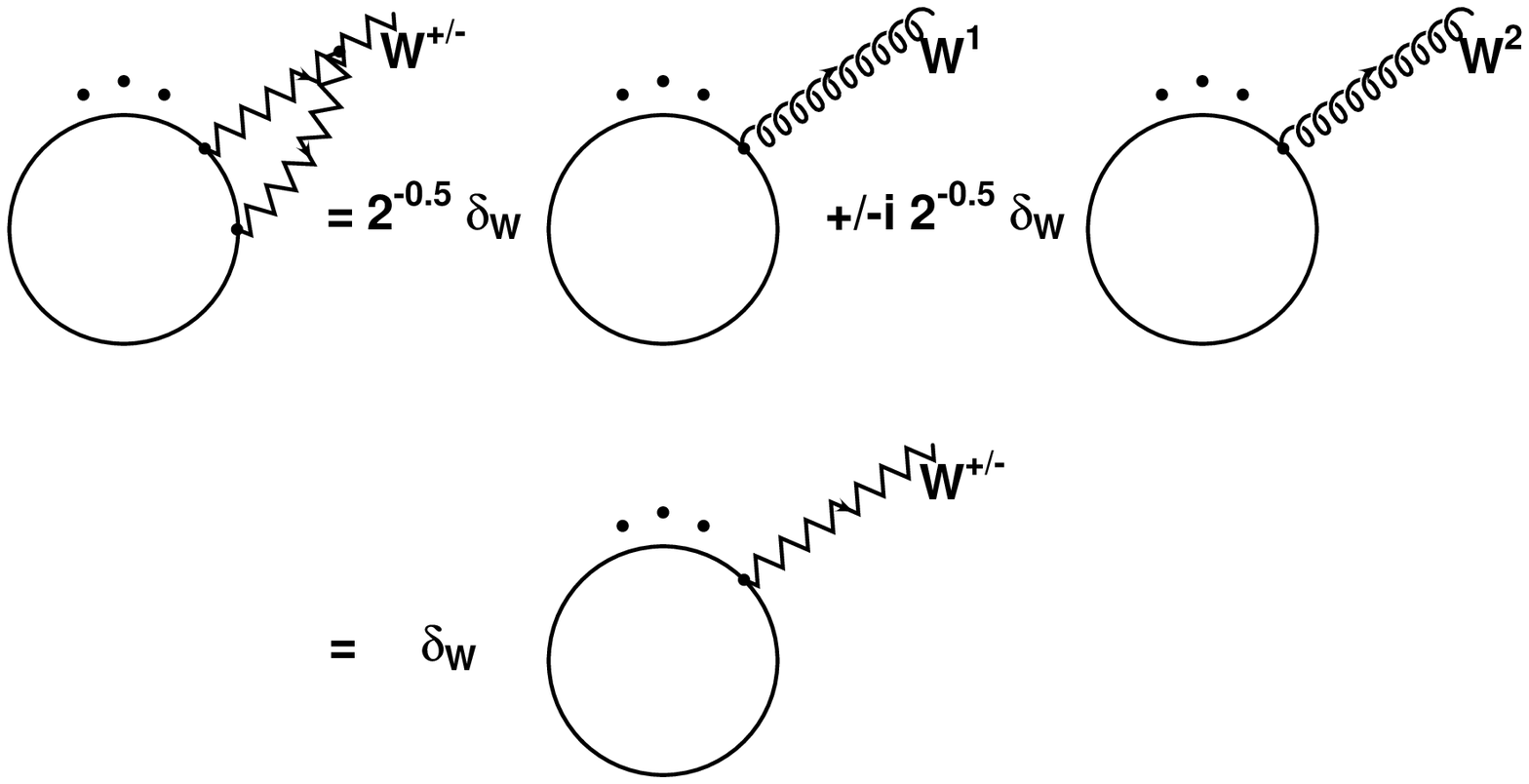,width=14cm}
\caption{The schematic corrections to external gauge boson emissions in terms of the fields in the unbroken phase
of the electroweak theory. There are no mixing terms between the $W^3_\nu$ and $B_\nu$ fields
for massless fermions. We denote $\cos \theta_{\rm w}$ by $c_{\rm w}$ and $\sin \theta_{\rm w}$ by
$s_{\rm w}$.
For $W^\pm$ final states, the corrections factorize with respect to the
physical amplitude. In general, one has to sum over all fields of the unbroken theory with each
amplitude being multiplied by the respective mixing coefficient as given in Eq. (\ref{eq:lc}).
At one loop, also the renormalization conditions must be included.}
\label{fig:mix}
\end{figure}

The results we obtain in this section are generally valid for spontaneously
broken gauge theories, however, for
definiteness we discuss only the electroweak Standard Model. The physical gauge bosons are thus a
massless photon (described by the field $A_\nu$) and massive $W^\pm$ and $Z$ bosons
(described correspondingly by fields $W_{\nu }^{\pm }$ and
$Z_{\nu }$).:
\begin{eqnarray}
W^\pm_\nu&=& \frac{1}{\sqrt{2}} \left( W^1_\nu \pm i W^2_\nu \right) \label{eq:wpm} \\
Z_\nu&=& \cos \theta_{\rm w}W^3_\nu + \sin \theta_{\rm w} B_\nu \label{eq:z} \\
A_\nu&=& -\sin \theta_{\rm w}W^3_\nu + \cos \theta_{\rm w} B_\nu \label{eq:p}
\end{eqnarray}
Thus, amplitudes containing physical fields will correspond to a linear combination of the
massless fields in the unbroken phase. The situation is illustrated schematically for a single
gauge boson external leg in Fig. \ref{fig:mix}. In case of the $W^\pm$ bosons, the corrections
factorize with respect to the physical amplitude.

In the general case 
let us denote physical particles (fields) by $f$
    and particles (fields) of the unbroken theory by $u$. Let the connection
    between
    them be denoted by $f=\sum_{u} C^{fu}u$, where the sum is performed over
    appropriate
    particles (fields) of the unbroken theory as in Eqs. (\ref{eq:wpm}),
(\ref{eq:z}) and (\ref{eq:p}).
Note that, in general,
    physical particles, having definite masses,  don't belong to
    irreducible representations of the symmetry
    group of the unbroken theory (for example, the photon and
    $Z$ bosons have no definite isospin). On the other hand, particles of
    the unbroken theory, belonging to irreducible representations of the
    gauge
    group, have no definite masses.
    Then for the amplitude
    ${\cal M}^{f_1,...f_n}(p_{1},...,p_{n};\mu ^{2})$ with $n$ physical
    particles $f_i$ with momenta $p_i$ and infrared cut-off
    $\mu^{2}$, the general case for virtual corrections is given by
\begin{equation}
    {\cal M}^{f_1,...f_n}(p_{1},...,p_{n};\mu ^{2}) =\sum_{u_1,...u_n }
    \prod_{i=1}^n C^{f_iu_i}
    {\cal M}^{u_1,...u_n}(p_{1},...,p_{n}, \mu^2) \label{eq:lc}
\end{equation}
On the one loop level and to subleading accuracy, Eq. (\ref{eq:lc}) must also include the
correct counterterms for the commonly chosen on-shell scheme. In this scheme the on-shell 
photon, for instance, does not mix with the $Z$-boson, thus including all mixing effects into
the massive neutral $Z$-boson sector. More details are given below.
In the following we give results only for the amplitudes ${\cal M}^{u_1,...u_n}(p_{1},...,p_{n}, 
\mu^2)$ keeping in mind that in general the physical amplitudes must be obtained
via Eq. (\ref{eq:lc}). For fermions, transverse $W^\pm$, longitudinal gauge bosons or Higgs bosons,
no linear combination arises, i.e. the universal corrections below factorize automatically
with respect to the physical Born process. Only photons and $Z$-bosons are affected by this
complication for the obvious reasons discussed above.
To logarithmic accuracy, all masses can be set equal:
\[
M_{Z}\sim M_{W}\sim M_{\rm Higgs}\sim M
\]
and the energy is considered to be much larger, $\sqrt{s}\gg M$.
The left and right handed
fermions are correspondingly doublets ($T=1/2$) and singlets ($T=0$) of the
$%
SU$(2) weak isospin group and have hypercharge $Y$ related to the electric
charge $Q$, measured in units of the proton charge, by the
Gell-Mann-Nishijima formula $Q=T^{3}+Y/2$.

The value for the infrared cutoff $\mu$ can be chosen in two different regimes (see Fig. 
\ref{fig:su2u1}):
I $\sqrt{s} \gg \mu \gg M$ and II $\mu \ll M$. The second case is universal in the sense
that it does not depend on the details of the electroweak theory and will be discussed below.
In the first region we can neglect spontaneous symmetry breaking effects (in particular
terms connected to the v.e.v.) and consider the theory with fields $B_{\nu }$ and $W_{\nu }^{a}$
as given by ${\cal L}_{\rm symm}$ in Eq. (\ref{eq:lsym}).
One could of course also calculate everything in terms of the physical fields, however, we
emphasize again that in this case we need to consider the photon also in region I). The
omission of the photon would lead to the violation of gauge invariance since the photon
contains a mixture of the $B_{\nu }$ and $W_{\nu }^{3}$ fields.

In region I), the renormalization group equation (or generalized infrared evolution equation)
(\ref{eq:mrg}) in the case of all $m_i < M$ reads\footnote{
Note, that the amplitude on the right hand
side is in general a linear combination of fields in the unbroken phase according to Eq. (\ref{eq:lc}).
In addition, in the electroweak theory matching will be required at the
scale $M$ and often on-shell renormalization of the couplings $e$ and $\sin \theta_{\rm w}$ is used. In
this case one has additional complications in the running coupling terms due to the different mass
scales involved below $M$. Details are presented in section \ref{sec:ewrg}.}
\begin{eqnarray}
&& \left( \frac{\partial}{\partial t} + \beta \frac{\partial}{\partial g} +
\beta^\prime \frac{\partial}{\partial g^\prime} + \sum_{i=1}^{n_g}
\Gamma^i_g(t) -n_{\rm W} \frac{1}{2} \frac{\alpha}{\pi} \beta_0
-n_{\rm B} \frac{1}{2} \frac{\alpha^\prime}{\pi} \beta^\prime_0 + \sum_{k=1}^{n_f} \left(
\Gamma^k_f(t) + \frac{1}{2}
\gamma^k_f \right) \right) \nonumber \\
&& \times {\cal M}^\perp (p_1,...,p_n,g,g^\prime,\mu) = 0
\label{eq:wmrg}
\end{eqnarray}
where the index $\perp$ indicates that we consider only $n_g$ transversely
polarized external gauge bosons with $n_{\rm W}+n_{\rm B}=n_g$ and $n_f$ denotes the 
number of external fermion
lines.
The two $\beta$-functions are given by:
\begin{eqnarray}
\beta (g(\overline{\mu}^2)) &=& \frac{\partial g(\overline{\mu}^2)}{\partial \log \overline{\mu}^2}
\approx - \beta_0 \frac{g^3(\overline{\mu}^2)}{8\pi^2} \label{eq:b0} \\
\beta^\prime (g^\prime(\overline{\mu}^2)) &=& \frac{\partial g^\prime
(\overline{\mu}^2)}{\partial \log \overline{\mu}^2}
\approx - \beta^\prime_0 \frac{{g^\prime}^3(\overline{\mu}^2)}{8\pi^2} \label{eq:b0p}
\end{eqnarray}
with the one-loop terms given by:
\begin{equation}
\beta_0=\frac{11}{12}C_A - \frac{1}{3}n_{gen}-\frac{1}{24}n_{h} \;\;\;,\;\;\;
\beta^\prime_0= - \frac{5}{9}n_{gen} -\frac{1}{24}n_{h} \label{eq:ewb}
\end{equation}
where $n_{gen}$ denotes the number of fermion generations \cite{wein,gross} and $n_h$ the number of
Higgs doublets.
Eq. (\ref{eq:wmrg}) describes the one loop RG corrections correctly. At higher orders the subleading
RG terms must be included according to the discussion in section \ref{sec:ewrg}.
The infrared singular anomalous dimensions read
\begin{equation}
\Gamma^i_{f,g} (t) = \left( \frac{\alpha}{4 \pi}T_i(T_i+1)+ \frac{\alpha^\prime}{4 \pi} \left( \frac{Y_i}{2}
\right)^2 \right) t
\label{eq:wad}
\end{equation}
where $T_i$ and $Y_i$ are the total weak isospin and hypercharge respectively of the particle emitting
the soft and collinear gauge boson. Analogously,
\begin{equation}
\gamma^i_{f} = -3 \left( \frac{\alpha}{4 \pi}T_i(T_i+1)+ \frac{\alpha^\prime}{4 \pi} \left( \frac{Y_i}{2}
\right)^2 \right) 
+ \frac{\alpha}{4 \pi} \left( \frac{1+\delta_{i,{\rm R}}}{4} \frac{m^2_i}{M^2} + \delta_{i,{\rm L}}
\frac{m^2_{i^\prime}}{4 M^2} \right)
\label{eq:qad}
\end{equation}
where the last two terms only contribute for quarks of the third generation. $i^\prime$ denotes the
isospin partner of $i$. The presence of Yukawa terms and also the Higgs contribution to the
$\beta$-functions in Eq. (\ref{eq:ewb}) are remnants of the spontaneous broken symmetry which
leads to differences even in the transverse sector compared to unbroken gauge theories as is
obvious from the form of ${\cal L}_{\rm symm}$ in Eq. (\ref{eq:lsym}).
In terms of the corresponding logarithmic probabilities we thus have the following expression
for fermions from the virtual splitting function approach:
\begin{eqnarray}
W_{f_i}(s,\mu^2) &=&  \frac{\alpha}{4 \pi} \!\! \left[ \! \left( T_i(T_i+1)+  \tan^2 \!
\theta_{\rm w}
\frac{Y^2_i}{4} \right) \!\!
\left( \log^2 \frac{s}{\mu^2}- 3 \log \frac{s}{\mu^2}
\! \right) \right. \nonumber \\ && \left.
+ \left( \frac{1+\delta_{i,{\rm R}}}{4} \frac{m^2_i}{M^2} + \delta_{i,{\rm L}}
\frac{m^2_{i^\prime}}{4 M^2} \right)
\log \frac{s}{\mu^2} \right] \label{eq:Wf}
\end{eqnarray}
For external transversely polarized gauge bosons:
\begin{eqnarray}
W_{g_i}(s,\mu^2) &=& \left( \frac{\alpha}{4 \pi}T_i(T_i+1)+ \frac{\alpha^\prime}{4 \pi}
\left( \frac{Y_i}{2} \right)^2 \right) \log^2 \frac{s}{\mu^2} \nonumber \\
&&
- \left( \delta_{i,{\rm W}} \frac{\alpha}{\pi} \beta_0 + \delta_{i,{\rm B}}
\frac{\alpha^\prime}{\pi} \beta^\prime_0 \right) \log \frac{s}{\mu^2} \label{eq:Wg}
\end{eqnarray}
The initial condition for Eq. (\ref{eq:wmrg}) is given by the requirement that for the infrared cutoff
$\mu^2=s$ we obtain the Born amplitude. The solution of (\ref{eq:wmrg}) is thus given by
\begin{eqnarray}
&& {\cal M}^\perp (p_1,...,p_n,g,g^\prime,\mu) = {\cal M}^\perp_{\rm Born}
(p_1,...,p_n,g, g^\prime)
\nonumber \\ && \times \exp \left\{ - \frac{1}{2} \sum^{n_g}_{i=1}
\left( \frac{\alpha}{4 \pi}T_i(T_i+1)+ \frac{\alpha^\prime}{4 \pi}
\left( \frac{Y_i}{2} \right)^2 \right) \log^2 \frac{s}{\mu^2} \right. \nonumber \\
&& \;\;\;\;\;\;\;\;\;\;\;\;
+ \left( n_{\rm W} \frac{\alpha}{2\pi} \beta_0 + n_{\rm B}
\frac{\alpha^\prime}{2\pi} \beta^\prime_0 \right) \log \frac{s}{\mu^2} \nonumber \\
&& -\frac{1}{2} \sum^{n_f}_{k=1} \left[ \left( \frac{\alpha}{4 \pi}T_k(T_k+1)+
\frac{\alpha^\prime}{4 \pi} \left( \frac{Y_k}{2} \right)^2 \right)
\left[ \log^2 \frac{s}{\mu^2} - 3 \log \frac{s}{\mu^2} \right] \right. \nonumber \\ && \left. \left.
+ \frac{\alpha}{4 \pi} \left( \frac{1+\delta_{k,{\rm R}}}{4} \frac{m^2_k}{M^2} + \delta_{k,{\rm L}}
\frac{m^2_{k^\prime}}{4 M^2} \right) \log \frac{s}{\mu^2} \right] \right\} \label{eq:mpsol1}
\end{eqnarray}
where we neglect RG corrections for now. These will be discussed thoroughly in section \ref{sec:ewrg}.
$n_{\rm W}$ and $n_{\rm B}$ denote the number of external $W$ and $B$ fields respectively.
The $SU(2) \times U(1)$ group factors in the exponential can be written in terms of
the parameters of the broken theory as follows:
\[
g^{2}T_i(T_i+1)+{g^{\prime }}^{2}\left( \frac{Y_i}{2}\right)
^{2}=e_i^{2}+g^{2}\left( T_i(T_i+1)-(T_i^{3})^{2}\right) +\frac{g^{2}}{\cos
^{2}\theta _{w}}\left( T_i^{3}-\sin ^{2}\theta _{w}Q_i\right) ^{2}
\]
where the three terms on the r.h.s. correspond to the contributions of the
soft photon (interacting with the electric charge $e_i=Q_ig\sin \theta _{w}$),
the $W^{\pm }$ and the $Z$ bosons, respectively. Although we may rewrite
solution (\ref{eq:mpsol1}) in terms of the parameters of the broken theory in
the form of a product of three exponents corresponding to the exchanges of
photons, $W^{\pm }$ and $Z$ bosons, it would be wrong to identify the
contributions of the diagrams without virtual photons with this expression
for the particular case $e_i^{2}=0$. This becomes evident when we note that if
we were to omit photon lines then the result would depend on the choice of
gauge, and therefore be unphysical. Only for $\theta _{w}=0$, where the
photon coincides with the $B$ gauge boson, would the identification of the
$%
e_i^{2}$ term with the contribution of the diagrams with photons be correct.

We now need to discuss the solution in the general case. In region I) we calculated the
scattering amplitude for the theory in the unbroken phase in the massless limit.
Choosing the cutoff $\mu$ in region II), $\mu \ll M$, we have to only consider the
QED contribution. In this region we cannot necessarily neglect all mass terms, so
we need to discuss the subleading terms for QED with mass effects.
If $m_i\ll \mu$, the results from massless QCD can
be used directly by using the Abelian limit $C_F =1$. In case $\mu \ll m_i$ we must use the
well known next to leading order QED results, e.g. \cite{yfs}, and the virtual probabilities take the
following form for fermions:
\begin{equation}
w_{f_i}(s,\mu^2) = \left\{ \begin{array}{lc} \frac{e_i^2}{(4 \pi)^2} \left( \log^2 \frac{s}{\mu^2}
- 3 \log \frac{s}{\mu^2} \right) & , \;\;\; m_i \ll \mu \\
\frac{e_i^2}{(4 \pi)^2} \left[ \left( \log \frac{s}{m_i^2}-1 \right) 2 \log \frac{m_i^2}{\mu^2} \right. \\
\left.\;\;\;\;\;\;\;\;\;\;+ \log^2 \frac{s}{m_i^2} - 3 \log \frac{s}{m_i^2} \right] & , \;\;\; \mu 
\ll m_i\end{array} \right. \label{eq:wf}
\end{equation}
Note, that in the last equation the full subleading collinear logarithmic term \cite{m}
is used in distinction
to Ref. \cite{yfs}. In the explicit two loop calculation presented in Ref.
\cite{bnb} it can be seen that the full collinear term also exponentiates
at the subleading level in massive QED.
For $W^\pm$ bosons we have analogously:
\begin{equation}
w_{{\rm w}_i}(s,\mu^2) =
\frac{e_i^2}{(4 \pi)^2} \left[ \left( \log \frac{s}{M^2}-1 \right)
2 \log \frac{M^2}{\mu^2}
+ \log^2 \frac{s}{M^2} \right] \label{eq:ww}
\end{equation}
In addition we have collinear terms for external on-shell photon lines
from fermions with mass $m_j$ and electromagnetic charge $e_j$
up to scale $M$:
\begin{equation}
w_{\gamma_i}(M^2,\mu^2) = \left\{ \begin{array}{lc}
\frac{n_f}{3} \frac{e_j^2}{4 \pi^2} N^j_C
\log \frac{M^2}{\mu^2} & , \;\;\; m_j \ll \mu \\
\frac{1}{3} \sum_{j=1}^{n_f} \frac{e_j^2}{4 \pi^2} N^j_C \log \frac{M^2}{m_j^2}
& , \;\;\; \mu \ll m_j\end{array} \right. \label{eq:wga}
\end{equation}
Note that automatically, $w_{\gamma_i}(M^2,M^2)=0$. At one loop order,
this contribution cancels against terms from the renormalization of the QED
coupling up to scale $M$. For external $Z$-bosons, however, there are no
such collinear terms since the mass is large compared to the $m_i$. Thus,
the corresponding RG-logarithms up to scale $M$ remain uncanceled.

The appropriate initial condition is given by Eq. (\ref{eq:mpsol1})
evaluated at the matching point $\mu=M$. Thus we find for the general solution in region II):
\begin{eqnarray}
&& {\cal M}^\perp (p_1,...,p_n,g,g^\prime,\mu) = {\cal M}^\perp_{\rm Born}
(p_1,...,p_n,g,g^\prime)
\nonumber \\ && \times \exp \left\{ - \frac{1}{2} \sum^{n_g}_{i=1}
\left( \frac{\alpha}{4 \pi}T_i(T_i+1)+ \frac{\alpha^\prime}{4 \pi}
\left( \frac{Y_i}{2} \right)^2 \right) \log^2 \frac{s}{M^2} \right. \nonumber \\
&& \;\;\;\;\;\;\;\;\;\;\;\;
+ \left( n_{\rm W} \frac{\alpha}{2\pi} \beta_0 + n_{\rm B}
\frac{\alpha^\prime}{2\pi} \beta^\prime_0 \right) \log \frac{s}{M^2} \nonumber \\
&& -\frac{1}{2} \sum^{n_f}_{k=1} \left[ \left( \frac{\alpha}{4 \pi}T_k(T_k+1)+
\frac{\alpha^\prime}{4 \pi} \left( \frac{Y_k}{2} \right)^2 \right)
\left[ \log^2 \frac{s}{M^2} - 3 \log \frac{s}{M^2} \right] \right. \nonumber \\ && \left. \left.
+ \frac{\alpha}{4 \pi} \left( \frac{1+\delta_{k,{\rm R}}}{4} \frac{m^2_k}{M^2} + \delta_{k,{\rm L}}
\frac{m^2_{k^\prime}}{4 M^2} \right) \log \frac{s}{m_t^2} \right] \right\} \nonumber \\
&&\times \exp \left[ - \frac{1}{2} \sum_{i=1}^{n_f} \left( w_{f_i}(s,\mu^2)
- w_{f_i}(s,M^2) \right)
- \frac{1}{2} \sum_{i=1}^{n_{\rm w}} \left( w_{{\rm w}_i}(s,\mu^2)
- w_{{\rm w}_i}(s,M^2) \right) \right. \nonumber \\
&& \;\;\;\;\;\;\;\;\;\;\; \left. -\frac{1}{2} \sum_{i=1}^{n_\gamma} w_{\gamma_i}(M^2,\mu^2)
\right] \nonumber \\
&& = {\cal M}^\perp_{\rm Born}
(p_1,...,p_n,g,g^\prime)
\nonumber \\ && \times \exp \left\{ - \frac{1}{2} \sum^{n_g}_{i=1}
\left( \frac{\alpha}{4 \pi}T_i(T_i+1)+ \frac{\alpha^\prime}{4 \pi}
\left( \frac{Y_i}{2} \right)^2 \right) \log^2 \frac{s}{M^2} \right. \nonumber \\
&& \;\;\;\;\;\;\;\;\;\;\;\;
+ \left( n_{\rm W} \frac{\alpha}{2\pi} \beta_0 + n_{\rm B}
\frac{\alpha^\prime}{2\pi} \beta^\prime_0 \right) \log \frac{s}{M^2} \nonumber \\
&& - \frac{1}{2} \sum^{n_f}_{k=1} \left[ \left( \frac{\alpha}{4 \pi}T_k(T_k+1)+
\frac{\alpha^\prime}{4 \pi} \left( \frac{Y_k}{2} \right)^2 \right)
\left[ \log^2 \frac{s}{M^2} - 3 \log \frac{s}{M^2} \right] \right. \nonumber \\ && \left. \left.
+ \frac{\alpha}{4 \pi} \left( \frac{1+\delta_{k,{\rm R}}}{4} \frac{m^2_k}{M^2} + \delta_{k,{\rm L}}
\frac{m^2_{k^\prime}}{4 M^2} \right) \log \frac{s}{m_t^2} \right] \right\} \nonumber \\
&&\times \exp \left[ - \frac{1}{2} \sum_{i=1}^{n} \! \left( \frac{e_{i}^{2}}{(4\pi )^{2}}%
\! \left( 2 \log \frac{s}{m_{i}\,M}\log \frac{M^{2}}{m_{i}^{2}}+ 2 \log \frac{s}{%
m_{i}^{2}}\log \frac{m_{i}^{2}}{\mu ^{2}} + 3 \log
\frac{m_i^2}{M^2} \right.  \right.  \right.  \nonumber \\
&& \left. \left. \left. -2 \log \frac{m_i^2}{\mu^2}
\right) \right) - \frac{1}{2} \sum_{i=1}^{n_\gamma} \frac{1}{3}
\frac{e_i^2}{4 \pi^2} N^i_C \log \frac{M^2}{m_i^2} \right] \label{eq:mpsol2}
\end{eqnarray}
The last equality holds for $\mu \ll m_i \ll M$ and we have replaced the matching scale $M$
by $m_t$ in the Yukawa enhenced subleading terms since the coefficients are unambiguously 
determined and the argument in the corresponding logarithm must be $m_t$ \cite{m3,dp}.
It is important to note again that, unlike the situation in QCD, in the electroweak theory we
have in general different mass scales determining the running of the couplings of the physical on-shell
renormalization scheme quantities. We have written the above result in such
a way that it holds for arbitrary chiral fermions and transversely polarized gauge bosons.
In order to include physical external photon states in the on-shell scheme,
the renormalization condition is given by the requirement that the physical
photon does not mix with the Z-boson. This leads to the condition that
the Weinberg rotations in Fig. \ref{fig:mix} at one loop receive no
RG-corrections. Thus, above the scale $M$ the subleading collinear and
RG-corrections cancel for physical photon and Z-boson states.

Since the Yukawa enhanced terms are novel features in broken gauge theories as compared
to the situation in QCD we use the non-Abelian generalization of the
Gribov theorem in the following to prove the correctness
of our splitting function approach for specific processes.
Since we are interested here in corrections to order ${\cal O}\left(\alpha^n \log^{2n-1} \frac{
s}{M^2} \right)$, each additional
loop correction to the universal subleading terms in the previous section must yield two logarithms,
i.e. we are considering DL-corrections to the basic process like the inner fermion loop
in Fig. \ref{fig:tyuk}.
It is of particular importance that all additional gauge bosons must couple to external legs, since
otherwise only a subleading term of order ${\cal O}\left(\alpha^n \log^{2n-2}
\frac{}{M^2} \right)$ would be generated.
All subleading corrections generated by the exchange of gauge bosons coupling both to
external Goldstone bosons and inner fermion lines cancel analogously to a mechanism found
in Ref. \cite{ms1} for terms in heavy quark production in $\gamma \gamma$-collisions in a
$J_z=0$ state. Formally this can be understood by noting that such terms contain an infrared
divergent correction. The sum of those terms, however, is given by the Sudakov form factor.
Thus any additional terms encountered in intermediate steps of the calculation cancel.

\begin{figure}[t]
\centering
\epsfig{file=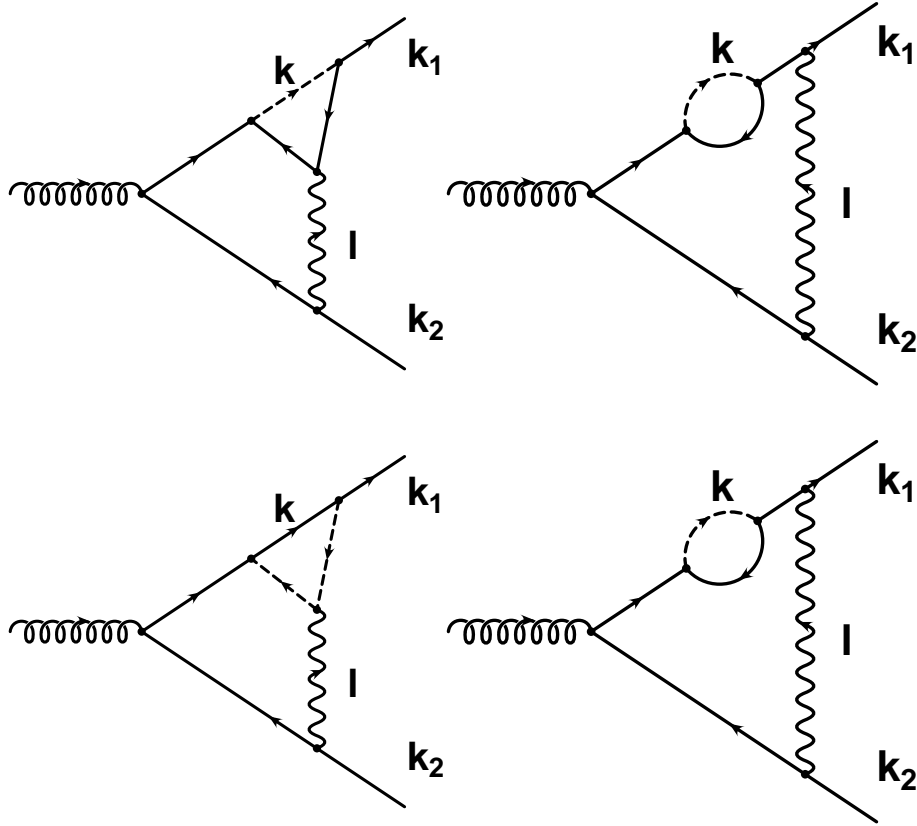,width=12cm}
\caption{Two loop Feynman diagrams yielding Yukawa enhanced logarithmic corrections to the third
generation of fermions in the final sate. 
The Ward identity in Eq. (\ref{eq:wi}) assures that in the Feynman gauge,
the sum
of all vertex and self energy diagrams does not lead to
additional SL logarithms at the two
loop level.
Only corrections to the original one loop vertex
need
to be considered and lead to the exponentiation of Yukawa terms in the
fermionic SM sector to SL accuracy.}
\label{fig:smfwi}
\end{figure}

This point can be understood also from the principle of gauge invariance. At the two loop
level for instance we have to consider diagrams of the type depicted in Fig. \ref{fig:smfwi}.
They involve Vertex corrections $\Gamma_\nu(k_1^2,l^2,(k_1-l)^2)$ and self energy
terms $\Sigma((k_1-l)^2)$ with the same overall Yukawa-term structure.
Writing the gauge coupling in the symmetric basis for clarity since
we are considering a regime where $s=(k_1-k_2)^2 \gg M^2$, where $M$ is the gauge boson mass. 
In any
case, local gauge invariance is not violated in the SM and for heavy particles in the
high energy limit, we can
perform the calculation in a basis which is more convenient.
For our purposes we need to
investigate terms containing three large logarithms in those diagrams. Since the would-be
Goldstone boson loops
at one loop only yield a single logarithm it is clear that the gauge boson loop momentum
$l$ must be soft. Thus we need to show that the UV logarithm originating from the $k$
integration is identical (up to the sign) in both diagrams.
We can therefore neglect the loop momentum $l$ inside the fermion loop.
It is then straightforward to see that
\begin{equation}
\frac{\partial}{\partial {k_1}_\mu} \Sigma (k_1^2)= \Gamma^\mu (k_1^2,0,k_1^2)
\label{eq:wi}
\end{equation}
where the full sum of all contributing self energy and vertex diagrams must be taken.
Thus, we have established a Ward identity for arbitrary Yukawa couplings of scalars to
fermions and thus, the identity of the UV singular contributions.
The relative sign is
such that the generated SL logarithms of the diagrams in Fig. \ref{fig:smfwi} 
cancel each
other. The existence of such an identity is not surprising since it expresses the fact
that also the Yukawa sector is gauge invariant. 
We are thus left with gauge boson corrections to the original vertices in the on-shell
renormalization scheme such as depicted in Fig. \ref{fig:tyuk}.
At high energies we can therefore employ the non-Abelian version
of Gribov's bremsstrahlung theorem. The soft photon corrections are included via
matching as discussed above.

For the one loop process in Fig. \ref{fig:tyuk}, for instance,
we include only corrections with top and bottom quarks and assume
on-shell renormalization.
Thus the corrections at higher orders factorize with respect to the one loop fermion amplitude
and ${\cal M}_{\rm ``Born''}(p_1,...,p_n) = {\cal M}_{\rm 1 loop}(p_1,...,p_n)$. Note that
the latter is also independent of the cutoff $\mu$ since the fermion mass serves as a
natural regulator. In principle we can choose the top-quark mass to be much larger than $\mu$
for instance. In our case we have for the electroweak DL corrections at the weak scale $\mu=M$:
\begin{equation}
W^{\rm ew}_l (s , M^2) = \left[
  \frac{ g^2}{16 \pi^2} T_i(T_i+1)+  \frac{ {g^\prime}^2}{16 \pi^2} \frac{Y^2}{4} \right]
 \log^2 \frac{s}{M^2}
 \label{eq:Wew}
\end{equation}
We now want to consider specific processes relevant at future $e^+e^-$ colliders and
demonstrate how to apply the non-Abelian version of Gribov's factorization theorem for
the higher order corrections. The subleading corrections are then compared to the
general splitting
function result in Eqs. (\ref{eq:Wf}) and (\ref{eq:mpsol1}). Below we use the
physical fields for the respective contributions.

From the arguments of section \ref{sec:qn} it is now straightforward
to include also top-Yukawa terms for chiral quark final states. These terms occur for
left handed bottom as well as top quark external lines. The situation for a typical Drell-Yan
process is depicted in Fig. \ref{fig:tyuk} where for the inner scattering amplitude we have
two contributions. We neglect all terms of order ${\cal O} \left( \frac{m_f^2}{s}, \frac{M^2}{s}
\right)$. Using on-shell renormalization we find for the inner amplitude on the left
in Fig. \ref{fig:tyuk} for a right handed electron in the initial and a left handed bottom quark
in the final state from the $\phi^\pm$ loop for the sum of the $\gamma$ and $Z$ contributions
according to the Feynman rules in appendix \ref{sec:fr}:
\begin{figure}
\centering
\epsfig{file=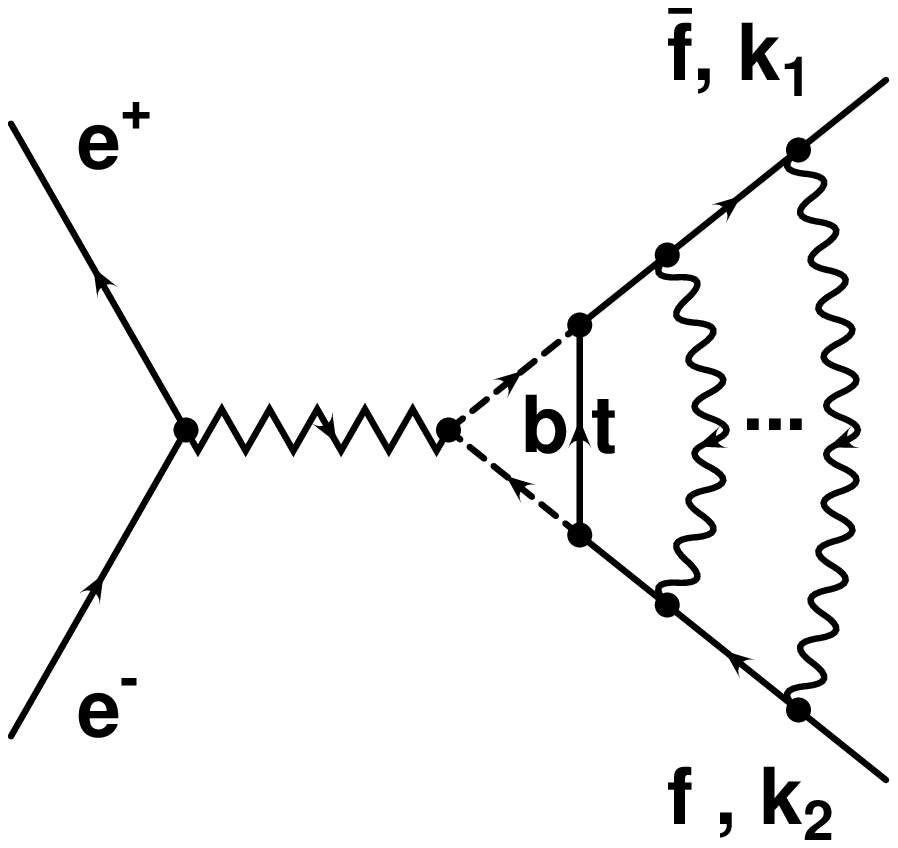,width=6cm} \hspace{2cm}
\epsfig{file=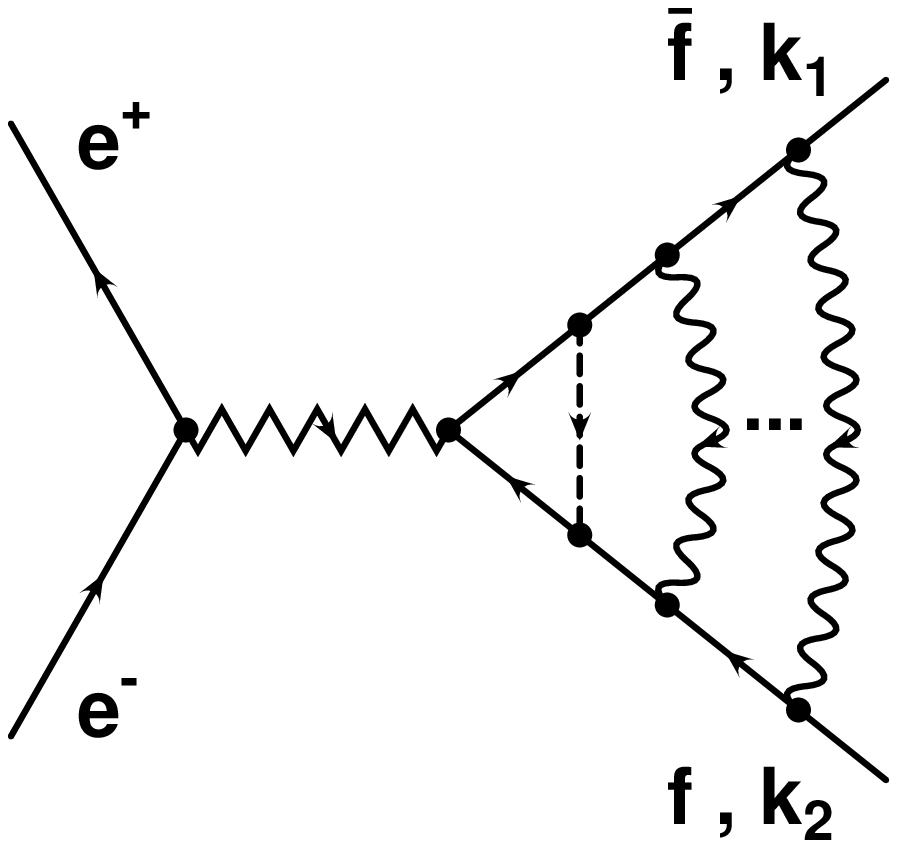,width=6cm}
\caption{Feynman diagrams yielding Yukawa enhanced logarithmic corrections to the third
generation of fermions in the final sate. The inner scattering amplitude
is taken on the mass shell. No DL-corrections originate from the inner loop.
At higher orders, the subleading corrections are given in factorized
form according to the non-Abelian generalization of Gribov's theorem as described in the
text. Corrections from gauge bosons inside the Goldstone-boson loop give only
sub-subleading contributions.
DL-corrections at two and higher loop order are given by gauge bosons coupling to (in principle
all) external legs as schematically indicated.}
\label{fig:tyuk}
\end{figure}
\begin{eqnarray}
{}^a\!\!{\cal A}^{\rm DY}_{\rm 1 loop} &=& - \frac{e^4m_t^2}{4 s M^2 s_{\rm w}^2 c_{\rm w}^2}
\langle e^+_{\rm L} | \gamma_\nu | e^-_{\rm R} \rangle \times \nonumber \\ &&
\int \frac{ d^nl}{(2 \pi)^n}
\frac{ \langle f_{\rm L} | {\rlap/ l} (2l-k_1-k_2)^\nu | f_{\rm R} \rangle}{(l^2-m_{f^\prime}^2+i \varepsilon
)((l-k_1)^2-M^2+i \varepsilon)((l-k_2)^2-M^2+i \varepsilon)}
+{}^a \delta^{\rm DY}_{\rm ct} \nonumber \\ &=&
-\frac{i}{32 \pi^2} \frac{e^4m_t^2}{4 s M^2 s_{\rm w}^2 c_{\rm w}^2}
\langle e^+_{\rm L} | \gamma_\nu | e^-_{\rm R} \rangle  \langle f_{\rm L} |
\gamma^\nu | f_{\rm R} \rangle (B_{23}-B^M_{23}) \label{eq:sfs}
\end{eqnarray}
The scalar functions at high energy
evaluate to $B_{23}-B^M_{23}=-\log \frac{s}{M^2}$. For the diagram on the
right in Fig. \ref{fig:tyuk} we have for the bottom again only the $\phi^\pm$ contribution.
Here we find for the sum of the $\gamma$ and $Z$ contributions:
\begin{eqnarray}
{}^b\!\!{\cal A}^{\rm DY}_{\rm 1 loop} &=& - \frac{e^4m_t^2Q_t}{2 s M^2 s_{\rm w}^2 c_{\rm w}^2}
\langle e^+_{\rm L} | \gamma_\nu | e^-_{\rm R} \rangle \times \nonumber \\ &&
\int \frac{ d^nl}{(2 \pi)^n}
\frac{ \langle f_{\rm L} | {\rlap/ l} \gamma^\nu {\rlap/ l}| f_{\rm R} \rangle}{(l^2-M^2+i \varepsilon
)((l-k_1)^2-m_t^2+i \varepsilon)((l-k_2)^2-m_t^2+i \varepsilon)}
+{}^b \delta^{\rm DY}_{\rm ct} \nonumber \\ &=&
\frac{i}{32 \pi^2} \frac{e^4m_t^2Q_t}{2 s M^2 s_{\rm w}^2 c_{\rm w}^2}
\langle e^+_{\rm L} | \gamma_\nu | e^-_{\rm R} \rangle  \langle f_{\rm L} |
\gamma^\nu | f_{\rm R} \rangle (B_{23}-B^M_{23}) \label{eq:fsf}
\end{eqnarray}
In all cases we renormalize on-shell, i.e. by requiring that the vertex vanishes when the momentum
transfer equals the masses of the external on-shell lines. All on-shell self energy contributions
don't contribute in this scheme.
For external left handed top quarks, the $\phi^\pm$ loop is mass suppressed and we only have
to consider the $\chi$ and $H$ corrections.
They are given by replacing $Q_t \longrightarrow 2 Q_t \left(T^3_t \right)^2$ and
$Q_t \longrightarrow \frac{1}{2} Q_t$ in Eq. (\ref{eq:fsf}). It turns out that the $Z \chi H$
contributions equal the corrections from the $\gamma \phi^\pm$ and $Z \phi^\pm$ in the case
of the bottom calculation.
The Born amplitude is given by:
\begin{eqnarray}
{\cal M}^{\rm DY}_{\rm Born} &=&
i \frac{e^2}{s c^2_{\rm w}} (Q_f - T^3_f) \langle e^+_{\rm L} | \gamma_\nu | e^-_{\rm R} \rangle
\langle f_{\rm L} | \gamma^\nu | f_{\rm R} \rangle \nonumber \\
&=& \left\{ \begin{array}{lc}
i \frac{e^2}{6 s c^2_{\rm w}} \langle e^+_{\rm L} | \gamma_\nu | e^-_{\rm R} \rangle
\langle f_{\rm L} | \gamma^\nu | f_{\rm R} \rangle  \;\;\;, f_{\rm L}=t_{\rm L}, b_{\rm L} \\
i \frac{e^2}{s c^2_{\rm w}} \frac{2}{3} \langle e^+_{\rm L} | \gamma_\nu | e^-_{\rm R} \rangle
\langle f_{\rm R} | \gamma^\nu | f_{\rm L} \rangle \;\;, f_{\rm R}=t_{\rm R} \end{array} \right.
\label{eq:LB}
\end{eqnarray}
for top and bottom quarks.
In all cases, $\log \frac{M^2}{m_t^2}$
terms can be savely neglected to the accuracy we are working.
Thus we find for left handed quarks of the third generation:
\begin{eqnarray}
{\cal M}^{{\rm DY}_{\rm L}}_{\rm 1 loop}(p_{1},...,p_{4})&=&{\cal M}^{\rm DY}_{\rm Born}(p_{1},...,p_{4}) \left\{
 1 -
\frac{g^2}{16 \pi^2} \frac{1}{4}
\frac{m_t^2}{M^2} \delta_{f,t_{\rm L}/b_{\rm L}} \log \frac{s}{M^2}
\right\} \label{eq:tyukL}
\end{eqnarray}
For right handed external top quarks we have $\phi^\pm$, $\chi$ and $H$ corrections. In that case we
observe that the $Z \chi H$, $\gamma \phi^\pm$ and $Z \phi^\pm$ loops have an opposite sign
relative to the left handed case.
For the corrections corresponding to the topology shown on the right in Fig. \ref{fig:tyuk}
we must replace $Q_t$ in Eq. (\ref{eq:fsf}) by $Q_f-T^3_f=\frac{1}{6}$ for the $\phi^\pm$ graph.
The same contribution is obtained by adding the $H$ and $\chi$ loops and we find:
\begin{equation}
{\cal M}^{{\rm DY}_{\rm R}}_{\rm 1 loop}(p_{1},...,p_{4})={\cal M}^{\rm DY}_{\rm Born}(p_{1},...,p_{4}) \left\{ 1
 -
\frac{g^2}{16 \pi^2} \frac{1}{2}
\frac{m_t^2}{M^2} \delta_{f,t_{\rm R}} \log \frac{s}{M^2}
\right\} \label{eq:tyukR}
\end{equation}
At higher orders
we note that the exchange of gauge bosons inside the one loop process is subsubleading
and we arrive at the factorized form analogous to the Yukawa corrections for external
Goldstone bosons.
Since these corrections are of universal nature we can drop the specific reference to the
Drell-Yan process and
the application of the generalized Gribov-theorem for external fermion lines
to all orders yields:
\begin{equation}
{\cal M} (p_{1},...,p_{n};\mu ^{2})={\cal M}_{\rm 1 loop}(p_{1},...,p_{n})\exp
\left( -\frac{1}{2}\sum_{l=1}^{n_f}W^{\rm ew}_{l}(s,M^{2})\right) \label{eq:gt}
\end{equation}
where $W^{\rm ew}_{l}(s,M^{2})$ is given in Eq. (\ref{eq:Wew}) and the quantum numbers are
those of the external fermion lines.
Since at high energies all fermions can be considered massless we can again absorb the chiral
top-Yukawa corrections into universal splitting functions as in Ref. \cite{m1}. Thus in the electroweak
theory we find to next to leading order the corresponding probability for the emission of
gauge bosons from chiral fermions subject to the cutoff $\mu$ are given by Eqs. (\ref{eq:Wf})
and (\ref{eq:mpsol1}). The corrections from below the scale $M$ need to be included via
matching as described above.

For physical observables, soft real photon emission must be taken into account in an inclusive (or
semi inclusive) way and the parameter $\mu^2$ in (\ref{eq:mpsol2}) will be replaced by parameters
depending on the experimental requirements. This will be briefly discussed in section \ref{sec:si}.
Next we turn to longitudinal degrees of freedom after first reviewing the Goldstone boson 
equivalence theorem.

\subsection{The equivalence theorem} \label{sec:et}

At high energies, the longitudinal polarization states can be described with the polarization
vector is given in Eq. (\ref{eq:lpv}).

The connection between S-matrix elements and Goldstone bosons
is provided by the equivalence theorem \cite{gb,gb2,gb3}. It states that at tree level for S-matrix
elements for longitudinal
bosons at the high
energy limit $M^2/s\longrightarrow 0$ can be expressed through matrix elements
involving their associated
would-be Goldstone bosons. We write schematically in case of a single gauge boson:
\begin{eqnarray}
{\cal M}(W^\pm_{L}, \psi_{{\rm phys}}) &=& {\cal M}(\phi^\pm, \psi_{{\rm phys}}
) + {\cal O} \left(
\frac{M_{\rm w}}{\sqrt{s}} \right)
\label{eq:wet} \\
{\cal M}(Z_{L}, \psi_{{\rm phys}}) &=& i {\cal M}(\chi, \psi_{{\rm phys}}) + {\cal
O} \left(
\frac{M_{\rm z}}{\sqrt{s}} \right)
\label{eq:zet}
\end{eqnarray}
The problem with this statement of the equivalence theorem is that it holds only
at tree level
\cite{yy,bs}. For
calculations at higher orders,
additional terms enter which change Eqs. (\ref{eq:wet}) and (\ref{eq:zet}).

Because of the gauge invariance of the physical theory and the associated BRST
invariance, a modified
version of Eqs. (\ref{eq:wet}) and (\ref{eq:zet}) can be derived \cite{yy} which
reads
\begin{eqnarray}
k^\nu {\cal M}(W^\pm_{\nu}(k), \psi_{{\rm phys}}) &=& C_{\rm w} M_{\rm w} {\cal
 M}(\phi^\pm (k),
 \psi_{{\rm phys}}) + {\cal O} \left( \frac{M_{\rm w}}{\sqrt{s}} \right) \label{
 eq:wetp} \\
 k^\nu{\cal M}(Z_{\nu}(k), \psi_{{\rm phys}}) &=& i C_{\rm z} M_{\rm z} {\cal M}(
 \chi (k), \psi_{{\rm phys}})
 + {\cal O} \left( \frac{M_{\rm z}}{\sqrt{s}} \right) \label{eq:zetp}
 \end{eqnarray}
 where the multiplicative factors $C_{\rm w}$ and $C_{\rm z}$ depend only on wave
 function renormalization
 constants and mass counterterms. Thus, using the form of the longitudinal
 polarization vector of
 Eq. (\ref{eq:lpv}) we can write
 \begin{eqnarray}
 {\cal M}(W^\pm_{L}(k), \psi_{{\rm phys}}) &=&  C_{\rm w} {\cal M}(\phi^\pm (k),
 \psi_{{\rm phys}}) + {\cal O} \left( \frac{M_{\rm w}}{\sqrt{s}} \right)
 \label{eq:wgs} \\
 {\cal M}(Z_{L}(k), \psi_{{\rm phys}}) &=& i C_{\rm z} {\cal M}(\chi (k), \psi_{{
 \rm phys}})
 + {\cal O} \left( \frac{M_{\rm z}}{\sqrt{s}} \right) \label{eq:zgs}
 \end{eqnarray}
 We see that in principle,
 there are logarithmic loop corrections to the tree level equivalence theorem.
 The important point in our approach, however, is that the correction coefficients
 are not functions of the energy variable $s$:
 \begin{equation}
 C_{\rm w}= C_{\rm w} ( \overline{\mu}, M, g, g^\prime) \;\;, \;
 C_{\rm z}= C_{\rm z} ( \overline{\mu}, M, g, g^\prime) \label{eq:etcc}
 \end{equation}
 \begin{figure}
 \centering
 \epsfig{file=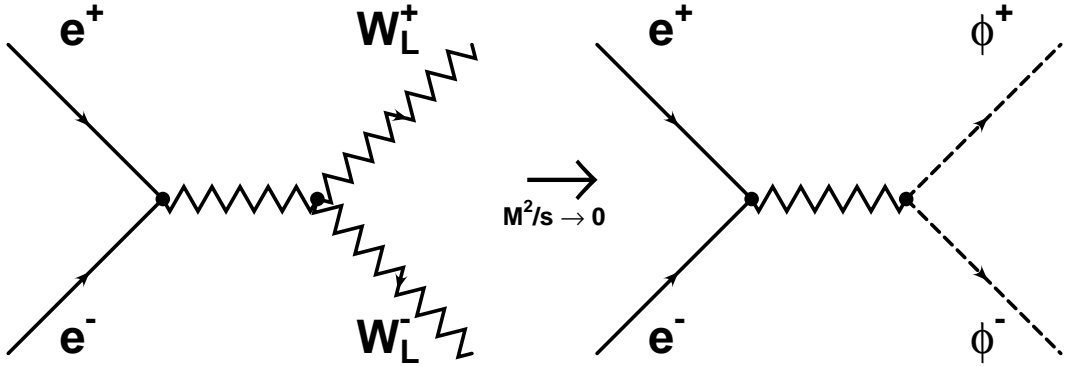,width=14cm}
 \caption{The pictorial Goldstone boson equivalence theorem for $W$-pair production in
 $e^+e^-$ collisions.
 The correct DL-asymptotics for longitudinally polarized bosons
 are obtained by using the quantum numbers of the charged would-be Goldstone scalars
 at high energies.}
 \label{fig:gsb}
 \end{figure}
 The pictorial form of the Goldstone boson equivalence theorem is depicted in Fig. \ref{fig:gsb}
 for longitudinal $W$-boson production at a linear $e^+e^-$ collider.
 In the following we denote the logarithmic variable $t\equiv \log \frac{s}{\mu^2}$, where $\mu$ is
 a cutoff on the transverse part of the exchanged virtual momenta $k$ of all involved particles, i.e.
 \begin{equation}
 \mu^2 \leq \mbox{\boldmath $k$}^2_\perp \equiv \min (2 (kp_l)(kp_j)/(p_lp_j))
 \label{eq:kplim}
 \end{equation}
 for all $j \neq l$. The non-renormalization group part of the evolution equation at high
 energies is given on the invariant matrix element level by Eq. (\ref{eq:mrg}):
\begin{equation}
\frac{\partial}{\partial t} {\cal M}(L(k), \psi_{{\rm phys}}) = K (t)
{\cal M}(L(k), \psi_{{\rm phys}}) \label{eq:leveq}
\end{equation}
and thus, after inserting Eqs. (\ref{eq:wgs}), (\ref{eq:zgs}) we find that the same evolution equation
also holds for ${\cal M}(\phi(k), \psi_{{\rm phys}})$. The notation here is $L=\{W_L^\pm, Z_L \}$
and $\phi=\{ \phi^\pm, \chi\}$, respectively.
Thus, the $\log \frac{s}{\mu^2}$ dependence in our approach is unrelated to the corrections
to the equivalence theorem, and in general, is unrelated to two point functions in a
covariant gauge at high energies where masses can be neglected.
This is a consequence of the physical on-shell renormalization scheme
where the $\overline{\rm MS}$ renormalization scale parameter ${\overline \mu} \sim M$.
Physically, this result can be understood by interpreting the correction terms $C_{\rm w}$
and $C_{\rm z}$ as corrections required by the gauge invariance of the theory in order
to obtain the correct renormalization group asymptotics of the physical Standard Model fields.
Thus, their origin is not related to Sudakov corrections.
In other words,
the results from the discussion of scalar QCD in section \ref{sec:ugt}
should be applicable to the subleading scalar sector
in the electroweak theory regarding a non-Abelian scalar gauge theory as the effective
description in this range according to ${\cal L}_{\rm symm}$ in Eq. (\ref{eq:lsym}). 
The only additional complication is
the presence of subleading Yukawa enhanced logarithmic corrections which will be discussed
below. It is also worth noticing, that at one loop, the authors of Ref. \cite{dp} obtain
the same result for the contributions from the terms of Eq. (\ref{eq:etcc}). In their approach,
where all mass-singular terms are identified and the renormalization scale ${\overline \mu}
=\sqrt{s}$,
these terms are canceled by additional corrections from mass and wave function counterterms.
At higher orders it is then clear that corrections from two point functions are subsubleading
in a covariant gauge.

\subsection{Longitudinal degrees of freedom} \label{sec:long}

According to the discussion of the previous section we can use Goldstone bosons in the high 
energy regime as the relevant degrees of freedom for longitudinal gauge boson production.
Thus, the Higgs boson and the would-be Goldstone bosons actually receive the same corrections
in high energy processes (up to purely electromagnetic terms).

Regulating the virtual infrared divergences with the transverse momentum cutoff as described above,
we find the virtual contributions to the splitting functions for external Goldstone and Higgs
bosons:
\begin{eqnarray}
&& P^V_{\phi^\pm \phi^\pm}(z) = P^V_{\chi \chi}(z) = P^V_{HH}(z) = \nonumber \\
&& \left[ \left( T_i(T_i+1)+  \tan^2 \theta_{\rm w} \left( \frac
{Y_i}{2}
\right)^2 \right) \left( - 2 \log \frac{s}{\mu^2} + 4 \right)
- \frac{3}{2} \frac{m^2_t}{M^2} \right] \delta(1-z)
\label{eq:pv}
\end{eqnarray}
The functions can be calculated directly from loop corrections to the elementary
processes in analogy to QCD \cite{aem,a,dot} and the logarithmic term corresponds to the
leading kernel of Ref. \cite{m1}.
We introduce virtual distribution functions which include only the effects of loop computations.
These fulfill the Altarelli-Parisi equations in analogy to Eq. (\ref{eq:apss}):
\begin{eqnarray}
\frac{\partial \phi(z,t)}{\partial t}&=& \frac{g^2}{8\pi^2} \int^1_z \frac{dy}{y} \phi(z/y,t)
P^V_{\phi \phi}(y)
\label{eq:app}
\end{eqnarray}
The splitting functions are related by
$P_{\phi \phi}=P^R_{\phi \phi}+P^V_{\phi \phi}$, where
$R$ denotes the contribution from real boson emission.
$P_{\phi \phi}$ is free of logarithmic
corrections and positive definite.

Inserting the virtual probability of Eq. (\ref{eq:pv}) into the Eq.
(\ref{eq:app}) we find:
\begin{eqnarray}
\phi(1,t)&=& \phi_0 \exp \left\{ - \frac{ g^2}{8 \pi^2} \left[ \left( T_i(T_i+1)+  \tan^2 \theta_{\rm w}
\left( \frac{Y_i}{2} \right)^2 \right) \left( \log^2 \frac{s}{\mu^2}
 - 4 \log \frac{s}{\mu^2} \right) \right. \right. \nonumber \\
 && \left. \left. \;\;\;\;\;\;\;\;\;\;\;\;\;\;\;\;\;\;\;\;\;\;\;\; +
 \frac{3}{2} \frac{m^2_t}{M^2} \log \frac{s}{\mu^2} \right] \right\} \label{eq:psol}
\end{eqnarray}
These functions describe the total contribution for the emission of virtual particles (i.e. $z=1$),
with all invariants large compared to the cutoff $\mu$, to the
densities $\phi(z,t)$ ($\phi = \{ \phi^\pm, \chi, {\rm H} \}$).
The normalization is not per line but on the level of the cross section.
For the invariant matrix element involving $n_\phi$ external
scalar particles we thus find at the subleading level:
\begin{eqnarray}
&& {\cal M} (p_1,...,p_n,g,\mu^2)={\cal M}_{\rm Born} (p_1,...,p_n,g)
\exp \left\{ - \frac{1}{2}
\sum^{n_\phi}_{i=1} W^\phi_i(s,\mu^2) \right\}
\label{eq:mgyuk}
\end{eqnarray}
where
\begin{equation}
 W^\phi_i(s,\mu^2) \!=\!  \frac{ g^2}{16 \pi^2} \!\! \left[ \! \left( T_i(T_i+1)+  \tan^2 \!
 \theta_{\rm w}
 \frac{Y^2_i}{4} \right) \!\! \left( \log^2 \frac{s}{\mu^2}- 4 \log \frac{s}{\mu^2}
 \! \right) \!\!
 + \frac{3}{2} \frac{m^2_t}{M^2} \log \frac{s}{\mu^2} \right] \label{eq:Wp}
\end{equation}
The functions $W^\phi_i$ correspond to the probability of emitting a virtual
soft and/or collinear gauge
boson from the particle $\phi$ subject to the infrared cutoff $\mu$. Typical diagrams contributing
to Eq. (\ref{eq:Wp}) in a covariant gauge are depicted in Fig. \ref{fig:nll}.
The universality of the splitting functions is crucial in obtaining the above result.

Again, since the Yukawa enhanced terms are novel features in broken gauge theories as compared
to the situation in QCD we use the non-Abelian generalization of the Gribov 
theorem in the following to prove the correctness
of our splitting function approach for specific processes using
on-shell renormalization of the external Goldstone bosons.

\begin{figure}
\centering
\epsfig{file=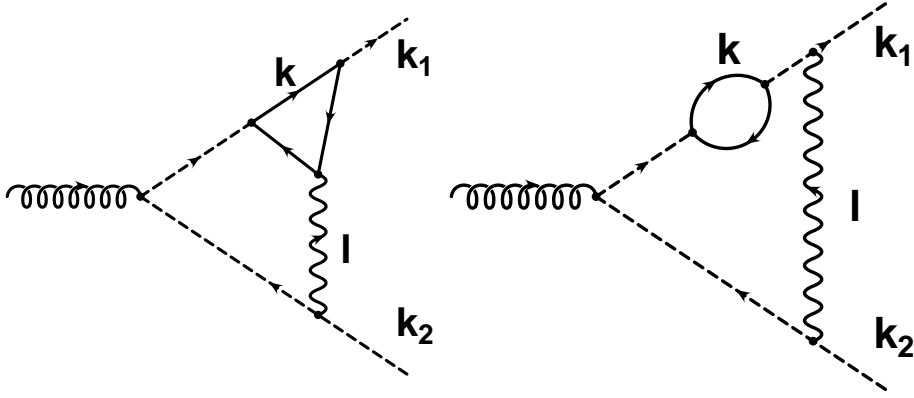,width=12cm}
\caption{Two loop corrections involving Yukawa couplings
of scalars to fermions.
The Ward identity in Eq. (\ref{eq:wi}) assures that in the Feynman gauge,
the sum
of both diagrams does not lead to
additional SL logarithms at the two
loop level.
Only corrections to the original one loop vertex
need
to be considered and lead to the exponentiation of Yukawa terms in the
SM to SL accuracy.}
\label{fig:wi}
\end{figure}

Since the three fermion loop is more complicated than the situation in
the fermionic sector above, we provide some more details in deriving the
respective Ward identity.
At the two loop level, we need to
consider the diagrams displayed in Fig. \ref{fig:wi}.
The corresponding two loop amplitudes read (neglecting $l$ outside the fermion loop):
\begin{eqnarray}
&&\!\!\!\!\!\!\!\!\!\!\!\!\!\! \int \!\! \frac{d^nl}{(4\pi)^n} \!\! \int \!\! 
\frac{d^nk}{(4\pi)^n} \frac{
(k_1-k_2)_\nu {\rm Tr} \left[
(G_r \omega_r + G_l \omega_l)( \rlap/ k - \rlap/ k_1 ) 2 \rlap/ k_2 ( \rlap/ k - \rlap/ k_1
+ \rlap/ l ) (G_r \omega_r + G_l \omega_l)
 \rlap/ k \right]}{(l^2-\lambda^2)(k_2+l)^2(k_1-l)^2k^2(k-k_1)^2(k-k_1+l)^2} \label{eq:ver} \\
&&\!\!\!\!\!\!\!\!\!\!\!\!\!\! \int \!\! \frac{d^nl}{(4\pi)^n} \!\! \int \!\! 
\frac{d^nk}{(4\pi)^n} \frac{
(k_1-k_2)_\nu {\rm Tr} \left[
(G_r \omega_r + G_l \omega_l) ( \rlap/ k - \rlap/ k_1 + \rlap/ l ) (G_r \omega_r + G_l \omega_l)
 \rlap/ k \right] 4 k_1 k_2}{(l^2-\lambda^2)(k_2+l)^2(k_1-l)^2k^2(k-k_1+l)^2(k_1-l)^2}
 \label{eq:se}
\end{eqnarray}
where we omit common factors and the scalar masses taking $M \sim \lambda$ for clarity.
The soft photon corrections must also be included via matching.
The $G_{r,l}$ denote the chiral Yukawa couplings and $\omega_{r,l}=\frac{1}{2} \left(1 \pm
\gamma_5 \right)$. The gauge coupling is again written in the symmetric basis.
For our purposes we need to
investigate terms containing three large logarithms in those diagrams. Since the fermion
loops at one loop only yield a single logarithm it is again clear that the gauge 
boson loop momentum
$l$ must be soft. Thus we need to show that the UV logarithm originating from the $k$
integration is identical (up to the sign) in both diagrams.
We can therefore neglect the loop momentum $l$ inside the fermion loop.
We find
for the fermion loop vertex $\Gamma^\mu(k_1^2,0,k_1^2)$ belonging to Eq. (\ref{eq:ver}):
\begin{eqnarray}
&& \frac{{\rm Tr} \left[(G_r \omega_r + G_l \omega_l)( \rlap/ k - \rlap/ k_1 ) \gamma^\mu ( \rlap/ k
- \rlap/ k_1 ) (G_r \omega_r + G_l \omega_l)\rlap/ k \right]}{k^2(k-k_1)^2(k-k_1)^2} \nonumber \\
&=& \frac{4G_rG_l \left(2 k_1^\mu (k^2-k_1k) +k^\mu (k_1^2-k^2) \right)}{k^2(k-k_1)^4}
\end{eqnarray}
This we need to compare with the self energy loop $\Sigma (k_1^2)$ from Eq. (\ref{eq:se}):
\begin{eqnarray}
&& \frac{\partial}{\partial {k_1}_\mu} \frac{{\rm Tr} \left[ (G_r \omega_r + G_l \omega_l)
( \rlap/ k - \rlap/ k_1 ) (G_r \omega_r + G_l \omega_l)  \rlap/ k \right]}{
k^2(k-k_1)^2} \nonumber \\
&=& \frac{\partial}{\partial {k_1}_\mu} \frac{4G_rG_l(k_1k-k^2)}{k^2(k-k_1)^2}
= 4 G_rG_l \frac{2k_1^\mu(k^2-k_1k)+k^\mu(k_1^2-k^2)}{k^2(k-k_1)^4}
\end{eqnarray}
In short we have established the analogous Ward identity in Eq. (\ref{eq:wi}) in
the longitudinal would-be Goldstone boson sector.
The relative sign is again
such that the generated SL logarithms of the diagrams in Fig. \ref{fig:wi} cancel each
other. The existence of such an identity expresses the fact
that also the longitudinal Yukawa
sector is gauge invariant. At higher orders this Ward identity ensures the 
corresponding cancellations to SL accuracy.
Also in an axial gauge the corrections can be seen to factorize accordingly since
in this gauge DL terms originate only from on-shell two point functions.

\begin{figure}
\centering
\epsfig{file=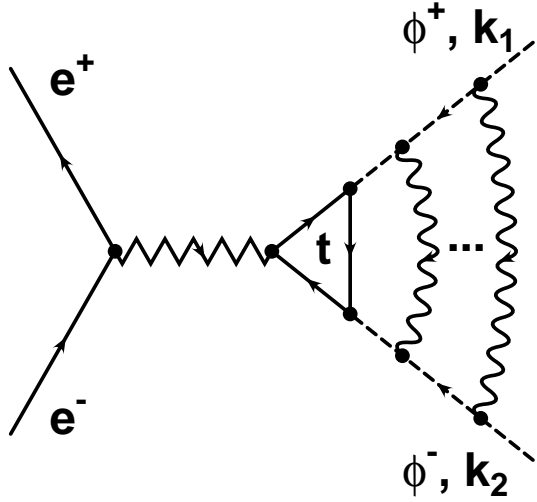,width=7cm}
\caption{A Feynman diagram yielding Yukawa enhanced logarithmic corrections in the
on-shell scheme. At higher orders, the subleading corrections are given in factorized
form according to the non-Abelian generalization of Gribov's theorem as described in the
text. Corrections from gauge bosons inside the top-loop give only sub-subleading contributions.
DL-corrections at two and higher loop order are given by gauge bosons coupling to (in principle
all) external legs as schematically indicated.}
\label{fig:yuk}
\end{figure}

We are thus left with gauge boson corrections to the original vertices in the on-shell
renormalization scheme such as depicted in Fig. \ref{fig:yuk}.
At high energies we can therefore employ the non-Abelian version
of Gribov's bremsstrahlung theorem.

Thus the corrections at higher orders factorize again with respect to the one loop amplitude
and ${\cal M}_{\rm ``Born''}(p_1,...,p_n) = {\cal M}_{\rm 1 loop}(p_1,...,p_n)$. Note that
the latter is also independent of the cutoff $\mu$ since the fermion mass serves as a
natural regulator.
The subleading corrections are then compared to the
general splitting
function approach yielding Eq. (\ref{eq:mgyuk}).

In the case of the amplitude of Fig. \ref{fig:yuk} we must use the quantum numbers of the
associated Goldstone bosons and we have the following Born amplitude
\begin{equation}
{\cal M}_{\rm Born}(p_{1},...,p_{4})=i \frac{e^2}{2s c^2_{\rm w}} \langle e^-_{\rm R}|
\gamma^\nu | e^+_{\rm L} \rangle (k_1-k_2)_\nu
\end{equation}
and at one loop we have two fermion loops contributing ($ttb$ and $bbt$).
The renormalization condition is provided by the requirement that the corrections
vanish at the weak scale, i.e. for $s=M^2$, which amounts to subtracting the
vertex for that case. The first diagram of the two is, according to the Feynman rules
of appendix \ref{sec:fr} using the physical fields, given by
\begin{eqnarray}
&&{\cal A}^{ttb}_{\rm 1 loop}(p_{1},...,p_{4})= 3 \sum_{\gamma, Z} \frac{e^4 m_t^2}{2 s M^2 s^2_{\rm w}}
\langle e^-_{\rm R}| \gamma^\nu | e^+_{\rm L} \rangle c_+^e \times \nonumber \\
&& \int \frac{d^nl}{(2 \pi)^n} \frac{ {\rm Tr} \left\{ \omega_- {\rlap/ l} \omega_+ (
{\rlap/ l}-{\rlap/ k_2}) \gamma_\nu (c^t_+ \omega_+ + c^t_- \omega_-) ( {\rlap/ l}+ {\rlap/ k_1})
\right\}}{(l^2-m_b^2+i \varepsilon)((l+k_1)^2-m_t^2+i \varepsilon)((l-k_2)^2-m_t^2+i \varepsilon)}
+ \delta^{ttb}_{\rm ct}
\nonumber \\ &&= \frac{3iQ_t}{16 \pi^2 c^2_{\rm w}} \frac{e^4 m_t^2}{2 s M^2 s^2_{\rm w}}
\langle e^-_{\rm R}| \gamma^\nu | e^+_{\rm L} \rangle (B_{23}-B_{23}^M)
(k_1-k_2)_\nu \label{eq:ttb}
\end{eqnarray}
where $\omega_\pm=\frac{1}{2} ( 1 \pm \gamma_5)$ and the chiral couplings are given by
$c^f_\pm=Q_f$ for the photon and $c^f_+=\frac{s_{\rm w}}{c_{\rm w}}Q_f$ and
$c^f_-= \frac{s_{\rm w}^2 Q_f-T^3_f}{s_{\rm w}c_{\rm w}}$ for Z-bosons respectively.
The counterterm $\delta^{ttb}_{\rm ct}$
is chosen such that the logarithmic corrections vanish for $s=M^2$.
Thus, the sum of the scalar functions is to logarithmic accuracy $B_{23}-B_{23}^M=-\log \frac{s}{M^2}$.
Analogously, we have for the $bbt$ quark loop:
\begin{eqnarray}
&&{\cal A}^{bbt}_{\rm 1 loop}(p_{1},...,p_{4})= 3 \sum_{\gamma, Z} \frac{e^4 m_t^2}{2 s M^2 s^2_{\rm w}}
\langle e^-_{\rm R}| \gamma^\nu | e^+_{\rm L} \rangle c_+^e \times \nonumber \\
&& \int \frac{d^nl}{(2 \pi)^n} \frac{ {\rm Tr} \left\{ \omega_+ (-{\rlap/ l}) \omega_- (-
{\rlap/ l}-{\rlap/ k_1}) \gamma_\nu (c^b_+ \omega_+ + c^b_- \omega_-) ( -{\rlap/ l}+ {\rlap/ k_2})
\right\}}{(l^2-m_t^2+i \varepsilon)((l+k_1)^2-m_b^2+i \varepsilon)((l-k_2)^2-m_b^2+i \varepsilon)}
+ \delta^{bbt}_{\rm ct}
\nonumber \\ &&= -\frac{3i(Q_b-T^3_b)}{16 \pi^2 c^2_{\rm w}} \frac{e^4 m_t^2}{2 s M^2 s^2_{\rm w}}
\langle e^-_{\rm R}| \gamma^\nu | e^+_{\rm L} \rangle (B_{23}-B_{23}^M)
(k_1-k_2)_\nu \label{eq:bbt}
\end{eqnarray}
Adding both results (\ref{eq:ttb}) and (\ref{eq:bbt}) we find
\begin{equation}
{\cal M}_{\rm 1 loop}(p_{1},...,p_{4}) =
{\cal M}_{\rm Born}(p_{1},...,p_{4}) \left\{ 1 -
\frac{g^2}{16 \pi^2} \frac{3}{2} \frac{m_t^2}{M^2} \log \frac{s}{M^2} \right\} \label{eq:yuk}
\end{equation}
and the all orders result to subleading accuracy is given by
\begin{equation}
{\cal M}(p_{1},...,p_{n};\mu ^{2})={\cal M}_{\rm 1 loop}(p_{1},...,p_{n})\exp
\left( -\frac{1}{2}\sum_{l=1}^{n}W^{\rm ew}_{l}(s,M^{2})\right) \label{eq:yukg}
\end{equation}
where $W^{\rm ew}_{l}(s,M^{2})$ is given in Eq. (\ref{eq:Wew}). 
The subleading Yukawa corrections from the Altarelli-Parisi in Eq. (\ref{eq:mgyuk}) agree with
the corresponding results from the application of the Gribov-theorem in Eq. (\ref{eq:yukg}).
\begin{figure}
\centering
\epsfig{file=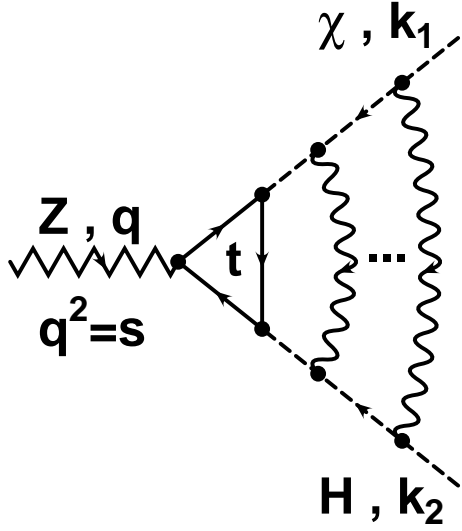,width=6cm}
\caption{A Feynman diagram yielding Yukawa enhanced logarithmic corrections to external
longitudinal Z-bosons and Higgs lines in the
on-shell scheme. At higher orders, the subleading corrections are given in factorized
form according to the non-Abelian generalization of Gribov's theorem as described in the
text. Corrections from gauge bosons inside the top-loop give only sub-subleading contributions.}
\label{fig:ZchiH}
\end{figure}
For longitudinal Z-boson and Higgs production, we note that there is only one non-mass suppressed
elementary vertex with two neutral scalars,
namely the $Z \chi H$ vertex. As mentioned above, universal terms are related
to the massless limit. For the ``Born amplitude'' of the Higgs-strahlung vertex we have
\begin{equation}
{\cal M}^{\rm Z \chi H}_{\rm Born} = \frac{e}{2 s_{\rm w} c_{\rm w}} (k_1^\nu - k_2^\nu)
\end{equation}
The universal Yukawa corrections to both external $\chi$ and $H$ states from an off shell
$Z$ line are then given by the corrections depicted in the inner fermion
loop of Fig. \ref{fig:ZchiH}. Here we find
\begin{eqnarray}
&&{\cal A}^{Z \chi H}_{\rm 1 loop}(p_{1},...,p_{3})= 3 \frac{e^3 m_t^2}{4 M^2 s^2_{\rm w}}
\times \nonumber \\
&& \int \frac{d^nl}{(2 \pi)^n} \frac{ {\rm Tr} \left\{ \gamma_5 ({\rlap/ l}) (
{\rlap/ l}-{\rlap/ k_2}) \gamma^\nu (c^t_+ \omega_+ + c^t_- \omega_-) ( {\rlap/ l}+ {\rlap/ k_1})
\right\}}{(l^2-m_t^2+i \varepsilon)((l+k_1)^2-m_t^2+i \varepsilon)((l-k_2)^2-m_t^2+i \varepsilon)}
+ \delta^{Z \chi H}_{\rm ct}
\nonumber \\ &&= \frac{6T^3_t}{16 \pi^2 s_{\rm w} c_{\rm w}} \frac{e^3 m_t^2}{4 M^2 s^2_{\rm w}}
(B_{23}-B_{23}^M)
(k^\nu_1-k^\nu_2)
\end{eqnarray}
and thus
\begin{equation}
{\cal M}^{Z \chi H}_{\rm 1 loop}(p_{1},...,p_{3}) = {\cal M}^{\rm Z \chi H}_{\rm Born} \left\{ 1
- \frac{3}{2} \frac{e^2 m^2_t}{ 16 \pi^2 s^2_{\rm w} M^2}
\log \frac{s}{M^2} \right\} \label{eq:ZchiH}
\end{equation}
From the same line of reasoning as for the charged Goldstone bosons we find that the
all orders result is given by Eq. (\ref{eq:yukg}). At the subleading level, this is equivalent to
the corresponding corrections obtained in Eq. (\ref{eq:mgyuk}).

In addition to the Sudakov corrections in Eq. (\ref{eq:Wp}) we also have to include terms
corresponding to the renormalization of the mass terms in the Yukawa coupling of the
Born amplitude $\left( \sim \frac{m_t^2}{M^2}, \frac{m_H^2}{M^2} \right)$
at the one loop level \cite{dp}. At
higher orders, mass renormalization terms are connected to two point functions and thus
subsubleading. The relevant higher order SL-RG terms, however, will be discussed section \ref{sec:ewrg}.

\subsection{Semi-inclusive cross sections} \label{sec:si}

In order to make predictions for observable cross sections, the unphysical infrared cutoff $\mu^2$
has to be replaced with a cutoff $\mu^2_{\rm expt}$, related to the lower bound of
$\mbox{\boldmath $k$}_\perp^2$ of the other virtual particles
of those gauge bosons emitted in the process which are not included in the cross section.
We assume that $\mu^2_{\rm expt}< M^2$, so that the non-Abelian component of the photon is not essential.
The case $\mu^2_{\rm expt}>M^2$ is much more complicated and is discussed in Ref. \cite{flmm} through
two loops at the DL level.

We can write
the expression for the semi-inclusive cross section in a compact way as follows:
\begin{eqnarray}
d\sigma (p_{1}, \ldots, p_{n},g,g^\prime,\mu_{\rm expt}) &=& d\sigma_{\rm elastic}(p_{1},
\ldots ,p_{n},g,g^\prime,\mu) \nonumber \\ && \times \exp
(w_{\rm expt}^{\gamma }(s,m_i,\mu,\mu_{\rm expt}))
\end{eqnarray}
The $\mu$ dependence in this expression cancels and the semi-inclusive cross section depends only on
the parameters of the experimental requirements.
In terms of all the virtual corrections discussed above the full result is given by:
\begin{eqnarray}
&& d\sigma (p_{1}, \ldots, p_{n},g,g^\prime,\mu_{\rm expt}) = d\sigma_{\rm Born} (p_{1},
\ldots ,p_{n},g,g^\prime )
\nonumber \\ && \times \exp \left\{ - \sum^{n_g}_{i=1} W_{g_i} (s,M^2)
- \sum^{n_f}_{i=1} W_{f_i} (s,M^2) - \sum^{n_\phi}_{i=1} W_{\phi_i} (s,M^2)
\right\} \nonumber \\
&&\times \exp \left[ - \sum_{i=1}^{n_f} \left( w_{f_i}(s,\mu^2)
- w_{f_i}(s,M^2) \right)
- \sum_{i=1}^{n_{\rm w}} \left( w_{{\rm w}_i}(s,\mu^2)
- w_{{\rm w}_i}(s,M^2) \right) \right. \nonumber \\
&& \;\;\;\;\;\;\;\;\;\;\; \left. - \sum_{i=1}^{n_\gamma} w_{\gamma_i}(M^2,m_j^2)
\right]
\times \exp \left( w_{\gamma_{\rm expt}} (s,m_i,\mu,\mu_{\rm expt})
\right) \label{eq:si}
\end{eqnarray}
where $n_g$ denotes the number of transversely polarized gauge bosons
and $n_f$ the number of {\it external} fermions.
This expression omits all RG corrections, even at the one loop level.
The functions $W$ and $w$ correspond to the logarithmic probability to emit
a soft and/or collinear particle per line, where the capital letters denote
the probability in the high energy effective theory and the lower case letter the
corresponding one from pure QED corrections below the weak scale. The matching
condition is implemented such that for $\mu=M$ only the high energy
solution remains.
For the contribution from scalar fields $\phi=\{\phi^\pm,\chi,H\}$ above the
scale $M$ we have $W_{\phi_i}(s,M^2)$ in Eq. (\ref{eq:Wp})
with $\alpha=g^2/4\pi$ and $\tan^2 \theta_{\rm w}= \alpha^\prime / \alpha$.
The last term is written as a logarithm containing the top quark mass $m_t$ rather than
the weak scale $M$ since these terms always contain $m_t$ as the heaviest mass in the
loop correction \cite{m3}.
$W_{f_i}(s,M^2)$ and $W_{g_i}$ are given in Eqs. (\ref{eq:Wf}) and (\ref{eq:Wg}) 
respectively.
Again we note that for external photon and Z-boson states we must include
the mixing appropriately as discussed in section \ref{sec:tr}.
For the terms entering from contributions below the weak scale we have 
$w_{f_i}(s,\mu^2)$ given in Eq. (\ref{eq:wf}), $w_{{\rm w}_i}$
in Eq. (\ref{eq:ww}) and $w_{\gamma_i}$ in Eq. (\ref{eq:wga}) 
for the virtual corrections.
For real photon emission we have in the soft
photon approximation: 
\begin{eqnarray}
w_{\gamma_{\rm expt}}(s,m_i,\mu,\mu_{\rm expt})
\!&=&\! \left\{ \begin{array}{lc}
\sum_{i=1}^n \frac{e_i^2}{(4 \pi)^2} \left[
- \log^2 \frac{s}{\mu^2_{\rm expt}}
+ \log^2 \frac{s}{\mu^2}- 3 \log \frac{s}{\mu^2} \right]
& , m_i \ll \mu \\
\sum_{i=1}^n \frac{e_i^2}{(4 \pi)^2} \left[ \left( \log
\frac{s}{m_i^2} -  1 \right)
2 \log \frac{m_i^2}{\mu^2} + \log^2 \frac{s}{m_i^2}
\right. \\ \left. \;\;\;\;\;\;\;\;\;\;\;\;\;\;\;\;\;
 - 2 \log \frac{s}{\mu^2_{\rm expt}} \left( \log \frac{s}{m_i^2} -
1 \right) \right]
& ,
\mu \ll m_i \end{array} \right.  
\end{eqnarray}
where $n$ is the number of external lines
and the upper case applies only to fermions since for $W^\pm$
we have $\mu < M$. Note that in all contributions from the regime $\mu<M$ we have
kept mass terms inside the logarithms. This approach is valid in the entire Standard
Model up to terms of order ${\cal O} \left( \log \frac{m_t}{M} \right)$.
Note, however, that in the Yukawa enhanced terms the replacement $M \longrightarrow m_t$
is legitimate.
The overall $\mu$-dependence in the semi-inclusive cross section cancels and we only have
a dependence on the parameter $\mu_{\rm expt}$ related to the experimental energy resolution.
All universal electroweak Sudakov corrections at DL and SL level exponentiate.

\subsection{Electroweak RG corrections} \label{sec:ewrg}

The way to implement the SL-RG corrections is clear from the discussion in section
\ref{sec:rg}. At high energies,
the DL phase space is essentially described by
an unbroken $SU(2)\times U(1)$ theory in which we can calculate the high energy contributions.
In this regime, all particle masses can be neglected and we have to consider the following
virtual electroweak DL phase space integral with running couplings in each gauge group:
\begin{eqnarray}
{\widetilde W}^{RG}_{i_V} \left(s,\mu^2 \right) &=&  \frac{1}{2\pi}
\int^s_{\mu^2} \frac{d {\mbox{\boldmath $k$}_{\perp }^{2}}}{
{\mbox{\boldmath $k$}_{\perp }^{2}}} \int^1_{{\mbox{\boldmath $k$}_{\perp }^{2}}/s}
\frac{d v}{v}  \left\{
 \frac{T_i(T_i+1)\alpha(\mu^2)}{1+c \; \log \frac{{\mbox{\boldmath $k$}_{\perp }^{2}}}{\mu^2}}
 + \frac{(Y^2_i/4) \alpha^\prime(\mu^2)}{1+c^\prime \; \log
 \frac{{\mbox{\boldmath $k$}_{\perp }^{2}}}{\mu^2}} \right\} \nonumber \\
&=& \frac{\alpha(\mu^2) T_i(T_i+1)}{2 \pi } \left\{ \frac{1}{c} \log \frac{s}{\mu^2}
\left( \log \frac{\alpha(\mu^2)}{\alpha
(s)} - 1 \right) + \frac{1}{c^2}
\log \frac{\alpha(\mu^2)}{\alpha(s)} \right\} \nonumber \\
&& +\frac{\alpha^\prime(\mu^2) Y^2_i}{8 \pi } \left\{ \frac{1}{c^\prime} \log \frac{s}{\mu^2}
\left( \log \frac{\alpha^\prime(\mu^2)}{\alpha^\prime
(s)} - 1 \right) + \frac{1}{{c^\prime}^2}
\log \frac{\alpha^\prime(\mu^2)}{\alpha^\prime(s)} \right\} \label{eq:vewrg}
 \end{eqnarray}
where $\alpha(\mu^2)=g^2(\mu^2)/4 \pi$, $\alpha^\prime(\mu^2)={g^\prime}^2(\mu^2)/4 \pi$,
$c=\alpha(\mu^2)\beta_0 / \pi$
and analogously, $c^\prime=\alpha^\prime(\mu^2)\beta^\prime_0 / \pi$.
In each case, the correct non-Abelian or Abelian limit is reproduced by letting the corresponding
couplings of the other gauge group approach zero. In this way it is easy to see that the
argument of the running couplings can only be what appears in Eq. (\ref{eq:vewrg}).

The form of Eq. (\ref{eq:vewrg}) is valid for fermions, transversely and longitudinally polarized
external lines but (omitted) subleading terms as well as the quantum numbers of the weak isospin $T_i$
and the weak hypercharge $Y_i$ differ.
In order to implement the missing soft photon contribution, we choose the
analogous form of solution in Eq. (\ref{eq:sp}) and have to implement it in such a way that
for $\mu=M$ Eq. (\ref{eq:vewrg}) is obtained.
The full result for the respective semi-inclusive cross sections is then given
by:
\begin{eqnarray}
&& d\sigma^{\rm RG} (p_{1}, \ldots, p_{n},g,g^\prime,\mu_{\rm expt}) = d\sigma_{\rm Born} (p_{1},
\ldots ,p_{n},g(s),g^\prime (s))
\nonumber \\ && \times \exp \left\{ - \sum^{n_g}_{i=1} W^{\rm RG}_{g_i} (s,M^2)     
- \sum^{n_f}_{i=1} W^{\rm RG}_{f_i} (s,M^2) - \sum^{n_\phi}_{i=1} W^{\rm RG}_{\phi_i} (s,M^2)
\right\} \nonumber \\
&& \times \exp \left[ - \sum_{i=1}^{n_f} \left( w^{\rm RG}_{f_i}(s,\mu^2)
- w^{\rm RG}_{f_i}(s,M^2) \right)
- \sum_{i=1}^{n_{\rm w}} \left( w^{\rm RG}_{{\rm w}_i}(s,\mu^2)
- w^{\rm RG}_{{\rm w}_i}(s,M^2) \right) \right. \nonumber \\
&& \;\;\;\;\;\;\;\;\;\;\; \left. - \sum_{i=1}^{n_\gamma} w_{\gamma_i}(M^2,m_j^2)
\right]
\times \exp \left( w^{\rm RG}_{\gamma_{\rm expt}} (s,m_i,\mu,\mu_{\rm expt})
\right) \label{eq:sirg}
\end{eqnarray}
where $n_f$ denotes here again the number of {\it external} fermions.
The argument of the gauge couplings in the Born cross section indicate the
one loop renormalization of the couplings which is not included in the exponential expressions
but which at one loop is genuinely subleading:
\begin{eqnarray}
\alpha (s) &=& \alpha (M^2) \left( 1 - \beta_0 \frac{\alpha (M^2)}{\pi} \log \frac{s}{M^2} \right) \\
\alpha^\prime (s) &=& \alpha^\prime (M^2) \left( 1 - \beta^\prime_0 \frac{\alpha^\prime
(M^2)}{\pi}\log \frac{s}{M^2} \right)
\end{eqnarray}
where $\alpha (M^2)= e^2(M^2) / 4 \pi s_{\rm w}^2$ and $\alpha^\prime (M^2)=
e^2(M^2) / 4
\pi c_{\rm w}^2$ with
\begin{equation}
e^2(M^2)=e^2 \left(1+ \frac{1}{3} \frac{e^2}{4 \pi^2} \sum_{j=1}^{n_f
} Q_j^2 N^j_C \log
\frac{M^2}{m_j^2} \right)
\end{equation}
and $e^2/ 4 \pi = 1/137$.
If there are non-suppressed mass ratios in the Born term, also these terms need to be renormalized
at one loop (see Ref. \cite{dp}).
Higher order mass renormalization terms would then be sub-subleading.
The function $W^{\rm RG}_{\phi_i} (s,M^2)$ is given by
\begin{eqnarray}
 W^{\rm RG}_{\phi_i}(s,M^2)\!&=&\! \frac{\alpha(M^2)
 T_i(T_i+1)}{2 \pi } \left\{ \frac{1}{c} \log \frac{s}{M^2}
\left( \log \frac{\alpha(M^2)}{\alpha
(s)} - 1 \right) + \frac{1}{c^2}
\log \frac{\alpha(M^2)}{\alpha(s)} \right\} \nonumber \\
\!&&\!\!\!\!\!\!\!\!\!\!\!\!\!\!\!\!\!\!\!\! +\frac{\alpha^\prime(M^2) Y^2_i}{8 \pi } \left\{ \frac{1}{c^\prime} \log \frac{s}{M^2}
\left( \log \frac{\alpha^\prime(M^2)}{\alpha^\prime
(s)} - 1 \right) + \frac{1}{{c^\prime}^2}
\log \frac{\alpha^\prime(M^2)}{\alpha^\prime(s)} \right\} \nonumber \\
\!&&\!\!\!\!\!\!\!\!\!\!\!\!\!\!\!\!\!\!\!\! - \left[ \left( \frac{ \alpha(M^2)}{4 \pi}  T_i(T_i+1)+
\frac{ \alpha^\prime(M^2)}{4 \pi} \frac{Y^2_i}{4} \right)  4 \log \frac{s}{M^2}
- \frac{3}{2} \frac{ \alpha(M^2)}{4 \pi} \frac{m^2_t}{M^2} \log \frac{s}{m_t^2} \right] \label{eq:WpRG}
 \end{eqnarray}
where we again have $m_t$ in the argument of the Yukawa enhanced correction \cite{m3}.
Analogously for fermions we have:
\begin{eqnarray}
W^{\rm RG}_{f_i}(s,M^2)&=& \frac{\alpha(M^2)
 T_i(T_i+1)}{2 \pi } \left\{ \frac{1}{c} \log \frac{s}{M^2}
\left( \log \frac{\alpha(M^2)}{\alpha
(s)} - 1 \right) + \frac{1}{c^2}
\log \frac{\alpha(M^2)}{\alpha(s)} \right\} \nonumber \\
&& +\frac{\alpha^\prime(M^2) Y^2_i}{8 \pi } \left\{ \frac{1}{c^\prime} \log \frac{s}{M^2}
\left( \log \frac{\alpha^\prime(M^2)}{\alpha^\prime
(s)} - 1 \right) + \frac{1}{{c^\prime}^2}
\log \frac{\alpha^\prime(M^2)}{\alpha^\prime(s)} \right\} \nonumber \\
&& - \left[ \left( \frac{ \alpha(M^2)}{4 \pi} T_i(T_i+1)+
\frac{ \alpha^\prime(M^2)}{4 \pi}\frac{Y^2_i}{4} \right)  3 \log \frac{s}{M^2}
\right. \nonumber \\ && \left.
- \frac{ \alpha(M^2)}{4 \pi} \left( \frac{1+\delta_{i,{\rm R}}}{4} \frac{m^2_i}{M^2} + \delta_{i,{\rm L}}
\frac{m^2_{i^\prime}}{4 M^2} \right)
\log \frac{s}{m_t^2} \right] \label{eq:WfRG}
\end{eqnarray}
The last term contributes only for left handed bottom and for top quarks as mentioned above and
$f^\prime$ denotes the corresponding isospin partner for left handed fermions.
\begin{eqnarray}
&&  W^{\rm RG}_{g_i}(s,M^2)= \frac{\alpha(M^2)
 T_i(T_i+1)}{2 \pi } \left\{ \frac{1}{c} \log \frac{s}{M^2}
\left( \log \frac{\alpha(M^2)}{\alpha
(s)} - 1 \right) + \frac{1}{c^2}
\log \frac{\alpha(M^2)}{\alpha(s)} \right\} \nonumber \\
&& +\frac{\alpha^\prime(M^2) Y^2_i}{8 \pi } \left\{ \frac{1}{c^\prime} \log \frac{s}{M^2}
\left( \log \frac{\alpha^\prime(M^2)}{\alpha^\prime
(s)} - 1 \right) + \frac{1}{{c^\prime}^2}
\log \frac{\alpha^\prime(M^2)}{\alpha^\prime(s)} \right\} \nonumber \\
&&- \left( \delta_{i,{\rm W}} \frac{\alpha(M^2)}{\pi} \beta_0 + \delta_{i,{\rm B}}
\frac{\alpha^\prime(M^2)}{\pi} \beta^\prime_0 \right) \log \frac{s}{M^2} \label{eq:WgRG}
\end{eqnarray}
Again we note that for external photon and Z-boson states we must include
the mixing appropriately as discussed in Ref. \cite{m1}.
For the terms entering from contributions below the weak scale we have for fermions:
\begin{equation}
w^{\rm RG}_{f_i}(s,\mu^2)\! = \!\left\{ \begin{array}{lc} \!
\frac{e^2_i}{8 \pi^2 } \left\{ \frac{1}{c} \log \frac{s}{\mu^2}
\left( \log \frac{e^2(\mu^2)}{e^2
(s)} - 1 \right) + \frac{1}{c^2}
\log \frac{e^2(\mu^2)}{e^2(s)}
- \frac{3}{2} \log \frac{s}{\mu^2} \right\} \!\!& , \; m_i \ll \mu \\ \!
\frac{e_i^2}{8 \pi^2 } \left\{ \frac{1}{c} \log \frac{s}{m^2}
\left( \log \frac{e^2(\mu^2)}{e^2
(s)} - 1 \right) -\frac{3}{2} \log \frac{s}{m^2} - \log \frac{ m^2}{\mu^2} \right. \\
\left. \;\;\;\;\;\; + \frac{1}{c^2}
\log \frac{e^2(m^2)}{e^2(s)} \left( 1- \frac{1}{3}\frac{e^2}{4 \pi^2}
\sum_{j=1}^{n_f } Q_j^2 N^j_C \log \frac{m^2}{m_j^2}
\right) \right\}
\!\!& , \; \mu
\ll m_i\end{array} \right.
\end{equation}
where $c=-\frac{1}{3}\frac{e^2}{4 \pi^2} \sum^{n_f}_{j=1}Q^2_j N_C^j$.
Analogously, for external W-bosons and photons we find:
\begin{eqnarray}
w^{\rm RG}_{{\rm w}_i}(s,\mu^2) &=&
\frac{e_i^2}{8 \pi^2 } \left\{ \frac{1}{c} \log \frac{s}{M^2}
\left( \log \frac{e^2(\mu^2)}{e^2
(s)} - 1 \right) - \log \frac{ M^2}{\mu^2} \right. \\
&& \left. \;\;\;\;\;\;+ \frac{1}{c^2}
\log \frac{e^2(M^2)}{e^2(s)} \left( 1- \frac{1}{3}\frac{e^2}{4 \pi^2}
\sum_{j=1}^{n_f } Q_j^2 N^j_C \log \frac{M^2}{m_j^2}
\right) \right\}
\end{eqnarray}
\begin{equation}
w_{\gamma_i}(M^2,\mu^2) = \left\{ \begin{array}{lc}
\frac{1}{3} \sum_{j=1}^{n_f} \frac{e_j^2}{4 \pi^2} N^j_C
\log \frac{M^2}{\mu^2} & , \;\;\; m_j \ll \mu \\
\frac{1}{3} \sum_{j=1}^{n_f} \frac{e_j^2}{4 \pi^2} N^j_C \log \frac{M^2}{m_j^2}
& , \;\;\; \mu \ll m_j\end{array} \right.
\end{equation}
Note that the function $w_{\gamma_i}(M^2,\mu^2)$ does not receive
any RG corrections to the order we are working since it contains only SL terms.
For the virtual corrections and for real photon emission we have in the soft
photon approximation:
\begin{eqnarray}
w^{\rm RG}_{\gamma_{\rm expt}}(s,m_i,\mu,\mu_{\rm expt})
\!&=&\! \left\{ \begin{array}{lc}
\sum_{i=1}^n \frac{e_i^2}{8 \pi^2}
 \left\{ \frac{1}{c} \log \frac{s}{\mu^2} \left( \log \frac{
e^2(\mu^2)}{e^2 (\mu^2_{\rm expt})} - 1 \! \right) - \frac{1}{c} \log \frac{\mu^2_{\rm expt}}{
s} \right. \\ \left.
\;\;\;\;\;\;\;\;\;\;\;\;\;\;\;\; + \frac{1}{c^2} \log \frac{e^2 (\mu^2)}{e^2 (\mu^2_{\rm expt})}
 -\frac{3}{2} \log \frac{s}{\mu^2}
\right\}
\;\;\;\;\;\;\;\;\;\;\;\;\;\;\; , m_i \ll \mu \\
\sum_{i=1}^n \frac{e_i^2}{8 \pi^2}
 \left\{ \frac{1}{c} \log \frac{s}{m^2}
 \left( \log \frac{ e^2 (\mu^2)}{e^2 (\mu_{\rm expt}^2m^2/s)} - 1  \right) \right.
\\ \;\;\;\;\;\;\;\;\;\;\;\;\;\;\;\;
  + \frac{1}{c^2} \log \frac{e^2(\mu^2_{\rm expt}m^2/s)}{e^2(\mu_{\rm expt}^2)}
\left( 1-
 \frac{1}{3} \frac{e^2}{4 \pi^2} \sum^{n_f}_{j=1}
 Q^2_j N_C^j
\log \frac{
\mu_{\rm expt}^2}{m_j^2} \right)
\\ \left. \;\;\;\;\;\;\;\;\;\;\;\;\;\;\;\;
- \log \frac{m^2}{\mu^2} + \log \frac{s}{\mu_{\rm expt}^2}
\right\}
\;\;\;\;\;\;\;\;\;\;\;\;\;\;\;\;\;\;\;\;\;\;\;\;\;
,\mu \ll m_i \end{array} \right. 
\end{eqnarray}
where $n$ is the number of external lines and $n_f$ fermions propagating in
the loops folded with the DL integrals.
The upper case applies only to fermions since for $W^\pm$
we have $\mu < M$. Note that in all contributions from the regime $\mu<M$ we have
kept mass terms inside the logarithms.
For the running above the weak scale $M$ we use only the massless $\beta_0$, $\beta_0^\prime$
terms with $n_{gen}=3$. This approach is valid in the entire Standard
Model up to terms of order ${\cal O} \left( \log \frac{m_t}{M} \right)$.

\begin{figure}
\centering
\epsfig{file=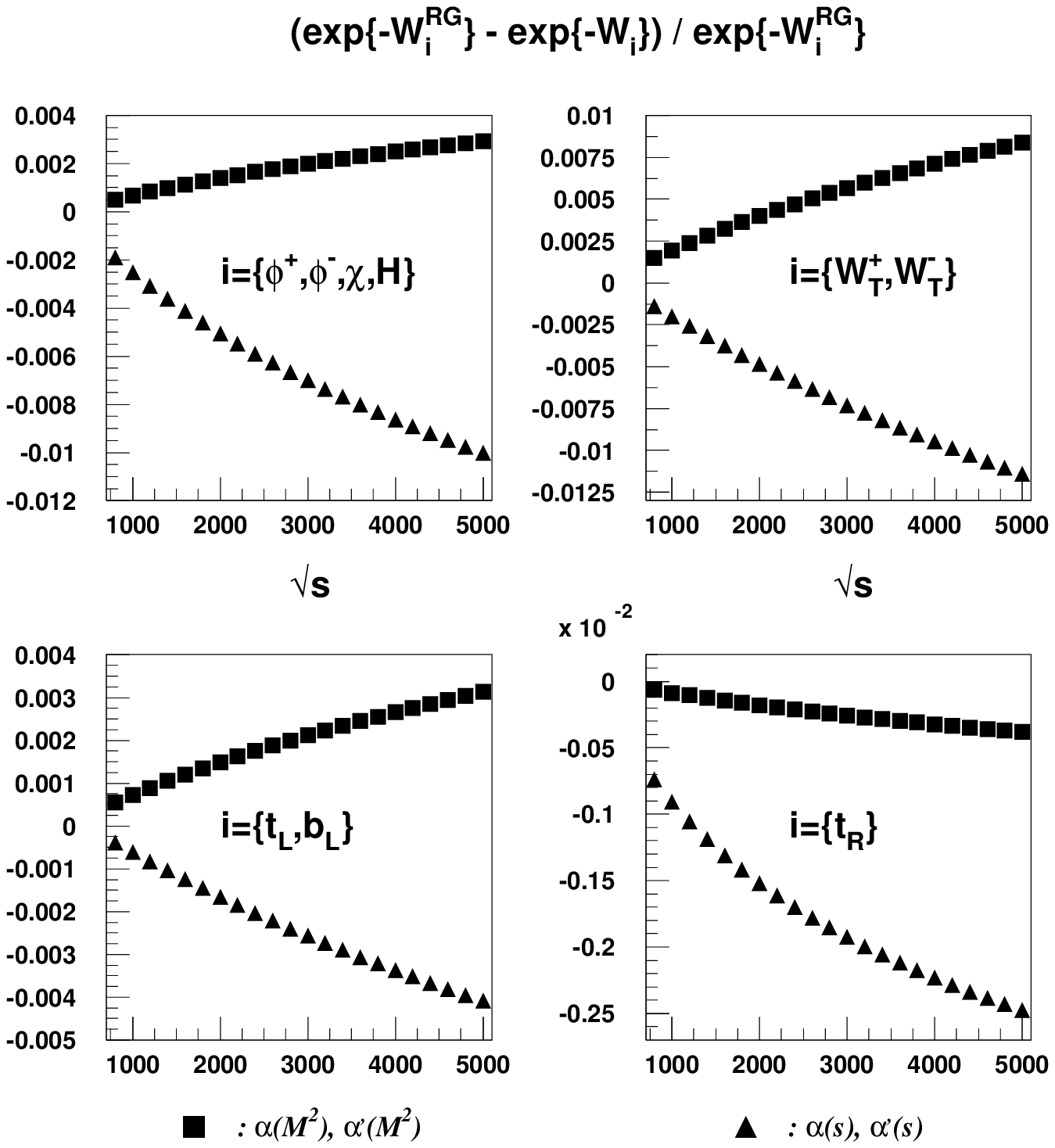,width=15.5cm}
\caption{This figure compares the renormalization group improved probabilities $W^{\rm RG}_i$
with the conventional Sudakov exponentials $W_i$ for various external particle lines. The
comparison is made with the indicated scale choices for the functions $W_i$ and takes into
account only the RG corrections from the scale $M$ to $\sqrt{s}$.
Taking the difference between the two curves is a measure of the uncertainty removed
by the RG effects. The variations in
the scale of the coupling in the $W_i$ functions is largest in the scalar (Goldstone and
Higgs boson) sector
and for transverse $W^\pm$ where the effect is
about $0.8\%$ at 2 TeV per line on the level of the cross section. In general, the RG
improved form factors differ by fractions of one percent per line and need to be taken into
account at future colliders if the experimental accuracy is in the percentile regime.}
\label{fig:wrel}
\end{figure}

In order to briefly discuss the size of the SL-RG corrections obtained 
here, we present a numerical comparison. For this purpose we will only compare the terms which are
new in the RG analysis, i.e. the running from the weak scale $M$ to $\sqrt{s}$. We
are thus interested in effects starting at the two loop level and want to compare the relative
size of the RG-improved form factors to the pure Sudakov terms. It is therefore of interest
to compare the ratios $\left(e^{\{-W^{\rm RG}_i\}}-e^{\{-W_i\}} \right)/ e^{\{-W^{\rm RG}_i\}}$
for the various particle labels $i$. Since the physical scales in the problem are given
by $M$ and $\sqrt{s}$, the lower and upper limits of the couplings are given accordingly
by these scales for the functions $W_i$. Fig. \ref{fig:wrel} compares the respective
ratios for various SM particles.
For definiteness we take $M=80$ GeV, $m_t=174$ GeV, $s_{\rm w}^2=0.23$,
$\alpha (M^2)=1/128/s_{\rm w}^2$, $\alpha^\prime (M^2)=1/128/c_{\rm w}^2$, $\beta_0
=19/24$ and $\beta_0^\prime=-41/24$.
The difference between the curves using $M^2$ and those
using $s$ as the scales in the conventional Sudakov form factors is a measure of the
inherent scale uncertainty which is removed by the RG-improved Sudakov form factors
$W^{\rm RG}_i$. The largest effect is obtained in the gauge boson sector \cite{m4}.
For external $\{ \phi^+,
\phi^-, \chi, H \}$ particles we have at 1 TeV a difference between the curves of
about $0.35\%$ per line on the level of the cross section, growing to $0.65\%$ at 2 TeV.
The situation is very similar for transversely polarized $W^+, W^-$ particles where
it reaches about 0.4\% at 1 TeV and 0.8\% at 2 TeV per line on the cross section
level.
For left handed quarks of the third generation the size of the corrections is about
0.15\% at 1 TeV per line on the level of the cross section and 0.33\% at 2 TeV. These corrections
are thus considerably smaller and only needed if precisions below the one percent level are
necessary from the theory side. For right handed top quarks the effect is even smaller
since only the running of $\alpha^\prime$ enters and it is
thus negligible for most applications. The form of the two curves in case of right handed
tops differs markedly from the other three cases because at the energies displayed, the
dominant effect is actually due to subleading Yukawa enhanced corrections ($\sim
\alpha$) since the DL
terms are proportional to $\alpha^\prime$ and since the ratio $m_t^2/M^2$ is of the size of
an additional logarithm for these values of $\sqrt{s}$.

In general it can be seen that-where the DL terms dominate-the renormalization group
improved results are indeed in-between the upper and lower bounds given by the respective
scale choices in the conventional Sudakov form factors. Indeed also for right handed top
quarks this pattern is observed if only DL corrections are taken into account.

It should be emphasized again that also the QED-RG corrections can be sizable since large mass
ratios with light particles occur. These should of course also be implemented in a full
SM prediction at TeV energies keeping in mind that one must always be far above any
particle thresholds.

\subsection{Angular dependent corrections} \label{sec:at}

\begin{figure}[t]
\centering
\epsfig{file=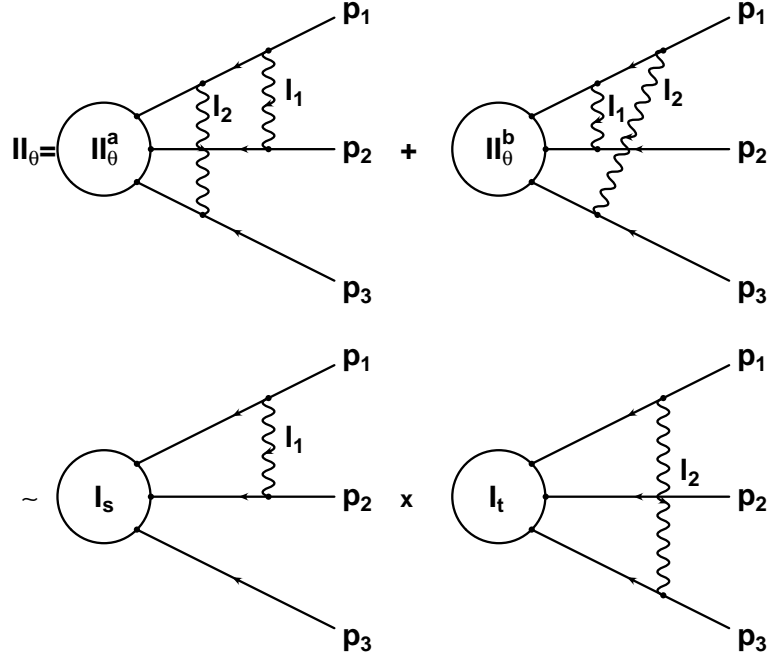,width=10cm}
\caption{Angular dependent two loop on-shell diagrams. The sum of II$^a_\theta$ and
II$^b_\theta$ factorizes in massive QED and the right handed SM into the product of the two one loop
corrections (each with a different invariant and mass terms) in leading order.}
\label{fig:qedang}
\end{figure}

In this section we discuss the important contribution from angular dependent 
logarithmic corrections. In massless four fermion processes these were
first given by Ref. \cite{kps} at the SL level and in Ref. \cite{kmps} at
the SSL level. Since the size of the observed corrections is large, these
terms are important for future collider phenomenology. In Ref. \cite{mang}
a general way of treating SL angular terms for arbitrary processes including
mass terms was presented. The regime of validity assumes that all $2p_ip_j \gg M^2$. 
In order to better understand the origin of angular dependent corrections
in the electroweak theory, consider the case of massive QED and the right handed
SM depicted in Fig.
\ref{fig:qedang}.
We will show that the mass terms do not lead to new effects compared to those in the
massless case.

The two scalar integrals of
Fig. \ref{fig:qedang}, regularized with gauge boson mass terms $\lambda_1$ and $\lambda_2$,
are given in massive QED or in the case of right handed (massive) fermions by
\begin{eqnarray}
{\mbox{II}}^a_{\theta} &=& 4 s t \int \frac{d^4 l_1}{(2 \pi)^4} \int \frac{d^4 l_2}{(2 \pi)^4}
\left\{ \frac{1}{(l_1^2-\lambda_1^2)((p_1-l_1)^2-m_1^2)((p_1-l_1-l_2)^2-m_1^2)} \times
\right. \nonumber \\
&& \left. \,\;\;\;\;\;\;\;\;\;\;\;\;\;\;\;\;\;\;\;\;\;\;\;\;\;\;\;\;\;\;\;\;\;
\frac{1}{((p_2+l_1)^2-m_2^2)
(l_2^2-\lambda_2^2)((p_3+l_2)^2-m_3^2)} \right\} \\
{\mbox{II}}^b_{\theta} &=& 4 s t \int \frac{d^4 l_1}{(2 \pi)^4} \int \frac{d^4 l_2}{(2 \pi)^4}
\left\{ \frac{1}{(l_1^2-\lambda_1^2)((p_1-l_2)^2-m_1^2)((p_1-l_1-l_2)^2-m_1^2)} \times
\right. \nonumber \\
&& \left. \,\;\;\;\;\;\;\;\;\;\;\;\;\;\;\;\;\;\;\;\;\;\;\;\;\;\;\;\;\;\;\;\;\;
\frac{1}{((p_2+l_1)^2-m_2^2)
(l_2^2-\lambda_2^2)((p_3+l_2)^2-m_3^2)} \right\}
\end{eqnarray}
denoting $s=2p_1p_2$, $t=2p_1p_3$ and
where the $m_i$ are the masses of the external charged particles on their mass shell.
Thus, it is straightforward to see that the sum of the two diagrams factorizes to leading
order:
\begin{eqnarray}
{\mbox{II}}^a_{\theta}+{\mbox{II}}^b_{\theta} &=& 4 s t \int \frac{d^4 l_1}{(2 \pi)^4} \int
\frac{d^4 l_2}{(2 \pi)^4}
\left\{ \frac{l_1^2+l_2^2-2p_1(l_1+l_2)}{(l_1^2-\lambda_1^2)((p_1-l_1)^2-m_1^2)((p_1-l_1-l_2)^2-m_1^2)} \times
\right. \nonumber \\
&& \left. \,\;\;\;\;\;\;\;\;\;\;
\frac{1}{((p_2+l_1)^2-m_2^2)
(l_2^2-\lambda_2^2)((p_1-l_2)^2-m_1^2)((p_3+l_2)^2-m_3^2)} \right\} \nonumber \\
&\approx&  \int \frac{d^4 l_1}{(2 \pi)^4} \frac{2s}{(l_1^2-\lambda_1^2)
((p_1-l_1)^2-m_1^2)
((p_2+l_1)^2-m_2^2)} \times \nonumber \\
&& \int \frac{d^4 l_2}{(2 \pi)^4} \frac{2t}{(l_2^2-\lambda_2^2)((p_1-l_2)^2-m_1^2)
((p_3+l_2)^2-m_3^2)}
\end{eqnarray}
The omitted cross term $2l_1l_2$ leads only to corrections containing three logarithms
at the two loop level. It is thus on the same level as the approximation in the beginning
of our discussion which only considers scalar integrals and can therefore be neglected.
To DL accuracy we can employ the Sudakov technique, parametrizing the loop momenta
along the external four momenta as
\begin{eqnarray}
l_1 &\equiv& v_1 \left( p_1 - \frac{m^2_1}{s} p_2 \right) + u_1 \left( p_2 - \frac{m^2_2}{s} p_1 \right)
+ {l_1}_\perp \\
l_2 &\equiv& v_2 \left( p_1 - \frac{m^2_1}{t} p_3 \right) + u_2 \left( p_3 - \frac{m^2_3}{t} p_1 \right)
+ {l_2}_\perp
\end{eqnarray}
Thus, after rewriting the measure and integrating over the perpendicular components we find for
the case of two photons with mass $\lambda$ (omitting
the principle value parts):
\begin{eqnarray}
{\mbox{II}}^a_{\theta}+{\mbox{II}}^b_{\theta} &\sim& \frac{1}{8 \pi^2} \left[ \int^1_0 \frac{dv_1}{v_1}
\int^1_0 \frac{du_1}{u_1} \theta \left(su_1v_1-\lambda^2 \right) \theta \left(u_1-\frac{m_1^2}{s}
v_1 \right)  \theta \left(v_1-\frac{m_2^2}{s} u_1 \right) \right] \times \nonumber \\
&& \frac{1}{8 \pi^2} \left[ \int^1_0 \frac{dv_2}{v_2}
\int^1_0 \frac{du_2}{u_2} \theta \left(tu_2v_2-\lambda^2 \right) \theta \left(u_2-\frac{m_1^2}{t}
v_2 \right)  \theta \left(v_2-\frac{m_3^2}{t} u_2 \right) \right] \nonumber \\
&=& \frac{1}{8 \pi^2} \left[ \int^1_\frac{\lambda^2}{s} \frac{dv_1}{v_1} \int^1_\frac{\lambda^2}{sv_1}
\frac{du_1}{u_1} - \int^\frac{\lambda m_2}{s}_\frac{\lambda^2}{s} \frac{dv_1}{v_1} \int^1_\frac{\lambda^2}{
s v_1} \frac{du_1}{u_1} - \int^\frac{m_2^2}{s}_\frac{\lambda m_2}{s} \frac{dv_1}{v_1}
\int^1_\frac{s v_1}{m_2^2}
\frac{du_1}{u_1} \right. \nonumber \\
&& \;\;\;\;\;\;\;\;\; \left. - \int^\frac{\lambda m_1}{s}_\frac{\lambda^2}{s} \frac{du_1}{u_1}
\int^1_\frac{\lambda^2}{
s u_1} \frac{dv_1}{v_1} - \int^\frac{m_1^2}{s}_\frac{\lambda m_1}{s} \frac{du_1}{u_1}
\int^1_\frac{s u_1}{m_1^2}
\frac{dv_1}{v_1} \right] \times \nonumber \\
&& \frac{1}{8 \pi^2} \left[ \int^1_\frac{\lambda^2}{t} \frac{dv_2}{v_2} \int^1_\frac{\lambda^2}{tv_2}
\frac{du_2}{u_2} - \int^\frac{\lambda m_3}{t}_\frac{\lambda^2}{t} \frac{dv_2}{v_2} \int^1_\frac{\lambda^2}{
t v_2} \frac{du_2}{u_2} - \int^\frac{m_3^2}{t}_\frac{\lambda m_3}{t} \frac{dv_2}{v_2}
\int^1_\frac{t v_2}{m_3^2}
\frac{du_2}{u_2} \right. \nonumber \\
&& \;\;\;\;\;\;\;\;\; \left. - \int^\frac{\lambda m_1}{t}_\frac{\lambda^2}{t} \frac{du_2}{u_2}
\int^1_\frac{\lambda^2}{
t u_2} \frac{dv_2}{v_2} - \int^\frac{m_1^2}{t}_\frac{\lambda m_1}{t} \frac{du_2}{u_2}
\int^1_\frac{t u_2}{m_1^2}
\frac{dv_2}{v_2} \right] \nonumber \\
&=& \frac{1}{8 \pi^2} \left[ -\frac{1}{4} \log^2 \frac{s}{m_1^2} -\frac{1}{4} \log^2 \frac{s}{m_2}
+ \frac{1}{2} \log \frac{s}{m_1^2} \log \frac{s}{\lambda^2} + \frac{1}{2} \log \frac{
s}{m_2^2} \log \frac{s}{\lambda^2} \right] \times \nonumber \\
&& \frac{1}{8 \pi^2} \left[ -\frac{1}{4} \log^2 \frac{t}{m_1^2} -\frac{1}{4} \log^2 \frac{t}{m_3}
+ \frac{1}{2} \log \frac{t}{m_1^2} \log \frac{t}{\lambda^2} + \frac{1}{2} \log \frac{
t}{m_3^2} \log \frac{t}{\lambda^2} \right] \label{eq:qedang}
\end{eqnarray}
The important point about the result in Eq. (\ref{eq:qedang}) is not only the factorized form in terms of
the two massive one loop form factors but also the fact that the fermion mass terms correspond to
each external on shell line in the amplitude.
Thus, by rewriting the term in the bracket of the last
line in Eq. (\ref{eq:qedang}) as
\begin{eqnarray}
&& -\frac{1}{4} \log^2 \frac{t}{m_1^2} -\frac{1}{4} \log^2 \frac{t}{m_3}
+ \frac{1}{2} \log \frac{t}{m_1^2} \log \frac{t}{\lambda^2} + \frac{1}{2} \log \frac{
t}{m_3^2} \log \frac{t}{\lambda^2}  \nonumber \\ &=& \frac{1}{2}
\log^2 \frac{t}{\lambda^2}-\frac{1}{4} \left( \log^2 \frac{m^2_1}{\lambda^2} + \log^2 \frac{m^2_3}{\lambda^2}
\right) \nonumber \\ &\approx& \frac{1}{2} \log^2 \frac{s}{\lambda^2} + \log \frac{s}{\lambda^2}
\log \frac{t}{s}
-\frac{1}{4} \left( \log^2 \frac{m^2_1}{\lambda^2} + \log^2 \frac{m^2_3}{\lambda^2} \right)
\end{eqnarray}
we see that the SL angular terms are indeed independent of the fermion mass terms.
For the corrections involving the invariant $u\equiv 2 p_2p_3$
the situation is analogous.

In the case of the SM with
right handed massive fermions we need to consider in addition the exchange of $Z$-bosons.
The results read
\begin{eqnarray}
&& {\mbox{II}}^a_{\theta}+{\mbox{II}}^b_{\theta} \sim \frac{1}{8 \pi^2} \left[ \frac{1}{2} \log^2
\frac{s}{M^2} \right] \times \nonumber \\ &&
\frac{1}{8 \pi^2} \left[ \frac{1}{2} \log^2 \frac{s}{\lambda^2} + \log \frac{s}{\lambda^2}
\log \frac{t}{s}
-\frac{1}{4} \left( \log^2 \frac{m^2_1}{\lambda^2} + \log^2 \frac{m^2_3}{\lambda^2} \right) \right]
\label{eq:gZ}
\end{eqnarray}
for the case of a photon and a $Z$-boson with mass $M$, and for the case of two $Z$'s we have
\begin{equation}
{\mbox{II}}^a_{\theta}+{\mbox{II}}^b_{\theta} \sim \frac{1}{8 \pi^2} \left[ \frac{1}{2} \log^2
\frac{s}{M^2} \right] \times
\frac{1}{8 \pi^2} \left[ \frac{1}{2} \log^2 \frac{s}{M^2} + \log \frac{s}{M^2}
\log \frac{t}{s} \right]
\label{eq:ZZ}
\end{equation}
Again we see the independence of the SL angular terms on the fermion mass terms and in addition,
the fact that the gauge boson mass gap does not spoil the type of factorization in the right
handed SM.

This type of factorization can be generalized on theoretical grounds to the situation in the
general SM. It should be noted that all fermion mass singularities in the SM only arise through
photon radiation or coupling renormalization. The latter is not important in our discussion here and
is anyhow sub-subleading at higher orders.
The exchange of the heavy gauge bosons does not lead to fermion mass singular terms assuming that
all $m_i \leq M$. This case is analogous to QCD where angular terms factorize in matrix form \cite{cat}.
Thus, only corrections where one heavy gauge boson and one photon are involved are novel features
in the SM. In this case, however, the type of factorization analogous to Eq. (\ref{eq:qedang})
for fermion mass and soft terms follows from the factorized form of real emission corrections.
They are of factorized form and the KLN theorem then leads to the analogous situation for
the sum of all virtual corrections. The soft terms must also factorize since we can always
define an observable by only allowing soft photon radiation, i.e. chosing $\Delta E \leq M$.

From the arguments presented in Ref. \cite{mang} it then follows that
\begin{eqnarray}
&& \!\!\!\!\!\!\!\!\! {\cal M}_{\rm SL}^{u_{i_1},...,
u_{i_n}} \left( \{ p_k \}; \{m_l \}; M, \lambda \right) =
\exp \left\{ - \frac{1}{2} \sum^{n_g}_{i=1} W^{\rm RG}_{g_i} (s,M^2)
 - \frac{1}{2} \sum^{n_f}_{i=1} W^{\rm RG}_{f_i} (s,M^2) \right. \nonumber \\ && \!\!\!\!\!\!\!\!\! \left.
 - \frac{1}{2} \sum^{n_\phi}_{i=1} W^{\rm RG}_{\phi_i} (s,M^2)+
 \frac{1}{8 \pi^2} \sum_{k=1}^n \sum^n_{l < k} \sum_{V_a=B,W^a} \!\!\! {\tilde I}^{V_a}_{i^\prime_k,i_k}
 {\tilde I}^{
 {\overline V}_a}_{i^\prime_l,
 i_l} \log \frac{s}{M^2} \log \frac{2 p_lp_k}{s}
 \right\} \nonumber \\
 && \!\!\!\!\!\!\!\!\! \times \exp \left[ - \frac{1}{2} \sum_{i=1}^{n_f} \left( w^{\rm RG}_{f_i}(s,\lambda^2)
 - w^{\rm RG}_{f_i}(s,M^2) \right)
 - \frac{1}{2} \sum_{i=1}^{n_{\rm w}} \left( w^{\rm RG}_{{\rm w}_i}(s,\lambda^2)
 - w^{\rm RG}_{{\rm w}_i}(s,M^2) \right) \right. \nonumber \\
 && \!\!\!\!\!\!\!\!\! \left. - \frac{1}{2} \sum_{i=1}^{n_\gamma} w_{\gamma_i}(M^2,m_j^2)
 + \sum_{k=1}^n \sum^n_{l < k} \left( w^\theta_{kl} \left(s,\lambda^2 \right) -
w^\theta_{kl} \left(s,M^2 \right) \right)
 \right] \times \nonumber \\ &&
{\cal M}^{u_{i_1},...,u_{i^\prime_k},...,u_{i^\prime_l},...,u_{i_n}}_{\rm Born}
(\{p_{k}\};\{m_{l}\})
 \label{eq:angr}
\end{eqnarray}
where $n_g$ denotes the number of external gauge bosons (in the symmetric basis), $n_f$ the number
of external fermions and $n_\phi$ the number of external scalars (including Higgs particles).
The notation used in the regime below the scale $M$ is analogous; note that we use a photon mass
regulator in this case. The fields $u$ have a well defined isospin, but for angular dependent terms involving
CKM mixing effects, one has to include the extended isospin mixing appropriately in
the corresponding couplings ${\tilde I}^{V_a}_{i^\prime_k,i_k}$ of the symmetric basis.
The expressions for the virtual probabilities $W^{\rm RG}$ are given in section \ref{sec:rg}.
In addition we denote 
\begin{equation}
w^\theta_{kl} \left(s,\lambda^2 \right)=\frac{e^2}{8 \pi^2}Q_k Q_l\log \frac{s}{\lambda^2}
\log \frac{2 p_lp_k}{s} \label{eq:wang}
\end{equation}
for the angular dependent corrections from the soft QED regime,
where the $Q_j$ denote the electrical charges of the {\it external} lines.
These terms are not manifestly positive and thus not related to emission probabilities 
and also process dependent.

The diagonal terms do not involve (CKM-extended) isospin rotated Born matrix elements.
These occur only from the angular
terms above the scale $M$.
Thus, the way the electroweak angular terms factorize is also in 
an exponentiated operator form as in QCD. Eq. (\ref{eq:angr}) is valid for arbitrary external lines.

\subsection{Sudakov logarithms in softly broken supersymmetric models} \label{sec:susy}

In the introduction we already discussed the motivations for considering
supersymmetric extensions of the SM. Since supersymmetry must be broken in
nature (if it is at all relevant to physics at the electroweak scale), the
original simplicity is lost and in the case of the MSSM over one-hundred
parameters are needed to describe the model. In order for supersymmetry to
stabilize the hierarchy problem, though, two conditions need to be fulfilled.
Firstly, the masses of the superpartner particles must not be much larger
than the electroweak scale and secondly, it must be a ``softly broken''
symmetry, i.e. broken by mass terms and couplings with a positive mass dimension. 

Thus, at energies in the TeV regime, these two conditions (assumed to be fulfilled
in the MSSM or the NMSSM for instance) lead to the following consequences for
radiative corrections of electroweak origin.
If we assume that the mass scale of the superpartner particles are not much larger
than the weak scale, say less than $500$ GeV, and energies in the TeV regime,
the results for the DL corrections outlined above for the SM
can be applied straightforwardly to the MSSM. The reason is that
the gauge couplings are preserved under supersymmetry and no additional spin 1 particles
are exchanged. The appropriate quantum numbers in the eigenvalues of the casimir operators
are the same as those of the SM partners.

In case the superpartner masses are larger than $500$ GeV, additional double logarithms
need to be taken into account in a way outlined in Ref. \cite{flmm}.
In the following we assume that we can neglect such
terms, i.e. that all particles in the MSSM have a mass below $500$ GeV.

At the subleading level, the situation in general is less
clear at higher orders.
For SL angular dependent terms, the same reasoning as above goes through since they
originate only from the exchange of spin 1 gauge bosons and can thus be resummed
as in the SM (see section \ref{sec:at}).
Box-type diagrams exchanging supersymmetric particles in the s-channel do not contribute
to SL angular terms.
The same holds for all universal
SL corrections which involve the exchange of SM particles since they are
properties of the external particles only.

New types of
SL Sudakov corrections are, however, involved in the exchange of supersymmetric particles as
discussed in Refs. \cite{brv3,brv4,bmrv} at the one loop level in the on-shell production of superpartner
particles in $e^+e^-$ collisions.
In the following we discuss the results obtained in those works.
We begin with the corrections
contributing in particular the
Yukawa terms from the final state corrections. The final result of calculating all
terms contributing to the Yukawa sector includes terms depending on
$\tan \beta$, which is the ratio of the two v.e.v.'s from the two Higgs doublet sector
of the MSSM. Since $\tan \beta$ could be as large as $40$, the bottom Yukawa terms are
crucial in supersymmetric models. Also higher order terms need to be considered
as in the SM and the exponentiation of the MSSM Yukawa terms follows from the same
arguments as in the SM. In particular the Ward identity Eq. (\ref{eq:wi}) holds \cite{bmrv}.
Again, here we assume that the susy masses are close to the electroweak
scale. These corrections apply to both the SM as well as the superpartner production.
In particular the process of charged Higgs production seems to be well suited for
an indirect measurement of $\tan \beta \geq 14$ of better than ${\cal O} \left(25 \%
\right)$, and a few percent for $\tan \beta \geq 25$ \cite{bmrv}. The important point
to note here is not only the precision but in particular the fact that this determination
of $\tan \beta$ is independent of soft breaking terms to SL accuracy and thus model
independent. This is due to the fact that the soft breaking contributions
are constants and can be eliminated via subtraction if a series of
precise measurements is performed at various energy scales.
In addition this approach is scheme and gauge invariant.

Also the SL-gauge terms get modified by loops containing novel superpartner contributions.
If we neglect the mass difference between the SM fermionic particle and its scalar partner we observe
an exact supersymmetry relation\footnote{Here we take a common mass also for
the two chiral superpartners. Since 
in general the mass eigenstates are different from the flavor eigenstates if
their masses differ when supersymmetry breaking is included, mixing effects must
then also be considered.} described by (for $\lambda=M$):

\begin{eqnarray}
&& d \sigma^{\rm SL}_{e^+_{\alpha} e^-_{\alpha} \longrightarrow {\overline f}_{\beta}
f_{\beta}}
/ d \sigma^{\rm Born}_{e^+_{\alpha} e^-_{\alpha} \longrightarrow {\overline f}_{\beta}
f_{\beta}} =
d \sigma^{\rm SL}_{e^+_{\alpha} e^-_{\alpha} \longrightarrow {\overline {\tilde f}}_{\beta}
{\tilde f}_{\beta}}
/ d \sigma^{\rm Born}_{e^+_{\alpha} e^-_{\alpha} \longrightarrow {\overline {\tilde f}}_{\beta}
{\tilde f}_{\beta}} = \nonumber \\ && \exp \left\{ - \frac{g^2(m_s^2)
T_{e^-_{\alpha}}(T_{e^-_{\alpha}}+1)}{8 \pi^2 } \left[ \log^2 \frac{s}{M^2}
- \frac{1}{3} {\tilde \beta_0} \frac{g^2(m_s^2) }{4 \pi^2 } \log^3 \frac{s}{m_s^2} \right] \right. \nonumber \\
&& -\frac{{g^\prime}^2(m_s^2) Y^2_{e^-_{\alpha}}}{32 \pi^2 } \left[
\log^2 \frac{s}{M^2}- \frac{1}{3} {\tilde \beta_0}^\prime \frac{{g^\prime}^2(m_s^2)
}{4 \pi^2 } \log^3 \frac{s}{m_s^2} \right]
\nonumber \\
&& + \left( \frac{ g^2(m_s^2)}{8 \pi^2} T_{e^-_{\alpha}}(T_{e^-_{\alpha}}+1)+
\frac{ {g^\prime}^2(m_s^2)}{8 \pi^2}\frac{Y^2_{e^-_{\alpha}}}{4} \right)  2 \log \frac{s}{M^2}
\nonumber \\ &&
- \frac{g^2(m_s^2)
T_{{\tilde f}_{\beta}}(T_{{\tilde f}_{\beta}}+1)}{8 \pi^2 } \left[ \log^2 \frac{s}{M^2}
- \frac{1}{3} {\tilde \beta_0} \frac{g^2(m_s^2) }{4 \pi^2 } \log^3 \frac{s}{m_s^2} \right]  \nonumber \\
&& -\frac{{g^\prime}^2(m_s^2) Y^2_{{\tilde f}_{\beta}}}{32 \pi^2 } \left[
\log^2 \frac{s}{M^2}- \frac{1}{3} {\tilde \beta_0}^\prime \frac{{g^\prime}^2(m_s^2)
}{4 \pi^2 } \log^3 \frac{s}{m_s^2} \right]
\nonumber \\
&& + \left( \frac{ g^2(m_s^2)}{8 \pi^2} T_{{\tilde f}_{\beta}}(T_{{\tilde f}_{\beta}}+1)+
\frac{ {g^\prime}^2(m_s^2)}{8 \pi^2}\frac{Y^2_{{\tilde f}_{\beta}}}{4} \right)  2 \log \frac{s}{M^2}
\nonumber \\ &&
- \frac{ g^2(m_s^2)}{8 \pi^2} \left( \frac{1+\delta_{\beta,{\rm R}}}{2} \frac{{\hat m}
^2_{\tilde f}}{M^2} + \delta_{\beta,{\rm L}}
\frac{{\hat m}^2_{{\tilde f}^\prime}}{2 M^2} \right)
\log \frac{s}{m_s^2} \nonumber \\ &&
-\frac{g^2(m_s^2)}{8\pi^2} \log \frac{s}{M^2} \left[ \left( \tan^2 \theta_{\rm w} Y_{e^-_{\alpha}} Y_{
{\tilde f}_\beta}
+ 4 I^3_{e^-_{\alpha}} I^3_{{\tilde f}_\beta} \right) \log \frac{t}{u} \right. \nonumber \\ &&
\left. \left. + \frac{\delta_{\alpha, L} \delta_{\beta,L}}{\tan^2 \theta_{\rm w} Y_{e^-_{\alpha}} Y_{
{\tilde f}_\beta} /4
+ I^3_{e^-_{\alpha}} I^3_{{\tilde f}_\beta}} \left( \delta_{d,{\tilde f}} \log \frac{-t}{s} -
\delta_{u,{\tilde f}}
\log \frac{-u}{s} \right)  \right] \right\}
\label{eq:Wfsusy}
\end{eqnarray}
where $T_j$ denotes the total weak isospin of the particle $j$, $Y_j$ its weak hypercharge
and at high $s$ the invariants are given by
$t=-\frac{s}{2} \left( 1-\cos \theta \right)$ and $u=-\frac{s}{2} \left( 1+\cos \theta \right)$.
The helicities are those of the fermions ($f$) whose superpartner is produced.
In addition we denote ${\hat m}_{\tilde f}=m_t / \sin \beta$
if ${\tilde f}= {\tilde t}$ and ${\hat m}_{\tilde f}=m_b / \cos \beta$ if ${\tilde f}={\tilde b}$.
${\tilde f}^\prime$ denotes the corresponding
isopartner of ${\tilde f}$. For particles other than those belonging to the third family
of quarks/squarks, the Yukawa terms are negligible.
Eq. (\ref{eq:Wfsusy}) depends on the important parameter $\tan \beta = \frac{v_u}{v_d}$, the
ratio of the two vacuum expectation values, and displays
an exact supersymmetry in the sense that the same corrections are obtained for the fermionic
sector in the regime above the electroweak scale $M$.

Here we assume that the asymptotic MSSM $\beta$-functions can be used from the scale $m_s\sim
m_{\tilde f} \sim M$ with
\begin{eqnarray}
{\tilde \beta}_0&=& \frac{3}{4} C_A- \frac{n_g}{2}-\frac{n_h}{8} \;\;,\;
{\tilde \beta}_0^\prime=-\frac{5}{6}n_g-\frac{n_h}{8} \label{eq:bMSSM} \\
g^2(s) &=& \frac{g^2(m_s^2)}{1+{\tilde \beta}_0 \frac{g^2
(m_s^2)}{4\pi^2}
\ln \frac{s}{m_s^2}} \;\;,\;
{g^\prime}^2 (s) = \frac{{g^\prime}^2 (m_s^2)}{1+{\tilde \beta}^\prime_0
\frac{{g^\prime}^2 (m_s^2)}{4\pi^2}
\ln \frac{s}{m_s^2}} \label{eq:arunMSSM}
\end{eqnarray}
where $C_A=2$, $n_g=3$ and $n_h=2$. In practice, one has to use the relevant numbers of active
particles in the loops. These terms correspond to the RG-SL corrections just as in the case
of the SM as discussed in Ref. \cite{m4}
but now with the MSSM particle spectrum contributing.
They originate only from RG terms within loops which without the RG contribution
would give a DL correction.
It should be noted that the one-loop RG corrections do not exponentiate and are omitted
in the above expressions. They are, however, completely determined by the renormalization group
in softly broken supersymmetric theories such as the MSSM
and sub-subleading at higher than one loop order.

The relation (\ref{eq:Wfsusy}) is expected since in unbroken supersymmetry
both chiral fermions and the superpartner sfermions
are part of the same supermultiplet. In the diagrammatic evaluation, however, the identity
expressed in (\ref{eq:Wfsusy}) is a highly non-trivial check on the overall correctness
of the calculation as different particles and loops contribute in each case.
Since in the real world
$m_{\tilde f} \neq m_f$, the corresponding matching terms containing light fermion mass
need to be included as well. 

For charged Higgs production we have analogously
\begin{eqnarray}
&& \!\!\!\!\!\!\!\!\!\!\!\!\!\!\!\!\!\!\!\!
d \sigma^{\rm SL}_{e^+_{\alpha} e^-_{\alpha} \longrightarrow H^+ H^-}
\!=\!\! d \sigma^{\rm Born}_{e^+_{\alpha} e^-_{\alpha} \longrightarrow H^+ H^-}
\times \nonumber \\ && \!\!\!\!\!\!\!\!\!\!\!\!\!\!\!\!\!\!\!\!
\exp \left\{ - \frac{g^2(m_s^2)}{8\pi^2} T_{e^-_{\alpha}} \left( T_{e^-_{\alpha}}+1
\right) \left[
\log^2 \frac{s}{M^2}
 - \frac{1}{3} {\tilde \beta_0} \frac{g^2(m_s^2) }{4 \pi^2 } \log^3 \frac{s}{m_s^2}
 \right] \right. \nonumber \\
 &&\!\!\!\!\!\!\!\!\!\!\!\!\!\!\!\!\!\!\!\! -\frac{{g^\prime}^2(m_s^2) Y^2_{e^-_{\alpha}}}{32 \pi^2 }
 \left[
 \log^2 \frac{s}{M^2}- \frac{1}{3} {\tilde \beta_0}^\prime \frac{{g^\prime}^2(m_s^2)
 }{4 \pi^2 } \log^3 \frac{s}{m_s^2}
 \right] \nonumber \\
 &&\!\!\!\!\!\!\!\!\!\!\!\!\!\!\!\!\!\!\!\! + \left( \frac{ g^2(m_s^2)}{8 \pi^2}
 T_{e^-_{\alpha}} \left( T_{e^-_{\alpha}}+1 \right)+
 \frac{ {g^\prime}^2(m_s^2)}{8 \pi^2} \frac{Y^2_{e^-_{\alpha}}}{4} \right)  2 \log \frac{s}{M^2}
 \nonumber \\ &&\!\!\!\!\!\!\!\!\!\!\!\!\!\!\!\!\!\!\!\!
 - \frac{g^2(m_s^2)}{8\pi^2} T_H \left( T_H+1
 \right) \left[
 \log^2 \frac{s}{M^2}
  - \frac{1}{3} {\tilde \beta_0} \frac{g^2(m_s^2) }{4 \pi^2 } \log^3 \frac{s}{m_s^2}
  \right] \nonumber \\
  \!\!\!\!&&\!\!\!\!\!\!\!\!\!\!\!\!\!\!\!\!\!\!\!\! -\frac{{g^\prime}^2(m_s^2)
  Y^2_H}{32 \pi^2 }
  \left[
  \log^2 \frac{s}{M^2}- \frac{1}{3} {\tilde \beta_0}^\prime \frac{{g^\prime}^2(m_s^2)
  }{4 \pi^2 } \log^3 \frac{s}{m_s^2}
  \right] \nonumber \\
  \!\!\!\!&&\!\!\!\!\!\!\!\!\!\!\!\!\!\!\!\!\!\!\!\! + \left( \frac{ g^2(m_s^2)}{8 \pi^2}
  T_H \left( T_H+1 \right)+
  \frac{ {g^\prime}^2(m_s^2)}{8 \pi^2} \frac{Y^2_H}{4} \right)  2 \log \frac{s}{
  M^2}
  \nonumber \\ && \!\!\!\!\!\!\!\!\!\!\!\!\!\!\!\!\!\!\!\!
  - 3 \; \frac{g^2(m_s^2)}{32 \pi^2} \left[ \frac{m_t^2}{M^2} \cot^2 \beta + \frac{m_b^2}{M^2} \tan^2 \beta \right]
  \log \frac{s}{m_s^2}
  \nonumber \\ && \!\!\!\!\!\!\!\!\!\!\!\!\!\!\!\!\!\!\!\! \left.
  -\frac{g^2 (m_s^2)}{4\pi^2} \log \frac{s}{M^2} \left[
  \delta_{\alpha,{\rm L}} \left( \frac{1}{2 c^2_{\rm w}} \log \frac{t}{u} + 2 c^2_{\rm w}
  \log \frac{-t}{s} \right)
  + \delta_{\alpha, {\rm R}} \tan^2 \theta_{\rm w}
  \log \frac{t}{u}  \right] \right\} \label{eq:Hang}
  \end{eqnarray}
  It should be noted here that the Yukawa terms proportional to $\tan \beta$ are quite large
  due to the additional factor of $3=N_C$ from the quark loops \cite{bmrv}. Overall, both Yukawa contributions
  in Eq. (\ref{eq:Wfsusy}) and (\ref{eq:Hang}) reinforce the Sudakov suppression factor of the leading
double logarithmic terms.

Eq. (\ref{eq:Hang}) is also valid for charged Higgsino production (as
suggested by the relation (\ref{eq:Wfsusy})), however, the latter is not a mass eigenstate.

The higher order exponentiation of the universal terms uses the Ward identity (\ref{eq:wi})
for the respective corrections. The angular terms are of SM origin only as described above
and are thus treated as in section \ref{sec:at}.
In addition we have the matching terms stemming form mass terms of both the gauge bosons
and particles involved in the process. In any case it would also be helpful for collider
experiments to have a full one loop calculation in the full supersymmetric theory in order
to have a better understanding of the size of constant terms omitted in our high energy
approximation. These terms, however, are difficult to discuss at this point since they
depend strongly on how supersymmetry breaking is realized in nature.

Neutral Higgs production in the MSSM depends also on the 
parameter $\alpha$, the mixing angle between the CP even neutral Higgs particles. 
For these terms the corresponding dependence on $\tan \beta$ is thus not
as evident as in the charged Higgs case without independent
knowledge on $\alpha$ but results can be found in Ref. \cite{bmrv}.
For transverse gauge boson production it should be clear that the $\beta$ functions in Eqs. 
(\ref{eq:WgRG}) should be replaced by the corresponding MSSM expression given
in Eq. (\ref{eq:bMSSM}). It can also be expected that an analogous supersymmetry
relation should hold between the high energy splitting function expression of
gauge bosons and gauginos under the above assumptions (in particular the neglect
of the mass differences).

Longitudinal gauge boson production is analogous to Higgs production after the application
of the Goldstone equivalence theorem.

\subsection{Fully-inclusive cross sections} \label{sec:fi}

Physical observables in the electroweak theory depend on the
infrared cutoff, the gauge boson mass $M$, as was discussed in section \ref{sec:si}. This
feature is closely related to electroweak symmetry breaking itself since it indicates that
in the unbroken limit $M \longrightarrow 0$ cross sections become unobservable. Thus, we
would expect in this limit confining effects limiting the types of initial or final state
particles to be in an isospin singlet state. In the broken physical theory, this is
evidently not the case and physical particles carry a non-Abelian group charge. It is
for this reason that even fully inclusive cross sections are expected to depend on
$\log^2 \frac{s}{M^2}$. Also in QCD the Bloch-Nordsieck violating terms are present 
in fully inclusive cross sections
and only the initial state color averaging eliminates the infrared problem. Thus we see
that masslessness of the gauge bosons in a non-Abelian theory is intimately connected with
confinement and therefore color neutral initial and final states.

The realization of this phenomenon in the electroweak theory was emphasized in Ref. \cite{ccc1} 
where these ``Bloch-Nordsieck violating'' terms were calculated by means of the coherent state
operator formalism from QCD. The basic idea is provided by the observation that in
collider experiments one does not take the average over, say, the $e^-e^+$ and $\nu e^+$
initial states since the asymptotic states carry weak isospin quantum numbers. The authors
of Ref. \cite{ccc2} conclude that these $\log^2 \frac{s}{M^2}$ terms in fully inclusive
electroweak processes are due only from initial state $W^\pm$ corrections.
There are, however, possibly additional terms from the Abelian sector due to the fact that
the hypercharge is broken in the physical fields. These terms, however, can
be neglected for light fermions.
By then considering the leading coherent state operators for the soft and hard (Born) parts
of the scattering amplitude, the authors of Ref. \cite{ccc2}
find that the Bloch-Nordsieck violating terms exponentiate
in the form of the Sudakov form factor in the adjoint representation. For initial state fermions
with two isospin components the results for the fully inclusive cross sections are then given by
\begin{eqnarray}
\sigma_{11} &=& \sigma_{22} = \frac{1}{2} \left( \sigma^{\rm Born}_{11} + \sigma^{\rm Born}_{12}
+\left[ \sigma^{\rm Born}_{11} - \sigma^{\rm Born}_{12} \right] \exp \left(-2 {\cal F}_S \right) \right) \\
\sigma_{12} &=& \sigma_{21} = \frac{1}{2} \left( \sigma^{\rm Born}_{11} + \sigma^{\rm Born}_{12}
-\left[ \sigma^{\rm Born}_{11} - \sigma^{\rm Born}_{12} \right] \exp \left(-2 {\cal F}_S \right) \right) 
\end{eqnarray}
where
\begin{equation}
{\cal F}_S=\frac{g^2}{16 \pi^2} \log^2 \frac{s}{M^2}
\end{equation}
is the Sudakov form factor and the factor of 2 in the exponential argument corresponds to the Casimir
eigenvalue in the adjoint representation, i.e. $T(T+1)$ with $T=1$.
A generalization to transversely polarized initial states is given in Ref. \cite{ccc3}.
Denoting the triplet representation as $+,3,-$ the authors obtain for the case of fermion
boson scattering:
\begin{eqnarray}
\sigma_{1+} &=& \sigma_{2-} = \frac{1}{2} \left( \sigma^{\rm Born}_{1+} + \sigma^{\rm Born}_{1-}
+\left[ \sigma^{\rm Born}_{1+} - \sigma^{\rm Born}_{1-} \right] \exp \left(-2 {\cal F}_S \right) \right) \\
\sigma_{1-} &=& \sigma_{2+} = \frac{1}{2} \left( \sigma^{\rm Born}_{1+} + \sigma^{\rm Born}_{1-}
-\left[ \sigma^{\rm Born}_{1+} - \sigma^{\rm Born}_{1-} \right] \exp \left(-2 {\cal F}_S \right) \right) \\
\sigma_{13} &=& \sigma_{23} = \frac{1}{2} \left( \sigma_{1+} + \sigma_{1-} \right)
= \frac{1}{2} \left( \sigma^{\rm Born}_{1+} + \sigma^{\rm Born}_{1-} \right) 
\end{eqnarray}
In the case of transverse boson boson scattering we have
\begin{eqnarray}
\sigma_{++} &=& \sigma_{--} = \sigma^{\rm Born}_{++} \left( \frac{1}{3} + \frac{1}{2} \exp \left(-2 
{\cal F}_S \right) + \frac{1}{6} \exp \left(-6 {\cal F}_S \right) \right) 
+ \sigma^{\rm Born}_{- \, +} \left( \frac{1}{3} - \frac{1}{2} \exp \left(-2 
{\cal F}_S \right) \right. \nonumber \\ && \left. \;\;\;\;\;\;\;\;\;\;+ \frac{1}{6} \exp \left(-6 {\cal F}_S \right) \right) 
+ \sigma^{\rm Born}_{3+} \left( \frac{1}{3}- \frac{1}{3}  \exp \left(-6 {\cal F}_S \right) \right) \\
\sigma_{- \, +} &=& \sigma_{+ \, -} = \sigma^{\rm Born}_{++} \left( \frac{1}{3} - \frac{1}{2} \exp \left(-2 
{\cal F}_S \right) + \frac{1}{6} \exp \left(-6 {\cal F}_S \right) \right) 
+ \sigma^{\rm Born}_{- \, +} \left( \frac{1}{3} + \frac{1}{2} \exp \left(-2 
{\cal F}_S \right) \right. \nonumber \\ && \left. \;\;\;\;\;\;\;\;\;\;+ \frac{1}{6} \exp \left(-6 {\cal F}_S \right) \right) 
+ \sigma^{\rm Born}_{3+} \left( \frac{1}{3}- \frac{1}{3}  \exp \left(-6 {\cal F}_S \right) \right) \\
\sigma_{3+} &=& \sigma_{3-}= \frac{1}{3} \left\{ \left[ \sigma^{\rm Born}_{++} +\sigma^{\rm Born}_{-\,+} \right]
\left( 1- \exp \left(-6 
{\cal F}_S \right) \right) 
+ \sigma^{\rm Born}_{3+} \left( 1+2 \exp \left(-6 {\cal F}_S \right) \right) \right\} 
\end{eqnarray}
In Ref. \cite{ccc4} results for the longitudinal and Higgs sector are presented employing the
equivalence theorem. For a light Higgs the inclusive cross sections are related as follows:
\begin{eqnarray}
\sigma_{\phi^-\phi^-}&=&\sigma_{\phi^+\phi^+}\;\;;\; \sigma_{\phi^+h} = \sigma_{\phi^-h}=\sigma_{\phi^+3} \\
\sigma_{hh}&=&\sigma_{33}=\frac{1}{2} \left( \sigma_{\phi^+\phi^+}+\sigma_{\phi^+\phi^-} \right) + 
Re \;\sigma^{\rm Born}_{
\phi_0 \phi_0^* \rightarrow \phi_0^* \phi_0} \exp \left(-(2+\tan^2 \theta_{\rm w}) {\cal F}_S \right) \\ 
\sigma_{3h}&=&\frac{1}{2} \left( \sigma_{\phi^+\phi^+}+\sigma_{\phi^+\phi^-} \right) - Re \; \sigma^{\rm Born}_{
\phi_0 \phi_0^* \rightarrow \phi_0^* \phi_0} \exp \left(-(2+\tan^2 \theta_{\rm w}) {\cal F}_S \right) 
\end{eqnarray}
The individual results can be obtained from 
\begin{eqnarray}
\!\!\!\!\!\!\!\!\!\!
\sigma_{\phi^+\phi^+}+\sigma_{\phi^-\phi^+} + 2 \sigma_{3\phi^+} &=& \!\sigma^{\rm Born}_{\phi^+\phi^+}+
\sigma^{\rm Born}_{\phi^-\phi^+} + 2 \sigma^{\rm Born}_{3\phi^+} \\
\!\!\!\!\!\!\!\!\!\!
\sigma_{\phi^+\phi^+}+\sigma_{\phi^-\phi^+} - 2 \sigma_{3\phi^+} &=& \!\left( \sigma^{\rm Born}_{\phi^+\phi^+}+
\sigma^{\rm Born}_{\phi^-\phi^+} - 2 \sigma^{\rm Born}_{3\phi^+} \right) \exp \left(-2 {\cal F}_S \right) \\
\!\!\!\!\!\!\!\!\!\!
\sigma_{\phi^+\phi^+}-\sigma_{\phi^-\phi^+} + 2 Im \; \sigma_{3\phi^+\rightarrow h \phi^+} &=& \! 
\sigma^{\rm Born}_{\phi^+\phi^+}-
\sigma^{\rm Born}_{\phi^-\phi^+} + 2 Im \; \sigma^{\rm Born}_{3\phi^+\rightarrow h \phi^+} \\
\!\!\!\!\!\!\!\!\!\!
\sigma_{\phi^+\phi^+}-\sigma_{\phi^-\phi^+} - 2 Im \; \sigma_{3\phi^+\rightarrow h \phi^+} &=& \!
\left( \sigma^{\rm Born}_{\phi^+\phi^+}-
\sigma^{\rm Born}_{\phi^-\phi^+} - 2 Im \; \sigma^{\rm Born}_{3\phi^+\rightarrow h \phi^+} \right) 
\exp \left(-2 {\cal F}_S \right) 
\end{eqnarray}
In general, the effects are sizable. In particular for initial state leptons and for gauge 
bosons ${\cal O} \left( 10 \% \right)$ effects can be expected in the TeV range.
In Ref. \cite{ccc5} an initial step towards electroweak splitting functions is taken for a spontaneously
broken $SU(2)$ gauge group. In this model there is no gauge boson mass gap, however, the important difference
to the QCD case is that the initial state isospin averaging is omitted. A rigorous extension to the
full electroweak theory would be important since the size of SL terms in non-negligible. 

\subsection{Physical fields at fixed order} \label{sec:pf}

In this section we present results obtained in real fixed order perturbative calculations
performed with the physical SM fields in the on-shell scheme. It is pivotal to compare these
results with the corrections obtained with the IREE method expounded on above where the calculation
is performed in the high energy effective theory including appropriate matching conditions at
the weak scale in order to include pure QED effects. Fortunately, a lot of effort has been
spent in perturbative electroweak calculations at high energies in order to allow for a 
cross check for all degrees of freedom at one loop. At the two loop level calculations
at the DL level are available in the regime we are interested in. We begin, however, at
the lowest order. It should be mentioned that in order to obtain an accuracy in the
1 \% regime, it is not sufficient to include only logarithmic 
corrections since
for processes involving unsuppressed mass-ratios in the Born cross section (such as
$\phi^+ \phi^- \longrightarrow \phi^+ \phi^-$) constants can be much larger.
Also $\pi^2$ terms and Yukawa enhanced constants can be of the order of several
\% and thus, for the precision objectives of the
linear collider
a full one loop calculation is necessary in most cases. At higher orders, however, these
constant terms can be neglected. 

\subsubsection{One loop results}

There have been various calculations at the one loop level, however, for our purposes
it is convenient to discuss the results of Ref. \cite{dp} in more detail where general formulas
are derived for one loop logarithmic corrections at high energies. For all SL Sudakov
logarithms we find agreement with the results presented in Ref. \cite{dp}. The universality
of the corrections in the physical theory, however, is more difficult to see since the Ward
identities are more involved in the broken gauge theory.
We summarize results presented in Ref. \cite{dp} for DL and
one loop angular terms containing $\log \frac{s}{M^2}
\log \frac{u}{t}$ type contributions. In addition there are also the on-shell parameter renormalization
of the couplings, masses and mixing angles. All results are in the 't Hooft-Feynman gauge
and the techniques discussed in this section are valid only at the one loop level.

In order to obtain double logarithmic corrections one can drop the dependence of the numerator
on the loop momentum.
In this approximation the one-loop corrections give
\beq \label{eikonalappA}
\delta \M^{i_1 \ldots i_n}=\sum_{k=1}^n\sum_{l<
  k}\sum_{\GB_a=A,Z,W^\pm}\int \frac{\rd^4q}{(2\pi)^4} \frac{-4ie^2
  p_kp_lI^{\GB_a}_{i'_k i_k}(k) I^{\bar{\GB}_a}_{i'_l i_l}(l) \M_0^{i_1 \ldots i'_k
    \ldots i'_l \ldots i_n}}
{(q^2-M_{\GB_a}^2)[(p_k+q)^2-m_{k'}^2][(p_l-q)^2-m_{l'}^2]}
\eeq
and in leading order, using the high-energy expansion of the scalar three-point
function \cite{Denn3}, one obtains
\beq \label{eikonalapp}
\delta \M^{i_1 \ldots i_n}
=\frac{1}{2}\sum_{k=1}^n\sum_{l\neq k}\sum_{\GB_a=A,Z,W^\pm} I^{\GB_a}_{i'_k i_k}(k) I^{\bar{\GB}_a}_{i'_l i_l}(l) \M_0^{
i_1 \ldots i'_k \ldots i'_l \ldots i_n}[\LrMa-\de_{\GB_aA}\Lkla] 
\eeq
where the short hand notations
\beqar \label{dslogs}
\LrM &:=& \frac{e^2}{16\pi^2}\log^2{\frac{r_{kl}}{M^2}} \qquad
\lrM:=\frac{e^2}{16\pi^2}\log{\frac{r_{kl}}{M^2}} \\
\Ls &:=& \LsW \qquad
\ls:=\lsW
\eeqar
are used with
\beq \label{Sudaklim} 
r_{kl}=(p_k+p_l)^2\sim 2p_kp_l \gg \MW^2
\eeq
The DL term containing the invariant $r_{kl}$ depends on the angle
between the momenta $p_k$ and $p_l$. Writing
\beq \label{angsplit}
\LrM=\LsM+2\lsM\lrs+\Lrs
\eeq
the angular-dependent part is isolated in logarithms of $r_{kl}/s$,
and gives a subleading soft--collinear ($\SS$) contribution of order
$\ls\log(|r_{kl}|/s)$, whereas terms $\Lrs$ can be neglected in LA.
The remaining part, together with the
additional contributions from photon
loops in \refeq{eikonalapp}, gives the leading soft--collinear ($\SC$)
contribution and is angular-independent.  This contribution corresponds to the
universal part discussed in the previous sections in the symmetric basis.
The eikonal approximation
\refeq{eikonalappA} applies to chiral fermions, Higgs bosons, and
transverse gauge bosons, and depends on their gauge couplings
$I^{\GB_a}(k)$ which can be found in appendix \ref{sec:qn}.

Since the longitudinal polarization vectors
(\ref{eq:lpv}) grow with energy, matrix elements involving longitudinal gauge bosons have to be 
treated with the
equivalence theorem, \ie they have to be expressed by matrix elements involving the corresponding Goldstone bosons.
As discussed in section \ref{sec:et}, 
the equivalence theorem
for Born matrix elements receives no DL one-loop corrections. 
Therefore, the soft-collinear  
corrections for external longitudinal gauge bosons can be obtained using the simple relations
\beqar \label{DLeqtheor}
\de^{\mathrm{DL}}\M^{\ldots \PW^\pm_\rL \ldots} &=&\de^{\mathrm{DL}}\M^{\ldots \phi^\pm \ldots},\nl
\de^{\mathrm{DL}}\M^{\ldots \PZ_\rL \ldots} &=&\ri\de^{\mathrm{DL}}\M^{\ldots \chi \ldots}
\eeqar
from the corrections \refeq{eikonalapp} for external Goldstone bosons.

The invariance of the S matrix with respect to global
$\SUtwo\times\Uone$ transformations implies (up to mass terms):
\beq\label{eq:gi}
0= \delta_{\GB_a} 
\M^{i_1 \ldots i_n}=\ri e \sum_{k} I^{\GB_a}_{i'_k i_k}(k) \M^{i_1 \ldots i'_k \ldots i_n}
\eeq
which is the analogue of Eq. (\ref{eq:cnag}) for the electroweak theory.
For external Goldstone fields extra contributions proportional to the
Higgs vacuum expectation value appear, which are, however, irrelevant
in the high-energy limit.  Using \refeq{eq:gi}, the $\SC$ logarithms
in \refeq{eikonalapp} can be written as a single sum over external legs,
\beq \label{SCsum}
\de^{\SC} \M^{i_1 \ldots i_n} =\sum_{k=1}^n \delta^\SC_{i'_ki_k}(k)
\M_0^{i_1 \ldots i'_k\ldots i_n}.
\end{equation}
After evaluating the sum over  A, Z, and W, in \refeq{eikonalapp}, the
correction factors read
\beq \label{deSC} 
\de^\SC_{i'_ki_k}(k)=- \frac{1}{2}\left[ C^{\ew}_{i'_ki_k}(k)\Ls -2(I^Z(k))_{i'_ki_k}^2 \log{\frac{\MZ^2}{\MW^2}}\, \ls+\
de_{i'_ki_k} Q_k^2\Lemk \right]
\eeq
The first term represents the DL symmetric-electroweak part and is
proportional to the electroweak Casimir operator $\cew$ defined in
\refeq{CasimirEW}. This is always diagonal in the $\SUtwo$ indices,
except for external transverse
neutral gauge bosons in the physical basis
\refeq{adjointCasimir2}, where it gives rise to mixing between
amplitudes involving photons and Z bosons. The second term originates
from Z-boson loops, owing to the difference between $\MW$ and $\MZ$,
and
\beq
\Lemk:= 2\ls\log{\left(\frac{\MW^2}{\la^2} \right)}+\LWla-\Lkla
\eeq
contains all logarithms of pure electromagnetic origin.
The $\SC$ corrections for external longitudinal gauge bosons are directly obtained from \refeq{deSC} by using the quantum
 numbers of the corresponding Goldstone bosons.

The contribution of the second term of \refeq{angsplit} to
\refeq{eikonalapp} remains a sum over pairs of external legs,
\beq \label{SScorr}
\de^\SS \M^{i_1 \ldots i_n} =\sum_{k=1}^n
\sum_{l<k}\sum_{\GB_a=A,Z,W^\pm}\delta^{\GB_a,\SS}_{i'_ki_k i'_li_l}(k,l)
\M_0^{i_1\ldots i'_k\ldots i'_l\ldots i_n}
\eeq
with angular-dependent terms. The exchange of soft, neutral gauge
bosons contributes with
\beqar \label{subdl1} 
\de^{A,\SS}_{i'_ki_k i'_li_l}(k,l)&=&
2 \left[\ls+\lWla\right]\lrs I_{i'_ki_k}^A(k)I_{i'_li_l}^A(l),\nl
\delta^{Z,\SS}_{i'_ki_k i'_li_l}(k,l)&=&
2\ls \lrs I_{i'_ki_k}^Z(k)I_{i'_li_l}^Z(l)
\eeqar
and, except for $I^Z$ in the neutral scalar sector $H,\chi$ (see
\ref{sec:qn}),
the couplings $I^\NB$ are diagonal matrices.  
The exchange of charged gauge bosons yields
\beq \label{subdl2} 
\delta^{W^\pm,\SS}_{i'_ki_k i'_li_l} (k,l)=2\ls\lrs I_{i'_ki_k}^\pm(k)I_{i'_li_l}^{\mp}(l)
\eeq
and owing to the non-diagonal matrices $I^\pm(k)$ [\cf
\refeq{ferpmcoup}, \refeq{scapmcoup} and \refeq{gaupmcoup}],
contributions of $\SUtwo$-transformed Born matrix elements appear on
the left-hand side of \refeq{SScorr}.
In general,
these transformed Born matrix elements are not related to the
original Born matrix element and have to be evaluated explicitly.

The $\SS$ corrections for external longitudinal gauge bosons are obtained from \refeq{SScorr} with the equivalence theorem
\refeq{DLeqtheor}
, \ie the couplings and the Born matrix elements for Goldstone bosons\footnote{Note that for Goldstone bosons $\chi$, the
 equivalence theorem as well as the couplings \refeq{scapmcoupB} and \refeq{ZHcoup} contain the imaginary constant $\ri$.
}  have to be used on the right-hand side of \refeq{SScorr}.

The application of the above formulas is illustrated in \refse{sec:app} for the case of 4-particle processes, where
owing to $r_{12}=r_{34}$,
$r_{13}=r_{24}$ and  $r_{14}=r_{23}$, 
\refeq{SScorr} reduces to
\beqar  \label{4fsubdl} 
\de^\SS \M^{i_1i_2i_3i_4} &=&\sum_{\GB_a=A,Z,W^\pm}2 \left[\ls+\lWa\right]\times \label{eq:at} \\
&&\left\{\log{\frac{|r_{12}|}{s}}\left[
I^{\GB_a}_{i_1'i_1}(1)
I^{\bar{\GB}_a}_{i_2'i_2}(2)
\M_0^{i'_1i'_2i_3i_4}
+I^{\GB_a}_{i_3'i_3}(3)I^{\bar{\GB}_a}_{i'_4i_4}(4)
\M_0^{i_1i_2i'_3i'_4}
\right]\right.\nl
&&\left.{}+\log{\frac{|r_{13}|}{s}}\left[
I^{\GB_a}_{i_1'i_1}(1)I^{\bar{\GB}_a}_{i_3'i_3}(3)
\M_0^{i'_1i_2i'_3i_4}
+I^{\GB_a}_{i_2'i_2}(2)I^{\bar{\GB}_a}_{i_4'i_4}(4)
\M_0^{i_1i'_2i_3i'_4}
\right]\right.\nl
&&\left.{}+\log{\frac{|r_{14}|}{s}}\left[
I^{\GB_a}_{i_1'i_1}(1)I^{\bar{\GB}_a}_{i_4'i_4}(4)
\M_0^{i'_1i_2i_3i'_4}
+I^{\GB_a}_{i_2'i_2}(2)I^{\bar{\GB}_a}_{i_3'i_3}(3)
\M_0^{i_1i'_2i'_3i_4}
\right]\right\}  \nonumber
\eeqar
and the logarithm with $r_{kl}=s$ vanishes.
Note that this formula applies to $4\rightarrow 0$ processes, where
all particles or antiparticles and their momenta are
incoming. Predictions for $2\rightarrow 2$ processes  are obtained by
substituting outgoing particles (antiparticles) by the corresponding incoming
antiparticles (particles).

In addition to the angular terms in Eq. (\ref{eq:at}) at one loop there are also SL terms
from logarithms connected to parameter renormalization. These include the RG corrections to
the mixing angles, electromagnetic charge, Yukawa couplings and scalar self couplings in
the on-shell scheme. The corresponding terms are given in Ref. \cite{dp}. 

In Refs. \cite{m1,m3} we have shown that the splitting function approach employed for
the electroweak radiative correction in the high energy regime described by ${\cal L}_{symm}$
is indeed justified at the one loop level. For this cross check it is crucial that the
DL and the non-Yukawa SL terms factorize with respect to the {\it same} group factor. As can be seen by
the explicit results in Eqs. (\ref{eq:Wf}) and (\ref{eq:Wp}), these terms are given by
\begin{equation}
\left[ \frac{ g^2}{16 \pi^2}  T_i(T_i+1)+  \frac{{g^\prime}^2}{16 \pi^2}
 \frac{Y^2_i}{4} \right]  \left( \log^2 \frac{s}{M^2}- 3 \log \frac{s}{M^2} \right)
\end{equation}
for fermions and
\begin{equation}
\left[ \frac{ g^2}{16 \pi^2}  T_i(T_i+1)+  \frac{{g^\prime}^2}{16 \pi^2}
 \frac{Y^2_i}{4} \right]  \left( \log^2 \frac{s}{M^2}- 4 \log \frac{s}{M^2} \right)
\end{equation}
for scalars. The correct reproduction of the explicit high energy approximation based on
the physical fields in Ref. \cite{bddms} constitutes a very strong check on the overall
approach. Since also the corrections from region II) in Fig. \ref{fig:su2u1} are found to
agree with Ref. \cite{bddms} (including soft photon bremsstrahlung), the matching condition
at the weak scale is confirmed at the subleading level for both transverse as well as
longitudinal degrees of freedom. The latter includes in particular the treatment of
the would-be Goldstone bosons in the on-shell scheme to SL accuracy at high energies
according to the effective Lagrangian ${\cal L}_{symm}$ in Eq. (\ref{eq:lsym}).

This concludes our discussion of the one loop case. More definitions and representations
in the physical basis are presented in appendix \ref{sec:qn}.

\subsubsection{Two loop results}

Also at the two loop level there have been checks performed with fixed order
calculations in terms of the physical SM fields. 
While the level of agreement demonstrated at the one loop level is impressive,
it is very important for the desired accuracy of future linear colliders that
there is agreement with real two loop DL calculations and the
IREE method in order to trust predictions at the SL level.

We follow Refs. \cite{m2,hkk}
to demonstrate the method used and to show how non-exponentiating terms cancel
in the electroweak theory in the 't Hooft-Feynman gauge. Analogous results have been
obtained in the Coulomb gauge in Ref. \cite{bw,bw2} for external fermion lines and
recently also for longitudinal and transverse gauge bosons.
In this gauge, the DL corrections are related to two point functions and
thus directly to external legs. The intermediate steps, however, are more
cumbersome and therefore we only discuss the 't Hooft-Feynman gauge. In this gauge,
a complete set of Feynman rules is listed in appendix \ref{sec:fr}.

We begin with the case for right handed fermions coupled to a gluon,
or in general a gauge singlet.
\begin{figure}
\centering
\epsfig{file=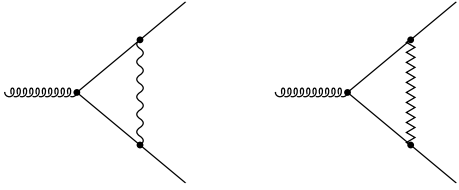,width=6cm}
\caption{The one loop electroweak SM Feynman diagrams leading to DL corrections
in the 't Hooft-Feynman gauge
for $g \longrightarrow f_{\rm R} {\overline f}_{\rm L}$.
Only the vertex corrections from the neutral Z-boson (zigzag-lines)
and the photon propagators contribute. At higher orders only corrections to
these two diagrams need to be considered in the DL approximation. The
photonic corrections are regulated by a fictitious mass terms $\lambda$.
In physical cross sections, the $\lambda$-dependence is canceled by the
effect of the emission of soft and collinear bremsstrahlung photons.}
\label{fig:ew1l}
\end{figure}
At the one loop level, the electroweak corrections are
depicted in Fig.~\ref{fig:ew1l}.
The fermion masses are
neglected for simplicity. They can, however, be added without changing the
nature of the higher order corrections.
For right handed fermions we only need to consider the neutral electroweak
gauge bosons, i.e. we are concerned with an $U(1)_{\rm R} \times
U(1)_{\rm Y}$ gauge theory which is spontaneously broken to yield
the Z-boson and photon fields.
The DL-contribution of a particular Feynman diagram is thus given by
\begin{equation}
{\cal M}_k = {\cal M}_{\rm Born} \; {\cal F}_k \label{eq:formfac}
\end{equation}
where the ${\cal F}_k$ are given by integrals over the remaining
Sudakov parameters
at the $n$-loop level:
\begin{equation}
\quad {\cal F}_k = \left( \frac{ e^2_f
}{8 \pi^2}
\right)^n \prod^n_{i=1} \int^1_0 \int^1_0 \frac{d u_i}{u_i} \frac{d
v_i}{v_i} \Theta_k \label{eq:SudQED}
\end{equation}
The $\Theta_k$ describe the regions of integration which lead to DL corrections.
At one loop, the diagrams of Fig.~\ref{fig:ew1l} lead to
\begin{eqnarray}
{\cal M}^{{\rm R}^{(1)}}_{\rm DL} &=& {\cal M}^{\rm R}_{\rm Born} \left( 1 -
\frac{e^2_f}{8 \pi^2} \int^1_0 \frac{du}{u}
\int^1_0 \frac{dv}{v} \left[ \theta (s u v- \lambda^2)+
\frac{\rm s_w^2}{\rm c_w^2} \theta ( s u v - M^2) \right] \right) \nonumber \\
&=&{\cal M}^{\rm R}_{\rm Born} \left( 1
- \frac{e^2_f}{16 \pi^2} \left[ \log^2 \frac{s}{\lambda^2}+
\frac{\rm s_w^2}{\rm c_w^2} \log^2 \frac{s}{M^2} \right] \right) \label{eq:ew1l}
\end{eqnarray}
which is the well known result from QED plus the same term with a rescaled
coupling (see the Feynman rules in appendix \ref{sec:fr}) and infrared cutoff. The restriction
to right handed fermions allows us to focus solely on the mass gap of
the neutral
electroweak gauge bosons. The $W^\pm$ only couples to left handed doublets.
\begin{figure}
\centering
\epsfig{file=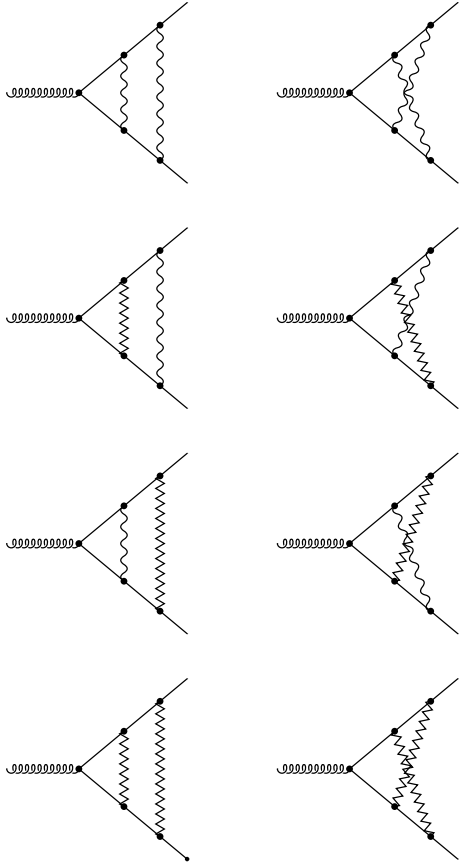,width=6cm}
\caption{The two loop electroweak SM Feynman diagrams leading to DL corrections
in the 't Hooft-Feynman gauge
for $g \longrightarrow f_{\rm R} {\overline f}_{\rm L}$.
The neutral Z-boson (zigzag-lines) and the photon propagators possess
different on-shell regions due to the mass gap.} \label{fig:ewfd}
\end{figure}
At the two loop level we have to consider more diagrams than in the
QED case. The relevant Feynman graphs that give DL corrections in the
't Hooft-Feynman gauge are depicted in Fig.~\ref{fig:ewfd}.
Only these corrections can yield four logarithms at the two loop level
in the 't Hooft-Feynman gauge. Otherwise one cannot obtain the required pole terms
(as is well known in QED \cite{ms1}).
They contain diagrams
where the exchanged gauge bosons enter with differing on-shell regions,
i.e. differing integration regions which give large DL corrections.
It is instructive to revisit
the case of pure QED corrections, since the topology of the graphs
yielding DL contributions in the 't Hooft-Feynman gauge is unchanged.
In QED at the two loop level, the scalar integrals corresponding
to the first row of Fig.~\ref{fig:ewfd} are given by:
\begin{eqnarray}
 S^{\rm QED}_1 \!\! &\equiv& \!\! \int^1_0 \! \frac{du_1}{u_1} \!
\int^1_0 \! \frac{dv_1}{v_1} \!
\int^1_0 \! \frac{du_2}{u_2} \!
\int^1_0 \frac{dv_2}{v_2} \theta ( s u_1 v_1 - \lambda^2) \theta (s u_2 v_2
- \lambda^2) \theta ( u_1\!\!-\!\!u_2) \theta (v_1\!\!-\!\!v_2) \;\;\;\;\; \label{eq:qed1} \\
S^{\rm QED}_2 \!\!
&\equiv&\!\!
\int^1_0 \! \frac{du_1}{u_1} \! \int^1_0 \! \frac{dv_1}{v_1} \! \int^1_0 \!
\frac{du_2}{u_2} \!
\int^1_0 \! \frac{dv_2}{v_2} \theta ( s u_1 v_1 - \lambda^2) \theta (s u_2 v_2
- \lambda^2) \theta ( u_1\!\!-\!\!u_2) \theta (v_2\!\!-\!\!v_1) \;\;\;\; \label{eq:qed2}
\end{eqnarray}
Thus, in QED we find the familiar result
\begin{eqnarray}
S^{\rm QED}_1+S^{\rm QED}_2 &=&
\int^1_0 \! \frac{du_1}{u_1} \!
\int^1_0 \! \frac{dv_1}{v_1} \!
\int^1_0 \! \frac{du_2}{u_2} \!
\int^1_0 \frac{dv_2}{v_2} \theta ( s u_1 v_1 - \lambda^2) \theta (s u_2 v_2
- \lambda^2) \theta ( u_1-u_2) \nonumber \\
&=& \int^1_\frac{\lambda^2}{s} \frac{du_1}{u_1} \int^{1}_\frac{\lambda^2}{
s u_1} \frac{dv_1}{v_1} \frac{1}{2} \log^2 \frac{s u_1}{\lambda^2}
\nonumber \\
&=& \frac{1}{2} \left( \frac{1}{2} \log^2 \frac{s}{\lambda^2} \right)^2
\label{eq:qedsud}
\end{eqnarray}
whichs yields the second term of the exponentiated one loop result in
Eq.~(\ref{eq:ew1l})
for ${\rm s_w} \longrightarrow 0$.
In the electroweak theory, we also need to consider the remaining diagrams
of Fig.~\ref{fig:ewfd}. The only differences occur because of the rescaled
coupling according to the Feynman rules in appendix \ref{sec:fr} and the
fact that the propagators have a different mass. Thus the second row
of Fig.~\ref{fig:ewfd} leads to
\begin{eqnarray}
S^{\rm M, \lambda}_1+S^{\rm M, \lambda}_2 &=& \frac{\rm s^2_w}{\rm c^2_w}
\int^1_0 \! \frac{du_1}{u_1} \!
\int^1_0 \! \frac{dv_1}{v_1} \!
\int^1_0 \! \frac{du_2}{u_2} \!
\int^1_0 \frac{dv_2}{v_2} \theta ( s u_1 v_1 - M^2) \theta (s u_2 v_2
- \lambda^2) \theta ( u_1-u_2) \nonumber \\
&=& \frac{\rm s^2_w}{\rm c^2_w}
\int^1_\frac{M^2}{s} \frac{du_1}{u_1} \int^{1}_\frac{M^2}{
s u_1} \frac{dv_1}{v_1} \frac{1}{2} \log^2 \frac{s u_1}{\lambda^2}
\nonumber \\
&=& \frac{\rm s^2_w}{\rm c^2_w}
\left[ \frac{1}{8} \log^4 \frac{s}{\lambda^2} - \frac{1}{6} \log^3
\frac{s}{\lambda^2} \log \frac{M^2}{\lambda^2}
+ \frac{1}{24} \log^4 \frac{ M^2}{\lambda^2} \right]
\label{eq:2sud}
\end{eqnarray}
where we indicate the gauge boson masses of the two propagators in the scalar
functions. Analogously we find for the remaining two rows
\begin{eqnarray}
S^{\rm \lambda, M}_1+S^{\rm \lambda, M}_2 &=& \frac{\rm s^2_w}{\rm c^2_w}
\int^1_0 \! \frac{du_1}{u_1} \!
\int^1_0 \! \frac{dv_1}{v_1} \!
\int^1_0 \! \frac{du_2}{u_2} \!
\int^1_0 \frac{dv_2}{v_2} \theta ( s u_1 v_1 - \lambda^2) \theta (s u_2 v_2
- M^2) \theta ( u_1-u_2) \nonumber \\
&=& \frac{\rm s^2_w}{\rm c^2_w}
\int^1_\frac{M^2}{s} \frac{du_2}{u_2} \int^{1}_\frac{M^2}{
s u_2} \frac{dv_2}{v_2} \frac{1}{2} \left(\log^2 \frac{s}{\lambda^2} -
\log^2 \frac{s u_2}{\lambda^2} \right)
\nonumber \\
&=& \frac{\rm s^2_w}{\rm c^2_w}
\left[ \frac{1}{4} \log^2
\frac{s}{M^2} \log \frac{ s}{\lambda^2} - \frac{\rm c^2_w}{\rm s^2_w} \left(
S^{\rm M, \lambda}_1+S^{\rm M,
\lambda}_2 \right) \right]
\label{eq:3sud} \\
{\rm and} \;\;\;\;\;\;\;\;\;\;\;\;\;\;\; &&  \nonumber \\
S^{\rm M, M}_1+S^{\rm M, M}_2 &=&
\frac{1}{2} \frac{\rm s_w^4}{\rm c_w^4} \left( \frac{1}{2} \log^2 \frac{s}{M^2} \right)^2
\label{eq:4sud}
\end{eqnarray}
Thus, we find for the full two loop electroweak DL-corrections
\begin{eqnarray}
{\cal M}^{{\rm R}^{(2)}}_{\rm DL} &\equiv& {\cal M}^{\rm R}_{\rm Born}\left( 1 +
\delta_{\rm R}^{(1)} + \delta_{\rm R}^{(2)} \right) \label{eq:2lres} \\
{\rm with} \;\;\;\;\;\;\;\;\;\; && \nonumber \\
\delta_{\rm R}^{(1)} &=&
- \frac{e^2_f}{16 \pi^2} \left( \log^2 \frac{s}{\lambda^2} + \frac{\rm s^2_w}{\rm c^2_w}
\log^2 \frac{s}{M^2} \right) \\
{\rm and} \;\;\;\;\;\;\;\;\;\; && \nonumber \\
\delta_{\rm R}^{(2)}&=& \left(\frac{e^2_f}{8 \pi^2} \right)^2 \left[
S^{\rm QED}_1+S^{\rm QED}_2+S^{\rm M, \lambda}_1+S^{\rm M,\lambda}_2
+S^{\rm \lambda, M}_1+S^{\rm \lambda, M}_2 + S^{\rm M, M}_1+S^{\rm M, M}_2
\right] \nonumber \\
&=& \left(\frac{e^2_f}{8 \pi^2} \right)^2 \left[
\frac{1}{8} \log^4 \frac{s}{\lambda^2} +
\frac{\rm s^2_w}{4 \rm c^2_w} \log^2 \frac{s}{M^2} \log^2 \frac{ s}{\lambda^2}
+\frac{\rm s^4_w}{8 \rm c^4_w} \log^4 \frac{s}{M^2} \right] \nonumber \\
&=& \frac{1}{2} \left[- \frac{e^2_f}{16 \pi^2} \left( \log^2 \frac{s}{\lambda^2} + \frac{\rm s^2_w}{\rm c^2_w}
\log^2 \frac{s}{M^2} \right) \right]^2 \label{eq:ewres}
\end{eqnarray}
which is precisely the second term of the exponentiated one loop result
in the process of $g \longrightarrow f_{\rm R} {\overline f}_{\rm L}$.
In Ref. \cite{m2} we showed that by using the appropriate quantum numbers listed
in appendix \ref{sec:qn}, that the result of the IREE method
of section \ref{sec:tr} gives indeed the same result. 

For left handed fermions
the calculation was performed in the 't Hooft-Feynman gauge in Ref. \cite{hkk}.
It is instructive to list the intermediate steps of the calculation
in order to identify terms not present in the case of unbroken gauge theories.
We write the form of the correction as
\begin{equation}
{\cal M}^{{\rm L}^{(2)}}_{\rm DL} \equiv {\cal M}^{\rm L}_{\rm Born}\left( 1 +
\delta_{\rm L}^{(1)} + \delta_{\rm L}^{(2)} \right) \label{eq:2lresl} 
\end{equation}
where the one loop result is given by
\begin{equation}
\delta_{\rm L}^{(1)} =
- \frac{e^2_f}{16 \pi^2} \log^2 \frac{s}{\lambda^2} - 
\left(  \frac{g^2}{16 \pi^2} T_f \left(T_f+1 \right) + \frac{{g^\prime}^2}{16 \pi^2}
\frac{Y^2}{4} - \frac{e^2_f}{16 \pi^2} \right)
\log^2 \frac{s}{M^2} 
\end{equation}
The two loop contribution can be written as a sum of the straight ladder diagrams,
the crossed ladder and the three boson diagrams, which also enter due to the charged
gauge boson exchange. Respectively we denote these terms as
\begin{equation}
\delta_{\rm L}^{(2)} = sl+cl+tb
\end{equation}
The reason for this is that in QCD, the non-Abelian contribution of $cl$ ($\sim
C_A$) is canceled by the $tb$ terms. The explicit results are
\begin{eqnarray}
&& \!\!\!\!\!\!\!\!\!\!\!\!\!\!\!\!\!\! sl = \frac{1}{64 \pi^4} \left\{ \frac{e^4_f}{24} \log^4 \frac{s}{\lambda^2}+
\frac{e^2_f}{4} \left(g^2 T_f \left(T_f+1 \right) + {g^\prime}^2 \frac{Y_f^2}{4} - e^2_f \right)
\log^2 \frac{s}{\lambda^2} \log^2 \frac{s}{M^2} \right. \nonumber \\ && \!\!\!\!\!\!\!\!\!\!\!\!\!\!\!\!\!\!
+ \left(g^2 T_f \left(T_f+1 \right) + {g^\prime}^2 \frac{Y_f^2}{4} - e^2_f \right)^2 
\frac{1}{24} \log^4 \frac{s}{M^2} + e^2_f \left(g^2 T_f \left(T_f+1 \right) + {g^\prime}^2
 \frac{Y_f^2}{4} - e^2_f \right) \nonumber \\ && \!\!\!\!\!\!\!\!\!\!\!\!\!\!\!\!\!\! \left. \times \left[
\frac{1}{6} \log^4 \frac{s}{M^2}-\frac{1}{3} \log^3 \frac{s}{M^2}\log \frac{s}{\lambda^2} \right]
\right\} \label{eq:sl} \\ 
&& \!\!\!\!\!\!\!\!\!\!\!\!\!\!\!\!\!\! cl = \frac{1}{64 \pi^4} \left\{ \frac{e^4_f}{12}  \log^4 \frac{s}{\lambda^2} +
\left[ \left(g^2 T_f \left(T_f+1 \right) + {g^\prime}^2 \frac{Y_f^2}{4} - e^2_f \right)^2-g^4
T_f \left( T_f+1 \right) \right] \times \right. \nonumber \\ && \!\!\!\!\!\!\!\!\!\!\!\!\!\!\!\!\!\!\frac{1}{12} \log^4 \frac{s}{M^2} 
+ e_f^2 \left(g^2 T_f \left(T_f+1 \right) + {g^\prime}^2 \frac{Y_f^2}{4} - e^2_f \right)
\left[ - \frac{1}{6} \log^4 \frac{s}{M^2} + \frac{1}{3} \log^3 \frac{s}{M^2} \log \frac{s}{\lambda^2}
\right] \nonumber \\ && \!\!\!\!\!\!\!\!\!\!\!\!\!\!\!\!\!\! \left.
+ 2 g^2 e^2 Q_f T_f^3 \frac{1}{6} \left[ \log^4 \frac{s}{M^2}- \log^3 \frac{s}{M^2} \log \frac{s}{\lambda^2}
\right] \right\}  \label{eq:cl} \\ && \!\!\!\!\!\!\!\!\!\!\!\!\!\!\!\!\!\!
tb = \frac{1}{64 \pi^4} \left\{ g^4 T_f \left( T_f+1 \right) \frac{1}{12} \log^4 \frac{s}{M^2}
- 2 g^2 e^2 Q_f T_f^3 \frac{1}{6}\left[ \log^4 \frac{s}{M^2}- \log^3 \frac{s}{M^2} \log \frac{s}{\lambda^2}
\right] \right\} \label{eq:tb}
\end{eqnarray}
Thus we see that we have
\begin{eqnarray}
\!\!\!\!\!\!\!\!\!\!\!\!\!\!\! \delta_{\rm L}^{(2)} &=& 
\frac{1}{64 \pi^4} \left\{ \frac{e^4_f}{8} \log^4 \frac{s}{\lambda^2}+
\frac{e^2_f}{4} \left(g^2 T_f \left(T_f+1 \right) + {g^\prime}^2 \frac{Y_f^2}{4} - e^2_f \right)
\log^2 \frac{s}{\lambda^2} \log^2 \frac{s}{M^2} \right. \nonumber \\  \!\!\!\!\!\!\!\!\!\!\!\!\!\!\!&&
+ \left. \left(g^2 T_f \left(T_f+1 \right) + {g^\prime}^2 \frac{Y_f^2}{4} - e^2_f \right)^2 
\frac{1}{8} \log^4 \frac{s}{M^2} \right\} \nonumber \\ \!\!\!\!\!\!\!\!\!\!\!\!\!\!\!&=&
\frac{1}{2} \left\{- \frac{e^2_f}{16 \pi^2} \log^2 \frac{s}{\lambda^2} - 
\left(  \frac{g^2}{16 \pi^2} T_f \left(T_f+1 \right) + \frac{{g^\prime}^2}{16 \pi^2}
\frac{Y^2}{4} - \frac{e^2_f}{16 \pi^2} \right)
\log^2 \frac{s}{M^2} \right\}^2 \label{eq:lres}
\end{eqnarray}
which is precisely half the square of the first order correction. The novel feature here is that
the way the cancellation happens is more complicated in the electroweak theory as already the
straight ladder diagrams possess non-exponentiating terms. Only the full sum displays the familiar
exponentiation. In the framework of the IREE method, however, these complications are naturally
accounted for by carefully examining the kernel in both the high energy regime according to
${\cal L}_{\rm symm}$ in Eq. (\ref{eq:lsym}) and the region
where only QED effects enter. At the scale $M$ the appropriate matching conditions have to be
employed. 
On the other hand, this complication is the reason for the wrong
result in Ref. \cite{cc} which is based on QCD factorization properties.

\section{Applications} \label{sec:app}

In this section we apply the results presented in section \ref{sec:bgt} to specific processes at a future
$e^+ e^-$ collider. As mentioned in the introduction, such a machine must be a precision tool to disentangle
and clarify the physics discovered at the LHC or the Tevatron. 
For our purposes we define precision as effects that change
cross sections at the 1 \% level. We would like to emphasize that at that level it is not sufficient
to calculate only logarithmic corrections to all orders. At one loop, constant terms 
such as $\frac{m_t^2}{M^2}$ can lead to non-negligible effects and thus, a full one loop analysis 
is needed in these cases. In this work we want to emphasize the effect of the higher order corrections where
constant terms are negligible. For all SL terms we can use the results summarized in Eq. (\ref{eq:sirg})
and the angular terms as described in section \ref{sec:at}. 
In the following we will display
only the ``pure'' electroweak corrections originating above the weak scale $M$ relative to the Born
cross sections.
The importance of the angular dependent 
corrections is also discussed in Refs. \cite{kps,kmps,mang} at higher orders and in
\cite{brv2} 
at the one loop level.
While we focus below on the linear collider case, also for hadronic machines the electroweak corrections
in the TeV range are important and must be included. 
In Ref. \cite{adp} the effect of one-loop logarithmic electroweak radiative corrections on WZ and $W\gamma$
production processes at the LHC was studied for instance. 
Using the leading-pole approximation these
corrections were implemented into Monte Carlo programs for 
$pp\to l\nu_l l'\bar l', l\nu_l\gamma$. The authors find that electroweak
corrections lower the predictions by 5-20 \% in the physically interesting region of large transverse momentum and
small rapidity separation of the gauge bosons. 

We begin in section \ref{sec:gbp} with transverse and longitudinal gauge boson production, followed by
Higgs boson production in section \ref{sec:hp} and heavy quark production in \ref{sec:hqp}. 
For the latter two cases we also include supersymmetric Sudakov effects, in particular the 
dependence on $\tan \beta$.
For all two to two processes we denote
\begin{equation}
t = - \frac{s}{2} \left( 1 - \cos \theta \right) \;, \;\; 
u = - \frac{s}{2} \left( 1 + \cos \theta \right)  
\end{equation}
which is valid at high energies and where $\theta$ is the scattering angle between the initial
and final state particles.
For definiteness, we use the following parameters
\begin{equation}
m_t=174 \; {\rm GeV}, \;\; M=80.35 \; {\rm GeV}, \;\; s_{\rm w}^2=0.22356, \;\; \frac{e^2(M^2)}{4 \pi}=\frac{1}{128}
\end{equation}
in all numerical results displayed. 
Furthermore we use $g=e/s_{\rm w}$ and $g^\prime=e/c_{\rm w}$ below. 
The angular terms at one loop are obtained from Eq. (\ref{eq:at})
which was derived in Ref. \cite{dp} except for the case of Higgs-strahlung.
All energy units in the figures are in GeV.

\subsection{Gauge boson production} \label{sec:gbp}

\begin{figure}
\centering
\epsfig{file=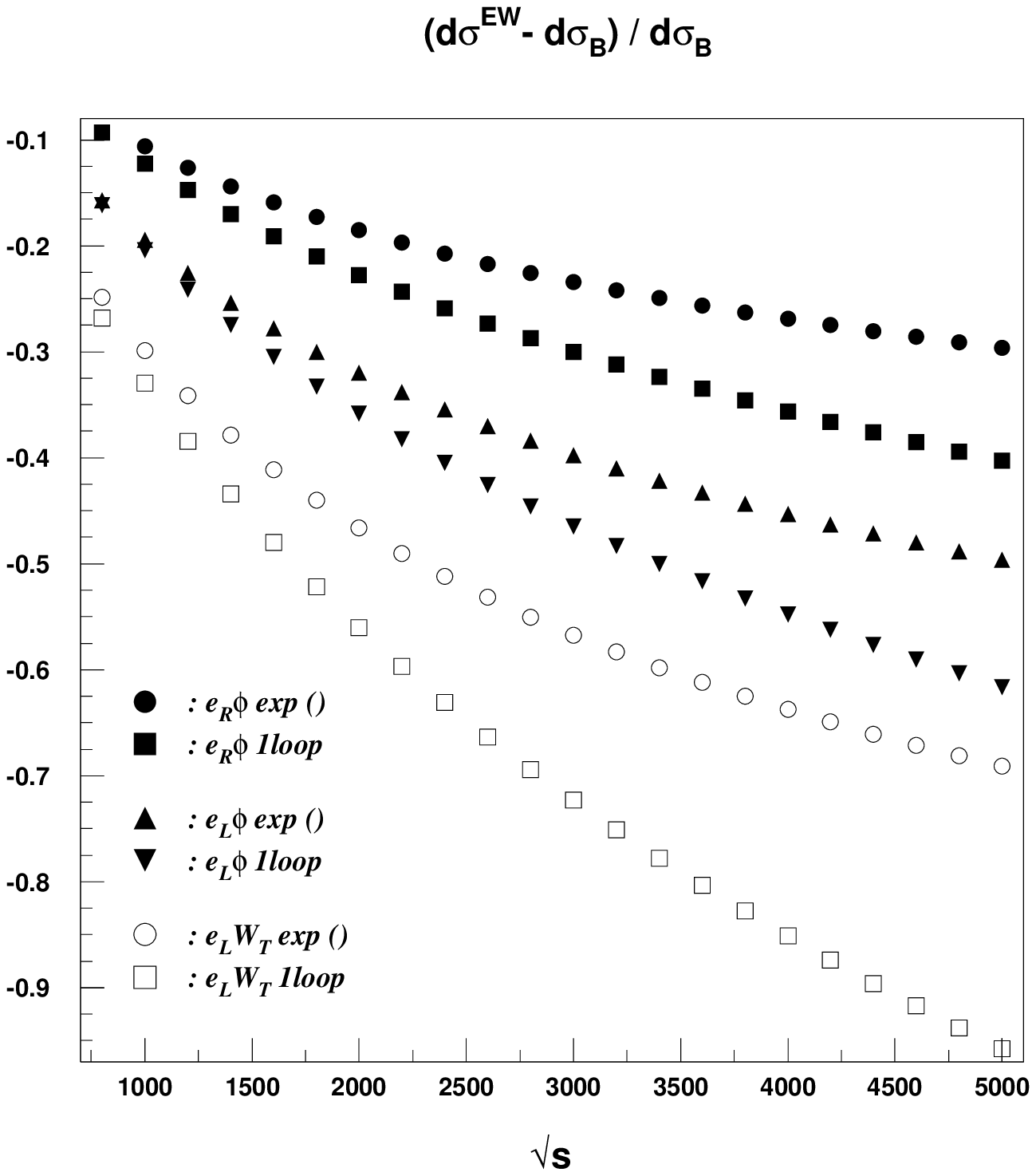,width=12cm}
\caption{The purely electroweak virtual
corrections relative to the Born cross section in transverse and
longitudinal $W^\pm$ production in $e^+ e^-$ collisions at 90$^o$ scattering
angle as a function of the c.m. energy. The polarization
is indicated in the figure for each symbol. Given are the one loop 
and the resummed corrections to SL accuracy in each case. 
It is clearly visible that the
difference between the two approaches is non-negligible at TeV energies and
necessitates the inclusion of the higher order terms. Pure QED corrections from below the
weak scale
are omitted.}
\label{fig:ewws}
\end{figure}              

In this section we discuss the effect of the higher order pure electroweak corrections to transverse
and longitudinal gauge bosons production at the cross section level. These processes, in particular the
longitudinal ones, are important at high energies for the unitarity of the theory. If no Higgs boson 
should be found, the $W^\pm$ sector has to become strongly interacting in the TeV range in order to
preserve unitarity. It would therefore be important to know when and how the new dynamics would deviate
from the perturbative SM predictions.

Fig. \ref{fig:ewws} displays the energy dependence of the respective cross sections in $W^\pm$ production 
for the various polarizations. All universal terms are given by Eqs. (\ref{eq:WfRG}), (\ref{eq:WgRG})
and (\ref{eq:WpRG}). In addition we have angular terms which read relative to the Born amplitude
in the high energy effective regime described by ${\cal L}_{symm}$ in Eq. (\ref{eq:lsym}):
\begin{eqnarray}
\sum_{B,W^a} \delta^\theta_{e^+_Re^-_L \longrightarrow W^+_{\rm T} W^-_{\rm T}} &=&
-\frac{g^2(M^2)}{8\pi^2} \log \frac{s}{M^2} \left( \log \frac{t}{u} + \left(1-\frac{t}{u}\right) \log
\frac{-t}{s} \right) \label{eq:amt} \\
\sum_{B,W^a} \delta^\theta_{e^+_Re^-_L \longrightarrow \phi^+ \phi^-} &=&
-\frac{g^2(M^2)}{8\pi^2} \log \frac{s}{M^2} \left(  \frac{1}{2c_{\rm w}^2}
\log \frac{t}{u} + 2 c_{\rm w}^2 \log
\frac{-t}{s} \right) \label{aml} \\
\sum_{B,W^a} \delta^\theta_{e^+_Le^-_R \longrightarrow \phi^+ \phi^-} &=&
-\frac{{g^\prime}^2(M^2)}{8\pi^2} \log \frac{s}{M^2}
\log \frac{t}{u}
\label{apl}
\end{eqnarray}
In terms of their numerical coefficients relative to the Born cross section we have
\begin{eqnarray}
d \sigma^{e_L,W_T}_\theta &=& d \sigma^{e_L,W_T}_{\rm Born} \left\{ - 8.95 \left[ \log \frac{t}{u} +
\left( 1 - \frac{t}{u} \right) \log \frac{-t}{s} \right] \frac{e^2(M^2)}{8 \pi^2} \log \frac{s}{M^2}
\right\} \label{eq:aelwt} \\
d \sigma^{e_L,W_L}_\theta &=& d \sigma^{e_L,W_L}_{\rm Born} \left\{ - \left[ 5.76 \log \frac{t}{u} +
13.9 \log \frac{-t}{s} \right] \frac{e^2(M^2)}{8 \pi^2} \log \frac{s}{M^2}
\right\} \label{eq:aelwl} \\
d \sigma^{e_R,W_L}_\theta &=& d \sigma^{e_R,W_L}_{\rm Born} \left\{ - 2.58 \log \frac{t}{u} 
\;\; \frac{e^2(M^2)}{8 \pi^2} \log \frac{s}{M^2}
\right\} \label{eq:aerwl}
\end{eqnarray}
and RG corrections which at one loop are given by
\begin{eqnarray}
d \sigma^{e_L,W_T}_{\rm RG} &=& d \sigma^{e_L,W_T}_{\rm Born} \left\{- 2 \beta_0 \;\; \frac{g^2 (M^2)}{4 \pi^2}
\log \frac{s}{M^2} \right\} \\
d \sigma^{e_L,W_L}_{\rm RG} &=& d \sigma^{e_L,W_L}_{\rm Born} \left\{ \frac{41-82 c^2_{\rm w}
+22 c^4_{\rm w}}{12 c^2_{\rm w}} \frac{g^2(M^2)}{4 \pi^2}
\log \frac{s}{M^2} \right\} \\
d \sigma^{e_R,W_L}_{\rm RG} &=& d \sigma^{e_R,W_L}_{\rm Born} \left\{ \frac{41
}{12} \;\; \frac{{g^\prime}^2(M^2)}{4 \pi^2}
\log \frac{s}{M^2} \right\} \\
\end{eqnarray}
The corrections are written relative to the Born cross sections and Fig. \ref{fig:ewws} shows the
one loop and all orders results in each case to SL accuracy. The largest effect
can be seen in the transverse sector for left handed electrons where the resummed terms are of order
30 \% at 1 TeV, 55 \% at 3 TeV
and 65 \% at 5 TeV. The respective cross section for right handed electrons
is mass suppressed \cite{bddms}.
Also for longitudinal $W^\pm$ ( $\sim \phi^\pm$ via the equivalence
theorem) we have significant effects. For left handed electrons the resummed terms are of order
20 \% at 1 TeV, 38 \% at 3 TeV
and 46 \% at 5 TeV.
For right handed electrons the resummed corrections are of order
11 \% at 1 TeV, 22 \% at 3 TeV
and 26 \% at 5 TeV.

In each case it can be seen that the one loop contributions alone are insufficient in the
TeV regime and differ from the resummed results in the percentile range at 1 TeV, of order
10 \% at 3 TeV and even more beyond. 

\begin{figure}
\centering
\epsfig{file=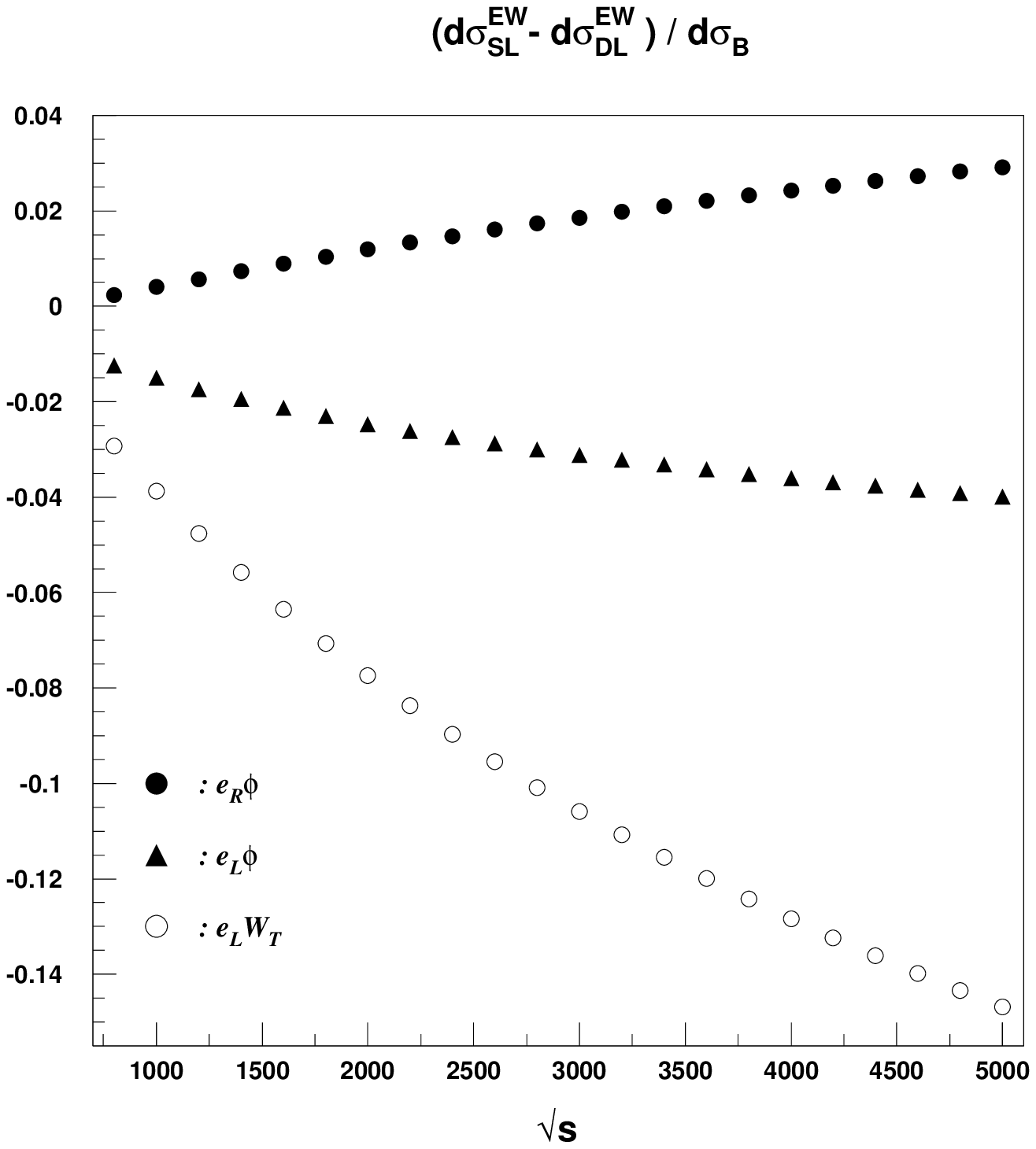,width=12cm}
\caption{The purely electroweak virtual
corrections relative to the Born cross section in transverse and
longitudinal $W^\pm$ production in $e^+ e^-$ collisions at 90$^o$ scattering
angle as a function of the c.m. energy. The polarization
is indicated in the figure for each symbol. Given are the full one loop 
and the resummed corrections to DL and SL accuracy in each case. 
It is clearly visible that the
difference between the two approaches, originating from SL terms at the
two loop level, is non-negligible at TeV energies and
necessitates the inclusion of the higher order SL terms. Pure QED corrections
from below the weak scale are omitted.}
\label{fig:ewwr}
\end{figure}              

It is, however, not only important to take
higher order DL corrections into account but also the higher order SL terms. This can be seen
in Fig. \ref{fig:ewwr} where the one loop terms agree (to SL accuracy) and at higher orders
in one case only the DL and in the other also the SL terms are kept. 
Again the largest effect occurs in the transverse $W^\pm$ sector. The difference between the
higher order SL and DL corrections is of order 4 \% at 1 TeV, 10.5 \% at 3 TeV
and 14.5 \% at 5 TeV. Also for longitudinal $W^\pm$ production the effects are 
significant. For left handed electrons the difference between the higher order SL and DL terms
changes cross sections by 1.8 \% at 1 TeV, 3 \% at 3 TeV 
and 3.5 \% at 5 TeV.  
For right handed electrons we have differences of
0.5 \% at 1 TeV, 2 \% at 3 TeV 
and 3 \% at 5 TeV.  

Thus, at these energies the SL terms can be as large as the leading terms and must not be omitted. 
For the longitudinal gauge bosons, there is a partial cancellation between the subleading Sudakov and
Yukawa enhanced terms. Therefore, the overall effect is larger in the transverse sector but still large
in the longitudinal one.

\begin{figure}
\centering
\epsfig{file=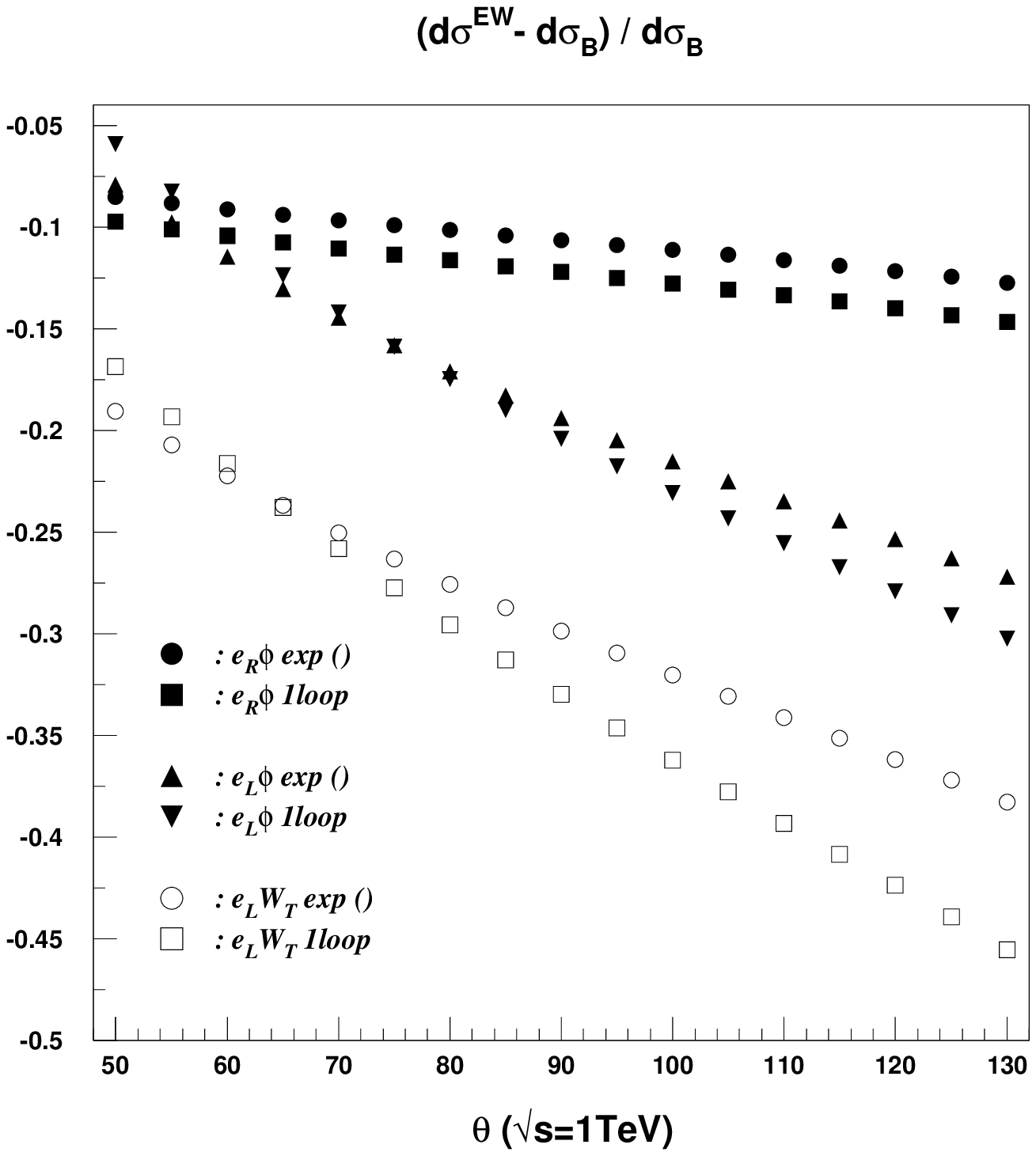,width=8.5cm}
\epsfig{file=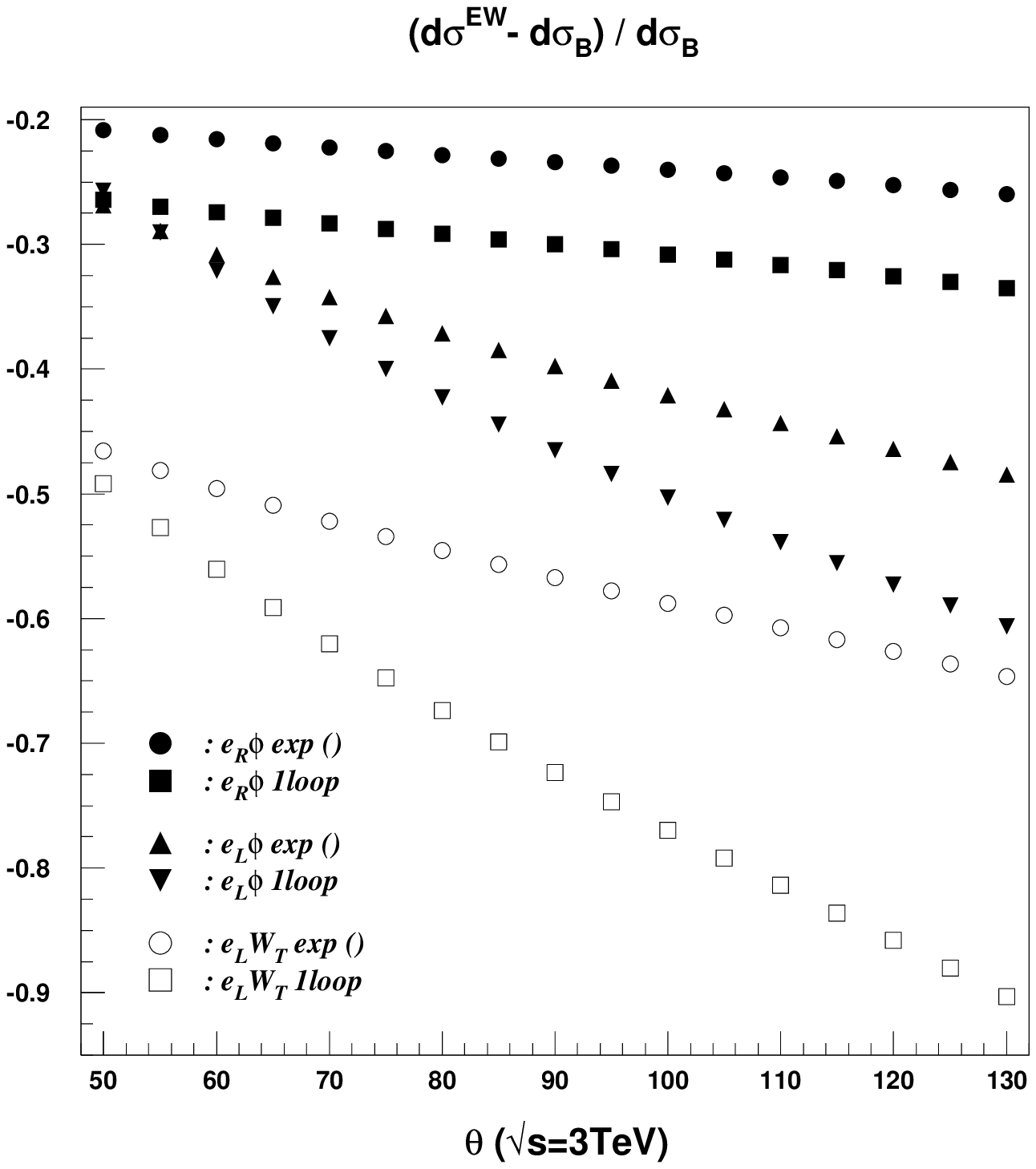,width=8.5cm} \vspace{-0.5cm}
\caption{The purely electroweak virtual
corrections relative to the Born cross section in transverse and
longitudinal $W^\pm$ production in $e^+ e^-$ collisions at 1 and 3 TeV 
c.m. energy as a function of the scattering angle. The polarization
is indicated in the figure for each symbol. Given are the one loop 
and the resummed corrections to SL accuracy in each case. It is clearly visible that the
difference between the two approaches is non-negligible at TeV energies and
necessitates the inclusion of the higher order terms. Pure QED corrections
from below the weak scale are omitted.}
\label{fig:ewwt}
\end{figure}              

Also the angular terms are significant as can be seen in Fig. \ref{fig:ewwt}. We treat the angular
terms at higher orders as described in section \ref{sec:at} and the one loop terms from Eqs. 
(\ref{eq:aelwt}), (\ref{eq:aelwl}) and (\ref{eq:aerwl}).
The figure displays the effect for fixed c.m. energy of 1 and 3 TeV and we have written all angular
terms in such a way that they are proportional to the Born cross section. This is always possible but
involves factor of $\frac{t}{u}$ etc. It can be seen that the angular corrections are large and vary
for the resummed contributions by almost 20 \% at 1 TeV and 17 \% at 3 TeV 
for transverse $W^\pm$ production for scattering angles
between 50$^o$ and 130$^o$. Also for longitudinal $W^\pm$ production the corrections are large.
For left handed electrons the cross section changes by about 19 \% at 1 TeV and 20 \% at 3 TeV
in the same angular range. For the same range of scattering angles, the
right handed electrons the resummed angular terms change cross sections by
about 4.5 \% at 1 TeV and 5 \% at 3 TeV. The one loop corrections are even larger and lead to 
significantly different results. Thus, the higher order terms are very important and it is mandatory to
investigate if the two loop angular corrections in the full electroweak theory are indeed given by
the product of the one loop terms and the Sudakov form factor. 

In $e^+ e^-$ collisions, the only non-mass suppressed longitudinal $Z$ process is the
Higgs-strahlung process. As such we discuss it in the next section together with other Higgs
production processes.

\subsection{Higgs production} \label{sec:hp}

In this section we discuss light SM Higgs production processes relevant to the linear collider program.
We begin with the Higgs-strahlung process $e^+ e^- \longrightarrow H \chi$. The cross section for this
process is smaller than the $W$-fusion process at TeV energies, however, it would still be utilized to 
collect more Higgs events. The electroweak corrections, however, will reduce the cross section.

\begin{figure}
\centering
\epsfig{file=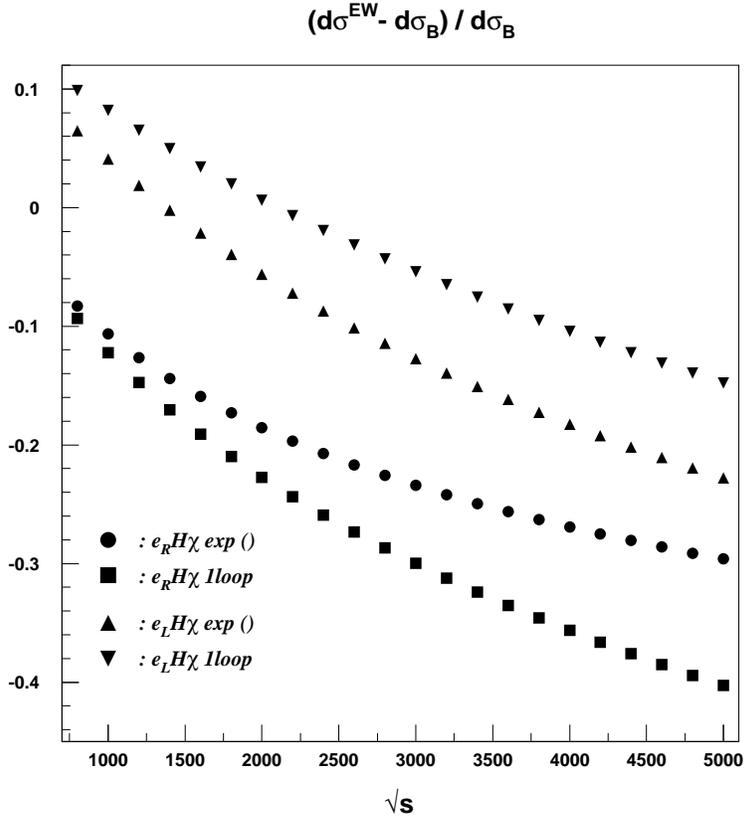,width=12cm}
\caption{The purely electroweak virtual
corrections relative to the Born cross section in Higgs-strahlung
in $e^+ e^-$ collisions at 90$^o$ scattering
angle as a function of the c.m. energy. The polarization
is indicated in the figure for each symbol. Given are the one loop 
and the resummed corrections to SL accuracy in each case. 
The difference between the two approaches is non-negligible at TeV energies and
necessitates the inclusion of the higher order terms. Pure QED corrections
from below the weak scale are omitted.}
\label{fig:ewhc}
\end{figure}

All universal corrections are given by Eqs. (\ref{eq:WfRG})
and (\ref{eq:WpRG}). In addition we have angular terms which read relative to the Born amplitude
in the high energy effective regime described by ${\cal L}_{symm}$ in Eq. (\ref{eq:lsym}):
\begin{equation}
\sum_{B,W^a} \delta^\theta_{e^+_Re^-_L \longrightarrow H \chi} =
-\frac{g^2(M^2)}{8\pi^2} \frac{4 c^2_{\rm w}}{c^2_{\rm w}-s^2_{\rm w}} \log \frac{s}{M^2} \left( \log \frac{-t}{s} + \log
\frac{-u}{s} \right) \label{eq:ahc} \\
\end{equation}
Numerically we have for relative to the Born cross section:
\begin{eqnarray}
d \sigma^{e_L,H}_\theta &=& d \sigma^{e_L,H}_{\rm Born} \left\{-50.25 \left[ \log \frac{-t}{s} +
\log \frac{-u}{s} \right] \frac{e^2(M^2)}{8 \pi^2} \log \frac{s}{M^2}
\right\} \label{eq:eHa}
\end{eqnarray}
and RG corrections which at one loop are given by
\begin{eqnarray}
d \sigma^{e_L,H}_{\rm RG} &=& d \sigma^{e_L,H}_{\rm Born} \left\{ \frac{41-82 c^2_{\rm w}
+60 c^4_{\rm w}}{12 c^2_{\rm w}(s^2_{\rm w}-c^2_{\rm w})} \frac{g^2(M^2)}{4 \pi^2}
\log \frac{s}{M^2} \right\} \\
d \sigma^{e_R,H}_{\rm RG} &=& d \sigma^{e_R,H}_{\rm Born} \left\{ \frac{41
}{12} \;\; \frac{{g^\prime}^2(M^2)}{4 \pi^2}
\log \frac{s}{M^2} \right\} \\
\end{eqnarray}
Fig. \ref{fig:ewhc} depicts the changes of the cross section as a function of c.m. energy for
the two electron polarizations at 90$^o$ scattering angle. 
Only $Z$-exchange contributes. The resummed corrections reduce the cross section
by about 11 \% at 1 TeV, 23 \% at 3 TeV and about 28 \% at 5 TeV for left handed electrons.
For right handed electrons, the effect is actually an increase in the cross section at energies
up to about 2 TeV due to the large positive angular terms.
The reduction at 3 TeV is about 5 \% and 14 \% at 5 TeV for the resummed cross sections.
Again we can see that the one loop predictions differ in the percentile regime at 1 TeV and by
about 7 \% at 3 TeV. At higher energies the difference grows even more rapidly. Thus, for the linear
collider in the TeV range, higher order contributions are necessary.

\begin{figure}
\centering
\epsfig{file=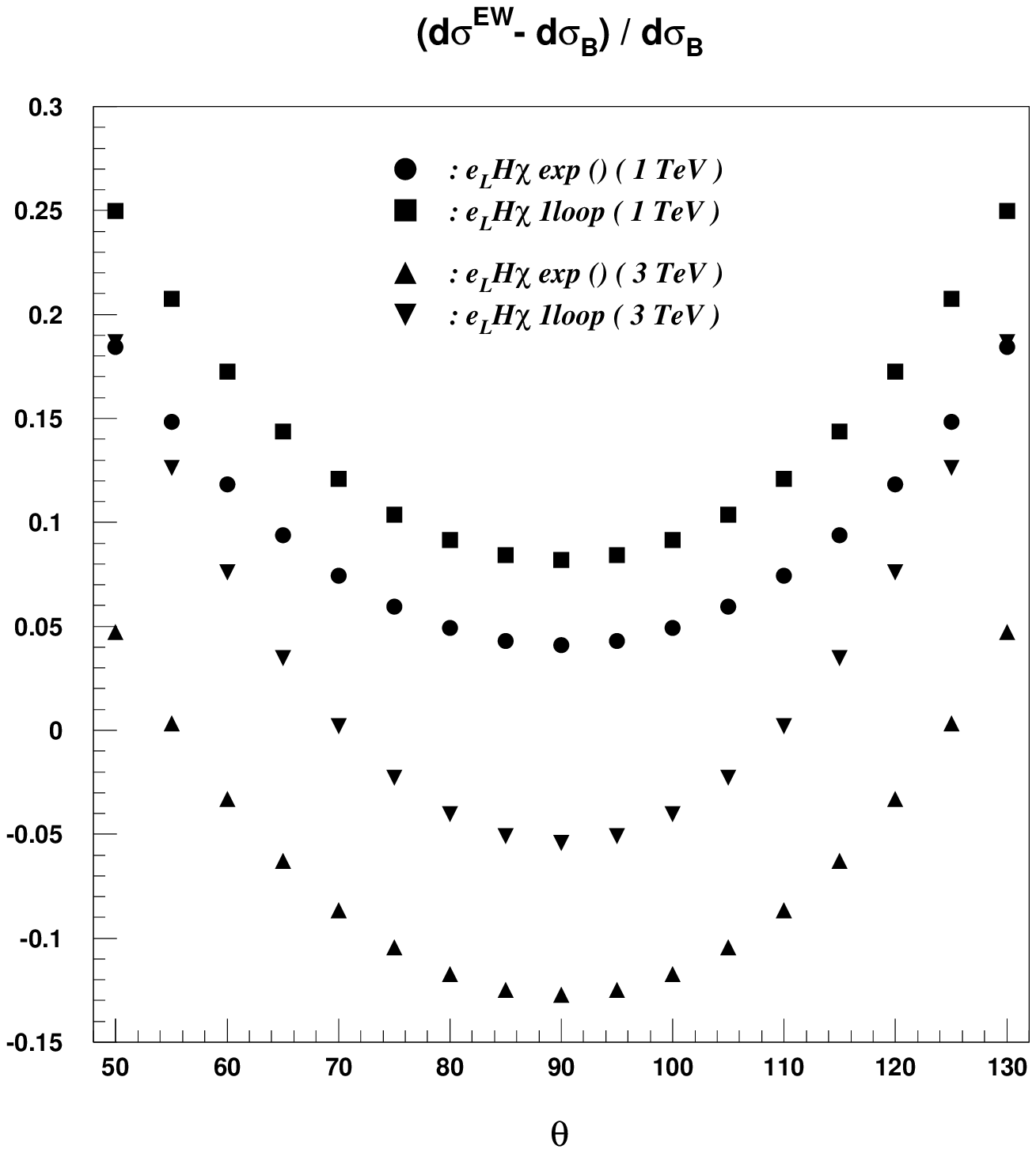,width=12cm}
\caption{The purely electroweak virtual
corrections relative to the Born cross section in Higgs-strahlung
in $e^+ e^-$ collisions at 1 and 3 TeV 
c.m. energy as a function of the scattering angle.
Only cross sections involving
left handed electrons receive SL angular corrections. Given are the one loop 
and the resummed corrections to SL accuracy in each case. The
difference between the two approaches is non-negligible at TeV energies and
necessitates the inclusion of the higher order terms. Pure QED corrections
from below the weak scale are omitted.}
\label{fig:ewhct}
\end{figure}              
In Fig. \ref{fig:ewhct} the angular dependence for the Higgs-strahlung process is displayed.
Only the cross section involving left handed electrons possess angular dependent terms.
The corrections are symmetric with respect to the central scattering angle (i.e. symmetric
in $u \leftrightarrow t$). The resummed cross sections are consistently lower due to
the large positive angular one loop result in Eq. (\ref{eq:eHa}). At 1 TeV, the corrections
change by about 14 \% and at 3 TeV by about 18 \% over the displayed angular range.
Again we see how important those terms are for future collider experiments in the TeV range.

\begin{figure}
\centering
\epsfig{file=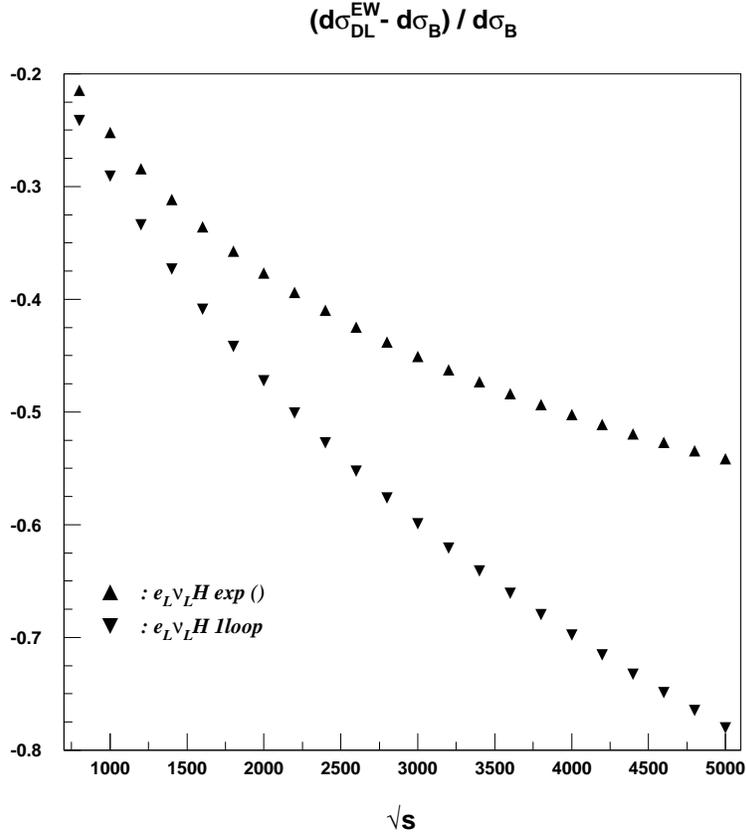,width=12cm}
\caption{The purely electroweak virtual
corrections relative to the Born cross section in Higgs radiation off
$W^\pm$ (fusion)
in $e^+ e^-$ collisions 
as a function of the c.m. energy. The polarization
is indicated in the figure for each symbol. Given are the one loop 
and the resummed corrections to DL accuracy in each case. 
It is clearly visible that the
difference between the two approaches is non-negligible at TeV energies and
necessitates the inclusion of the higher order terms. Pure QED corrections
from below the weak scale are omitted. The SL terms are omitted for simplicity, however, are needed
for a full treatment.}
\label{fig:ewnnh}
\end{figure}              

We now want to briefly discuss two other important Higgs production processes at 
$e^+ e^-$ colliders, namely the $W$ fusion process and Higgs-strahlung off top quarks. 
Both of these processes involve three final state particles and given the multiplicity
of final state scattering states, we restrict ourselves here to a discussion only of the
energy-dependence at DL accuracy.
All universal DL corrections are given by the DL terms in Eqs. (\ref{eq:Wf})
and (\ref{eq:Wp}). 

We begin with the $W$-fusion process $e^+ e^- \longrightarrow \nu_e {\overline \nu}_e H$
in Fig. \ref{fig:ewnnh}. 
Only left handed electrons contribute. This process yields the largest cross section at high
energies for Higgs production. The resummed DL 
corrections reduce the cross section by about 25 \% at 1 TeV, 46 \% at 3 TeV and 53 \%
at 5 TeV. The one loop DL predictions are off by about 5 \% at 1 TeV, 15 \% at 3 TeV and
20 \% at 5 TeV relative to the Born cross section! 

While the SL terms can change this picture significantly, the typical size of the corrections
should be correctly depicted in Fig. \ref{fig:ewnnh}.

\begin{figure}
\centering
\epsfig{file=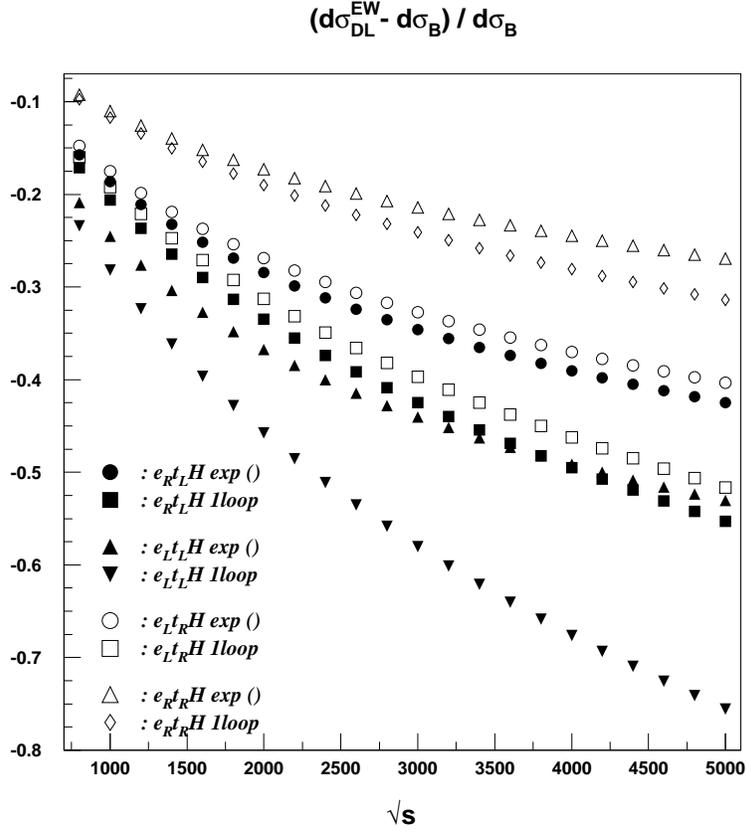,width=12cm}
\caption{The purely electroweak virtual
corrections relative to the Born cross section in Higgs radiation off
final state top-quarks
in $e^+ e^-$ collisions
as a function of the c.m. energy. The polarization
is indicated in the figure for each symbol. Given are the one loop 
and the resummed corrections to DL accuracy in each case. 
It is clearly visible that the
difference between the two approaches is non-negligible at TeV energies and
necessitates the inclusion of the higher order terms. Pure QED corrections
from below the weak scale are omitted. The SL terms are omitted for simplicity, however, are needed
for a full treatment.}
\label{fig:ewth}
\end{figure}              

In Fig. \ref{fig:ewth} we display the electroweak DL corrections to Higgs productions off
top quarks for the various polarizations. This process is crucial in order to determine the
top-Higgs Yukawa coupling \cite{daw}.
The DL terms reduce the cross section for $e_Lt_L$
by about 24 \% at 1 TeV, 43 \% at 3 TeV and 52 \% at 5 TeV. 
For $e_Rt_L$ by about 18 \% at 1 TeV, 35 \% at 3 TeV and 42 \% at 5 TeV.
For $e_Lt_R$ by about 17 \% at 1 TeV, 33 \% at 3 TeV and 40 \% at 5 TeV and finally 
for $e_Rt_R$ by about 11 \% at 1 TeV, 21 \% at 3 TeV and 27 \% at 5 TeV.
Again the one loop DL corrections differ significantly, especially for left handed polarizations.

In all Higgs production processes discussed in this section, the electroweak radiative corrections
are important and can reduce the cross sections considerably. Even for the top-Yukawa measurement at
800 GeV at TESLA or the NLC, corrections are of ${\cal O} \left( 20 \% \right)$ and the difference
between one loop and resummed and be a few percent. Therefore a full higher order SL analysis is
warranted for this process in addition to the QCD corrections.

\subsubsection{Charged MSSM Higgs production}

\begin{figure}
\centering
\epsfig{file=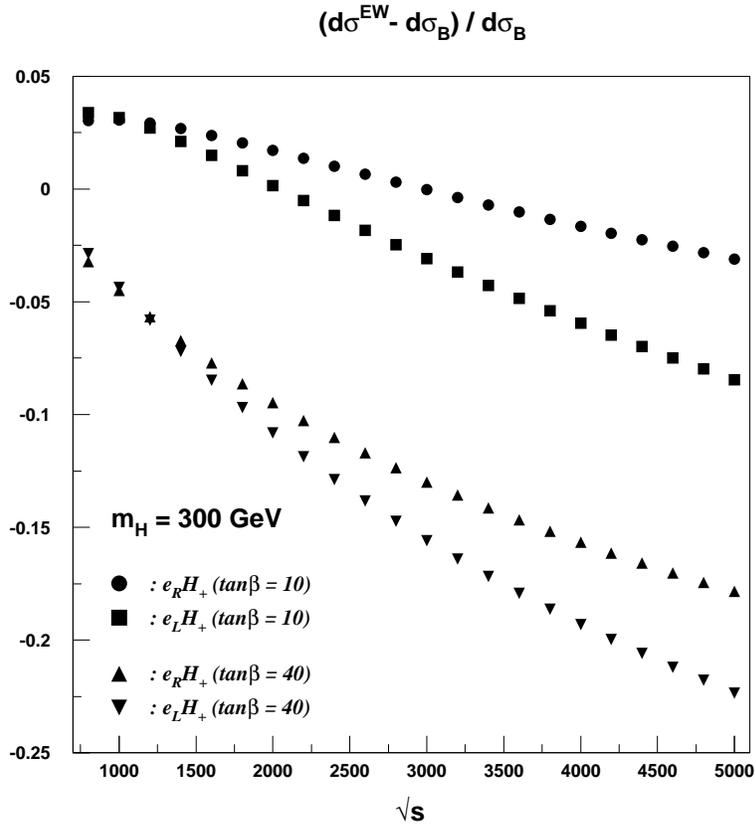,width=12cm}
\caption{The purely electroweak MSSM virtual corrections to charged Higgs production
above the susy scale set by $m_H=300$ GeV
to SL accuracy.
The dependence in shown for two characteristic choices of $\tan \beta$ for both right and
left handed electron polarizations.}
\label{fig:ewhphm}
\end{figure}

In this section we discuss the effect of MSSM Sudakov effects in charged Higgs production.
The relevant radiative corrections are given in Eq. (\ref{eq:Hang}) for the universal
and angular dependent Sudakov
terms above the susy scale set by $m_s=m_H$ under the assumptions stated in section \ref{sec:susy}.
The angular dependent corrections can also be obtained form Eqs. (\ref{eq:aelwl}) and  (\ref{eq:aerwl})
via the replacement $M \rightarrow m_H$, which for our purposes, however, is of SSL accuracy
in our ``light susy'' mass assumption. In addition we have RG contributions at one loop
which read
\begin{eqnarray}
d \sigma^{e_L,H_+}_\theta &=& d \sigma^{e_L,H_+}_{\rm Born} \left\{ - 2 \frac{\alpha^2(m_H^2)
{\tilde \beta}_0 + {\alpha^\prime}^2(m_H^2) {\tilde \beta}^\prime_0}{ \pi \left( \alpha(m_H^2)
+\alpha^\prime(m_H^2) \right)} \right\} \log \frac{s}{m_H^2} \\
d \sigma^{e_R,H_+}_\theta &=& d \sigma^{e_R,H_+}_{\rm Born} \left\{ - 2 
\frac{ \alpha^\prime(m_H^2)}{\pi} {\tilde \beta}^\prime_0 \right\} \log \frac{s}{m_H^2}
\end{eqnarray}
where ${\tilde \beta}_0$ and ${\tilde \beta}^\prime_0$ are given in Eq. (\ref{eq:arunMSSM}).
In Fig. \ref{fig:ewhphm} the energy dependence for the process $e^+e^- \rightarrow H^+H^-$
is depicted for two typical values of $\tan \beta = 10$ and $\tan \beta = 40$. 
The scattering angle is held fixed at $\theta=90^o$ since the angular dependence
is analogous to the case of longitudinal $W$ production in Fig. \ref{fig:ewwt}.
The helicity of the electrons is indicated and only the resummed all orders SL results
are presented. The value
of the heavy charged Higgs particles is $m_H=300$ GeV. For a different heavy Higgs mass,
the displayed results shift accordingly.

\begin{figure}
\centering
\epsfig{file=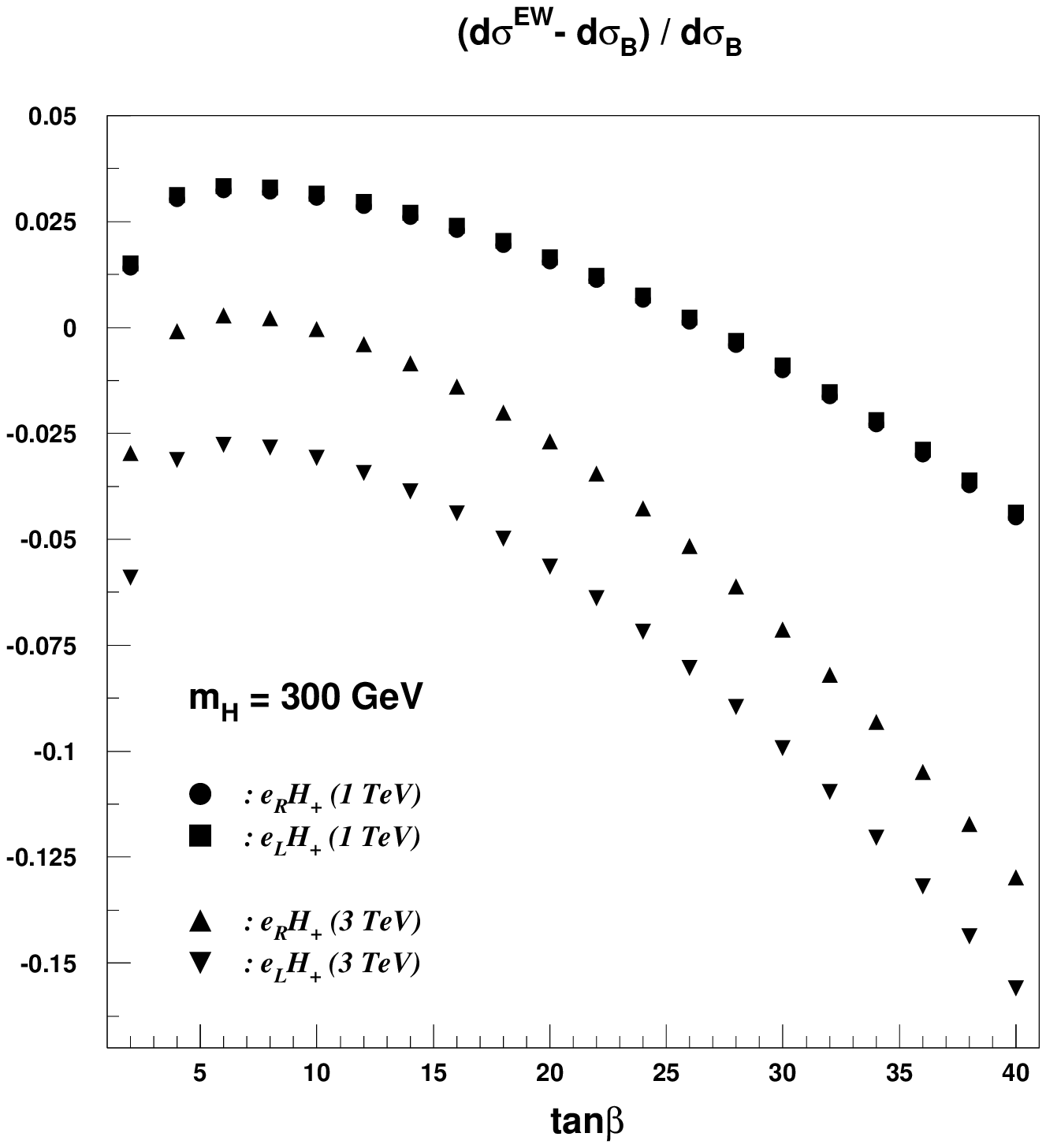,width=12cm}
\caption{The purely electroweak MSSM virtual corrections for charged Higgs production
above the susy scale set by $m_H=300$ GeV
to SL accuracy as a function of $\tan \beta$.
The dependence in shown for two characteristic choices of $\sqrt{s}$ for both right and
left handed electron polarizations.}
\label{fig:ewhphmb}
\end{figure}

Fig. \ref{fig:ewhphm} shows that the effect of the MSSM Sudakov corrections is somewhat
reduced compared to the SM case. This is almost entirely due to the different mass scale
used. The left handed cross sections are more suppressed than the right handed ones
due to the larger DL group factors. In addition, larger values for $\tan \beta$ lead
to an enhanced suppression since the overall sign of the Yukawa terms is negative. 
The dependence on $\tan \beta$ is explicitly shown in Fig. \ref{fig:ewhphmb} for c.m. energies
of $\sqrt{s}=1$ TeV and $\sqrt{s}=3$ TeV. At $1$ TeV the right and left handed cross sections
display the same overall dependence on $\tan \beta$, differing by about 7 \% in the range
displayed.
At $3$ TeV, the left handed cross section is more suppressed but the overall $\tan \beta$ dependence
is similar. The cross sections can differ by about 13 \% in the range between $\tan \beta=6$
to $\tan \beta=40$. This strong dependence, especially for larger values on $\tan \beta$
originates from both, the size of the Yukawa terms as well as from the number of colors in
the quark loops. In Ref. \cite{bmrv} this was utilized to suggest a measurement of $\tan \beta$ at CLIC
with a relative precision better than 25 \% (a few percent for large values $\tan \beta \geq 25$). 
More importantly, this determination of $\tan \beta$ to SL accuracy does not depend on the soft
breaking terms (which are constants) and is scheme and gauge invariant.

While we focussed here only on the case of heavy charged Higgs production, also the neutral Higgs,
fermion and
sfermion processes have important information on MSSM parameters contained in the coefficients
of large Sudakov logarithms and should be fully exploited at such a collider.

\subsection{Heavy quark production} \label{sec:hqp}

In this section we discuss heavy quark production at the linear collider. These processes
can be used to measure $\alpha_s$ above the production threshold and should be fully understood.
A general result for the angular terms relative to the Born amplitude is given by
\begin{eqnarray}
\sum_{B,W^a} \delta^\theta_{e^+_{\alpha} e^-_{\alpha} \longrightarrow {\overline f}_{\beta} f_{\beta}}
&=&
-\frac{g^2(M^2)}{16\pi^2} \log \frac{s}{M^2} \left\{ \left[ \tan^2 \theta_{\rm w} Y_{e^-_{\alpha}} Y_{f_\beta}
+ 4 T^3_{e^-_{\alpha}} T^3_{f_\beta} \right] \log \frac{t}{u} \right. \nonumber \\ && 
\left. + \frac{\delta_{\alpha, L} \delta_{\beta,L}}{\tan^2 \theta_{\rm w} Y_{e^-_{\alpha}} Y_{f_\beta} /4
+ T^3_{e^-_{\alpha}} T^3_{f_\beta}} \left( \delta_{d,f} \log \frac{-t}{s} - \delta_{u,f}
\log \frac{-u}{s} \right) \!\! \right\} \label{eq:efang}
\end{eqnarray}
where the last line only contributes for left handed fermions and the $d,u$ symbols denote the
corresponding isospin quantum number of $f$.

\begin{figure}
\centering
\epsfig{file=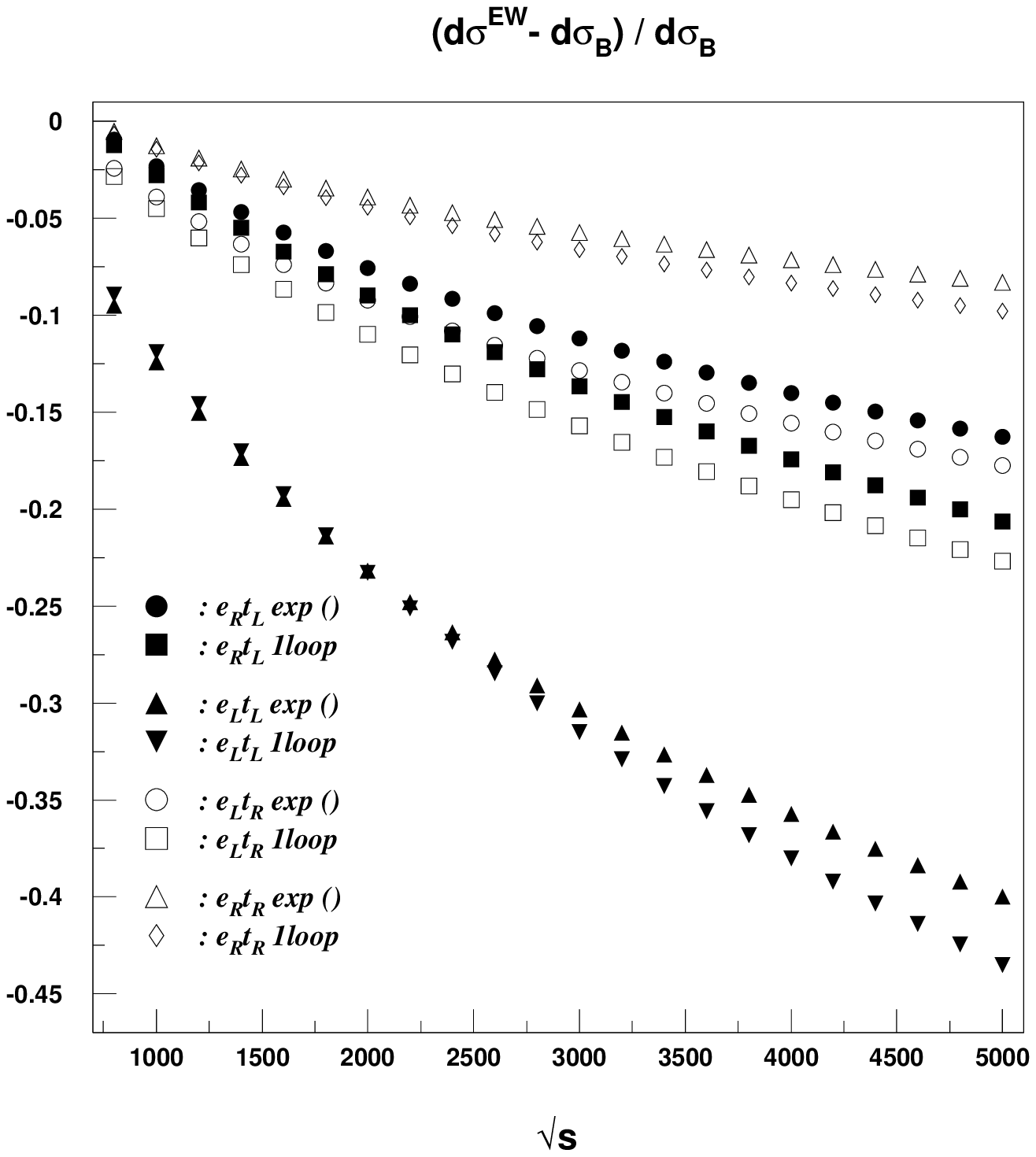,width=12cm}
\caption{The purely electroweak virtual 
corrections relative to the Born cross section in top quark production
in $e^+ e^-$ collisions at 90$^o$ scattering
angle as a function of the c.m. energy. The polarization
is indicated in the figure for each symbol. Given are the one loop 
and the resummed corrections to SL accuracy in each case. 
The difference between the two approaches is non-negligible at TeV energies and
necessitates the inclusion of the higher order terms. Pure QED corrections
from below the weak scale are omitted.}
\label{fig:ewt}
\end{figure}

We begin with top production.  
All universal terms are given by Eqs. (\ref{eq:WfRG})
and (\ref{eq:WpRG}). The angular terms and RG corrections which at one loop are given by
\begin{eqnarray}
d \sigma^{e_L,t_L}_\theta &=& d \sigma^{e_L,t_L}_{\rm Born} \left\{ \left[ 4.9 \log \frac{t}{u} -
16.3  \log \frac{-u}{s} \right] \frac{e^2(M^2)}{8 \pi^2} \log \frac{s}{M^2}
\right\} \\
d \sigma^{e_R,t_L}_\theta &=& d \sigma^{e_R,t_L}_{\rm Born} \left\{ 0.86 \log \frac{t}{u} 
\;\; \frac{e^2(M^2)}{8 \pi^2} \log \frac{s}{M^2}
\right\} \\
d \sigma^{e_L,t_R}_\theta &=& d \sigma^{e_L,t_R}_{\rm Born} \left\{ 1.72 \log \frac{t}{u} 
\;\; \frac{e^2(M^2)}{8 \pi^2} \log \frac{s}{M^2}
\right\} \\
d \sigma^{e_R,t_R}_\theta &=& d \sigma^{e_R,t_R}_{\rm Born} \left\{ 3.43 \log \frac{t}{u} 
\;\; \frac{e^2(M^2)}{8 \pi^2} \log \frac{s}{M^2}
\right\} 
\end{eqnarray}
and
\begin{eqnarray}
d \sigma^{e_L,t_L}_{\rm RG} &=& d \sigma^{e_L,t_L}_{\rm Born} \left\{- 12.2 \; \frac{e^2(M^2)}{8 \pi^2}
\log \frac{s}{M^2} \right\} \\
d \sigma^{e_R,t_L}_{\rm RG} &=& d \sigma^{e_R,t_L}_{\rm Born} \left\{ 8.8 \; \frac{e^2(M^2)}{8 \pi^2}
\log \frac{s}{M^2} \right\} \\
d \sigma^{e_L,t_R}_{\rm RG} &=& d \sigma^{e_L,t_R}_{\rm Born} \left\{ 8.8 \; \frac{e^2(M^2)}{8 \pi^2}
\log \frac{s}{M^2} \right\} \\
d \sigma^{e_R,t_R}_{\rm RG} &=& d \sigma^{e_R,t_R}_{\rm Born} \left\{ 8.8 \; \frac{e^2(M^2)}{8 \pi^2}
\log \frac{s}{M^2} \right\} 
\end{eqnarray}
Fig. \ref{fig:ewt} displays the energy dependence of the 
corrections for central
scattering angles for the various polarizations. The difference between the resummed and
the one loop contributions is in the several percent range only above 1 TeV. The largest
corrections are again obtained in the case where both fermions are left handed. The overall
corrections are large and non-negligible.

\begin{figure}
\centering
\epsfig{file=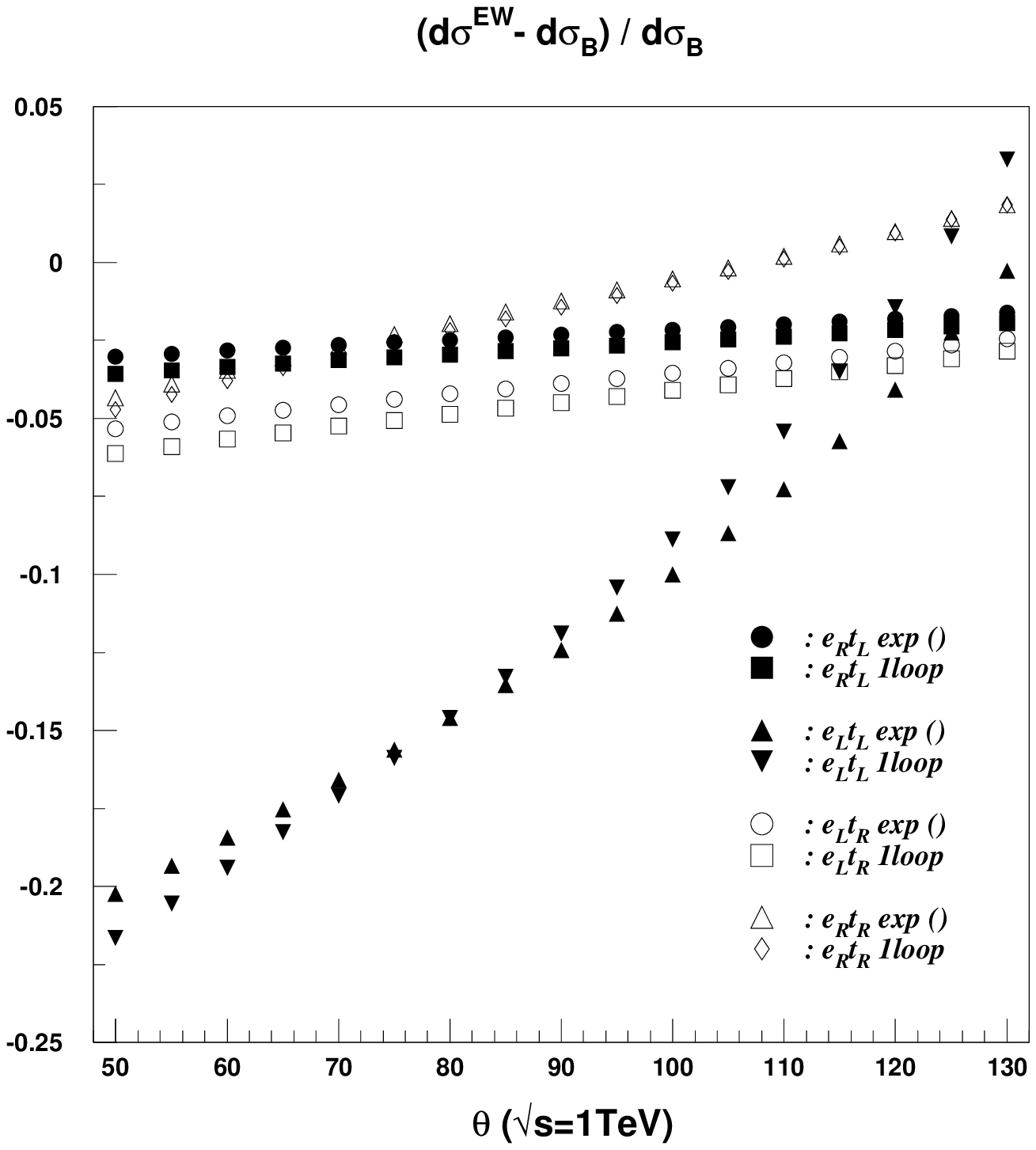,width=8.5cm}
\epsfig{file=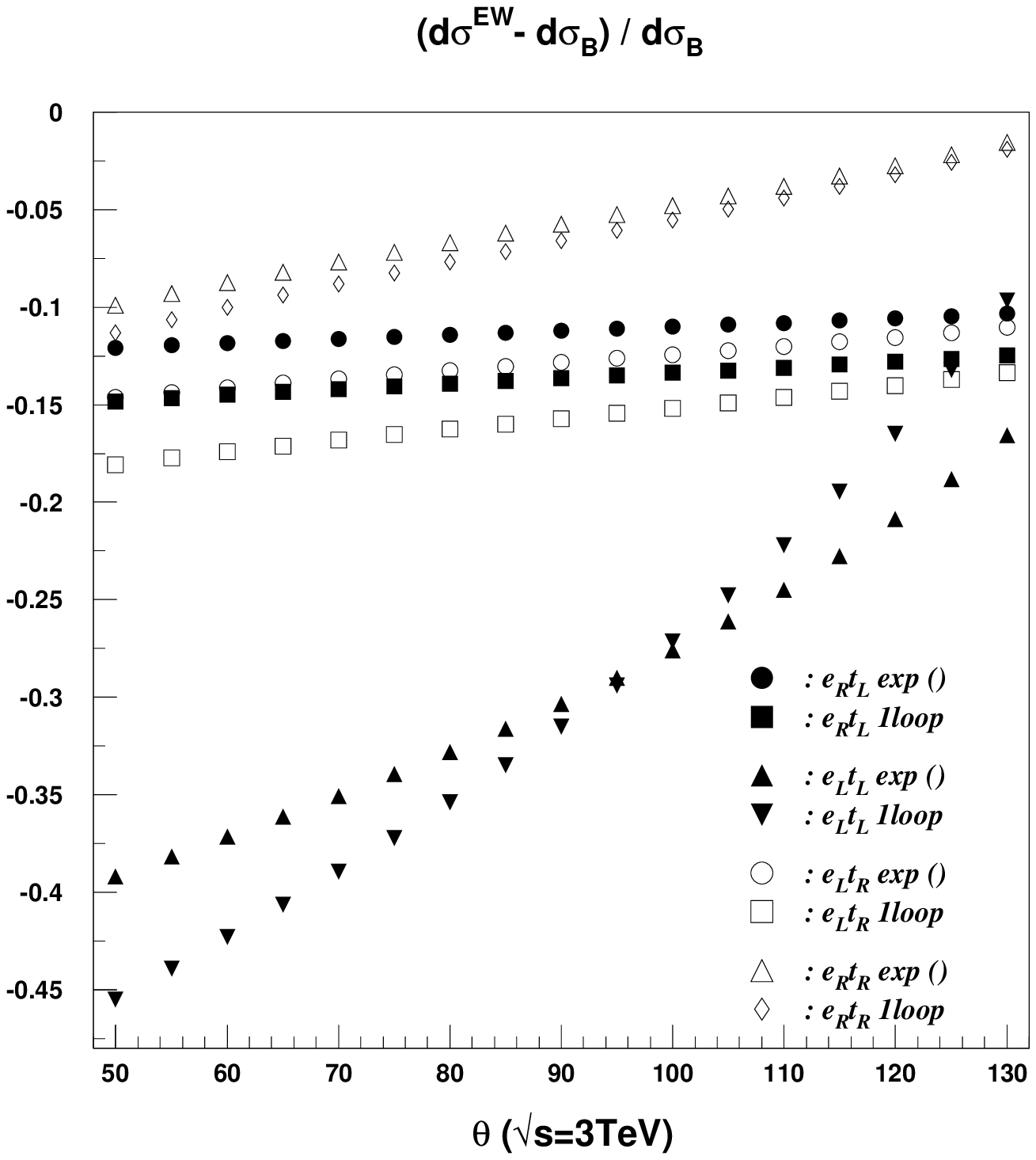,width=8.5cm} \vspace{-0.5cm}
\caption{The purely electroweak virtual
corrections relative to the Born cross section in top quark production
in $e^+ e^-$ collisions at 1 and 3 TeV 
c.m. energy as a function of the scattering angle. The polarization
is indicated in the figure for each symbol. Given are the one loop 
and the resummed corrections to SL accuracy in each case. 
The difference between the two approaches is non-negligible at TeV energies and
necessitates the inclusion of the higher order terms. Pure QED corrections
from below the weak scale are omitted.}
\label{fig:ewtt}
\end{figure}              

Also the angular terms, depicted in Fig. \ref{fig:ewtt}, 
are significant, especially for $e_L t_L$ where they
change cross sections by about 20 \% for 1 and 3 TeV over the angular range
between 50$^o$ and 130$^o$. This is another indication that the higher order
angular terms need to be understood in the full electroweak theory.

\begin{figure}
\centering
\epsfig{file=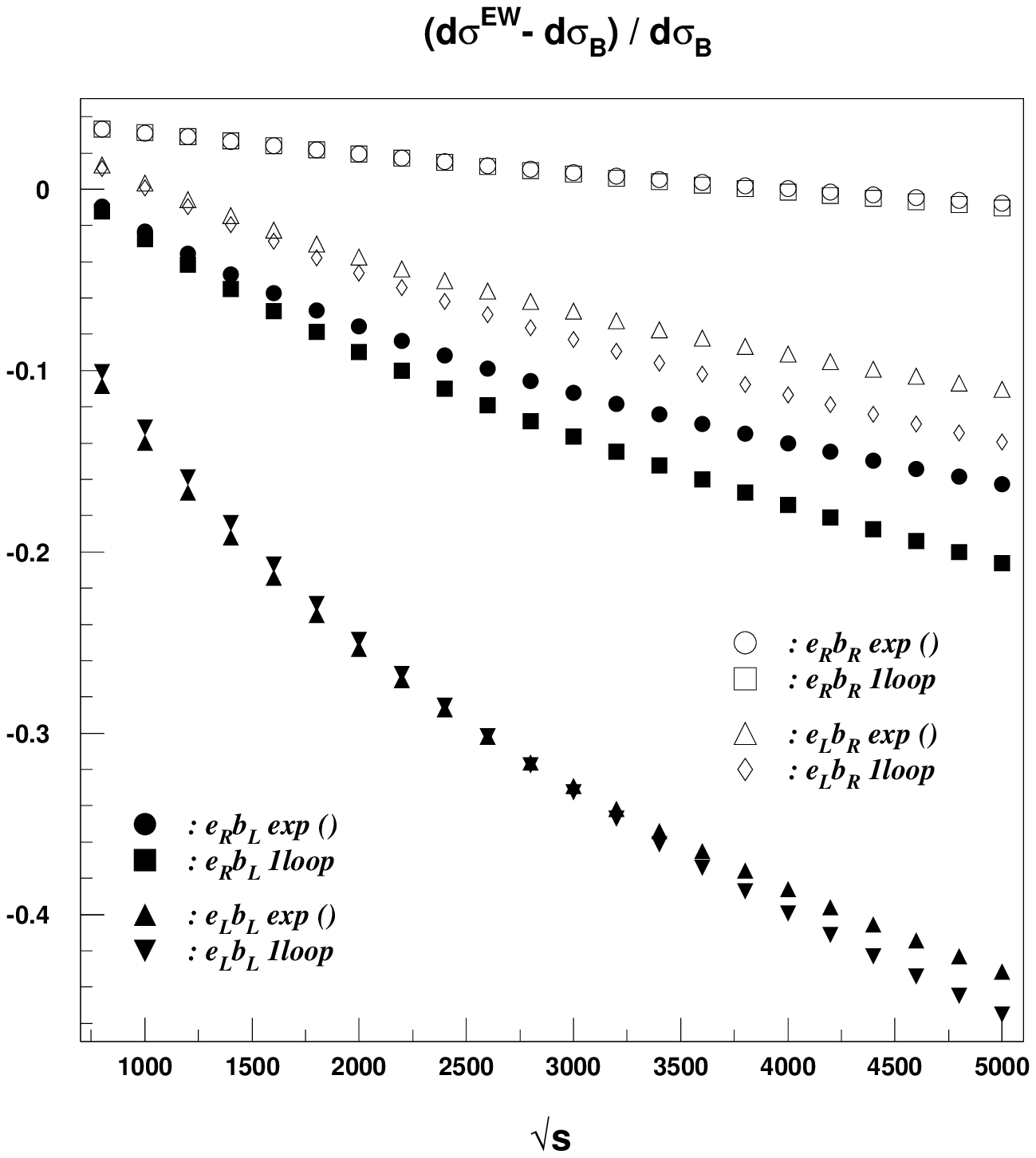,width=12cm}
\caption{The purely electroweak virtual 
corrections relative to the Born cross section in 
bottom quark production
in $e^+ e^-$ collisions at 90$^o$ scattering
angle as a function of the c.m. energy. The electron polarization
is indicated in the figure for each symbol. Given are the one loop 
and the resummed corrections to SL accuracy in each case. 
The difference between the two approaches is non-negligible at TeV energies and
necessitates the inclusion of the higher order terms. Pure QED corrections
from below the weak scale are omitted.}
\label{fig:ewb}
\end{figure}              

For bottom quark production $b_R$ does not contain SL-Yukawa terms. 
All universal corrections are given by Eqs. (\ref{eq:WfRG})
and (\ref{eq:WpRG}). In addition we have angular terms and RG corrections which at one loop are given by
\begin{eqnarray}
d \sigma^{e_L,b_L}_\theta &=& d \sigma^{e_L,b_L}_{\rm Born} \left\{ - \left[ 4.04 \log \frac{t}{u} +
19.8  \log \frac{-t}{s} \right] \frac{e^2(M^2)}{8 \pi^2} \log \frac{s}{M^2}
\right\} \\
d \sigma^{e_R,b_L}_\theta &=& d \sigma^{e_R,b_L}_{\rm Born} \left\{ 0.86 \log \frac{t}{u} 
\;\; \frac{e^2(M^2)}{8 \pi^2} \log \frac{s}{M^2}
\right\} \\
d \sigma^{e_L,b_R}_\theta &=& d \sigma^{e_L,b_R}_{\rm Born} \left\{ -0.86 \log \frac{t}{u} 
\;\; \frac{e^2(M^2)}{8 \pi^2} \log \frac{s}{M^2}
\right\} \\
d \sigma^{e_R,b_R}_\theta &=& d \sigma^{e_R,b_R}_{\rm Born} \left\{ -1.72 \log \frac{t}{u} 
\;\; \frac{e^2(M^2)}{8 \pi^2} \log \frac{s}{M^2}
\right\} 
\end{eqnarray}
and
\begin{eqnarray}
d \sigma^{e_L,b_L}_{\rm RG} &=& d \sigma^{e_L,b_L}_{\rm Born} \left\{- 16.6 \; \frac{e^2(M^2)}{8 \pi^2}
\log \frac{s}{M^2} \right\} \\
d \sigma^{e_R,b_L}_{\rm RG} &=& d \sigma^{e_R,b_L}_{\rm Born} \left\{ 8.8 \; \frac{e^2(M^2)}{8 \pi^2}
\log \frac{s}{M^2} \right\} \\
d \sigma^{e_L,b_R}_{\rm RG} &=& d \sigma^{e_L,b_R}_{\rm Born} \left\{ 8.8 \; \frac{e^2(M^2)}{8 \pi^2}
\log \frac{s}{M^2} \right\} \\
d \sigma^{e_R,b_R}_{\rm RG} &=& d \sigma^{e_R,b_R}_{\rm Born} \left\{ 8.8 \; \frac{e^2(M^2)}{8 \pi^2}
\log \frac{s}{M^2} \right\} 
\end{eqnarray}
Fig. \ref{fig:ewb} demonstrates that in the energy range displayed, the electroweak
corrections are actually positive for $e_R b_R$. This is mainly due to
the RG corrections for the right handed coupling ($g^\prime$) which is Abelian
and therefore increases with energy. Secondly it is due to the fact that the
Yukawa terms are absent and the DL terms are partially offset by the SL terms.

The remaining corrections are similar to the top production discussion.

\begin{figure}
\centering
\epsfig{file=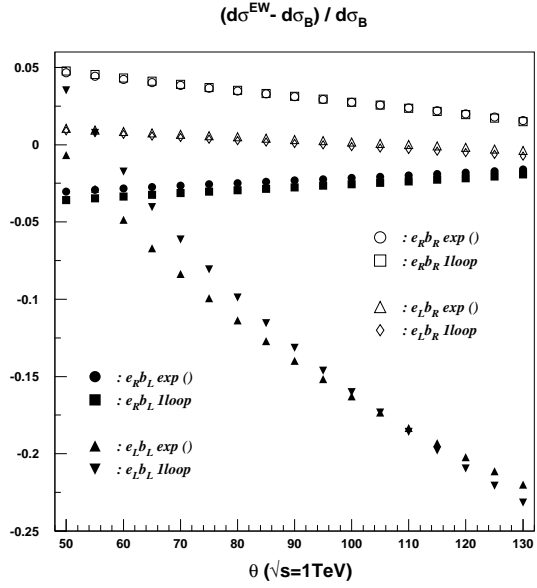,width=8.5cm}
\epsfig{file=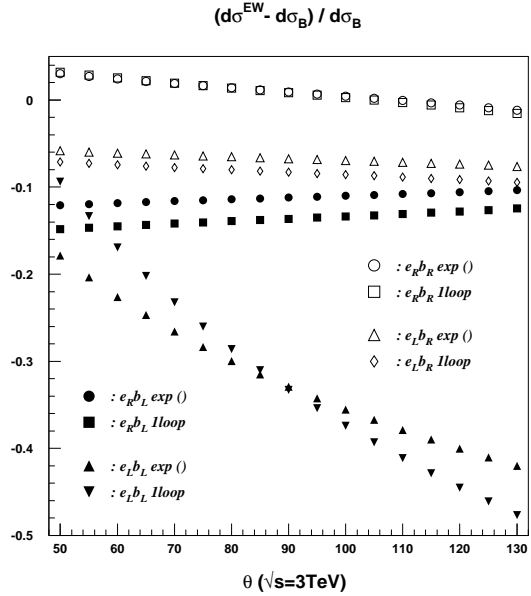,width=8.5cm}
\caption{The purely electroweak virtual
corrections relative to the Born cross section in bottom 
quark production
in $e^+ e^-$ collisions at 1 and 3 TeV 
c.m. energy as a function of the scattering angle. The electron polarization
is indicated in the figure for each symbol. Given are the one loop 
and the resummed corrections to SL accuracy in each case. 
The difference between the two approaches is non-negligible at TeV energies and
necessitates the inclusion of the higher order terms. Pure QED corrections
from below the weak scale are omitted.}
\label{fig:ewbt}
\end{figure}              

The angular terms, depicted in Fig. \ref{fig:ewbt}, show a similar behavior to the
case of top quark production in Fig. \ref{fig:ewtt} in that the largest
contribution is again for the purely left handed case. The overall size of the corrections differs
and the $e_R b_R$ contributions stay positive for most of the displayed angular range.

\section{Outlook} \label{sec:out}

Electroweak radiative corrections at high energies have received much attention recently
due to their importance at experiments in the TeV regime. It is not only the phenomenological
importance, however, that has led to a surge in interest into the high energy behavior of the
SM but also the fact that conceptually new effects enter due to EWSB.
The main differences to the case of unbroken gauge theories originate from the
longitudinal sector and the fact that three 
of the gauge bosons acquire masses, while the photon stays massless.
In addition, the asymptotic states carry a non-Abelian group charge, the weak isospin
and are superpositions of the fields of the unbroken phase. 

As a consequence, fully and semi-inclusive cross sections show double and single logarithmic
dependencies on the ratio of the energy and the gauge boson mass, and longitudinal
degrees of freedom are not mass suppressed. Thus, observables in the
SM depend on the infrared cutoff, the weak scale, which in this case, however, is a physical
parameter. At one loop, general methods exist which allow to calculate corrections relevant
to precision measurements at future colliders. Also at higher orders, a general approach
to SL accuracy is available, based on the high energy approximation of the symmetric
part of the SM Lagrangian and is phenomenologically necessary.

Focusing on techniques to calculate 
the higher order corrections, we have summarized the present status of virtual
electroweak radiative corrections to SM and MSSM
high energy processes. In the framework of the IREE method, the high energy effective theory is
based on the high energy limit of the SM Lagrangian in the symmetric limit where all terms connected
to the v.e.v. can be neglected to SL accuracy.
The QED corrections below the weak scale are incorporated with the appropriate matching conditions.
This approach is so far the only one able to allow for a two loop calculation of the DL and SL 
terms which are relevant for TeV experiments at future colliders. We have shown that the
one loop terms are insufficient when the c.m. energy is larger than 1 TeV and that both, DL and
SL terms at the two loop level
are necessary at the several percent level. Also angular dependent corrections cannot be 
neglected and it should be investigated if their calculation at the SSL level is 
needed for some observables. It should also be investigated if
large Yukawa constants can be treated in a systematic manner to SSL accuracy.
In this context it is also important to consider the emission of real gauge bosons
above the weak scale even at the SL level. 

In summary, there exists a way to calculate all higher order virtual SL electroweak radiative 
corrections to high energy processes. The approach is in agreement with all available one loop
calculations in terms of the physical SM or MSSM fields and at the two loop level to DL accuracy. 
These terms are crucial for the experiments at future colliders in the TeV regime since
the effects of new physics expected in this range can be rather small.

\section*{Acknowledgements}  

I would like to thank A.~Denner, S.~Pozzorini, E.~Accomando for valuable discussions
and D.~Wyler for his support of this work.

\section{Appendix} \label{sec:apd}

In this appendix we list the relevant quantum numbers of the physical SM fields and the corresponding
Feynman rules of the full SM in the 't Hooft-Feynman gauge.

\subsection{Operators and quantum numbers} \label{sec:qn}

In this section we present the
generators of the physical gauge group and various group-theoretical matrices used in section \ref{sec:pf}.
We follow the discussion of Ref. \cite{dp}.
The notation for the components of such matrices is
\beq
M_{\varphi_i\varphi_{i'}}(\varphi)
\eeq
where the argument $\varphi$ represents a multiplet and fixes the
representation for the matrix $M$, whereas $\varphi_i$ are the
components of the multiplet. Explicit
representations for left- and right-handed fermions
($\varphi=f^L,f^R,\bar{f}^L,\bar{f}^R$), for gauge bosons
($\varphi=\GB$) and for the scalar doublet ($\varphi=\Phi$) are given below.  Where the
representation is implicit, the argument $\varphi$ is omitted. For the
eigenvalues of diagonal matrices one has
\beq
M_{\varphi_i\varphi_{i'}}=\de_{\varphi_i\varphi_{i'}}M_{\varphi_i}
\eeq

\subsection*{Symmetric and physical gauge fields and gauge couplings}
For physical gauge bosons one needs to take special care of the effect of Weinberg
rotation (mixing). The symmetric basis $\tilde{\GB}_a=B,W^1,W^2,W^3$, is
formed by the $\Uone$ and $\SUtwo$ gauge bosons, which transform as a
singlet and a triplet, respectively, and quantities in this basis are
denoted by a tilde. The physical basis is given by the charge and mass
eigenstates $\GB_a=A,Z,W^+,W^-$. The physical charged gauge bosons,
\begin{equation}
W^\pm=\frac{W^1\mp\ri W^2}{\sqrt{2}},
\end{equation}
are pure $\SUtwo$ states, whereas in the neutral sector the $\SUtwo$
and $\Uone$ components mix, and the physical fields $\NB=A,Z$  are
related to the symmetric fields $\sNB=B,W^3$ by the Weinberg rotation,
\beq \label{weinbergrotation}
\NB=U_{\NB\sNB}(\thw)\sNB,\qquad  U(\thw)=\left(\begin{array}{c@{\;}c}\cw & -\sw \\ \sw & \cw \end{array}\right)
\eeq
with $\cw=\cos{\thw}$ and $\sw=\sin{\thw}$. In the on shell
renormalization scheme the Weinberg angle is fixed by
\beq \label{mixingangle}
\cw=\frac{\MW}{\MZ}
\eeq

The gauge couplings are given by the generators of global gauge transformations.
In the symmetric basis, they read
\beq
\tilde{I}^B=-\frac{1}{\cw}\frac{Y}{2},\qquad \tilde{I}^{W^a}=\frac{1}{\sw}T^a,\qquad a=1,2,3
\eeq
where $Y$ is the weak hypercharge and $T^a$ are the components of the weak isospin.  In the physical basis one has 
\beq
I^A=-Q,\qquad I^Z=\frac{T^3-\sw^2 Q}{\sw\cw},\qquad
I^\pm=\frac{1}{\sw}T^\pm=\frac{1}{\sw}\frac{T^1\pm\ri T^2}{\sqrt{2}}
\eeq
with $Q=T^3+Y/2$.

\subsection*{Casimir operators}
The  $\SUtwo$ Casimir operator is defined by
\beq
C=\sum_{a=1}^3(T^a)^2
\eeq
Loops involving charged gauge bosons are often associated with the
product of the non-Abelian charges
\begin{equation}
(I^W)^2:=\sum_{\sigma=\pm}\left[ I^\sigma I^{-\sigma} \right]=\left[\frac{C-(T^3)^2}{\sw^2}\right]
\end{equation}
and if one includes the contributions of neutral gauge bosons, one obtains the effective electroweak Casimir operator
\begin{equation}\label{CasimirEW} 
\cew:=\sum_{\GB_a=A,Z,W^\pm} I^{\GB_a}I^{\bar{\GB}_a}=\frac{1}{\cw^2}\left(\frac{Y}{2}\right)^2+\frac{1}{\sw^2}C
\end{equation}
For irreducible representations (fermions and scalars) with isospin
$T_\varphi$, the $\SUtwo$ Casimir operator is proportional to the
identity and reads
\beq
C_{\varphi_i\varphi_{i'}}(\varphi)=\de_{\varphi_i \varphi_{i'}}C_\varphi,\qquad C_\varphi=T_\varphi[T_\varphi+1]
\eeq
Physical gauge bosons have a reducible representation as already discussed in section \ref{sec:bgt}. 
In the symmetric basis $\tilde{C}(\GB)$ is a diagonal $4\times 4$ 
matrix
\beq \label{44diagonal}
\tilde{C}_{\tilde{\GB}_a\tilde{\GB}_{b}}=\de_{ab}\tilde{C}_{\tilde{\GB}_a}
\eeq
with  $\Uone$ and $\SUtwo$ eigenvalues
\beq \label{adjointCasimir1}
\tilde{C}_{B}=0,\qquad \tilde{C}_{W^a}=2
\eeq
The transformation of a matrix like \refeq{44diagonal} to the physical
basis, yields a $4\times 4$ matrix with diagonal $2\times 2$ block
structure, \ie without mixing between the charged sector ($W^\pm$) and
the neutral sector ($\NB=A,Z$). In the neutral sector $C(\GB)$ becomes
non-diagonal owing to mixing of the $\Uone$ and $\SUtwo$ eigenvalues,
\beq  \label{adjointCasimir2} 
C_{\NB\NB'}= \left[U(\thw) \tilde{C} U^{-1}(\thw)\right]_{\NB\NB'}=
2\left(\begin{array}{c@{\;}c}\sw^2 & -\sw\cw \\ -\sw\cw & \cw^2 \end{array}\right)
\eeq
whereas in the charged sector it remains diagonal,
\beq  \label{adjointCasimir3} 
C_{W^\si W^{\si'}}=2\de_{\si\si'}
\eeq

\subsection*{Explicit values for $Y$, $Q$,  $T^3$, $C$, $(I^A)^2$,
  $(I^Z)^2$, $(I^W)^2$, $C^\ew$, and $I^\pm$}
Here a list of the eigenvalues (or components) of the operators $Y$,
$Q$, $T^3$, $C$, $(I^A)^2$, $(I^Z)^2$, $(I^W)^2$, $C^\ew$, and
$I^\pm$ is given following Ref. \cite{dp}, that have to be inserted in the general results. 
For incoming
particles or outgoing antiparticles the values for the particles have
to be used, for incoming antiparticles or outgoing particles the
values of the antiparticles.

\subsubsection*{Fermions}
The fermionic doublets $f^\kappa=(f^\kappa_+, f^\kappa_-)^\rT$
transform according to the fundamental or trivial representations,
depending on the chirality $\kappa=\rL,\rR$. Except for $I^\pm$, the
above operators are diagonal. For lepton and quark doublets,
$L^\kappa= (\nu^\kappa, l^\kappa)^\rT$ and $Q^\kappa=(u^\kappa,
d^\kappa)^\rT$, their eigenvalues are
\beq \label{Llept}
\renewcommand{\arraystretch}{1.5}
\begin{array}{c@{\quad}|@{\quad}c@{\quad}c@{\quad}c@{\quad}c@{\quad}c@{\quad}c@{\quad}c@{\quad}c@{\quad}}
& {Y}/{2}& Q & T^3 &C &(I^A)^2&(I^Z)^2&(I^W)^2 & C^\ew \\
\hline
 \nu^{\rL},\bar{\nu}^{\rL} &\mp \frac{1}{2}  &0 & \pm \frac{1}{2} & \frac{3}{4}&  0  & \frac{1}{4\sw^2\cw^2} & \frac{1}{2
\sw^2}  & \frac{1+2\cw^2}{4\sw^2\cw^2}  \\
 l^{\rL},\bar{l}^{\rL} &\mp \frac{1}{2} &\mp 1 & \mp \frac{1}{2} & \frac{3}{4}   & 1  &  \frac{(\cw^2-\sw^2)^2}{4\sw^2\cw
^2} & \frac{1}{2\sw^2} & \frac{1+2\cw^2}{4\sw^2\cw^2}  \\
 l^{\rR},\bar{l}^{\rR}  & \mp 1  & \mp 1 & 0 & 0  & 1  & \frac{\sw^2}{\cw^2} & 0  & \frac{1}{\cw^2} \\
 u^{\rL},\bar{u}^{\rL} &\pm \frac{1}{6} & \pm \frac{2}{3} & \pm \frac{1}{2} & \frac{3}{4}  &  \frac{4}{9}  &  \frac{(3\cw
^2-\sw^2)^2}{36\sw^2\cw^2} & \frac{1}{2\sw^2}   & \frac{\sw^2+27\cw^2}{36\cw^2\sw^2} \\
 d^{\rL},\bar{d}^{\rL} &\pm \frac{1}{6} & \mp \frac{1}{3} & \mp \frac{1}{2} & \frac{3}{4}  &  \frac{1}{9}  &  \frac{(3\cw
^2+\sw^2)^2}{36\sw^2\cw^2} & \frac{1}{2\sw^2}  & \frac{\sw^2+27\cw^2}{36\cw^2\sw^2} \\
 u^{\rR},\bar{u}^{\rR}  & \pm \frac{2}{3}  & \pm \frac{2}{3} & 0 & 0  & \frac{4}{9}  &  \frac{4}{9}\frac{\sw^2}{\cw^2} &
0 & \frac{4}{9\cw^2}  \\
 d^{\rR},\bar{d}^{\rR}  & \mp \frac{1}{3}  & \mp \frac{1}{3} & 0 & 0  &  \frac{1}{9}  & \frac{1}{9}\frac{\sw^2}{\cw^2} &
0 & \frac{1}{9\cw^2}  \\
\end{array}
\eeq
For left-handed fermions, $I^\pm(f^\rL)$ have  the non-vanishing components
\beq \label{ferpmcoup}
I^{\si}_{f_{\si'}f_{-\si'}}(f^\rL)=-I^{\si}_{\bar{f}_{-\si'}\bar{f}_{\si'}}(\bar{f}^\rL)=\frac{\de_{\si\si'}}{\sqrt{2}\sw
}
\eeq
whereas for right-handed fermions $I^\pm(f^\rR)=0$.

\subsubsection*{Scalar fields}
The symmetric scalar doublet,
$\Phi= (\phi^+,\phi_0)^\rT$, $\Phi^*= (\phi^-,\phi_0^*)^\rT$,
transforms according to the fundamental representation, and its
quantum numbers correspond to those of left-handed leptons
\refeq{Llept} with
\beq
\phi^+ \leftrightarrow \bar{l}^\rL, \qquad \phi_0 \leftrightarrow \bar{\nu}^\rL, \qquad
\phi^- \leftrightarrow {l}^\rL, \qquad \phi_0^* \leftrightarrow {\nu}^\rL
\eeq
After symmetry breaking
the neutral scalar fields are parametrized by the mass eigenstates
\beq \label{Higgschi}
\phi_0=\frac{1}{\sqrt{2}} (v+ H + \ri\chi)
\eeq
With respect to this basis $S=(H,\chi)$ the operators $Q,C,(I^\NB)^2$,
and $\cew$ remain unchanged, while $T^3$ and $Y$ become non-diagonal
in the neutral components
\begin{equation}
T^3_{SS'}=-\left(\frac{Y}{2}\right)_{SS'}=
-\frac{1}{2}\left(\begin{array}{c@{\;}c}0 & -\ri \\  \ri & 0 \end{array}\right)
,
\end{equation}
and
\beq \label{ZHcoup}
I^Z_{H\chi}=-I^Z_{\chi H}=\frac{\ri}{2\sw\cw}
\eeq
The $\PW^\pm$ couplings read
\beq \label{scapmcoup}
I^{\si}_{S\phi^{-\si'}}=-I^{\si}_{\phi^{\si'}S}=\de_{\si\si'}I^{\si}_S
\eeq
with
\beq \label{scapmcoupB}
I^\si_{H}:=-\frac{\si}{2\sw},\qquad
I^\si_{\chi}:=-\frac{\ri}{2\sw}
\eeq
\subsubsection*{Gauge fields}
For transversely polarized external gauge bosons one has to use the
adjoint representation. In the symmetric basis the diagonal operators have eigenvalues
\begin{equation}\label{gaugeeigenvalues}
\renewcommand{\arraystretch}{1.5}
\begin{array}{c@{\quad}|@{\quad}c@{\quad}c@{\quad}c@{\quad}c@{\quad}c@{\quad}c@{\quad}c@{\quad}c@{\quad}}
& {Y}/{2}&Q & T^3 &C &(I^A)^2&(I^Z)^2&(I^W)^2 & \cew  \\
\hline

 W^\pm  & 0  & \pm 1  & \pm 1 & 2   & 1  & \frac{\cw^2}{\sw^2} &\frac{1}{\sw^2}  & \frac{2}{\sw^2}   \\

 W^3  &  0  &  0  & 0  & 2  & 0  & 0 & \frac{2}{\sw^2}  & \frac{2}{\sw^2}  \\

 B  &  0  & 0  & 0 & 0  & 0  & 0 & 0  & 0  \\
\end{array}
\end{equation}
In the neutral sector, owing to the Weinberg rotation, the non-trivial
operators $\cew,C$ and $(I^W)^2$ become non-diagonal in the physical
basis $\NB=A,Z$,
with components
\beq
\cew_{\NB\NB'}=\frac{1}{\sw^2}C_{\NB\NB'}=(I^W)_{\NB\NB'}^2=
\frac{2}{\sw^2}\left(\begin{array}{c@{\;}c} \sw^2 & -\sw\cw \\ -\sw\cw & \cw^2 \end{array}\right)
\eeq
whereas the trivial operators ${Y}/{2}=Q=T^3=(I^A)^2=(I^Z)^2=0$ remain unchanged.
In the physical basis the non-vanishing components of the $I^\pm$ couplings are
\beq \label{gaupmcoup}
I^{\si}_{\NB W^{-\si'}}=-I^{\si}_{W^{\si'}\NB}=\de_{\si\si'}I^\si_\NB
\eeq
with
\beq
I^\si_A=-\si,\qquad
I^\si_Z=\si\frac{\cw}{\sw}
\eeq

\subsection{Electroweak Feynman rules} \label{sec:fr}

\textwidth 160 mm
\textheight 235 mm
\oddsidemargin 1 mm
\footskip 40 pt
\itemsep 2pt

\newsavebox{\blobdz}
\newsavebox{\blobza}
\newsavebox{\Vt}
\newsavebox{\Vr}
\newsavebox{\St}
\newsavebox{\Sr}
\newsavebox{\Fr}
\newsavebox{\Fl}
\newsavebox{\Fup}
\newsavebox{\Vtr}
\newsavebox{\Vbr}
\newsavebox{\Str}
\newsavebox{\Sbr}
\newsavebox{\Fupr}
\newsavebox{\Fbr}
\newsavebox{\Vtbr}
\newsavebox{\Fupbr}
\newsavebox{\Stbr}
\newsavebox{\Gr}
\newsavebox{\Gtbr}
\newsavebox{\Lt}
\newsavebox{\Lr}
\newsavebox{\Ltr}
\newsavebox{\Lbr}
\newsavebox{\Lb}
\newsavebox{\Lbl}
\newsavebox{\Ltl}
\newsavebox{\Vp}

\savebox{\Gr}(48,0)[bl]
{ \put(23,0){\vector(1,0){3}} \multiput(0,0)(4.8,0){11}{\circle*1{1}} }
\savebox{\Gtbr}(32,48)[bl]
{ \multiput(0,24)(4,3){9}{\circle*{1}}
 \multiput(0,24)(4,-3){9}{\circle*{1}}
 \put(16,12){\vector(-4,3){3}} \put(16,36){\vector(4,3){3}} }

\savebox{\Vr}(48,0)[bl]
{\multiput(3,0)(12,0){4}{\oval(6,4)[t]}
\multiput(9,0)(12,0){4}{\oval(6,4)[b]} }
\savebox{\Vtr}(32,24)[bl]
{\multiput(4,0)(8,6){4}{\oval(8,6)[tl]}
\multiput(4,6)(8,6){4}{\oval(8,6)[br]} }
\savebox{\Vbr}(32,24)[bl]
{\multiput(4,24)(8,-6){4}{\oval(8,6)[bl]}
\multiput(4,18)(8,-6){4}{\oval(8,6)[tr]}}
\savebox{\Vtbr}(32,48)[bl]
{\put(00,24){\usebox{\Vtr}}
\put(00,00){\usebox{\Vbr}}}

\savebox{\Sr}(48,0)[bl]
{ \multiput(0,0)(12.5,0){4}{\line(4,0){10}} }
\savebox{\Str}(32,24)[bl]
{ \multiput(-2,-1.5)(12,9){3}{\line(4,3){10}} }
\savebox{\Sbr}(32,24)[bl]
{\multiput(-2,25.5)(12,-9){3}{\line(4,-3){10}} }
\savebox{\Stbr}(32,48)[bl]
{\put(00,24){\usebox{\Str}}
\put(00,00){\usebox{\Sbr}}}

\savebox{\Fr}(48,0)[bl]
{ \put(0,0){\vector(1,0){26}} \put(24,0){\line(1,0){24}} }
\savebox{\Fupr}(32,24)[bl]
{ \put(0,0){\vector(4,3){18}} \put(16,12){\line(4,3){16}} }
\savebox{\Fbr}(32,24)[bl]
{ \put(32,0){\vector(-4,3){19}} \put(16,12){\line(-4,3){16}} }
\savebox{\Fupbr}(32,48)[bl]
{\put(00,24){\usebox{\Fupr}}
\put(00,00){\usebox{\Fbr}}}

In this appendix we list the Feynman rules of the SM in the
't~Hooft-Feynman gauge including the counterterms in a
way appropriate for the concept of generic diagrams \cite{dhab}. I.e.\ we write down
generic Feynman rules obtained from the classic Lagrangian in Eq. (\ref{eq:LSM})
and give the possible actual insertions.
We omit any field renormalization constants for the unphysical
fields. For brevity we introduce the shorthand notation
\beq
c = \cw,\qquad s= \sw
\eeq
In the vertices all momenta are considered as incoming.
\vspace{2mm}

Propagators:
\vspace{2mm}

for gauge bosons $V = \gamma,\,\PZ,\,\PW$
in the 't Hooft-Feynman gauge ($\xi_{i}=1$)
\bma
\barr{l}
\framebox{
\begin{picture}(90,24)
\put(33,16){\makebox(20,10)[bl] {$k$}}
\put(-1,9){\makebox(12,10)[bl] {$\PV_{\mu}$}}
\put(67,9){\makebox(12,10)[bl] {$\PV_{\nu}$}}
\put(15,12){\usebox{\Vr}}
\put(15,12){\circle*{4}}
\put(63,12){\circle*{4}}
\end{picture} }
\earr
\barr{l}
\disp = \frac{-\ri g_{\mu \nu }}{k^{2}-\MV^{2}}
\earr
\ema

for Faddeev--Popov ghosts $G = u^{\gamma},\,u^{\PZ},\,u^{\PW}$
\bma
\barr{l}
\framebox{
\begin{picture}(90,24)
\put(34,16){\makebox(20,10)[bl] {$k$}}
\put(0,9){\makebox(12,10)[bl] {$G$}}
\put(67,9){\makebox(12,10)[bl] {$\bar{G}$}}
\put(15,12){\usebox{\Gr}}
\put(15,12){\circle*{4}}
\put(63,12){\circle*{4}}
\end{picture} }
\earr
\barr{l}
\disp = \frac{\ri}{k^{2}-M_{G}^{2}}
\earr
\ema

for scalar fields $S = H,\,\chi,\,\phi$
\bma
\barr{l}
\framebox{
\begin{picture}(90,24)
\put(34,16){\makebox(20,10)[bl] {$k$}}
\put(0,9){\makebox(12,10)[bl] {$S$}}
\put(67,9){\makebox(12,10)[bl] {$S$}}
\put(15,12){\usebox{\Sr}}
\put(15,12){\circle*{4}}
\put(63,12){\circle*{4}}
\end{picture} }
\earr
\barr{l}
\disp= \frac{\ri}{k^{2}-M_{S}^{2}}
\earr
\ema

and for fermion fields $F = f_{i}$
\bma
\rule{2mm}{0mm}\barr{l}
\framebox{
\begin{picture}(90,24)
\put(34,17){\makebox(20,10)[bl] {$p$}}
\put(0,9){\makebox(12,10)[bl] {$F$}}
\put(67,9){\makebox(12,10)[bl] {$\bar{F}$}}
\put(15,12){\usebox{\Fr}}
\put(15,12){\circle*{4}}
\put(63,12){\circle*{4}}
\end{picture} }
\earr
\barr{l}
\disp= \frac{\ri(\ps + m_{F})}{p^{2}-m_{F}^{2}}
\earr
\ema

In the 't Hooft-Feynman gauge we have the following relations:
\beq
M_{u^{\gamma}} = 0,\qquad
M_{u^{\PZ}} = M_{\chi} = \MZ,\qquad
M_{u^{\pm}} = M_{\phi} = \MW
\eeq

Tadpole:
\bma
\barr{l}
\framebox {
\begin{picture}(72,24)
\put(50,15){\makebox(10,10)[bl]{$S$}}
\put(7.5,6){\line(3,4){9}}
\put(7.5,18){\line(3,-4){9}}
\put(12,12){\usebox{\Sr}}
\end{picture} }
\earr
\barr{l}
\disp = \ri\delta t.
\earr
\ema
\newcommand{\dt}{\frac{e}{2s}\frac{\delta t}{\MW}}
\newcommand{\dth}{\frac{e}{2s}\frac{\delta t}{\MW\MH^2}}

{\samepage
VV counterterm:
\bma
\barr{l}
\framebox{
\begin{picture}(120,30)
\put(100,19){\makebox(10,20)[bl]{$V_{2,\nu}$\strut}}
\put(0,19){\makebox(10,20)[bl] {$V_{1,\mu},k$}}
\put(55.5,6){\line(3,4){9}}
\put(55.5,18){\line(3,-4){9}}
\put(12,12){\usebox{\Vr}}
\put(60,12){\usebox{\Vr}}
\end{picture} }
\earr
\barr{l}
= -\ri g_{\mu\nu}\Bigl[C_{1}k^{2} - C_{2} \Bigr] + \ri k_\mu k_\nu C_3
\earr
\ema
with the actual values of $V_{1},\:V_{2}$ and $C_{1},\:C_{2}$

\nobreak
without renormalization of the gauge-fixing term
\begin{subequations}
\beq
\begin{array}[b]{l@{\quad : \quad}ll}
\PWp\PWm & C_{1} = C_3 =\DZW, & C_{2} = \MW^{2}\DZW + \DMWS \co \\[1ex]
\PZ\PZ & C_{1} = C_3 =\DZZ, & C_{2} = \MZ^{2}\DZZ + \DMZS \co \\[1ex]
\PA\PZ & C_{1} = C_3 =\frac{1}{2}\DZAZ + \frac{1}{2}\DZZA , \qquad &
C_{2} =\MZ^{2}\frac{1}{2}\DZZA \co \\[1ex]
\PA\PA & C_{1} = C_3 = \DZA, & C_{2} = 0
\earr
\eeq

with renormalization of the ('t~Hooft--Feynman) gauge-fixing term
\beq
C_3 = 0, \qquad C_1, C_2 \mbox{ as above}
\eeq
\end{subequations}
}

{\samepage
VS counterterm:
\bma
\barr{l}
\framebox{
\begin{picture}(120,30)
\put(100,19){\makebox(10,20)[bl]{$S$\strut}}
\put(0,19){\makebox(10,20)[bl] {$V_{\mu},k$}}
\put(55.5,6){\line(3,4){9}}
\put(55.5,18){\line(3,-4){9}}
\put(12,12){\usebox{\Vr}}
\put(60,12){\usebox{\Sr}}
\end{picture} }
\earr
\barr{l}
= \ri k_\mu C
\earr
\ema
with the actual values of $V,\:S$ and $C$

without renormalization of the gauge-fixing term
\begin{subequations}
\beq
\begin{array}[b]{l@{\quad : \quad}ll}
\PW^\pm\phi^\mp & C = \pm\MW\frac{1}{2}(\DZW + \frac{\DMWS}{\MW^2}) \co \\[1ex]
\PZ\chi & C = \ri\MZ \frac{1}{2}(\DZZ + \frac{\DMZS}{\MZ^2}) \co \\[1ex]
\PA\chi & C = \ri\MZ \frac{1}{2}\DZZA  
\earr
\eeq

with renormalization of the ('t~Hooft--Feynman) gauge-fixing term
\beq
C = 0
\eeq
\end{subequations}
}

{\samepage
SS-counterterm:
\bma
\barr{l}
\framebox{
\begin{picture}(120,30)
\put(100,18){\makebox(10,20)[bl]{$S_{2}$}}
\put(0,18){\makebox(10,20)[bl] {$S_{1},k$}}
\put(55.5,6){\line(3,4){9}}
\put(55.5,18){\line(3,-4){9}}
\put(12,12){\usebox{\Sr}}
\put(60,12){\usebox{\Sr}}
\end{picture} }
\earr
\barr{l}
= \ri\Bigl[C_{1}k^{2} - C_{2} \Bigr]
\earr
\ema
with the actual values of $S_{1},\:S_{2}$ and $C_{1},\:C_{2}$

\nopagebreak
without renormalization of the gauge-fixing term
\begin{subequations}
\beq
\begin{array}[b]{l@{\quad : \quad}ll}
\PH\PH & C_{1} = \delta Z_H, \quad & C_{2} = \MH^{2}\DZH + \DMHS \co\\[1em]
\chi\chi & C_{1} = 0, \quad & C_{2} = -\dt  \co\\[1em]
\phi^+\phi^- & C_{1} = 0, \quad & C_{2} = -\dt 
\earr
\eeq

with renormalization of the ('t~Hooft--Feynman) gauge-fixing term
\beq
\begin{array}[b]{l@{\quad : \quad}ll}
\PH\PH & C_{1} = \delta Z_H, \quad & C_{2} = \MH^{2}\DZH + \DMHS \co\\[1em]
\chi\chi & C_{1} = 0, \quad & C_{2} = -\dt + \DMZS \co\\[1em]
\phi^+\phi^- & C_{1} = 0, \quad & C_{2} = -\dt + \DMWS 
\earr
\eeq
\end{subequations}
}

{\samepage
FF-counterterm:
\bma
\barr{l}
\framebox{
\begin{picture}(120,30)
\put(100,18){\makebox(10,20)[bl]{$\bar{F}_{2}$}}
\put(0,18){\makebox(10,20)[bl] {$F_{1},p$}}
\put(55.5,6){\line(3,4){9}}
\put(55.5,18){\line(3,-4){9}}
\put(12,12){\usebox{\Fr}}
\put(60,12){\usebox{\Fr}}
\end{picture} }
\earr
\barr{l}
= \ri\Bigl[C_{L}\ps\omega_{-} + C_{R}\ps\omega_{+} - C_{S}^{-}\omega_{-} -
C_{S}^{+}\omega_{+} \Bigr]
\earr
\ema\samepage
with the actual values of $F_{1},\:\bar{F}_{2}$ and
$C_{L},\:C_{R},\:C_{S}^{-},\:C_{S}^{+}$
\beq
\begin{array}[b]{l@{\quad : \quad}l}
f_{j} \bar{f}_{i} & \left\{
\barr{l} C_{L} = \frac{1}{2}\left(\delta Z_{ij}^{f,L}+ \delta
Z_{ij}^{f,L\dagger}\right),
\qquad C_{R} = \frac{1}{2}\left(\delta Z_{ij}^{f,R}+ \delta
Z_{ij}^{f,R\dagger}\right)\co\\[1em]
C_{S}^{-} = m_{f,i}\frac{1}{2}\delta Z_{ij}^{f,L}
+ m_{f,j}\frac{1}{2}\delta Z_{ij}^{f,R\dagger} + \delta_{ij}\delta
m_{f,i}\co\\[1em]
C_{S}^{+} = m_{f,i}\frac{1}{2}\delta Z_{ij}^{f,R}
+ m_{f,j}\frac{1}{2}\delta Z_{ij}^{f,L\dagger} + \delta_{ij}\delta
m_{f,i}
\earr \right.
\earr
\eeq
}

{\samepage
VVVV~coupling:
\bma
\barr{l}
\framebox{
\begin{picture}(96,77)(0,-2)
\put(6,65){\makebox(10,20)[bl]{$V_{1,\mu}$}}
\put(72,65){\makebox(10,20)[bl]{$V_{3,\rho}$}}
\put(6,0){\makebox(10,20)[bl]{$V_{2,\nu}$}}
\put(72,0){\makebox(10,20)[bl]{$V_{4,\sigma}$} }
\put(48,36){\circle*{4}}
\put(16,12){\usebox{\Vtr}}
\put(16,36){\usebox{\Vbr}}
\put(48,12){\usebox{\Vtbr}}
\end{picture} }
\earr
\barr{l}
= \ri e^{2}C \Bigl[2g_{\mu\nu}g_{\sigma \rho } -
g_{\nu\rho}g_{\mu \sigma } - g_{\rho\mu}g_{\nu \sigma }\Bigr]
\earr
\ema\samepage
with the actual values of $V_{1},\:V_{2},\:V_{3},\:V_{4}$ and $C$
\beq
\begin{array}[b]{l@{\quad : \quad}l}
\PWp \PWp \PWm \PWm & C = \frac{1}{s^{2}}
\Bigl[1+2\DZe -2\frac{\delta s}{s}+ 2\DZW
\Bigr]\co\\[1em]
\PWp \PWm \PZ \PZ & C = -\frac{c^{2}}{s^{2}}
\Bigl[1 + 2\DZe - 2\frac{1}{c^{2}}\frac{\delta s}{s}
+ \DZW + \DZZ \Bigr]
+ \frac{c}{s}\DZAZ\co\\[1em]
\PWp \PWm \PA \PZ & \left\{
\barr{lll} C &=& \frac{c}{s}
\Bigl[1 + 2\DZe -\frac{1}{c^{2}}\frac{\delta s}{s}
+ \DZW + \frac{1}{2}\DZZ + \frac{1}{2}\DZA \Bigr] \\[1ex]
&&\quad -\frac{1}{2}\DZAZ
-\frac{1}{2}\frac{c^{2}}{s^{2}}\DZZA \co\\[1ex]
\earr \right.\\[1em]
\PWp \PWm \PA \PA & C = - \Bigl[1 + 2\DZe + \DZW + \DZA \Bigr]
+ \frac{c}{s}\DZZA
\earr
\eeq
}

{\samepage
VVV~coupling:
\bma
\barr{l}
\framebox{
\begin{picture}(96,77)(0,-2)
\put(60,65){\makebox(10,20)[bl]{$V_{2,\nu},k_{2}$}}
\put(0,43){\makebox(10,20)[bl] {$V_{1,\mu},k_{1}$}}
\put(60,0){\makebox(10,20)[bl] {$V_{3,\rho},k_{3}$}}
\put(48,36){\circle*{4}}
\put(0,36){\usebox{\Vr}}
\put(48,12){\usebox{\Vtbr}}
\end{picture} }
\earr
\barr{l}
= -\ri eC \Bigl[g_{\mu \nu }(k_{2}-k_{1})_{\rho}
+g_{\nu\rho}(k_{3}-k_{2})_{\mu} +g_{\rho\mu}(k_{1}-k_{3})_{\nu}\Bigr]
\earr
\ema\samepage
with the actual values of $V_{1},\:V_{2},\:V_{3}$ and $C$
\beq
\begin{array}[b]{l@{\quad : \quad}l}
\PA \PWp \PWm & C = 1+\DZe+\DZW+\frac{1}{2}\DZA
-\frac{1}{2}\frac{c}{s}\DZZA\co\\[1ex]
\PZ \PWp \PWm & C = -\frac{c}{s}(1 + \DZe
- \frac{1}{c^{2}}\frac{\delta s}{s}
+ \DZW + \frac{1}{2}\DZZ ) + \frac{1}{2}\DZAZ
\earr
\eeq
}

{\samepage
SSSS~coupling:
\bma
\barr{l}
\framebox {
\begin{picture}(96,77)(0,-2)
\put(9,65){\makebox(10,20)[bl]{$S_{1}$}}
\put(75,65){\makebox(10,20)[bl]{$S_{3}$}}
\put(9,0){\makebox(10,20)[bl]{$S_{2}$}}
\put(75,0){\makebox(10,20)[bl]{$S_{4}$}}
\put(48,36){\circle*{4}}
\put(18,14){\usebox{\Str}}
\put(18,34){\usebox{\Sbr}}
\put(48,12){\usebox{\Stbr}}
\end{picture} }
\earr
\barr{l}
= \ri e^{2}C
\earr
\ema\samepage
with the actual values of $S_{1},\:S_{2},\:S_{3},\:S_{4}$ and $C$
\beq
\begin{array}[b]{l@{\quad : \quad}l@{\quad}l@{\qquad}l@{\quad}l}
\PH \PH \PH \PH & C = - \frac{3}{4s^{2}} \frac{\MH^{2}}{\MW^{2}}
\Bigl[1+2\DZe -2\frac{\delta s}{s}
+\frac{\DMHS}{\MH^{2}} +\dth -\frac{\delta \MW^{2}}{\MW^{2}}
+ 2\DZH \Bigr] \co\\[1ex]
\barr{l} \PH \PH \chi \chi \\ \PH \PH \phi \phi \earr \left.\rule[-
1.6ex]{0mm}{4ex}\right\}
& C = - \frac{1}{4s^{2}} \frac{\MH^{2}}{\MW^{2}}
\Bigl[1+2\DZe -2\frac{\delta s}{s}
+\frac{\DMHS}{\MH^{2}} +\dth -\frac{\delta \MW^{2}}{\MW^{2}}
+ \DZH \Bigr] \co\\[1ex]
\chi \chi \chi \chi & C = - \frac{3}{4s^{2}} \frac{\MH^{2}}{\MW^{2}}
\Bigl[1+2\DZe -2\frac{\delta s}{s}
+\frac{\DMHS}{\MH^{2}} +\dth -\frac{\delta \MW^{2}}{\MW^{2}}
\Bigr] \co\\[1ex]
\chi \chi \phi \phi & C = - \frac{1}{4s^{2}} \frac{\MH^{2}}{\MW^{2}}
\Bigl[1+2\DZe -2\frac{\delta s}{s}
+\frac{\DMHS}{\MH^{2}} +\dth -\frac{\delta \MW^{2}}{\MW^{2}}
\Bigr] \co\\[1ex]
\phi \phi \phi \phi & C = - \frac{1}{2s^{2}} \frac{\MH^{2}}{\MW^{2}}
\Bigl[1+2\DZe -2\frac{\delta s}{s}
+\frac{\DMHS}{\MH^{2}} +\dth -\frac{\delta \MW^{2}}{\MW^{2}}
\Bigr] 
\earr
\eeq
}

{\samepage
SSS~coupling:
\bma
\barr{l}
\framebox {
\begin{picture}(96,77)(0,-2)
\put(75,65){\makebox(10,20)[bl]{$S_{2}$}}
\put(0,42){\makebox(10,20)[bl]{$S_{1}$}}
\put(75,0){\makebox(10,20)[bl]{$S_{3}$}}
\put(48,36){\circle*{4}}
\put(0,36){\usebox{\Sr}}
\put(48,12){\usebox{\Stbr}}
\end{picture} }
\earr
\barr{l}
= \ri eC
\earr
\ema\samepage
with the actual values of $S_{1},\:S_{2},\:S_{3}$ and $C$
\beq
\begin{array}[b]{l@{\quad : \quad}l@{\quad}l@{\qquad}l@{\quad}l}
\PH \PH \PH & C = - \frac{3}{2s} \frac{\MH^{2}}{\MW}
\Bigl[1+\DZe -\frac{\delta s}{s}
+\frac{\DMHS}{\MH^{2}} +\dth
-\frac{1}{2}\frac{\delta \MW^{2}}{\MW^{2}}
+ \frac{3}{2}\DZH \Bigr] \co\\[1ex]
\barr{l} \PH\chi\chi \\ \PH\phi\phi \earr \left.
\rule[-1.5ex]{0mm}{4ex}\right\}
& C = - \frac{1}{2s} \frac{\MH^{2}}{\MW}
\Bigl[1+\DZe -\frac{\delta s}{s}
+\frac{\DMHS}{\MH^{2}} +\dth
-\frac{1}{2}\frac{\delta \MW^{2}}{\MW^{2}}
+ \frac{1}{2}\DZH \Bigr] 
\earr
\eeq
}

{\samepage
VVSS~coupling:
\bma
\barr{l}
\framebox{
\begin{picture}(96,77)(0,-2)
\put(6,65){\makebox(10,20)[bl]{$V_{1,\mu}$}}
\put(6,0){\makebox(10,20)[bl]{$V_{2,\nu}$}}
\put(75,65){\makebox(10,20)[bl]{$S_{1}$}}
\put(75,0){\makebox(10,20)[bl]{$S_{2}$}}
\put(48,36){\circle*{4}}
\put(16,12){\usebox{\Vtr}}
\put(16,36){\usebox{\Vbr}}
\put(48,12){\usebox{\Stbr}}
\end{picture} }
\earr
\barr{l}
= \ri e^{2}g_{\mu\nu}C
\earr
\ema\samepage
with the actual values of $V_{1},\:V_{2},\:S_{1},\:S_{2}$ and $C$
\beq
\begin{array}[b]{l@{\quad : \quad}l@{\quad}l@{\qquad}l@{\quad}l}
\PWp \PWm \PH \PH & C =  \frac{1}{2s^{2}}
\Bigl[1+2\DZe -2\frac{\delta s}{s}
+ \DZW + \DZH\Bigr] \co\\[1ex]
\barr{l} \PWp \PWm \chi \chi \\ \PWp \PWm \phi^{+} \phi^{-} \earr
\left.\rule[-1.5ex]{0mm}{4ex}\right\}
& C =  \frac{1}{2s^{2}}
\Bigl[1+2\DZe -2\frac{\delta s}{s}
+ \DZW \Bigr] \co\\[1ex]
\PZ \PZ \phi^{+} \phi^{-} & C =
\frac{(s^{2}-c^{2})^{2}}{2s^{2}c^{2}}
\Bigl[1+2\DZe
+\frac{2}{(s^{2}-c^{2})c^{2}}\frac{\delta s}{s} + \DZZ \Bigr]
+\frac{s^{2}-c^{2}}{sc} \DZAZ \co\\[1ex]
\PZ\PA\phi^{+} \phi^{-} & \left\{
\barr{l} C =\frac{s^{2}-c^{2}}{sc}  \Bigl[1+2\DZe
+\frac{1}{(s^{2}-c^{2})c^{2}}\frac{\delta s}{s}
+ \frac{1}{2}\DZZ + \frac{1}{2}\DZA  \Bigr] \\[1ex]
\qquad+\frac{(s^{2}-c^{2})^{2}}{2s^{2}c^{2}}\frac{1}{2}\DZZA
+ \DZAZ \co \earr \right.\\[1em]
\PA\PA\phi^{+} \phi^{-} & C =
2\Bigl[1+2\DZe + \DZA \Bigr]
+\frac{s^{2}-c^{2}}{sc} \DZZA \co\\[1ex]
\PZ\PZ\PH\PH& C = \frac{1}{2s^{2}c^{2}}
\Bigl[1+2\DZe
+ 2\frac{s^{2}-c^{2}}{c^{2}}\frac{\delta s}{s}
+ \DZZ+ \DZH \Bigr] \co\\[1ex]
\PZ\PZ\chi \chi & C = \frac{1}{2s^{2}c^{2}}
\Bigl[1+2\DZe
+ 2\frac{s^{2}-c^{2}}{c^{2}}\frac{\delta s}{s}
+ \DZZ \Bigr] \co\\[1ex]
\barr{l} \PZ\PA\PH\PH\\ \PZ\PA\chi \chi \earr
\left.\rule[-1.5ex]{0mm}{4ex}\right\}
& C =  \frac{1}{2s^{2}c^{2}} \frac{1}{2}\DZZA \co\\[1ex]
\PW^{\pm} \PZ \phi^{\mp} \PH & C = - \frac{1}{2c}
\Bigl[1+2\DZe -\frac{\delta c}{c}
 + \frac{1}{2}\DZW + \frac{1}{2}\DZH
+ \frac{1}{2}\DZZ \Bigr]
- \frac{1}{2s}\frac{1}{2}\DZAZ \co\\[1ex]
\PW^{\pm} \PA \phi^{\mp} \PH & C = - \frac{1}{2s}
\Bigl[1+2\DZe -\frac{\delta s}{s}
+ \frac{1}{2}\DZW + \frac{1}{2}\DZH
+ \frac{1}{2}\DZA \Bigr]
- \frac{1}{2c}\frac{1}{2}\DZZA \co\\[1ex]
\PW^{\pm} \PZ \phi^{\mp} \chi & C = \mp \frac{i}{2c}
\Bigl[1+2\DZe -\frac{\delta c}{c}
+ \frac{1}{2}\DZW + \frac{1}{2}\DZZ \Bigr]
\mp \frac{i}{2s}\frac{1}{2}\DZAZ \co\\[1ex]
\PW^{\pm} \PA \phi^{\mp} \chi & C =  \mp\frac{i}{2s}
\Bigl[1+2\DZe -\frac{\delta s}{s}
+ \frac{1}{2}\DZW + \frac{1}{2}\DZA \Bigr]
\mp \frac{i}{2c}\frac{1}{2}\DZZA 
\earr
\eeq
}

{\samepage
VSS~coupling:
\bma
\barr{l}
\framebox {
\begin{picture}(96,77)(0,-2)
\put(67,65){\makebox(10,20)[bl]{$S_{1},k_{1}$}}
\put(0,42){\makebox(10,20)[bl]{$V_{\mu}$}}
\put(67,0){\makebox(10,20)[bl]{$S_{2},k_{2}$}}
\put(48,36){\circle*{4}}
\put(0,36){\usebox{\Vr}}
\put(48,12){\usebox{\Stbr}}
\end{picture} }
\earr
\barr{l}
= \ri eC(k_{1}-k_{2})_{\mu}
\earr
\ema\samepage \samepage
with the actual values of $V,\:S_{1},\:S_{2}$ and $C$
\beq
\begin{array}[b]{l@{\quad : \quad}l@{\quad}l@{\qquad}l@{\quad}l}
\PA \chi \PH & C = - \frac{i}{2cs} \frac{1}{2}\DZZA\co\\[1ex]
\PZ \chi \PH & C = - \frac{i}{2cs}
\Bigl[1+\DZe +\frac{s^{2}-c^{2}}{c^{2}}\frac{\delta s}{s}
+ \frac{1}{2}\DZH + \frac{1}{2}\DZZ\Bigr]\co\\[1ex]
\PA\phi^{+}\phi^{-} & C = -\Bigl[1 + \DZe + \frac{1}{2}\DZA
+\frac{s^{2}-c^{2}}{2sc}\frac{1}{2}\DZZA\Bigl] \co\\[1ex]
\PZ \phi^{+} \phi^{-} & C = - \frac{s^{2}-c^{2}}{2sc}
\Bigl[1+\DZe +\frac{1}{(s^{2}-c^{2})c^{2}}\frac{\delta s}{s}
+\frac{1}{2}\DZZ\Bigr]
- \frac{1}{2}\DZAZ \co\\[1ex]
\PW^{\pm} \phi^{\mp} \PH & C = \mp \frac{1}{2s}
\Bigl[1+\DZe -\frac{\delta s}{s}
+ \frac{1}{2}\DZW + \frac{1}{2}\DZH \Bigr] \co\\[1ex]
\PW^{\pm} \phi^{\mp} \chi & C = - \frac{i}{2s}
\Bigl[1+\DZe -\frac{\delta s}{s}
+ \frac{1}{2}\DZW\Bigr] 
\earr
\eeq
}

{\samepage
SVV~coupling:
\bma
\barr{l}
\framebox {
\begin{picture}(96,77)(0,-2)
\put(72,65){\makebox(10,20)[bl]{$V_{1,\mu}$}}
\put(72,0){\makebox(10,20)[bl] {$V_{2,\nu}$}}
\put(0,42){\makebox(10,20)[bl]{$S$}}
\put(48,36){\circle*{4}}
\put(0,36){\usebox{\Sr}}
\put(48,12){\usebox{\Vtbr}}
\end{picture} }
\earr
\barr{l}
= \ri eg_{\mu\nu}C
\earr
\ema\samepage
with the actual values of $S,\:V_{1},\:V_{2}$ and $C$
\beq
\begin{array}[b]{l@{\quad : \quad}l@{\quad}l@{\qquad}l@{\quad}l}
\PH \PWp \PWm & C =  \MW\frac{1}{s}
\Bigl[1+\DZe -\frac{\delta s}{s}
+\frac{1}{2}\frac{\delta \MW^{2}}{\MW^{2}}
+ \frac{1}{2}\DZH+ \DZW \Bigr] \co\\[1ex]
\PH\PZ\PZ& C = \MW\frac{1}{sc^{2}}
\Bigl[1+\DZe +\frac{2s^{2}-c^{2}}{c^{2}}\frac{\delta s}{s}
+\frac{1}{2}\frac{\delta \MW^{2}}{\MW^{2}}
+ \frac{1}{2}\DZH+ \DZZ \Bigr] \co\\[1ex]
\PH \PZ\PA& C = \MW\frac{1}{sc^{2}} \frac{1}{2}\DZZA \co\\[1ex]
\phi^{\pm} \PW^{\mp} \PZ & C = - \MW\frac{s}{c}
\Bigl[1+\DZe +\frac{1}{c^{2}}\frac{\delta s}{s}
+\frac{1}{2}\frac{\delta \MW^{2}}{\MW^{2}}
+ \frac{1}{2}\DZW + \frac{1}{2}\DZZ \Bigr]
- \MW \frac{1}{2}\DZAZ \co\\[1ex]
\phi^{\pm} \PW^{\mp} \PA & C = - \MW
\Bigl[1+\DZe +\frac{1}{2}\frac{\delta \MW^{2}}{\MW^{2}}
+ \frac{1}{2}\DZW +  \frac{1}{2}\DZA \Bigr]
- \MW\frac{s}{c} \frac{1}{2}\DZZA
\earr
\eeq
}

{\samepage
VFF~coupling:
\bma
\barr{l}
\framebox {
\begin{picture}(96,77)(0,-2)
\put(75,65){\makebox(10,20)[bl]{$\bar{F}_{1}$}}
\put(0,42){\makebox(10,20)[bl]{$V_{\mu}$}}
\put(75,0){\makebox(10,20)[bl]{$F_{2}$}}
\put(48,36){\circle*{4}}
\put(0,36){\usebox{\Vr}}
\put(48,12){\usebox{\Fupbr}}
\end{picture} }
\earr
\barr{l}
= \ri e\gamma_{\mu}(C^{-}\omega_{-} + C^{+}\omega_{+})
\earr
\ema\samepage
with the actual values of $V,\:\bar{F}_{1}\:,F_{2}$ and $C^{+},\:C^{-}$
\beq
\begin{array}[b]{l@{\quad : \quad}l}
\gamma \bar{f}_{i} f_{j} & \left\{
\barr{l}
C^{+} = -Q_{f}\Bigl[\delta_{ij}\Bigl(1 + \DZe +
\frac{1}{2}\DZA\Bigl)
+ \frac{1}{2}(\delta Z^{f,R}_{ij}+\delta Z^{f,R\dagger}_{ij})\Bigr]
+ \delta_{ij}g_{f}^{+}\frac{1}{2}\DZZA\co
\\[1ex]
C^{-} = -Q_{f}\Bigl[\delta_{ij}\Bigl(1 + \DZe +
\frac{1}{2}\DZA\Bigl)
+ \frac{1}{2}(\delta Z^{f,L}_{ij}+\delta Z^{f,L\dagger}_{ij})\Bigr]
+ \delta_{ij}g_{f}^{-}\frac{1}{2}\DZZA\co
\earr \right.\\[1.6em]
\PZ \bar{f}_{i} f_{j} & \left\{\barr{l}
C^{+} = g_{f}^{+}\Bigl[\delta_{ij}\Bigl(1 + \frac{\delta
g_{f}^{+}}{g_{f}^{+}}+\frac{1}{2}\DZZ\Bigr)
+ \frac{1}{2}(\delta Z^{f,R}_{ij}+\delta Z^{f,R\dagger}_{ij})\Bigr]
-\delta_{ij}Q_{f}\frac{1}{2}\DZAZ \co\\[1ex]
C^{-} = g_{f}^{-}\Bigl[\delta_{ij}\Bigl(1 + \frac{\delta
g_{f}^{-}}{g_{f}^{-}}+\frac{1}{2}\DZZ\Bigr)
+ \frac{1}{2}(\delta Z^{f,L}_{ij}+\delta Z^{f,L\dagger}_{ij})\Bigr]
-\delta_{ij}Q_{f}\frac{1}{2}\DZAZ \co
\earr \right.
\\[1.8em]
\PWp \bar{u}_{i} d_{j} & \left\{ \barr{l}
C^{+} = 0,   \qquad
C^{-} = \frac{1}{\sqrt{2}s}\Bigl[V_{ij}\Bigl(1 + \DZe -
\frac{\delta s}{s}+\frac{1}{2}\DZW\Bigr) + \delta V_{ij} \\[1ex]
\hspace{4cm}
+\frac{1}{2}\sum_{k}(\delta Z_{ik}^{u,L\dagger}V_{kj}+V_{ik}\delta
Z_{kj}^{d,L})\Bigr]\co
\earr \right.
\\[1.6em]
\PWm \bar{d}_{j} u_{i} & \left\{ \barr{l}
C^{+} = 0,   \qquad
C^{-} = \frac{1}{\sqrt{2}s}\Bigl[V_{ji}^\dagger\Bigl(1 + \DZe -
\frac{\delta s}{s}+\frac{1}{2}\DZW\Bigr) + \delta V_{ji}^\dagger \\[1ex]
\hspace{4cm}
+\frac{1}{2}\sum_{k}(\delta Z_{jk}^{d,L\dagger}V_{ki}^\dagger
+V_{jk}^\dagger\delta Z_{ki}^{u,L})\Bigr]\co
\earr \right. \\[1.6em]
\PWp \bar{\nu}_{i} l_{j} &
C^{+} = 0,   \qquad
C^{-} = \frac{1}{\sqrt{2}s}\delta_{ij}\Bigl[1 + \DZe -
\frac{\delta s}{s}+\frac{1}{2}\DZW
+\frac{1}{2}(\delta Z_{ii}^{\nu,L\dagger}+\delta Z_{ii}^{l,L})\Bigr]\co
\\[1em]
\PWm \bar{l}_{j} \nu_{i} &
C^{+} = 0,  \qquad
C^{-} = \frac{1}{\sqrt{2}s}\delta_{ij}\Bigl[1 + \DZe -
\frac{\delta s}{s}+\frac{1}{2}\DZW
+\frac{1}{2}(\delta Z_{ii}^{l,L\dagger}+\delta Z_{ii}^{\nu,L})\Bigr]\co
\earr
\eeq
}
where
\beq \label{geZ}
\barr{ll}
g_{f}^{+} = -\frac{s}{c} Q_{f}, &
\delta g_{f}^{+} = -\frac{s}{c} Q_{f}\Bigl[\DZe +
\frac{1}{c^{2}}\frac{\delta s}{s}\Bigr]\co \\[1.2em]
g_{f}^{-} =  \frac{I_{W,f}^{3}-s^{2}Q_{f}}{sc},\qquad &
\delta g_{f}^{-} =  \frac{I_{W,f}^{3}}{sc}\Bigl[\DZe +
\frac{s^{2}-c^{2}}{c^{2}}\frac{\delta s}{s}\Bigr] + \delta g_{f}^{+}
\earr
\eeq
The vector and axial vector couplings of the $Z$-boson are given by
\beq \label{vfaf}
\barr{l}
v_{f} = \frac{1}{2}(g_{f}^{-} + g_{f}^{+}) = \frac{I_{W,f}^{3}-
2s^{2}Q_{f}}{2sc}, \quad
a_{f} = \frac{1}{2}(g_{f}^{-} - g_{f}^{+}) = \frac{I_{W,f}^{3}}{2sc}
\earr
\eeq

{\samepage
SFF~coupling:
\bma
\barr{l}
\framebox {
\begin{picture}(96,77)(0,-2)
\put(75,65){\makebox(10,20)[bl]{$\bar{F}_{1}$}}
\put(0,42){\makebox(10,20)[bl]{$S$}}
\put(75,0){\makebox(10,20)[bl]{$F_{2}$}}
\put(48,36){\circle*{4}}
\put(0,36){\usebox{\Sr}}
\put(48,12){\usebox{\Fupbr}}
\end{picture} }
\earr
\barr{l}
= \ri e(C^{-}\omega_{-} + C^{+}\omega_{+})
\earr
\ema\samepage
with the actual values of $S,\:\bar{F}_{1}\:,F_{2}$ and $C^{+},\:C^{-}$
\beq
\begin{array}[b]{l@{\quad : \quad}l}
\PH \bar{f}_{i} f_{j} & \left\{ \barr{l}
C^{+} = - \frac{1}{2s}\frac{1}{\MW}\Bigl[\delta_{ij}m_{f,i}
\Bigl(1 + \DZe -\frac{\delta s}{s} +
\frac{\delta m_{f,i}}{m_{f,i}} - \frac{\delta
\MW}{\MW}+\frac{1}{2}\DZH\Bigr) \\[1ex]
\hspace{3cm}
{}+ \frac{1}{2}(m_{f,i}\delta Z^{f,R}_{ij}+\delta Z^{f,L\dagger}_{ij}m_{f,j})
\Bigr]\co \\[1ex]
C^{-} = - \frac{1}{2s}\frac{1}{\MW}\Bigl[\delta_{ij}m_{f,i}
\Bigl(1 + \DZe -\frac{\delta s}{s} +
\frac{\delta m_{f,i}}{m_{f,i}} - \frac{\delta
\MW}{\MW}+\frac{1}{2}\DZH\Bigr) \\[1ex]
\hspace{3cm}
{}+ \frac{1}{2}(m_{f,i}\delta Z^{f,L}_{ij}+\delta Z^{f,R\dagger}_{ij}m_{f,j})
\Bigr]\co \earr\right.\\[3.7em]
\chi \bar{f}_{i} f_{j} & \left\{ \barr{l}
C^{+} = \ri\frac{1}{2s}2I_{W,f}^{3}\frac{1}{\MW}\Bigl[\delta_{ij}m_{f,i}
\Bigl(1 + \DZe -\frac{\delta s}{s} +
\frac{\delta m_{f,i}}{m_{f,i}} - \frac{\delta \MW}{\MW}\Bigr)
\\[1ex] \hspace{3cm}
{}+ \frac{1}{2}(m_{f,i}\delta Z^{f,R}_{ij}+\delta Z^{f,L\dagger}_{ij}m_{f,j})
\Bigr]\co
\\[1ex]
C^{-} = -\ri\frac{1}{2s}2I_{W,f}^{3}\frac{1}{\MW}\Bigl[\delta_{ij}m_{f,i}
\Bigl(1 + \DZe -\frac{\delta s}{s} +
\frac{\delta m_{f,i}}{m_{f,i}} - \frac{\delta \MW}{\MW}\Bigr)
\\[1ex] \hspace{3cm}
{}+ \frac{1}{2}(m_{f,i}\delta Z^{f,L}_{ij}+\delta Z^{f,R\dagger}_{ij}m_{f,j})
\Bigr]\co \earr\right.\\[3.7em]
\phi^{+} \bar{u}_{i} d_{j} &
\left\{ \barr{l}
C^{+} = -\frac{1}{\sqrt{2}s}\frac{1}{\MW}
\Bigl[V_{ij}m_{d,j}\Bigl(1 + \DZe -\frac{\delta s}{s} +
\frac{\delta m_{d,j}}{m_{d,j}} - \frac{\delta \MW}{\MW}\Bigl) +
\delta V_{ij}m_{d,j}\\[1ex]
\hspace{4cm}
+\frac{1}{2}\sum_{k}(\delta Z_{ik}^{u,L\dagger}V_{kj}m_{d,j}+V_{ik}m_{d,k}
\delta Z_{kj}^{d,R})\Bigr]\co \\[1ex]C^{-} = \frac{1}{\sqrt{2}s}\frac{1}{\MW}
\Bigl[m_{u,i}V_{ij}\Bigl(1 + \DZe -\frac{\delta s}{s} +
\frac{\delta m_{u,i}}{m_{u,i}} - \frac{\delta \MW}{\MW}\Bigl) +
m_{u,i}\delta V_{ij} \\[1ex]
\hspace{4cm}
+\frac{1}{2}\sum_{k}(\delta Z_{ik}^{u,R\dagger}m_{u,k}V_{kj}
+m_{u,i}V_{ik}\delta Z_{kj}^{d,L})\Bigr]\co
\earr\right.\\[3.8em]
\phi^{-} \bar{d}_{j} u_{i} &
\left\{ \barr{l}
C^{+} = \frac{1}{\sqrt{2}s}\frac{1}{\MW}
\Bigl[V_{ji}^\dagger m_{u,i}\Bigl(1 + \DZe -\frac{\delta s}{s} +
\frac{\delta m_{u,i}}{m_{u,i}} - \frac{\delta \MW}{\MW}\Bigl) +
\delta V_{ji}^\dagger m_{u,i}\\[1ex]
\hspace{4cm}
+\frac{1}{2}\sum_{k}(\delta Z_{jk}^{d,L\dagger}V_{ki}^\dagger m_{u,i}
+V_{jk}^\dagger m_{u,k}\delta Z_{ki}^{u,R})\Bigr]\co\\[1ex]
C^{-} = -\frac{1}{\sqrt{2}s}\frac{1}{\MW}
\Bigl[m_{d,j}V_{ji}^\dagger \Bigl(1 + \DZe -\frac{\delta s}{s} +
\frac{\delta m_{d,j}}{m_{d,j}} - \frac{\delta \MW}{\MW}\Bigl) +
m_{d,j}\delta V_{ji}^\dagger \\[1ex]
\hspace{4cm}
+\frac{1}{2}\sum_{k}(\delta Z_{jk}^{d,R\dagger}m_{d,k}V_{ki}^\dagger
+m_{d,j}V_{jk}^\dagger \delta Z_{ki}^{u,L})\Bigr]\co
\earr\right. \\[3.8em]
\phi^{+} \bar{\nu}_{i} l_{j} & \left\{\barr{l}
C^{+} = -\frac{1}{\sqrt{2}s}\frac{m_{l,i}}{\MW}
\delta_{ij}\Bigl[ 1 + \DZe -\frac{\delta s}{s} +
\frac{\delta m_{l,i}}{m_{l,i}} - \frac{\delta \MW}{\MW}
+\frac{1}{2}(\delta Z_{ii}^{\nu,L\dagger}+\delta Z_{ii}^{l,R})\Bigr]\co
\\[1ex]
C^{-} = 0\co
\earr \right.
\\[1.5em]
\phi^{-} \bar{l}_{j} \nu_{i} & \left\{\barr{l}
C^{+} = 0\co\\[1ex]
C^{-} =  -\frac{1}{\sqrt{2}s}\frac{m_{l,i}}{\MW}
\delta_{ij}\Bigl[1 + \DZe -\frac{\delta s}{s} +
\frac{\delta m_{l,i}}{m_{l,i}} - \frac{\delta \MW}{\MW}
+\frac{1}{2}(\delta Z_{ii}^{l,R\dagger}+\delta Z_{ii}^{\nu,L})\Bigr]
\earr \right.
\earr
\eeq
}

{\samepage
VGG~coupling:
\bma
\barr{l}
\framebox {
\begin{picture}(96,77)(0,-2)
\put(67,65){\makebox(10,20)[bl]{$\bar{G_{1}},k_{1}$}}
\put(0,42){\makebox(10,20)[bl]{$V_{\mu}$}}
\put(75,0){\makebox(10,20)[bl]{$G_{2}$}}
\put(48,36){\circle*{4}}
\put(0,36){\usebox{\Vr}}
\put(48,12){\usebox{\Gtbr}}
\end{picture} }
\earr
\barr{l}
= \ri ek_{1,\mu}C
\earr
\ema\samepage
with the actual values of $V,\:\bar{G}_{1}\:,G_{2}$ and $C$
\beq
\begin{array}[b]{l@{\quad : \quad}l@{\quad}l@{\qquad}l@{\quad}l}
\PA \bar{u}^{\pm} u^{\pm} & C = \pm
\Bigl[1 + \DZe + \frac{1}{2}\DZA\Bigl]
\mp \frac{c}{s}\frac{1}{2}\DZZA \co\\[1ex]
\PZ \bar{u}^{\pm} u^{\pm} & C = \mp \frac{c}{s}
\Bigl[1 + \DZe - \frac{1}{c^{2}}\frac{\delta s}{s}
+ \frac{1}{2}\DZZ\Bigl]
\pm \frac{1}{2}\DZAZ \co\\[1ex]
\PW^{\pm} \bar{u}^{\pm} u^{\PZ} & C = \pm \frac{c}{s}
\Bigl[1+\DZe  - \frac{1}{c^{2}}\frac{\delta s}{s} +
\frac{1}{2}\DZW\Bigr] \co\\[1ex]
\PW^{\pm} \bar{u}^{\PZ} u^{\mp} & C = \mp \frac{c}{s}
\Bigl[1+\DZe  - \frac{1}{c^{2}}\frac{\delta s}{s} +
\frac{1}{2}\DZW\Bigr] \co\\[1ex]
\PW^{\pm} \bar{u}^{\pm} u^{\gamma} & C = \mp
\Bigl[1+\DZe  + \frac{1}{2}\DZW\Bigr] \co\\[1ex]
\PW^{\pm} \bar{u}^{\gamma} u^{\mp} & C = \pm
\Bigl[1+\DZe  + \frac{1}{2}\DZW\Bigr] 
\earr
\eeq
}

{\samepage
SGG~coupling   :
\bma
\barr{l}
\framebox {
\begin{picture}(96,77)(0,-2)
\put(75,65){\makebox(10,20)[bl]{$\bar{G_{1}}$}}
\put(0,42){\makebox(10,20)[bl]{$S$}}
\put(75,0){\makebox(10,20)[bl]{$G_{2}$}}
\put(48,36){\circle*{4}}
\put(0,36){\usebox{\Sr}}
\put(48,12){\usebox{\Gtbr}}
\end{picture} }
\earr
\barr{l}
= \ri eC
\earr
\ema\samepage
with the actual values of $S,\:\bar{G}_{1}\:,G_{2}$ and $C$
\beq
\begin{array}[b]{l@{\quad : \quad}l@{\quad}l@{\qquad}l@{\quad}l}
\PH \bar{u}^{\PZ} u^{\PZ} & C = -\frac{1}{2sc^{2}}\MW
\Bigl[1 + \DZe + \frac{2s^{2}-c^{2}}{c^{2}}
\frac{\delta s}{s} +\frac{1}{2}\frac{\delta \MW^{2}}{\MW^{2}}
+ \frac{1}{2}\DZH \Bigl] \co\\[1ex]
\PH \bar{u}^{\pm} u^{\pm} & C = -\frac{1}{2s}\MW
\Bigl[1 + \DZe - \frac{\delta s}{s}
+\frac{1}{2}\frac{\delta \MW^{2}}{\MW^{2}}
+ \frac{1}{2}\DZH \Bigl] \co\\[1ex]
\chi \bar{u}^{\pm} u^{\pm} & C = \mp i \frac{1}{2s}\MW
\Bigl[1 + \DZe - \frac{\delta s}{s}
+\frac{1}{2}\frac{\delta \MW^{2}}{\MW^{2}} \Bigl] \co\\[1ex]
\phi^{\pm} \bar{u}^{\PZ} u^{\mp} & C =  \frac{1}{2sc}\MW
\Bigl[1 + \DZe + \frac{s^{2}-c^{2}}{c^{2}}
\frac{\delta s}{s}
+\frac{1}{2}\frac{\delta \MW^{2}}{\MW^{2}} \Bigl] \co\\[1ex]
\phi^{\pm} \bar{u}^{\pm} u^{\PZ} & C =
\frac{s^{2}-c^{2}}{2sc}\MW
\Bigl[1 + \DZe + \frac{1}{(s^{2}-c^{2})c^{2}}
\frac{\delta s}{s}
+\frac{1}{2}\frac{\delta \MW^{2}}{\MW^{2}} \Bigl] \co\\[1ex]
\phi^{\pm} \bar{u}^{\pm} u^{\gamma} & C =  \MW
\Bigl[1 + \DZe
+\frac{1}{2}\frac{\delta \MW^{2}}{\MW^{2}} \Bigl] 
\earr
\eeq
}
   
\end{document}